\documentclass[12pt,a4paper]{book}

\usepackage{epsf}
\usepackage{graphics}
\usepackage{amsmath}
\usepackage{amssymb}
\usepackage{latexsym}
\usepackage{tabularx}
\usepackage{natbib}
\usepackage{fullpage}
\newcommand{\fl}{\hspace{3mm}}

\def\pt(#1){({\it #1\/})}
\begin{document}
\pagenumbering{Roman}
\begin{center}
\vspace{5cm}

{\Large Ph.D. Dissertation}

\vspace{3cm}

{\LARGE \bf Studies on Entanglement 

in Nuclear and Electron Spin Systems for 

Quantum Computing}

\vspace{6cm}

{\Large Robabeh Rahimi Darabad}

\vspace{4cm}

{\Large Department of Systems Innovation

Graduate School of Engineering Science

Osaka University

\vspace{2cm}

March, 2006}
\end{center}\thispagestyle{empty} 
\newpage

\thispagestyle{empty} 
\newpage
\section*{}
\begin{center}
\vspace{8cm}

{\Large To my Love, MEHDI

      My Mother and Father}
\end{center}    
\thispagestyle{empty} 
\newpage                    
\newpage
\setcounter{page}{1} 

\vspace{4cm}
\section*{Abstract}
\vspace{1cm}

The enormous impact of computers on everyday life is an incredible fact. The current technology of the computers is based on th technology of transistors, which is an appreciated synergy between computer science and quantum physics. Smaller transistors can work with less power, can be highly densed and can be switched on and off very fast. While the miniaturization is necessary to increase the computation power, this miniaturization also provides us with an intuitive way of understanding why, in a near future, quantum laws will become important for computation.

Gordon Moore \cite{p1} in 1965 codified his law, known as Moore's law, on the growth of computers. It states that the power of computers will double for constant cost roughly once every two years. This statement amazingly has come true for decades, since the 1960s. This growth is achieved through miniaturizing the size of the computers, by compacting the number of transistors. There is of course a limit for the number of transistors per chip and at present time it is almost $10^{18}$ for a typical size of circuit components to be of the order of $100$ nanometers. Then this is a fact that around $2020$ we will reach the atomic size for storing a single bit of information, hence quantum effects will become incredibly dominant.

So far, we have considered the increase of the computational power of the current computers, so called classical computers. These nowadays commonly being used computers go back into not a well-known origin in their very old history. Up to now, the technology of computers has been improved incredibly. However, Alan Turing, a great mathematician, pioneered the concept of what is called now programmable computer \cite{p2} and this concept yields to the strong Church-Turing thesis \cite{p3}\cite{p4}.

 {\it ^^ ^^ Any model of computation can be simulated on a probabilistic Turing machine with at most a polynomial increase in the number of elementary operations required."}
 
Even although the polynomial difference in speed might be still significant, this statement verifies the equivalency of all the computers, from the most pioneer ones, to the current supercomputers. Whereas, intuitively, if a quantum computer works truly based on quantum postulates namely superposition of states, then an exponential enhancement would be achieved by massive operations on a total quantum state.

The situation on our desire for a quantum computer should be clear in the sense that it's not just our intuition for a far future but it is somehow our current demand, too. Particularly, for some special cases, the atomic scales have already been achieved \cite{p5}. Then classical physics cannot explain the evolution of the corresponding system any more.

One considerable step through our desire for a quantum computer has been established by Feynman,  1980's \cite{p6}. He suggested that a quantum computer based on quantum logic would be ideal for simulating quantum-mechanical system by making a hint on how to get involved to these problems.

 {\it ^^ ^^ So, as we go down and fiddle around with the atoms down there, we are working with different laws, and we can expect to do different things. We can manufacture in different ways. We can use, not just circuits, but some system involving the quantized energy levels, or the interactions of quantized spins, etc."}
 
 Apart from the serious requirement on quantum algorithms, one of the most important points that has to be considered for any quantum physical system is that {\it the system should be stable, while the operations and processing are going on.} This statement explains the requirement of the maintaining coherent superposition of quantum states.
 
 Generally speaking, decoherence is one of the ultimate obstacle to the practical realization of a quantum computer. For any quantum computer, which has inevitable interaction with environment, the quantum information is decaying and this means that the quantum system is suffering from decoherences. Then, errors are provided by decoherence. Hence, there are also another sources of errors, such as the imperfections in the quantum computer hardware. By the way, it is inevitable to have an appropriate quantum error correction code, which can eliminate the errors and make the quantum processing stable.
 
 For a while it has been assumed that quantum error correction codes would not be accessible due to some quantum fundamental difficulties in interchanging the classical error correction code to the quantum counterpart. However, quantum error correction code has been invented by Peter Shor \cite{p7} and Andrew Stean \cite{p8}, provided that the probability of error per computational step is not large. This field has been extended much after these pioneer works and we will revisit some of the following works in this thesis. Therefore, employing quantum error correction codes, the problem of the random error due to decoherence is assumed to be solved out. Nevertheless, the threshold on the error probability is still so strict that it is not trivially accessible.
 
 In this work, the objects of quantum algorithms and quantum error corrections will be reviewed in separate sections.
 \begin{description}
 \item[Chapter 1]
 Classical computers, as much as being advanced, still can give only polynomial enhancement compared to the previous not advanced versions. A quantum computer can be much more faster, exponentially, as it involves superposition of states. Then, any process is applied to the whole massive quantum states. The result of the processing is however hidden in between the superposed states. Then a large number of outputs would be only potentially possible and totally useless. Nevertheless, it is the task of quantum algorithms to exploit the inherent quantum parallelism of quantum mechanics and to highlight the desired output. Therefore, quantum computers, in order to be useful, require the development of appropriate quantum software, that is of efficient quantum algorithms. As we will see in the following sections, entanglement is introduced as an essential prerequisite for some of these quantum algorithms specially those with exponential enhancement over classical processing. Furthermore, quantum information is decaying, which means that quantum system is suffering from decoherence. Errors are provided by decoherence. Hence, there are also another sources of errors, such as the imperfections in the quantum computer hardwares. By the way, it is inevitable to have an appropriate quantum error correction code which can eliminate errors and make the quantum processing stable. As we will see, in the following sections, quantum teleportation or entanglement plays a crucial role in a fault tolerant quantum computer. Therefore, in order to realize a quantum computer in its real meaning as to be quantum digital computer and to be different from the classical analog computers, furthermore in order to get the advantages of quantum computer by running the quantum algorithms, entanglement should be provided among the quantum states. 
 
 \item[Chapter 2]
 Therefore, we conclude that entanglement might be a resource for a quantum computations's power through the theoretical point of view. Entanglement is important for a number of quantum algorithms, also it is an essential requirement for fault tolerant quantum computing. Then, one challenge would be understanding the quantum entanglement in more details. Enormous studies on the concept of quantum entanglement exist. We study the problem in its corresponding chapter. We will emphasize parts, which will be used for our introduced new measure of entanglement, applicable for the particular experimental system.
 
 \item[Chapter 3]
 After all, the main goal of this study would be reasonably focused on generating or even manipulating quantum entanglement. Nuclear magnetic resonance (NMR) based quantum information processing seems to be a good candidate for this idea as there are variety of seemingly successful implementations of quantum algorithms with NMR. We first explain NMR system and NMR quantum information processing and quantum computation, as much as we need for this work. Then, we will prove, by studying a particular quantum algorithm, that NMR at high spin temperature does not give a true quantum processing . The experimental conditions in order to fulfill all the requirements for generalizing an entangled state will be shown to be very strict, almost out of reach. However, there are still some indirect approach to solve this problem. This will be discussed extremely in its corresponding part.
 
 \item[Chapter 4]
 In order to generalize a true quantum entanglement, while getting benefits of the inevitable advantages of the previously studied NMR systems, we move to a rather different physical system, so called Electron Nuclear DOuble Resonance (ENDOR). We study ENDOR system with several experiments and finally we go to our final goal of realization of a true quantum entanglement.This will be shown by several experiments on introducing ENDOR system, which is a rather more complicated experimental scheme as compared to NMR because that ENDOR is a double resonance experiment and needs to have simultaneously controls of radio frequency and microwave frequency. After all, we will show {\it ENDOR for quantum computing} experiments under different conditions, magnetic fields from $9.5$ GHz up to $95$ GHz and temperature from $\sim 300$ K down to $\sim 3$ K. Results demonstrate truly establishment of (pseudo-)entangled state. This will be discussed to be particularly important because it demonstrates our ability to manipulate electron and nuclear spins on our arbitrary desired values and conditions. Also, we will introduce an almighty molecular sample, which will be discussed to be a proper sample for quantum computing even for a larger number of qubits or under very different experimental conditions of magnetic field and temperature. Finally, we will go much closer to the realization of quantum entanglement as it is based on experimental conditions just to be in hand.
 \end{description}

\newpage
\vspace{4cm}
\section*{Preface}
\vspace{1cm}


Once in 1850, William Gladstone, the British minister of finance asked Faraday, ^^ ^^ If electricity had any practical value". He replied, ^^ ^^ One day, sir, you may tax it".


Soon or later, this is a dream though, we may be asked to pay monthly also, for purified entanglement supplied by Entanglement Purification Centre of our city. Entanglement would be purified there and sent to our houses via teleportation similar way that can exchange the used entangled states with refresh and pure entangled ones. We may use it then for running the quantum part of our computers for computing also for information processing. Let's go further in such a sweet dream, those days people may know all about the dynamics of entanglement, and how entanglement plays role for macroscopic worlds of physical properties, down to its explanation for gauge field theory. I am but still not so brave to say that it will be a key for unification in physics but I'd like to say it will work for small systems of course and even large systems as then we know how to use it in place for Partition functions of large number of molecules. Such a world would be a wonderful place and I wish I could choose my life time to enjoy the smell of knowledge in such an era.
\newpage

\vspace{4cm}
\section*{Acknowledgments}
\vspace{1cm}

This research has been carried out during the author's tenure of Ph.D. degree (2003-2006) at the Division of Advanced Electronics and Optical Science, Department of Systems Innovation, Graduate School of Engineering Science, Osaka University. During these years, many people have helped me and it is my great pleasure to take this opportunity to express my gratitude to them all even if their name might not be listed here.

First of all, I would like to express my deepest sense of gratitude to Professor Masahiro Kitagawa for invaluable guidance and continuous support throughout my study in the Department of Systems Innovation, Graduate School of Engineering Science, Osaka University.

I extend my sincere appreciation and gratitude to Professor Takeji Takui, Departments of Chemistry and Materials Science, Graduate School of Science, Osaka City University, for his patient guidance and careful advice. I am grateful for his highly respected friendship and encouragement.

Thanks to Professor Masanao Ozawa, Graduate School of Information Sciences, Tohoku University, for his incredibly highly valuable helps and constructive discussions.

I present my thanks to Dr. Kazuyuki Takeda, Department of Systems Innovation, Graduate School of Engineering Science, Osaka University, for his guidance and helps, in his special manner, and appreciate him for all that he constantly taught me. I wish to express my sincere appreciation to Associate Professor Kazunobu Sato, Departments of Chemistry and Materials Science, Graduate School of Science, Osaka City University, for his brilliant guidances, kind attentions and valuable discussions. I owe him a lot for his constantly teaching me every even fundamentals.

I would like to thank Professor Nebuyuki Imoto and Professor Shinji Urabe, Graduate School of Engineering Science, Osaka University, for their kind attentions and serving as members of my dissertation committee.

My special thanks go to Professor Hiroshi Miyasaka, Graduate School of Engineering Science, for deeply understanding and excellent help at the worst time of my research. He introduced me to Professor Takeji Takui and actually opened a very wide area in my research career. Also, I specially thank Professor Mikio Nakahara, Department of Theoretical Physics, Kinki University, for his valuable friendship and careful advice.

I owe a lot of previous or current members of Department of Systems Innovation, specially my home laboratory members. I should firstly thank Ms. Madoka Konishi and Ms. Mutsuyo Hattori specially supporting me at any moment. I feel that I should respect all the students helping me but I have to be conscious of the place. I specially thank Mr. SaiToh Akira for all that he did and most that he planned to help me. I wish him success in his scientific researches.

I extend my appreciation to Dr. Kou Furukawa and Associate Professor Toshihiro Nakamura, Institute for Molecular Science, IMS, Dr. Kazuo Toyota, Associate Professor Daisuke Shiomi, Dr. Shinsuke Nishida, Departments of Chemistry and Materials Science, Graduate School of Science, Osaka City University, Dr. Hideyuki Hara, Bruker Biospin, Japan. Also, I would like specially to thank Dr. Peter Hoefer, Dr. Patrick Carl, and Dr. Dietz Schmalbein, Bruker Biospin, Germany, for providing me excellent opportunities for testing our ideas with excellent and most advanced Bruker machines. I appreciate their kind hospitality during my stay at Germany and specially their pure scientific attitude that helped me lots.

Thanks to our host family Mrs. Yuko Hattori and Mr. Hiroaki Hattori for their extreme kindness and special attentions. Thanks to Mr. Ueno Sadao, Rotary Yoneyama Suita Club, for his kind attentions and perfect supports. He prepared excellent opportunities for me to get closer to Japanese culture. Thanks to President Yaku Takahiro and all other Rotarian kind people in Rotary Yoneyama, Japan. Their attempts to realize the meaning of the sentence ^^ ^^ Service Above Self" is the most valuable thing that I found and hope to do the same in my beloved country, Iran. I should also thank my other Japanese friends, people at the University and other places, for their generosity and hospitality. I feel happy that I have finished my postgraduate studies with big respect and homage to Japan.

I would also like to thank financial supports. COE project of Osaka University, Professor Kazumasa Miyake, constantly supported me from september 2003 to March 2006. I received financial support from Osaka University, Honors Scholarship, from April 2004 to March 2005. I have been selected by International Soroptimist for research grant for the year 2005. Also, the research grant from COE Osaka University, September 2004 to March 2005, specially helped me introducing the new field of ENDOR Quantum Computation to my Ph.D. research. I specially enjoyed full support and kind attentions from Japan Yoneyama Rotary Club. Thanks to them all.

I would like to extend my deepest gratitude to my father, mother, brothers and sisters and their families for their deep understanding, encouragements, supports and invaluable assistance. They have always supported me and never let me down. I also sincerely appreciate the teachers and professors of my previous study degrees.

Above all, this thesis is specially dedicated to my love and best friend, Mehdi. His generosity and understanding go beyond any achievement that I have been able to reach and might not be spoiled down by being translated to the format of any language.  

\tableofcontents
\vspace{5cm}
\chapter{Classical Computing/Quantum Computing}
\pagenumbering{arabic}\setcounter{page}{1} 

Let's start with this question. What do we want from a quantum computer? In principle, all the calculations that can be performed in a classical computer can also be performed in a quantum computer, simply by replacing the irreversible gates of the classical computer with their reversible counterparts. The new circuit, which includes only the reversible gates, representing the unitary operations, can be implemented in a quantum computer. Then, the question remains that what is the advantage if we construct such a quantum computer only by interchanging from a classical computation to a quantum computer.

{\it The appeal of a quantum computer is the ability of quantum algorithms faster than classical ones.} This is the core statement for any attempt regarding improvement in quantum computation.

The most welcome advantage of a quantum computing compared to the classical one is for the case of having a polynomial complexity in solving a particular problem with a quantum computer versus an exponential complexity of solving the same problem with a classical computer. Shor's algorithm of integer factorization is a specific case of this class.  

In this chapter, we review the concept of quantum computing very briefly in order to find out the inherent advantages to the classical computation. In this regard, we review quantum algorithms to verify the advantages of each case compared to the classical ones. Quantum algorithms must use quantum features, which are not available in classical computers, in order to improve the computational complexities. The quantum algorithms, which have been defined up to now, can be separated to two main classes of the database search algorithms and the algorithms for finding the generators of a normal subgroup of a given group. We will see that quantum superposition is the critical quantum property for the former class including Grover's algorithm, which gives quadratic enhancement for an unsorted search problem. While for the latter class, which includes Shor's factoring algorithm, {\it entanglement plays a crucial role}.

However, most of the hard problems are NPC, NP complete, and it is most likely that a quantum computer also cannot solve them in polynomial time. Hence, building a very expensive quantum computer, with which only very rare quantum algorithms give exponential enhancement, seems not to be so wise. As a matter of fact, polynomial complexity class versus exponential complexity class is mostly a mathematical convenience. Sometimes, a polynomial case is even worse than exponential type of complexity and it is of course the nature of the problem itself which identifies a good or a bad class of complexity.

In order to look for more elaborated advantages of a quantum computer, we might look for more subtle realistic distinctions specially inside polynomial type of complexities. This concept is discussed in the part of measurement based quantum computer. We will see that {\it entanglement plays the conceptual role}, in this regard too.

Then the straight conclusion would be that quantum features are reasonably crucial for quantum computational advantages over classical computation. While quantum superposition, coherence, is an important feature, entanglement also might be very important in this sense. 

In this chapter our aiming would be on giving enough motivations for the study of quantum entanglement through emphasizing its critical role in quantum algorithms, parallelism and even more importantly its unique role for a complete fault tolerant quantum computing. Quantum algorithms, measurement based quantum computing and fault tolerant quantum computing are explained briefly in this regard only to emphasize the critical role of entanglement rather than to be the concrete explanation of each topic.
\section{Classical Computing}
In a classical computer data are stored in a sequence of bits that are represented with a string of $0$'s and $1$'s. Computational processing is applied on this sequence in discrete steps such as Boolean gates, \cite{Y1}.

A classical computational task is not a single task such as finding the primarity of a particular number but it is a family of similar tasks, such as ^^ ^^ given an integer N, is N prime?" Input size is defined to be the space in computer that is needed to store the particular input N.

Now, we want to know how hard is the particular task to be done on the computer. The computational effort or {\it complexity} is then given by finding out that how the computational effort grows as the input size grows.
 \subsection{Polynomial Versus Exponential Complexity}
 
 Growth in the number of computational steps as the input size grows is mainly divided to two classes of polynomial and exponential complexities. The former case is known as problems, which are feasible computation on a classical computer while the latter one is known as unfeasible computation. Very wellknown example is the problem of finding the prime factor for a given number. Up to now, there is no classical algorithm, which can solve this problem efficiently, meaning that the number of steps that are required to solve the problem is not bounded by any polynomial function. Then, the problem of factoring is known as a classical hard problem, whereas quantum Shor's algorithm gives an efficient way to solve this problem.
 \subsection{Query Complexity}
 
 The complexity of a classical task would also be explained in a query complexity. The assumption is that a ^^ ^^ black-box" is given, which computes a function $f:x \to y$. During the processing, each call of the query is counted as one step.
 
 For example, for $f: n\,bits \mapsto 1\,bit$, the problem is that whether $f$ is balance or constant. The promise is that $f$ is either balance or constant. For a classical computer the query complexity of this particular example is $2^{n-1} +1$, in a worst case. However, with a quantum computer, this problem can be solved with a single query to the black-box. The point is that the property, which has been asked is a global property. Therefore, there is no need to examine all the elements and compare them to each other, in order to find the answer. Quantum superposition makes it possible to manipulate all the elements of initial states, simultaneously and extract the desired global property with only one query to the black-box.
\section{Quantum Computing}
Generalization of the above concepts to the quantum computing, gives the corresponding quantum complexities. In quantum computation, bits are represented by qubits, two-level system with chosen computational bases. The computational steps are quantum operations on qubits and are represented by unitary gates or quantum measurement.
 
 There are various models possible for structure of a quantum computer. Each case is evaluated by a particular computational complexity, and entanglement is particularly an important quantum property for each class to represent the desired advantages of the quantum complexities compared to the classical counterparts, \cite{Y1}.
 \subsection{Gate Array Model}
 
This is the standard model for quantum computation. It starts with row of qubits in a simple state, generally initialized state. The computational steps are unitary gates and measurement that is performed at the end of the computation in $\{|0\rangle , |1\rangle \}$ basis.

Quantum algorithms prescribe the quantum processing and quantum computational steps. Quantum algorithms use the properties of quantum physics in order to provide new modes of computation and information processing, which are not available to classical computers. Because of the fundamental postulates of quantum physics, namely superposition, a quantum computer, which works based on the laws of quantum physics is able to store and process large volumes of information. All the information is represented in a(n) (entangled) quantum state. However, generally the access to the whole information is strictly restricted by quantum measurement theory. Then, only a relatively small amount of the whole stored information can be read out.

However, the extracted information maybe of a global nature and then gives an intuitive information on the whole quantum physical system. On the other hand, this process may be impossible for a classical computer to be performed efficiently to give a global feature of the total system. Because a classical computer may still need to examine each individual state in order to get the total information.

Quantum algorithms are actually to give a routine on a way how to extract the desired information from the total information stored in superposed quantum states, in a rather more efficient way compared to the classical approaches. Particularly, the most celebrated quantum algorithm is Shor's algorithm for integer factorization. While the best known classical algorithm for this problem runs in superpolynomial time of order $exp(n^{1/3}(\log n)^{2/3})$, Shor's algorithm provides a method for factoring an integer of $n$ digits in the order of $O(n^3)$. 

The significance of all quantum algorithms lies in the issue of computational complexity theory. Then, we classify our review based on the complexity classes for quantum algorithms.

The computational step for the gate array model of quantum computation is counted as the required unitary gates with dimension equal to or less than two qubits, e.g. CNOT, Hadamard and so on, in addition to the read-out step, which is the measurement at the end of the quantum circuit and only is performed in the computational basis.

\subsubsection{Polynomial Versus Exponential Complexity; Shor's Algorithm}

Suppose $\mathcal{H}_N$ to be the Hilbert space with dimension $N$ and with basis $|0\rangle, |1\rangle, ..., |N-1\rangle$, being labeled by integers module $N$.  For this space, quantum Fourier transformation ${\rm QFT}_N$ is defined as unitary transformation $\mathcal{H}_N \to \mathcal{H}_N$ as follows, \cite{Y1}, \cite{Y2}, \cite{Y3}, \cite{Y4}, \cite{Y5}, \cite{Y6}
\begin{equation}
{\rm QFT}_N |x\rangle =\frac{1}{\sqrt N}\sum_{j=0} ^{N-1} e^{\frac{2\pi ixy}{N}}|y \rangle,
\end{equation}
for $x, y= 0,1,...,N-1$. The matrix elements are represented as
\begin{equation}
[ {\rm QFT}] _{xy}=\frac{1}{\sqrt N} W^{xy},
\end{equation}
where
\begin{equation}
W=e^{\frac{2\pi i}{N}}.
\end{equation}

For example, ${\rm QFT}_2$ is just the Hadamard gate.

In a classical scheme with dimension $N=2^n$, Fourier transformation is an $N \times N$ matrix. So, calculation of the matrix needs $O(N^2)$ steps. The number of steps can be decreased to $O(N \log N)$, i.e. still exponential function of $n$.

However, it is proved that ${\rm QFT}_N$ can be implemented in $O((\log N)^2)=O(n^2)$ computational steps. If $N\neq 2^n$ then ${\rm QFT}_N$ cannot be exactly implemented in ${\rm poly}(\log)$ steps, but ${\rm QFT}_N$ can be still be closely approximated (to any accuracy $\epsilon$) by ${\rm QFT}_N$, where $M>N$, $M=2^m$ and $M=O(N\log N)$. i.e. we need only $O(\log \log N)$ additional qubits.

This is a remarkable fact that ${\rm QFT}_N$ can be implemented polynomially and this fact is fundamental step in Shor's algorithm.

The problem of factoring can be reduced to the problem of periodicity finding. This connection is entirely mathematical and does not have any quantum physical aspects. Therefore, if we could have an efficient classical algorithm for solving the periodicity problem then factoring problem also could be solved entirely classically.

Suppose that the integer number $N$ is given. Pick an integer $N^2\leq q\leq2N^2$. This step is clearly always possible. Also, pick a random number integer $x$, such that ${\rm gcd} (x,N)=1$. This step can be run using Euclid's algorithm for calculating the greatest common divisor, $b={\rm gcd} (x, N)$. If $b>1$ then $b$ is a factor of $N$ and actually we are done. Otherwise keep the choice of $x$.

Now, create two quantum registers. Register {\bf I} must contain enough qubits to represent numbers as large as $q-1$. Register {\bf II} must contain enough qubits to represent numbers as large as $N-1$. Then, load quantum register {\bf I} with an equally weighted superposition of all the integers from $0$ to $q-1$. Load register {\bf II} with zeros. The total state at this point will be 
\begin{equation}
\frac{1}{\sqrt q} \sum_{a=0}^{q-1} |a, 0\rangle .
\end{equation}

Now, apply the transformation $x^a {\rm mod} N$ to each number stored in register {\bf I} and store it in register {\bf II}. The total state is changed into
\begin{equation}
\frac{1}{\sqrt q} \sum_{a=0}^{q-1} |a, x^a {\rm mod} N \rangle .
\end{equation}

Measure the state of the second register. The measurement must yield one value of $x^a {\rm mod} N$, let's say $V$. The measurement on the register {\bf II} has the effect of collapsing the register {\bf I} into an equal superposition of each state between $0$ and $q-1$, such that $x^a {\rm mod} N=V$, because these are the only states, which could give that results.

By Euler's theorem, there is the lowest power $r$ such that 
\begin{equation}
x^r \equiv 1 \, {\rm mod}\,  N,
\end{equation}
where $r$ is the periodicity of $f(x)=x^a {\rm mod} N$.

Up to now, ${\rm QFT}_N$ is explained to be implementable efficiently. Also, factoring problem is shown to be a periodicity problem. Then, the remaining step in Shor's algorithm is to compute the ${\rm QFT}_N$ on the collapse register {\bf I}.

If the periodicity of ${\rm QFT}_N$ is $r$, then the ${\rm QFT}_N$ output has the periodicity $q/r$. With a classical computer the $r$ period is evaluated. Now, the factor of $N$ can be determined by taking the greatest common devisor, ${\rm gcd} (x^{r/2}+1,N)$ and ${\rm gcd} (x^{r/2}-1,N)$. If this does not work, the same process should be tried for a new number $x$.

The important point supporting the efficiency of Shor's factoring algorithm is the fact that QFT can be implemented efficiently, then exponentially faster than the classical case. The following example makes it clear how Shor's algorithm works.

Problem is to find the prime factors for $N=15$. Firstly, pick an integer $q$ such that $N^2\leq q\leq 2N^2$.
\begin{equation}
\label{30}
(N^2=(15)^2=225)\leq (q=2^8=256)\leq (2N^2=2(15)^2=450).
\end{equation}

This implies that $8$ qubits are sufficient for register {\bf I}. The second register must contain enough qubits to represent numbers up to $15-1=14$. Let register {\bf II} holds numbers up to $2^4=16$.

Therefore, we are working with two registers, ${\bf I,\, {\rm and},\, II}$.
\paragraph{Register {\bf I}}: given as
\begin{eqnarray}
\label{31}
|00000000\rangle +|00000001\rangle+|00000010\rangle+...\\ \nonumber
{\rm or}\, {\rm equivalently}\\ \nonumber
|0\rangle +|1\rangle +|2\rangle+...
\end{eqnarray}
\paragraph{Register {\bf II}}:  all zeros
\begin{equation}
|0000\rangle=|0\rangle.
\end{equation}

The total state would be then as follows 
\begin{equation}
\label{22}
\frac{1}{\sqrt {256}} \left[|0\rangle +|1\rangle +...\right]|0\rangle .
\end{equation}

Now, apply the transformation $x^a\,mod\,N$ to each number stored in register {\bf I} and store it in register {\bf II}. The number $x$ is picked at random here to be $x=7$, satisfying gcd$(x,N)=1$.

An important result from number theory is shown as follows.\\
$f(a)=x^a\,mod\,N$ {\it is a periodic function.} ($x$ and $a$ and $N$ are all integers; the greatest common divisor of $x$ and $N$ is $1$.)\\
{\bf Definition}: $B\,mod\,C$ is the remainder after division of $B$ by $C$.

This is known that $f(a)=x^a\,mod\,N$ has a period $P$. If $P$ is an even number then $(x^{P/2}-1)(x^{P/2}+1)$ is an integer multiple of N.

Hence, we have
\begin{eqnarray}
\label{33}
{\rm First}\,{\rm register}\hspace{2cm}&7^a\,mod\,15& \hspace{2cm} {\rm Second}\,{\rm register}\\ \nonumber
\hspace{1cm}|0\rangle      \hspace{3cm}&7^0\,mod\,15& \hspace{3cm}             1                    \\ \nonumber
\hspace{1cm}|1\rangle      \hspace{3cm}&7^1\,mod\,15& \hspace{3cm}             7                    \\ \nonumber
\hspace{1cm}|2\rangle      \hspace{3cm}&7^2\,mod\,15& \hspace{3cm}             4                    \\ \nonumber
\hspace{1cm}|3\rangle      \hspace{3cm}&7^3\,mod\,15& \hspace{3cm}             13                    \\ \nonumber
\hspace{1cm}|4\rangle      \hspace{3cm}&7^4\,mod\,15& \hspace{3cm}             1                    \\ \nonumber
\hspace{1cm}|5\rangle      \hspace{3cm}&7^5\,mod\,15& \hspace{3cm}             7                    \\ \nonumber
\hspace{1cm}|6\rangle      \hspace{3cm}&7^6\,mod\,15& \hspace{3cm}             4                    \\ \nonumber
\hspace{1cm}|7\rangle      \hspace{3cm}&7^7\,mod\,15& \hspace{3cm}             13                    \\ \nonumber
\hspace{1cm}|8\rangle      \hspace{3cm}&7^8\,mod\,15& \hspace{3cm}             1                    
\end{eqnarray}
Now, measure the second register. For this example, the possible outcomes are $|1\rangle, |4\rangle, |7\rangle, |13\rangle$. E.g. if the measurement on the second register gives $V=|1\rangle$, then the first register must collapse into $|0\rangle+|4\rangle+|8\rangle+|12\rangle+|16\rangle+|20\rangle \, ...$, which has the periodicity of $4$. Then apply the quantum Fourier transformation (QFT).

We know that the discrete classical Fourier transformation is as follows.\\
{\bf Input}: a series of $N$ complex numbers, $x_0, x_1, x_2,...,x_N$.\\
{\bf Output}: a second series of $N$ complex numbers, $y_0, y_1, y_2,...,y_N$, where
\begin{equation}
\label{35}
y_k=\frac{1}{\sqrt N} \sum^{N-1}_{j=0} x_j e^{2\pi ijk/N}.
\end{equation}

For the input series, if it has a period $P$, the discrete Fourier transform (DFT) output will repeat with a period $N/P$.
\begin{equation}
\label{36}
\begin{array}{@{\,}cccccc@{\,}}
            {\rm Input}&{\rm Period (p)}&{\rm DFT}  &{\rm Output}&{\rm Period (N/P)}\\
            10000000   &8               &\Rightarrow&11111111 &1              \\
            10001000   &4               &\Rightarrow&10101010 &2              \\
            10101010   &2               &\Rightarrow&10001000 &4              \\
            11111111   &1               &\Rightarrow&10000000 &8              \\           
\end{array}
\end{equation}
The quantum Fourier transform (QFT) is almost the same as the discrete Fourier transform, however it operates on the complex numbers which represent phase and amplitude of the various contributions of a superposition of states. For example consider three-qubit case.

\begin{eqnarray}
\label{37}
|\psi\rangle&=& \alpha_{000}|000\rangle +\alpha_{001}|001\rangle +\alpha_{010}|010\rangle +\alpha_{011}|011\rangle +\\ \nonumber
            & &\alpha_{100}|100\rangle +\alpha_{101}|101\rangle +\alpha_{110}|110\rangle +\alpha_{111}|111\rangle \\ \nonumber
            &=&\alpha_{000}|0\rangle +\alpha_{001}|1\rangle +\alpha_{010}|2\rangle +\alpha_{011}|3\rangle +\alpha_{100}|4\rangle +\alpha_{101}|5\rangle +\alpha_{110}|6\rangle +\alpha_{111}|7\rangle.
\end{eqnarray}
For this general case, suppose that the amplitudes are equal to each other. Then the Fourier transform would be written as follows.
\begin{equation}
\label{38}
\begin{array}{ccccc}
\fl {\rm Input}&{\rm Period (p)}&{\rm QFT}  &{\rm Output}                                     &{\rm Period (N/P)}    \\
\fl |0\rangle                                  &8&\Rightarrow&\{|0\rangle +|1\rangle +|2\rangle +|3\rangle + &1 \\
\fl                                           & &           &|4\rangle +|5\rangle +|6\rangle +|7\rangle \}   &  \\
\fl |0\rangle +|4\rangle                       &4&\Rightarrow&|1\rangle +|3\rangle +|5\rangle +|7\rangle   &2 \\
\fl |0\rangle +|2\rangle +|4\rangle +|6\rangle &2&\Rightarrow&|0\rangle +|4\rangle                         &4 \\
\fl \{|0\rangle +|1\rangle +|2\rangle +|3\rangle+&1&\Rightarrow&|0\rangle                                    &8 \\
\fl |4\rangle +|5\rangle +|6\rangle +|7\rangle \} & &           &                                             & \\            
\end{array}
\end{equation}

\begin{figure}
\begin{center}
{\includegraphics[2cm,21cm][10cm,23cm]{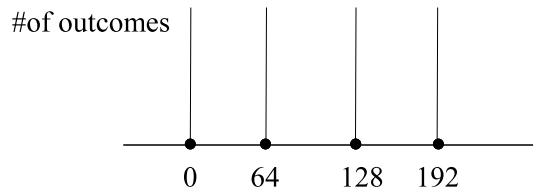}}
\end{center}
\caption{\label{YF1}The result of the measurement.} 
\end{figure}

As with the discrete Fourier transform, if the period of the $N$ input is $P$, then the period for the output of the quantum Fourier transform is $N/P$. Then in the above explained example for $N=15$, we expect the periodicity of $256/4$ in the output. The new state can be written as  $|0\rangle+|64\rangle+|128\rangle+|192\rangle$ for $q/r=256/4=64$. With a high probability, the measurement yields an integer which is an integer of $q/4$. If the measurement were to be repeated many times, the output then with high probability could be plotted as Figure \ref{YF1}. 

Now, it turns out that ${\gcd}(7^2+1, 15)$ and ${\gcd}(7^2-1, 15)$ or $5$ and $3$ are the factors. If this does not work, the process would be repeated for another random number $x$.

\subsubsection{Query Complexity; Grover Algorithm}

The black-box is a unitary operator, which can be applied as a single gate. Then the complexity is counted with the number of query to the box.

Grover's algorithm is for searching in an unstructured database for a single item, \cite{Y1}, \cite{Y8}, \cite{Y9}, \cite{Y10}, \cite{Y11}. The particular problem is among $N$ database, to find the special $x$ value for which $f(x=x_0)=1$, where $f$ is a black-box defined as 
\begin{equation}
f:\{1,2,...,N\}\mapsto \{0,1\}.
\end{equation}

Unstructured search simply means that if the answer for the trial is ^^ ^^ NO" then we get absolutely no idea where the answer could be. Instead, for an structured search the situation is completely different. For example, suppose an ordered search problem with a given $N=2^n$ number in ascending order. Suppose that one of them $(x_0)$ is in hand. Then try to find $x_0$ in the list. Classically, $\log_2 N =n$ queries are necessary and sufficient to solve the problem. With quantum computation, however it is different, hence representing the quantum benefits.

We assume an unstructured database search problem. Classically, $O(N)$ queries are required to solve this problem. But, in a quantum case, $O(\sqrt N)$ queries are necessary. This is given by Grover algorithm and it is proved that ^^ ^^ no quantum process can do this with fewer steps", \cite{Y10}.

The important point to be discussed here before starting with Grover algorithm is to make sure about the benefits of quantum searching. We try to go around this question with a very simple example. Assume that we are given a database including $N$ values, for which $f(x=x_0)=1$ and otherwise $f(x)=0$. The task is to find $x=x_0$. The important point is that we do not have different compositions of all the states. Instead, we have only an equal superposition of items. Then making an equal superposition of the items need no exponential effort. However, if we originally have all the different cases, then to make an equal superposition of all the items requires exponential steps and hence the benefits of Grover algorithm is lost, \cite{Y1}.

We start with a black box $f: n\,{\rm bits}\to 1\,{\rm bit}$ with $f(x\neq x_0)=0$ and $f(x=x_0)=1$. However, it turns out that it is more convenient to use equivalent box, phase box as follows
\begin{equation}
I_{x_0}|x\rangle=
\begin{cases}
|x\rangle  & \text{if $x \neq x_0$},\\
-|x\rangle & \text{if $x=x_0$}.
\end{cases}
\end{equation}

The equivalency becomes clear if we try ot for $\frac{1}{\sqrt 2}(|0\rangle -|1\rangle)$. In other words, one use of $I_{x_0}$ needs just one query to $U_f$. The Grover iteration is introduced as follow
\begin{equation}
Q=-H_n I_0 H_n I_{x_0},
\end{equation}
where
\begin{equation}
H_n=H\otimes ...\otimes H
\end{equation}
with {\it H} to be the Hadamard operation. The starting state is an equal superposition of all possible $x$-values as follows
\begin{equation}
|\psi_0\rangle =H_n |0...0\rangle=\frac{1}{\sqrt {2^n}}\sum_x|x\rangle.
\end{equation}

Then $(n+1)$th state is written based on the previous $n$th state as follows 
\begin{equation}
|\psi_{n+1}\rangle =Q|\psi_n\rangle.
\end{equation}

For $|\psi\rangle$, any state in space, we have 
\begin{equation}
I_{|\psi\rangle}=I-2|\psi\rangle\langle\psi|,
\end{equation}
indicating that $I_{|\psi\rangle}$ is a reflection in $\mathcal{H}_{\bot}(\psi)$, orthogonal subspace. The proof is clear because that for any chosen state $|\xi\rangle =a|\psi\rangle+b|\xi^\prime\rangle$, where $|\xi^\prime \rangle \bot |\psi\rangle$, then $I_{|\psi\rangle}|\xi\rangle=-a|\psi\rangle+b|\xi^\prime\rangle$.

$I_{|\psi\rangle}$ preserves the $2$D subspace spanned by $|\psi\rangle$ and $|\xi\rangle$. Also, for any unitary operation $U$, we have
\begin{equation}
UI_{|\psi\rangle}U^\dagger =I_{U|\psi\rangle},
\end{equation}
because
\begin{eqnarray}
U(I-2|\psi\rangle\langle\psi|)U^\dagger&=&I-2|U\psi\rangle\langle U\psi|\\ \nonumber
                                       &=&I_{U|\psi\rangle}.
\end{eqnarray}          
If all the above properties are applied on the iteration operator $Q$ and also recall that $H_n=H^\dagger_n$, then
\begin{eqnarray}
Q&=&-H_nI_0H_nI_{x_0}\\ \nonumber
 &=&-I_{H_n|0\rangle}I_{x_0}\\ \nonumber
 &=&-I_{|\psi_0\rangle}I_{x_0}.
\end{eqnarray}
In the same manner, both $I_{|\psi_0\rangle}$ and $I_{x_0}$ preserve the $2$D plane, spanned by $|\psi_0\rangle$ and $|x_0\rangle$.

\begin{figure}
\begin{center}
\scalebox{0.65}
{\includegraphics[3cm,21cm][15cm,27cm]{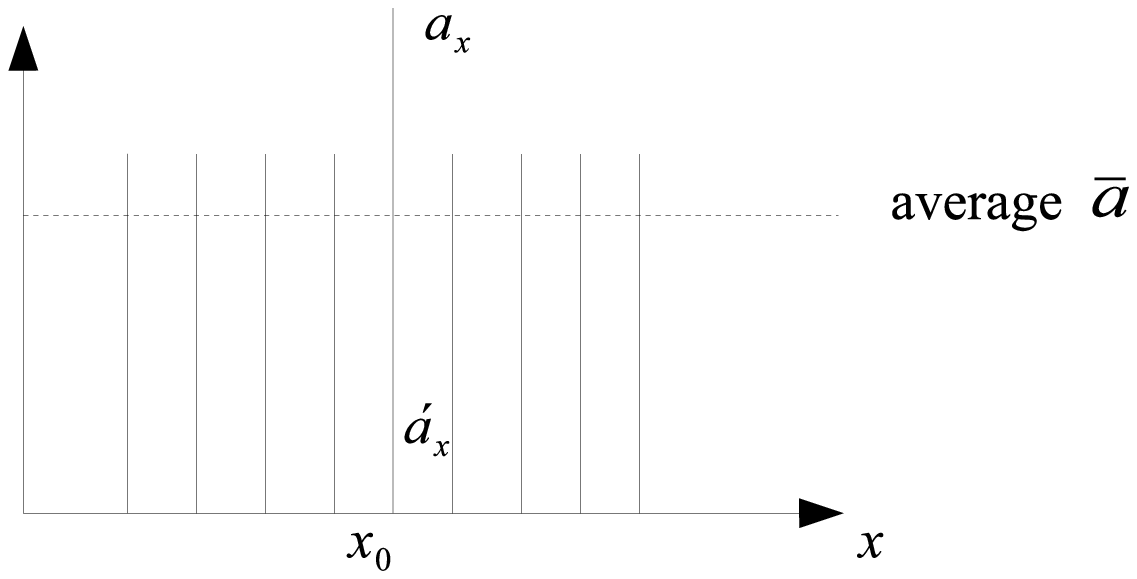}}
\end{center}
\caption{\label{YF2} $a^\prime_x$ is the inverse of $a_x$ as to the average $\overline a$.} 
\end{figure}

Now, we want to show, that $-I_{|\psi_0\rangle}$ is actually inversion in the average. Recall $|\psi_0\rangle=\frac{1}{\sqrt N}\sum |x\rangle$. For $|\alpha \rangle =\sum^{N-1}_{x=0} a_x|x\rangle$, with average amplitude $\overline a=\frac{\sum a_x}{N}$, then
\begin{eqnarray}
-I_{|\psi_0\rangle}|\alpha\rangle&=&-(I-2|\psi_0\rangle\langle\psi_0|)|\alpha\rangle\\ \nonumber
                                 &=&-\sum_{x^\prime}(I-\frac{2}{N}\sum_{xy}|x\rangle\langle y|)a_{x^\prime}|x^\prime\rangle\\ \nonumber
                                 &=&-\sum_x(a_x-2\overline a)|x\rangle\\ \nonumber
                                 &=&-\sum_x a_x-2(a_x-\overline a) |x\rangle \\ \nonumber
                                 &=&\sum_x a^\prime_x|x\rangle,
\end{eqnarray}
where $a^\prime_x=a_x-2(a_x-\overline a)$ and $(a_x-\overline a)$ is the distance of $a_x$ from average, see Figure \ref{YF2}.

The total steps of the Grover algorithm can be represented as follows
\begin{description}
\item[.]
Start with a uniform superposition, Figure \ref{YF3}.
\begin{figure}
\begin{center}
\scalebox{0.65}
{\includegraphics[3cm,21cm][15cm,27cm]{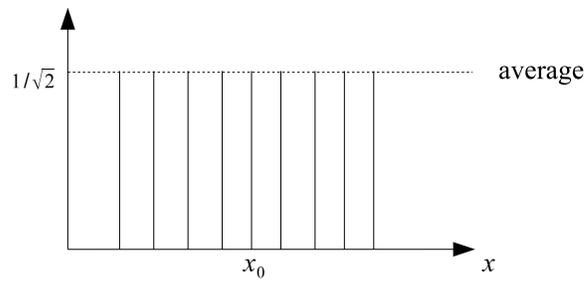}}
\end{center}
\caption{\label{YF3} The initial states are in a uniform superposition.} 
\end{figure}

\item[.]
Apply $I_{x_0}$, hence the inverse of $x_0$, Figure \ref{YF4}.
\begin{figure}
\begin{center}
\scalebox{0.65}
{\includegraphics[3cm,18cm][15cm,26.6cm]{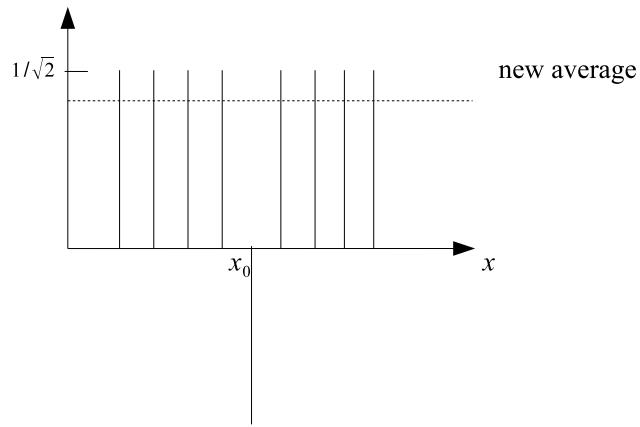}}
\end{center}
\caption{\label{YF4} Applying $I_{x_0}$ gives the inverse of $x_0$ and makes the average less than the previous one.} 
\end{figure}

\item[.]
Apply $-I_{|\psi_0\rangle}$, which gives the inverse in average, Figure \ref{YF5}.
\begin{figure}
\begin{center}
\scalebox{0.65}
{\includegraphics[3cm,21cm][15cm,29cm]{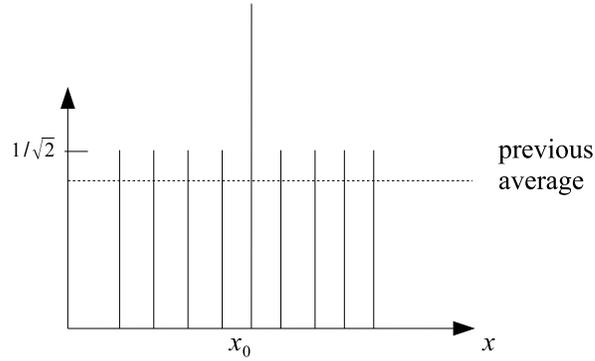}}
\end{center}
\caption{\label{YF5} $-I_{|\psi_0\rangle}$ gives the inverse in average.} 
\end{figure}
\end{description}

Amplitude for $x=x_0$ much increased, while amplitude for $x\neq x_0$ has been slightly decreased. Now, repeat the iteration operation.

It is proved that in $O(\sqrt N)$ the value $x_0$ can be extracted and no other quantum algorithm can solve this problem faster. Clearly, the scheme is based on quantum properties such as superposition of states. 

\subsubsection{Discussion}

Our review on quantum algorithms up to now has been mainly on the quantum computational algorithms with the main idea of emphasizing the quantum benefits compared to the classical counterparts. Shor's algorithm is introduced as a quantum algorithm, which represents polynomial complexity compared to any classical known algorithms, generally with exponential complexities. Entanglement is an essential quantum property, which supports exponential speed-up for Shor's algorithm. Grover algorithm represents quadratic query complexity compared to classical algorithms. Original algorithm represented by Grover, involved entanglement in order to represent quantum enhancement, however quantum coherence or superposition seems to be the required quantum property, \cite{Y11}.

There are some other quantum algorithms, but we are not going to get involved for their explanations here. Through our study on quantum entanglement, we emphasize that quantum enhancement is available if entanglement exists between the states even though the vice explanation is not correct, generally. Both quantum superposition and quantum entanglement play crucial role for quantum algorithms. Then, serious study on the concept of entanglement still seems to be reasonable.

We should maybe go around explanation on quantum information processing, too. Namely quantum teleportation and superdense coding. But, here our main idea of studying quantum algorithms has been finding out the complexity advantages due to quantum properties such as superposition and entanglement, compared to classical counterpart algorithms for solving the similar problems in quantum gate array model. Quantum teleportation is an essential part of the measurement based model. Then, we postpone it to the next section, when speaking about measurement based model and corresponding complexities. Superdense coding turns out to be very important for introducing a new measure of entanglement, as we proposed for ensemble quantum computer. Then we will go around superdense coding in the following chapter when we discuss on entanglement properties and different measures of entanglement. Superdense coding will be explained precisely in order to introduce a new entanglement measure.
                    
 \subsection{Measurement Based Model}
 
Measurement based models are particularly important for several reasons. Firstly, they do not have any evident classical analogous. This means they can be regarded to be more intrinsically quantum. Secondly and more importantly for us, measurement based models offer new perspective on the role of entanglement in computation. Finally, they might be shown to be more interesting for experimental implementations, then to be in our interest. Therefore, we study this model of quantum computation. 

In gate array model, measurement is the last part of quantum circuit for converting quantum information into classical information in order to read out classical answers. In contrast to unitary gates, measurements are very destructive. In other words, we carry large amount of information, all during the calculation in quantum circuit whereas in the last step we spoil most part of the whole information. It can be then surprisingly beneficial to perform measurement at the most beginning part of calculation or even use several measurements for steps of calculation.

In this formalism of measurement based quantum computation, \cite{Y1}, \cite{Y12}, \cite{Y13}, \cite{Y14}, \cite{Y15}, \cite{Y16}, \cite{Y17}, \cite{Y18}, fixed entangled state of many qubits are used to start up the computation. Therefore, entanglement plays a crucial role. Two very general schemes exist, teleportation quantum computing TQC and Cluster model or one-way quantum computer (1WQC). The latter one is based on multi-partite entanglement among all the cluster state, while the former one is based on bipartite entanglement. We will discuss TQC and give up 1WQC for future because that these two models are shown to be equivalent. Though the role of entanglement is so clear in TQC through quantum teleportation, as a non-local quantum algorithm.

After all, we will conjecture that the model can basically give an exponential enhancement, however on polynomial classical algorithms, by parallelism and due to the manipulation of entangled states. 

\subsubsection{Teleportation Quantum Computation}

Quantum teleportation, \cite{Y19}, is a non-local quantum algorithm for transferring quantum state between two parts. This algorithm is based on the entangled state preshared between the involved parties.

Suppose that Alice posses a quantum state $|\psi\rangle$ and wants to send it to Bob. Each of the parties, Alice and Bob, posses a part of an entangled state $|\beta_{00}\rangle$, where $|\beta_{zx}\rangle$ is maximally entangled Bell state, defined as follows
\begin{equation}
|\beta_{zx}\rangle\equiv\frac{|0,x\rangle+(-1)^z|1,\overline x\rangle}{\sqrt 2}.
\end{equation}

Totally, Alice has two qubits in her disposal, one to be in the state $|\psi\rangle$ and the other that is entangled to the qubit in Bob's disposal. Alice applies Bell measurement on qubits and gets the experimental outcomes $c$ and $d$. The state is then teleported to Bob, as $X^dZ^c|\psi\rangle$. She sends the classical information, outcomes of the measurement to Bob. He applies corresponding unitary operations on the state and corrects it into $|\psi\rangle$.

The rotated Bell basis is defined as
\begin{equation}
|\beta(u)_{cd}\rangle={\rm U}^\dagger \otimes {\rm I} |\beta_{cd}\rangle,
\end{equation}
for any $1$-qubit ${\rm U}$. Then, consider the situation, where we want to apply the unitary transformation ${\rm U}$ on a state $|\psi \rangle$. One way is as follows
\begin{equation}
 _{12}\langle \beta(u)_{cd}|\psi\rangle_1|\beta_{00}\rangle_{23}=_{12}\langle \beta_{cd}|U\psi\rangle_1|\beta_{00}\rangle_{23}=\frac{1}{2}{\rm X}^d {\rm Z}^c |U\psi\rangle_3.
\end{equation}

 The last equation is written in a similar manner as the standard teleportation for $|\psi\rangle$ as follows
 \begin{equation}
 _{12}\langle \beta_{cd}|\psi\rangle_{1}|\beta_{00}\rangle_{23}=\frac{1}{2} {\rm X}^d {\rm Z}^c |\psi \rangle_3.
 \end{equation}
 
 The interpretation of the above equations is that given a resource of $|\beta_{00}\rangle$, we can apply any $1$-qubit unitary transformation (up to known Pauli operators) to any state $|\psi\rangle$, by measurement alone. Generalization for $2$-qubit is done with more elaborated teleportation, which we do not mention here but may be found in some review papers, \cite{Y1}.
 
 Generalization of teleportation for dimension $d$ of space gives 
 \begin{equation}
 _{12} \langle \phi|\alpha\rangle_1|\phi\rangle_{23}=\frac{1}{d}|\alpha\rangle_3,
 \end{equation}
 where $|\alpha\rangle_1$ is teleported into $|\alpha\rangle_3$. Here, the maximally entangled state in $d$ dimension is shown by
 \begin{equation}
 |\phi\rangle=\frac{1}{\sqrt d}\sum^{d-1}_{i=0}|i\rangle|i\rangle.
 \end{equation}
 
 In a same manner, for rotated maximally entangled state, we have
 \begin{equation}
 |\phi(u)\rangle_{12}=U_1^\dagger\otimes I_2|\phi\rangle_{12},
 \end{equation}
 and hence
 \begin{eqnarray}
 _{12}\langle\phi(u)|\alpha\rangle_1|\phi\rangle_{23}&=&_{12} \langle \phi|U\alpha\rangle_1|\phi\rangle_{23}\\ \nonumber
                                                    &=&\frac{1}{d}U|\alpha\rangle_3.
\end{eqnarray}
The unitary operation $U$ is applied on the state by only measurement.
 
 The unitary transformations $(U_i)$ are chosen such that $|\phi(u_i)\rangle$'s are orthonormal basis in $12$-space. In the standard teleportation, $|\phi\rangle$ is the Bell basis and $U_{i}$'s are Pauli operators ${\rm I, X, Y, Z}$ and the output state is (Pauli)$|\alpha\rangle_3$. For a general $1$-qubit gate $V_i$ then $U_i$'s are $V^\dagger$ Pauli's and the output state is (Pauli)$V|\alpha\rangle_3$. For a general $2$-qubit gate $V$, just apply the same but with $d=4$ and use 
 \begin{equation}
 |\phi\rangle=1/2 (|00\rangle+|11\rangle)_{12}(|11\rangle+|11\rangle)_{1^\prime2^\prime},
 \end{equation}
 which is the maximally entangled state in four dimensions.
 
 For a universal computation, we should know, how to perform arbitrary sequence of gates from a universal set, by measurement. One set of universal gates is controlled-Z, X-rotation and Z-rotation. Hence, we know how to implement the required quantum gates. Then, we are almost done if only we can get rid of unpleasant extra Pauli operations premium the measurement outcomes. The reason for these extra Pauli operations is the randomness of the measurement.
 
 So, in order to get rid of the extra Pauli operators, we use unrandom measurement or adaptive measurements. This means that the result of a measurement determines and modifies the basis for the following measurement. Simple computational relationship between the Pauli operators and the rotational gates are useful in this regard.
 
 Suppose we want to do
 \begin{equation}
 U_l...U_1|\psi\rangle={\rm X}_l^{m_l}{\rm Z}_l^{n_l}...{\rm X}_2^{m_2}{\rm Z}_2^{n_2}{\rm X}_1^{m_1}{\rm Z}_1^{n_1}(U_l...U_1)|\psi\rangle,
 \end{equation}
 where $n_i$ and $m_i$ are the outcomes of $i$th measurement. So, we find out the Pauli corrections for the sequence of the unitary transformations.
 
 After all, measurement is done in Z-basis for all qubits. The Z Pauli operators do not have any effect on Z-measurement outcomes. For X Pauli operators also the situation is not complicated. Recall that $U|\psi\rangle=a|0\rangle+b|1\rangle$, then ${\rm X}U|\psi\rangle=a|1\rangle+b|0\rangle$. Therefore, probability of outcome $S_i=0,1$ for $Z_i$ measurement on $U|\psi\rangle$ is equal to the probability of outcome $S_i^\prime=S_i+m_i$ for ${\rm Z}_i$ measurement on ${\rm X}_i^{m_i} U|\psi\rangle$. The discussion above can be easily generalized for controlled Z operation supporting the universality of the gates in this model.
 
 Up to here, we gave a very brief sketch on the TQC. The idea of this studying is to show that how this model, which is based on entanglement can give quantum advantages over classical ones. The answer is not cleared up yet. But in some cases we can get the desired advantages as described in the followings.  
 
 In fact in TQC, we require the ^^ ^^ adaptive measurement ". However, if at least for some cases the measurement basis choices are not adaptive then we can do all the measurements in parallel as they all commute as quantum operations. The special operators, in this regard, are known as Clifford operations on $n$ qubits. We recall definitions as follows
  \begin{description}
 \item[.]
 Pauli group $P_n$ on $n$ qubits is generated by $\pm {\rm I}, \pm i, {\rm X, Y, Z}$ on each qubit.
 \item[.]
 C is a Clifford group operation on $n$ qubits if
 \begin{equation}
 CP_nC^{-1}=P_n\hspace{1cm} i.e. \hspace{1cm} CP_n=P_nC.
 \end{equation}
  \end{description}
 The Clifford group on $n$ qubits is generated by X, Z, H, CX, ${\rm p}=\left(\begin{array}{@{\,}cc@{\,}}0&1\\1&i\\ \end{array}\right)$.

Any n-qubit Clifford operator C can be represented by an $O(\log n)$ depth circuit. For any given general Poly$(n)$ sized gate, there is generally Poly$(n)$ depth so as a special parallelisability property of Clifford array are appropriate. 

Therefore, in measurement based model, to apply sequence of Clifford operations on a state, it is equivalently correct to apply all measurements, simultaneously and in parallel, i.e. Clifford operations are done in depth 1 process. However, Pauli exponents $m_i$ and $n_i$ require further $O(n)$ classical computation. Adding up the steps, then we conclude that any Clifford operation on n qubits can be represented as constant depth quantum $O(\log(n))$ depth computation followed by purely classical $O(n)$ computation.

We try to re-organize the quantum algorithm into a {\it quantum part} and a {\it classical part}, then minimize the quantum part in its depth. In this sense, entanglement inherent in measurement based model has been shown to be useful to come up with quantum advantage.

\begin{figure}
\begin{center}
\scalebox{0.99}
{\includegraphics[3cm,24cm][9cm,27.5cm]{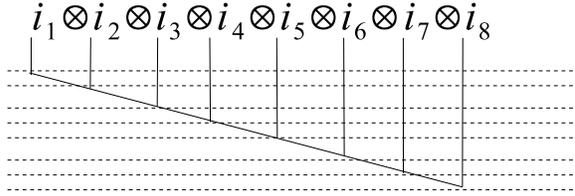}}
\end{center}
\caption{\label{YF6} The depth of calculation is shown for a particular example for a sequential adding, where the elements are added step by step to each other.} 
\end{figure}

\begin{figure}
\begin{center}
\scalebox{0.99}
{\includegraphics[3cm,23.5cm][9cm,27cm]{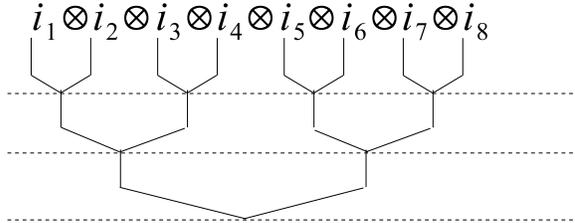}}
\end{center}
\caption{\label{YF7} For the particular example as shown on Figure \ref{YF6}. Adding one by one at each level decreases the depth of calculation.} 
\end{figure}

Suppose the problem for adding up 8 digits. This simple problem may be solved by sequential addition as shown in Figure \ref{YF6}.
The sequential addition requires $O(n)$ depth. However, if we add up one by one at each level then the same problem would be solved in $\log(n)$ depth, which verifies the logarithmic enhancement, see Figure \ref{YF7}. 

We note that the quantum gate array model is equivalent to the measurement based model. But the equivalency is not generalized to parallelisibility. In that sense, measurement based model seems more adequate to find out whether quantum algorithms exhibit better parallelisibility properties. There are very interesting research outcomes for parallelisability for special circuits with special gates, however for the general case of problems, still this is an open problem.

The conjecture is that any Poly$(n)$ sized quantum gate array can be done with Poly$(\log(n))$ layer, and Poly$(n)$ classical computation. This is a remarkable result if can be proved. It means that any polynomial time quantum computation can have its quantum part squeezed into logarithmic depth with supplementary polynomial time classical computation along the way. Clearly, the advantage is based on the quantum bipartite entanglement in TQC and multipartite entanglement on the whole state in 1WQC. This is then another motivation for us to study entanglement and to try to realize true entangled state in the experiment.

\section{Fault Tolerant; Classical Computing}
Ideally the attempt is that to implement algorithms, classical or quantum algorithms, with computers far from any unpleasant attack of noise. However, this is only possible in a theoretical study and in reality even in a very clean experiment in a laboratory, this idea does not come true. Always there are some noises, through environment or the interaction inside the system or what so ever, which make the system unstable. Therefore, in order to get benefits of theory, there might be some possibility to overcome the undesired effects of the noise. 

Old classical computers encountered problems due to the existence of noise. The computational elements, on those old days, had been unstable against noise. Then the efforts had been made on making the fault tolerant computations. Nowadays, the classical computation is approximately well stable, however the situation for quantum computation is still unstable.

The final aiming of the study of quantum information theory and quantum computation is of course the possibility of realizing a true quantum computer, which can realize the enhancement of quantum algorithms. Still nobody knows about the most appropriate hardware for constructing a quantum computer. Nevertheless, we are quite sure that any practical trial for making a quantum computer will be so fragile against the noise. Then it is a formidable task to study the noise and the ways to overcome the effects of the noise, error correction, and to make a fault tolerant computation.

In this regards, the experience through the classical computation can be useful, only to some extent. In this section, we study the fault tolerant quantum computation after a brief overview on the fault tolerant classical computation. We will get it that in case of fault tolerant quantum computation, entanglement is a prerequisite property which makes it possible to get the fault tolerant computation. {\it Then, we conclude that for a reliable quantum computation from the first point of making the quantum algorithms to the end idea of correctly realizing the quantum algorithms, entanglement should exist to assure the quantum nature of the processing.}

In the current technology of classical digital computers, because of that the existing silicon-based elements are remarkably reliable, fault tolerancy is not essential to the operation of modern digital computers. However, in the early days of the classical computers, when gates were realized using vacuum tubes, the situation was completely different. Indeed the gate failure was quite high and a great deal of thought went into how to design fault tolerant systems. Von Neumann, \cite{Y20}, suggested improving the reliability of a circuit with noisy gates by executing each gate many times, and using majority vote. Based on his conclusion, if the gate failures are statistically independent, and the probability of failure per gate is not so high, then any computation can be performed with a reasonable reliability.

In communication through noisy channel, Shannon, \cite{Y21}, proved that as far as the error probability is provided to be less than a certain threshold, so called the {\it capacity} of the noisy channel, a message can be sent with error probability  $\epsilon$ using only poly(log($1/\epsilon$)) noisy messages, using suitable codes. The threshold or the capacity is an intrinsic property of the channel and does not depend on the message size or on $\epsilon$.

A similar scenario goes to the classical computation, namely computational threshold theorem. A circuit with $N$ error free gates can be simulated with error probability less than $\epsilon$ using $O({\rm poly}(\log(N/\epsilon).N))$ gates, each gate fails with probability $p$, as long as $p<p_{th}$, where $p_{th}$ is independent of $N$ and $\epsilon$. 

If the circuit is made with only a single element, with error probability $p$, then the processing fails with probability $p$. However, we may use recursive codes. Encoding, meaning that each gate is copied to three gates, and majority vote is taking on the total gates. This is the level-1 circuit and fails with probability of $\binom{6}{2}=15$. Therefore the failure probability is $\sim 15p^2$ and if $15p^2<p$ then the process is successful. This gives $p<1/15$.

For higher level, say level-2, repeat the encoding once more. As a result, it turns out that output is incorrect if two or more blocks, each block containing three encoded gates, fail. Then, $p_{fail}\sim15(15p^2)^2=\frac{(15p)^4}{15}$. If we make notation as the number of faulty paths to be represented by $c\equiv\frac{1}{p_{th}}$, then
\begin{eqnarray}
{\rm level\, 1}: cp_{fail}=(cp)^2\\ \nonumber
{\rm level\, 2}: cp_{fail}=(cp)^4\\ \nonumber
{\rm level\, k}: cp_{fail}=(cp)^{2^k}
\end{eqnarray}
then 
\begin{equation}
\frac{p_{fail}}{p_{th}}=\left(\frac{p}{p_{th}}\right)^{2^k}.
\end{equation}

The code, which we used above is the recursion code. The computation is performed on encoded data, which is never decoded. By definition a procedure is fault tolerant if a single component causes at most one error in each encoded block of bits in the output.

Clearly, to make a circuit fault tolerant requires increase in the size of the circuit for each recursion. If $d$ stands for the size of the level-1 circuit, then the circuit size increases as $\sim d^k$ where $k$ is the recursion level. Therefore, there is an exponential increase in the recursion for fault tolerant. However, a double exponential decrease in error probability shows that the situation is not so bad. More explicitly, the size of a fault tolerant circuit can be computed as follows. To simulate an $N$ gate circuit, with error probability less than $\epsilon$, each gate should have error less than $\epsilon/N$. Thus, the fault tolerant procedure for each gate must have failure probability satisfying the following relation
\begin{equation}
\left( \frac{p}{p_{th}}\right)^{2{^k}}<\frac{\epsilon}{N p_{th}},
\end{equation}
the circuit size $Nd^k$ can be computed with solving the above inequalities
\begin{equation}
2^k\log\frac{p}{p_{th}}\sim \log\,\frac{\epsilon}{N p_{th}},
\end{equation}
then 
\begin{equation}
k\sim\log(\frac{\log\frac{Np_{th}}{\epsilon}}{\log\frac{p_{th}}{p}}).
\end{equation}

Then we have
\begin{equation}
Nd^k\sim N(\frac{\log\frac{Np_{th}}{\epsilon}}{\log\frac{p_{th}}{p}})^{\log d},
\end{equation}
which gives
\begin{equation}
Nd^k\sim {\rm poly} (\log \frac{N}{\epsilon}).N.
\end{equation}

This proves that the circuit size grows polynomially for making the fault tolerant circuit. Of course there have been several assumptions such as free of noise wires. The assumptions may make the threshold smaller than the estimated one.
\section{Fault Tolerant Quantum Computing}
The issue of fault tolerant quantum computing is more critical because that a quantum computer is more fragile against noises. Coherent states are inevitably required all during a quantum processing, \cite{Y22}, \cite{Y23}, \cite{Y24}, \cite{Y25}, \cite{Y26}, \cite{Y27}.
 \subsection{Quantum Error Correction Code}
 
Classical error correction codes might seem at a glance to be useful for introducing the quantum counterpart of error correction codes. However, there are several quantum essential properties preventing this idea.
 \begin{description}
 \item[.]
 Quantum states cannot be cloned. Therefore, encoding, meaning that making several copies of a particular state is not directly possible.
 \item[.]
 Errors are continuous. In addition to the bit errors there are also phase errors involved in a quantum system.
 \item[.]
 Quantum states collapse when are measured. Therefore, measurement can not give a proper way to repair the quantum state from the noises as the state is already destroyed. 
 \end{description}
 In 1995, it turned out that quantum error correction also is possible, \cite{Y24}. The objection of the quantum measurement destroying the state is resolved by an idea similar to what has been used for introducing the quantum algorithms. There is no need to measure the quantum state as far as we only need to know about the effect of the environment. Then, a measurement on the effect of the environment is sufficient and in this process the part of quantum state would be preserved. This is shown by the operator measurement. Actually, operator measurement is widely being used for defining measures of entanglement in addition to its usage for quantum algorithms. 
 
\begin{figure}
\begin{center}
\scalebox{0.777}
{\includegraphics[0cm,13cm][15cm,18cm]{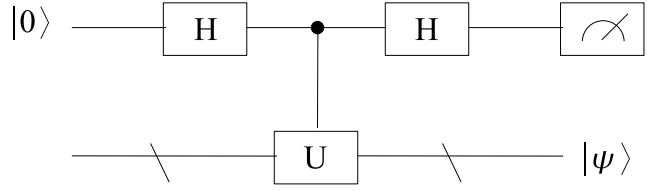}}
\end{center}
\caption{\label{100} Quantum circuit for operator measurement.} 
\end{figure}

 Assume a system, which is supposed to be measured. In a straight manner, one may measure the system directly, while in a rather elaborated scheme the system can be measured if we make the system close to the other system, any ancilla, and make them interact to each other and finally measure the state of the ancilla system. This process is shown in Figure \ref{100}.
 
 Given a unitary transformation $U$, with eigenvalues $\pm 1$ and eigenvectors $|u^{\pm}\rangle$, the operator measurement with ancilla state initially $|0\rangle$, as shown in Figure \ref{100}, can be described as follows
 \begin{eqnarray}
 & & (|0\rangle)(c_0|u_+\rangle+c_1|u_-\rangle) \overset{H}{\to}\\ \nonumber
 & & \left(\frac{|0\rangle+|1\rangle}{\sqrt 2}\right)(c_0|u_+\rangle+c_1|u_-\rangle) \overset{H}\to \\ \nonumber
 & & \frac{|0\rangle(c_0|u_+\rangle+c_1|u_-\rangle)}{\sqrt 2}+\frac{|1\rangle(c_0|u_+\rangle-c_1|u_-\rangle)}{\sqrt 2}\overset{H}{\to} \\ \nonumber
 & & \frac{|0\rangle+|1\rangle}{2}(c_0|u_+\rangle+c_1|u_-\rangle)+\frac{|0\rangle-|1\rangle}{2}(c_0|u_+\rangle-c_1|u_-\rangle)\equiv \\ \nonumber
 & & c_0|0\rangle|u_+\rangle+c_1|1\rangle|u_-\rangle.
 \end{eqnarray}
 Therefore, if the measurement result of the ancilla measurement is zero the $|\psi\rangle$ is at $|u_+\rangle$ and otherwise for the measurement outcome being one the state of $|\psi\rangle$ is $|u_-\rangle$. By the means of operator measurement we have got information on the desired state without spoiling the state itself.
 
 The objections of continuous errors and no-cloning, both can be circumvented by the concept of {\it entanglement}. Entangled states make the errors orthogonal and then distinguishable. Also, entangled states play the same role as the redundant copies in classical error corrections. Here, we would like to have an overview on quantum error correction codes and fault tolerant quantum computing in order to see the essential role of entanglement for this concept.
 
 A quantum code $C[[N,K]]$ is a $k$ qubit superposition of an $n$-qubit Hilbert space. For example $C[[3,1]]$ is a $3-$qubit quantum code, which is explained as follows. 
 
 \subsubsection{3-qubit Bit-flip Code}
 
 By definition, bit-flip error is the quantum operation defined as follows, \cite{Y28}
 \begin{equation}
 \varepsilon (\rho)=(1-p)\rho+pX\rho X,
 \end{equation}
 where $p$ is a probability for a bit-flip.
 
 In a $3$-qubit bit-flip error correction code, the state is encoded to $3$ qubits. Logical $0$ is $|0_L\rangle=|000\rangle$ and logical $1$ is $|1_L\rangle=|111\rangle$. Through out the circuit with bit-flip error, the state is changed as follows
\begin{equation}
a|0_L\rangle+b|1_L\rangle\overset{\epsilon}{\to}
\begin{cases}
\text{final\, state}       &\text{probability},\\
a|000\rangle+b|111\rangle  &\text{$(1-p)^3$},\\
a|001\rangle+b|110\rangle  &\text{$p(1-p)^2$},\\
a|010\rangle+b|101\rangle  &\text{$p(1-p)^2$},\\
a|100\rangle+b|011\rangle  &\text{$p(1-p)^2$},\\
a|011\rangle+b|100\rangle  &\text{$p^2(1-p)$},\\
a|101\rangle+b|010\rangle  &\text{$p^2(1-p)$},\\
a|110\rangle+b|001\rangle  &\text{$p^2(1-p)$},\\
a|111\rangle+b|000\rangle  &\text{$p^3$}.\\
\end{cases}
\end{equation}

Using the operator measurement as introduced above with the unitary transformations $U_1=IZZ$ and $U_2=ZZI$, the bit-flip error that is occurred on each element is given and is called {\it error syndrome}. For each error syndrome, then proper {\it recovery operation} can be applied in order to recover the state from the effect of the noise. 
\begin{equation}
a|0_L\rangle+b|1_L\rangle\overset{\epsilon}{\to}
\begin{cases}
State                & \text{$U_1\hspace{1.5cm}U_2\hspace{1.5cm}$Recovery Operation},\\
a|000\rangle+b|111\rangle  & \text{$0 \hspace{2cm} 0 \hspace{2cm}I$  },\\
a|001\rangle+b|110\rangle  & \text{$1 \hspace{2cm} 0 \hspace{2cm}X_1$},\\
a|010\rangle+b|101\rangle  & \text{$1 \hspace{2cm} 1 \hspace{2cm}X_2$},\\
a|100\rangle+b|011\rangle  & \text{$0 \hspace{2cm} 1 \hspace{2cm}X_3$},\\
a|011\rangle+b|100\rangle  & \text{$0 \hspace{2cm} 1 \hspace{2cm}X_3$},\\
a|101\rangle+b|010\rangle  & \text{$1 \hspace{2cm} 1 \hspace{2cm}X_2$},\\
a|110\rangle+b|001\rangle  & \text{$1 \hspace{2cm} 0 \hspace{2cm}X_1$},\\
a|111\rangle+b|000\rangle  & \text{$0 \hspace{2cm} 0 \hspace{2cm}I$  }.\\
\end{cases}
\end{equation}

 As far as there is no information on $a$ or $b$, then the measurement results leave the state unchanged. 
 
 Fidelity is a measure of distance between quantum states. The fidelity of states $\rho$ and $\sigma$ is defined to be
 \begin{equation}
 F(\rho,\sigma)\equiv {\rm tr} \sqrt{\rho^{1/2}\sigma\rho^{1/2}}.
 \end{equation}
 
 If one of the state is a pure state then the general definition is changed into
 \begin{equation}
 F(|\psi\rangle,\rho)=\sqrt{\langle \psi|\rho|\psi\rangle}.
 \end{equation}
 
\begin{figure}
\begin{center}
\scalebox{0.777}
{\includegraphics[3cm,20cm][14cm,28cm]{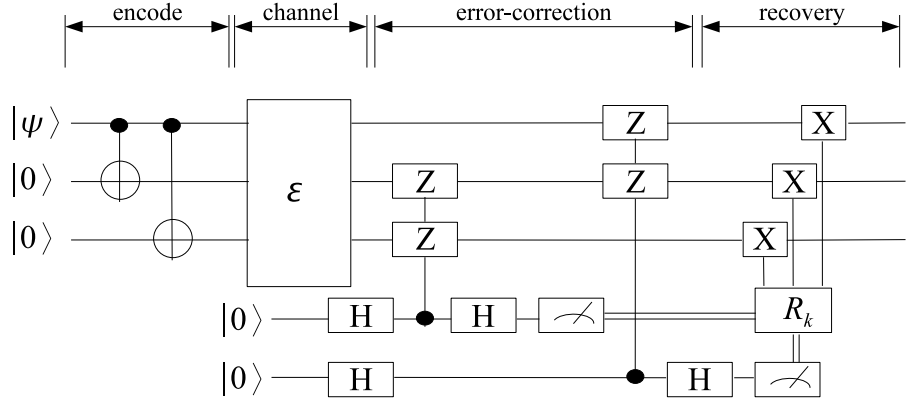}}
\end{center}
\caption{\label{Fig7} Quantum circuit for $3$-qubit bit-flip code.} 
\end{figure}

 Clearly, the final outcome $\rho$ is not perfect, because there is some probability of occurring two or more bit-flip error, that this code cannot correct them. Then, qualitatively, we measure the distance of the perfect corrected state and the state corrected with $3$-qubit bit-flip code.
 \begin{eqnarray}
  F(|\psi\rangle,\rho)&=&\sqrt{1-3p^2+2p^3}\\ \nonumber
                      &=&1-O(p^2).
  \end{eqnarray}
  Then the fidelity, even though is not perfect, still is better that $1-O(p)$, which is the fidelity for not using the code. The quantum circuit, which can realize the $3$-qubit bit-flip error code is shown in Figure \ref{Fig7}.
It is also shown that the above bit-flip error correction code corrects small rotation angle error.
 
 \subsubsection{$3$-qubit Phase-flip Code}
 
 Let the error for phase-flip channel be described as follows
 \begin{equation}
 \varepsilon_{pf}(\rho)=p\rho+(1-p)Z\rho Z.
 \end{equation}
 
 Correcting a phase flip error can be made in a same manner as the one for bit flip error. Recall that $HXH=Z$ and $HZH=X$, so $H\varepsilon_{pf}(H\rho H)H=\varepsilon_{bf}(\rho)$. The encoding of the states is as follows
 \begin{eqnarray}
 |0_L\rangle=|+++\rangle,\\ \nonumber
 |1_L\rangle=|---\rangle,
 \end{eqnarray}
 where $|\pm \rangle\equiv (|0\rangle\pm |1\rangle)/\sqrt2$. Then the syndrome measurement is performed by $XXI$ and $IXX$.

 The reason for the importance of the phase flip error is that the arbitrary qubit errors are proved to be combinations of bit-flip $(X)$, phase-flip $(Z)$ and bit-phase flip $(XZ)$ errors. The quantum code, which corrects the arbitrary single qubit error is Shor's $9$-qubit code.

\subsubsection{$9$-qubit Shor's Code}

 This code corrects any single qubit error. The encoding is as follows
 \begin{eqnarray}
 |0_L\rangle\equiv\frac{(|000\rangle+|111\rangle)^{\otimes 3}}{\sqrt 8},\\ \nonumber
 |1_L\rangle\equiv\frac{(|000\rangle-|111\rangle)^{\otimes 3}}{\sqrt 8}.
 \end{eqnarray}
 The syndrome measurement operators for bit-flip and phase-flip errors are
 \begin{eqnarray}
 Z_1Z_2,\, Z_2Z_3,\, Z_4Z_5,\, Z_5Z_6,\, Z_7Z_8,\, Z_8Z_9\\ \nonumber
 X_1X_2X_3X_4X_5X_6,\, X_4X_5X_6X_7X_8X_9
 \end{eqnarray}
It is simple, by direct examination, to check that any single bit flip or phase flip or bit-phase flip error gives a unique syndrome measurement result, then can be corrected. Let us give an explicit example. Suppose an error on fifth qubit. The initial state changes into the following state after the error $Y_5$
\begin{eqnarray}
& & a(|000\rangle+|111\rangle)(|000\rangle+|111\rangle)(|000\rangle+|111\rangle)+\\ \nonumber
& & b(|000\rangle-|111\rangle)(|000\rangle-|111\rangle)(|000\rangle-|111\rangle)\to\\ \nonumber
& & a(|000\rangle+|111\rangle)(|010\rangle-|101\rangle)(|000\rangle+|111\rangle)+\\ \nonumber
& & b(|000\rangle-|111\rangle)(|010\rangle+|101\rangle)(|000\rangle-|111\rangle)
\end{eqnarray}
Then the syndrome measurement with the operators defined above, gives result as follows
\begin{center}
\begin{tabular}{c|c}
Operator  & Syndrome\\\hline
 $Z_1Z_2$ & 0 \\
 $Z_2Z_3$ & 0 \\
 $Z_4Z_5$ & 1 \\
 $Z_5Z_6$ & 1 \\
 $Z_7Z_8$ & 0 \\
 $Z_8Z_9$ & 0 \\
 $X_1X_2X_3X_4X_5X_6$ & 1\\
 $X_4X_5X_6X_7X_8X_9$ & 1
\end{tabular}
\end{center}
Or if a phase flip error occurs on the second qubit, $Z_2$ then
\begin{center}
\begin{tabular}{c|c}
Operator  & Syndrome\\\hline
 $Z_1Z_2$ & 0 \\
 $Z_2Z_3$ & 0 \\
 $Z_4Z_5$ & 0 \\
 $Z_5Z_6$ & 0 \\
 $Z_7Z_8$ & 0 \\
 $Z_8Z_9$ & 0 \\
 $X_1X_2X_3X_4X_5X_6$ & 1\\
 $X_4X_5X_6X_7X_8X_9$ & 0
\end{tabular}
\end{center}
It is easy to check that the above syndrome is the same as if $Z_1$ or $Z_3$ happens. Then, this code is not an ideal code.

We realize an easier way for finding the error syndrome. Use the commutator relation as $[Z_i,Z_j]=0$ and $[X_i,X_j]=0$. Then, any $Z$ or $X$ error has a zero value through measurement of the syndrome operators which include the operators that commute with error operators. On the other hand, $[X_i,Z_j]$ is not zero if $i\neq j$. Then, the corresponding syndrome measurement would give one. This can be checked through the above examples, too.

After finding the error syndrome, the proper operator is applied on the state in order to remove the effect of the noise and recover the state.

\subsection{Quantum Error Correction Criteria}

Suppose a quantum channel as $\varepsilon=\sum_k E_k \rho E_k^\dagger$. Then, let $C$ be a quantum code defined by the orthonormal states $\{ |\psi_l\rangle\}$. There exists a quantum error correcting $\varepsilon$ on $C$, {\it iff}
\begin{description}
\item[.]
Orthogonality is satisfied: $\langle \psi_l|E_j^\dagger E_k |\psi_l\rangle=0,\,\forall j\neq k,\,\forall l$. This means that error should take code words to different points. 
\item[.]
Non-deformation: $\langle \psi_l|E_k^\dagger E_k|\psi_l\rangle=d_k \forall l$. This means that error should not be able to distinguish between the code words.
\end{description}
We escape the proof here. Instead we explain the stabilizer and quantum fault tolerant, which we can see clearly the role of entanglement. We will see that these conditions will be met again by stabilizers.
\subsection{Stabilizer and Quantum Fault Tolerant}
Up to now, we have introduced classical fault tolerant and quantum error correction codes, which define the basis for being able to build a quantum fault tolerant containing system of quantum gates. However, we have not yet seen how to perform operations on encoded data and how to control the properties of quantum error correction inside a quantum circuit. These will be met by the method of stabilizer.

Stabilizer is a mathematical method, however it is the Heisenberg representation of quantum computation, instead of the Schrodinger representation, which we are used to it and it works with wave functions. We first define the mathematics of stabilizer and then will use it to define quantum codes using stabilizers, \cite{Y22}, \cite{Y28}.

Suppose a Pauli group. The Pauli group $G_n$ on $n$ qubits is the group of all $n$-fold tensor products of $\{ X, Y, Z\}$ and $\{ \pm 1, \pm i\}$. For example
\begin{equation}
G_1\equiv \{ \pm I, \pm iI, \pm X, \pm iX, \pm Y, \pm iY, \pm Z, \pm iZ\}.
\end{equation}

By definition, a stabilizer for a vector space $V_s$ with basis as $\{|\psi_1\rangle\}$ is a set 
\begin{equation}
S=\{g\in G|\hspace{0.2cm} \,g|\psi\rangle=|\psi\rangle, \forall |\psi\rangle\in V_s\}.
\end{equation}

By convention $-I\notin S$. Note that $S$ is Abelian, meaning that all the elements of this group commute to each other because that they have the same eigenstate.

For example, for the vector space as follows
\begin{equation}
V_s=\frac{|00\rangle+|11\rangle}{\sqrt 2},
\end{equation}
the stabilizer is 
\begin{equation}
S=\langle XX, ZZ \rangle=\{XX,II,ZZ,YY\}.
\end{equation}

Another example is that the vector space for the following stabilizer
\begin{equation}
S=\{X, Z \},
\end{equation}
is the empty space, clearly because that the defined stabilizer elements do not commute to each other.

For the vector space
\begin{equation}
V_s=\{ |000\rangle, |111\rangle \},
\end{equation}
the stabilizer is
\begin{equation}
S=\langle IZZ, ZZI \rangle.
\end{equation}

As we had for the syndrome measurement.

Another more interesting example, might be the one corresponding to the stabilizer for the vector space
\begin{equation}
V_s=\{ (000+111)^{\otimes 3}, (000-111)^{\otimes 3}\}.
\end{equation}

The stabilizer for this case is 
\begin{eqnarray}
S&=&\{Z_1Z_2, Z_2Z_3, Z_4Z_5, Z_5Z_6, Z_7Z_8, Z_8Z_9,\\ \nonumber
 & &\, X_1X_2X_3X_4X_5X_6, X_4X_5X_6X_7X_8X_9 \}
          \end{eqnarray}
This is the familiar Shor code for error correction. Then it turns out that these stabilizers are actually codes, that have been introduced, part of them, as error correction codes.

\subsubsection{Stabilizer Code} 

A $[[n, k]]$ stabilizer code $C(S)$ is the vector space stabilized by $S\in G_n$, where $S$ is the set of stabilizers $S=\langle g_1,..., g_{n-k}\rangle$. Here, $n$ is the number of physical qubits and $k$ is the number of logical qubits. The minimal generator for the stabilizer is $(n-k)$. For example Shor's code $[[9,1]]$ has $8$ generators.

Given a certain set of errors $\{ E_a \} \in G$, if there exists an element in the stabilizer, $\exists g \in S$, such that $Eg=-gE$, for all $E=E_a^\dagger E_b$, then the set of errors $\{ E_a \}$, can be corrected by the stabilizer code, $C(s)$. 

The proof comes in the following for all $|\psi \rangle \in C(s)$, but we emphasize that the explanation is also the same as the previously shown quantum error correction criteria.
\begin{eqnarray}
\langle \psi|E|\psi\rangle&=&\langle\psi|Eg|\psi\rangle\\ \nonumber
                          &=&-\langle\psi|gE|\psi\rangle\\ \nonumber
                          &=&-\langle\psi|E|\psi\rangle\\ \nonumber
                          &=&0
\end{eqnarray}
This is equivalent to the quantum error correction criterion as explained by, $\langle\psi_l|E^\dagger_iE_j|\psi_l\rangle=0, \forall l, i\neq j$. Hence, the above explanation says that the error can be corrected, but what can be the syndrome of errors?

For $S=\langle g_1g_2...g_k\rangle$, and error $E$, the error syndrome is
\begin{equation}
\begin{cases}
0  & \text{if $[E,g_l]=0\,|\,l=1 \to k$},\\
1  & \text{otherwise}.
\end{cases}
\end{equation}

For example, consider the bit error, which has been explained before. The stabilizer $S=\langle IZZ, ZZI \rangle$. For error $E=IXI$, the error syndrome is $\{1, 1\}$, showing that an error has been occurred on the second qubit. However, the same stabilizer elements for an error $E=IIZ$, which is a phase-flip error, gives the error syndrome as $\{0, 0\}$. Then, phase flip error is not an error or in other words is not a detectable error with this stabilizer. This statement is coincident with our previous outcome for the bit flip error correction code. 

Lets assume the stabilizer defined as follows
\begin{equation}
S=\{ XZZXI, IXZZX, XIXZZ, ZXIXZ \}.
\end{equation}

Clearly all the elements commute to each other as this should be an Abelian group to be a stabilizer. This is the $5$-qubit quantum error correction. Suppose a bit flip error on the third qubit, $E=X_3$. The syndrome would be $\{ 1, 1, 0, 0 \}$. The error syndrome for a phase flip error e.g. $E=Z_4$ would be $\{ 0, 1, 0, 1\}$. This code corrects any $\{ X_i, Z_i, Y_i\}$ errors. 

$5$-qubit error correction code is called {\it perfect code}. The number of possible error for five qubits are $\binom{5}{1}\times 3 +1 = 16$, addition to one is for no error case. The number of syndrome strings is $2^4=16$. Meaning that the number of syndrome strings is equal to the number of possible errors. However, this code has been known to have some bad property. It does not give the fault tolerant for some important quantum gates namely, Hadamard gate, as will be explained later.

The other example which can be explained nicely with stabilizer is the Steane $7$-qubit code. The stabilizer elements $S$ is defined as follows
\begin{eqnarray}
g_1&=&IIIXXXX \\ \nonumber
g_2&=&IXXIIXX \\ \nonumber
g_3&=&XIXIXIX \\ \nonumber
g_4&=&IIIZZZZ \\ \nonumber
g_5&=&IZZIIZZ \\ \nonumber
g_6&=&ZIZIZIZ 
\end{eqnarray}
This stabilizer is a representation  for Steane code, $[[7, 1]]$, and corrects any single error.

\subsubsection{Encoded Operation; Normalizer}

The complete fault tolerant quantum computation requires the corresponding fault tolerant operation, too. In this sense, unitary operation on the stabilizer state is called {\it Normalizer}. Suppose, a unitary operation $U$ applied on the state $|\psi\rangle$. Then
\begin{equation}
|\psi\rangle\to U|\psi\rangle.
\end{equation}

If $|\psi\rangle$ is a stabilizer state, $g$, then in the Heisenberg picture it is equivalently written as
\begin{equation}
g\to UgU^\dagger.
\end{equation}

The proof is simple as follows
\begin{eqnarray}
U|\psi\rangle&=&Ug|\psi\rangle \\ \nonumber
             &=&[UgU^\dagger]U|\psi\rangle
             \end{eqnarray}
where $[UgU^\dagger]$ is the new stabilizer of the transformed state in the Heisenberg picture. We want to use this picture in order to understand how to perform operations on the encoded state. Instead of following the state around, we follow the stabilizer around to have a fault tolerant quantum computation with fault tolerant operations on encoded states.

For example, suppose the state $|\psi\rangle=a|000\rangle+|111\rangle$, which is the encoded state. For a fault tolerant quantum computation, operations should be applied on this encoded state. For example, for an operation $X$, then the encoded operation would be the logical $X=\bar X$, which changes the state into $|\psi^\prime\rangle=a|111\rangle+b|000\rangle$. For important quantum gates, the operations are as follows
\begin{eqnarray}
\bar X &:& a|000\rangle+|111\rangle\to a|111\rangle+b|000\rangle \\ \nonumber
\bar Z &:& a|000\rangle+|111\rangle\to a|000\rangle-b|111\rangle \\ \nonumber
\bar R_Y(\theta) &:& |000\rangle\to \cos \frac{\theta}{2}|000\rangle+\sin \frac{\theta}{2}|111\rangle \\ \nonumber
\bar H &:& |000\rangle\to \frac{|000\rangle+|111\rangle}{\sqrt 2}
\end{eqnarray}
But the question is; Are these fault tolerant operations possible? The requirement for all of these to be possible is that, the transformed state after the operation still should be an element in the stabilizer. This is explained as follows
\begin{equation}
\forall g\in S, UgU^\dagger \in S.
\end{equation}

Meaning that for any unitary transformation $U$ the element of the stabilizer should stay in the same stabilizer. Satisfying this requirement limits the set of operations to the normalizer, $N(S)$, defined as follows
\begin{equation}
 N(S)=\{g\in G|\hspace{0.5cm} ghg^\dagger \in S,\, \forall h\in S \}.
 \end{equation}
 
 The state may changes into other elements of the stabilizer but would not leave it. 
 
 Now, recall that $gh=\pm hg$, so $ghg^\dagger=\pm hgg^\dagger=\pm h$, but $-h\notin S$, because by definition $-I \notin S$, thus $gh=hg$. Therefore, for any normalizer $g$, $[g,h]=0$, for all the stabilizer elements $h$. As a result, the important fact is that the normalizer of the Pauli group is generated by the following elements
 \begin{equation}
 N(G)=\langle H, S, CNOT \rangle,
 \end{equation}
 where $S=\sqrt Z$. This group is called Clifford group and includes all the operations that are involved in the stabilizer. However, the group does not include some important gate such as Toffoli gate or any arbitrary operation. Therefore, the group is not universal for classical computation, because of not including Toffoli gate, and then for quantum computation, as it is the subset of the classical computation.
 
 For example, consider the $5$-qubit code, with the stabilizer elements as explained before. It is possible to check that for the normalizer elements $N(S)$, $\bar X=XXXXX$ and $\bar Z=ZZZZZ$. However, it is not possible to have the same for Hadamard, namely $\bar H\neq HHHHH$. This is the bad property for the $5$-qubit code, as mentioned before.
 
 Now, recall the Steane $7$-qubit code. It is possible to check that in addition to the logical $X$ and $Z$, Hadamard gate also can be written as $\bar H =HHHHHHH$. So, all the Clifford group can be normalized with Steane $7$-qubit code. This is the important property for this code because the operations are important for fault tolerant quantum computation.
 
\subsubsection{Complete Fault Tolerant Quantum Computation; Entanglement}

 In order to have a complete fault tolerant quantum computation, there are two main requirements which should be satisfied. Firstly, all the operations should be performed on the encoded data. Meaning that there should be a way to construct the fault tolerant operation for any quantum gate. Secondly, the error propagation should be controlled in a proper manner.
 
 In order to satisfy the required conditions and to get the correct fault tolerant quantum computation, we see that entanglement is inevitably required. Suppose the first requirement, namely the universal normalizers. Universal set of gates is possible if we could make the fault tolerant Toffoli gate. As this is a $3$-qubit gate, so it is difficult to work with it. Lets make it rather bit simple, by introducing the $T$ gate as follows
 \begin{eqnarray}
 T=\pi/8&=&\left(\begin{array}{@{\,}cc@{\,}}1&0\\0&\sqrt{i}\\ \end{array}\right)\\ \nonumber
        &\sim&\sqrt S \\ \nonumber
        &\sim&\sqrt[4] Z
\end{eqnarray}        
 Therefore, it is possible to construct the Toffoli gate from $T$ gate, if some non-unitary steps, such as measurement, are involved.
 
\begin{figure}
\begin{center}
\scalebox{0.8}
{\includegraphics[3cm,22cm][15cm,26.5cm]{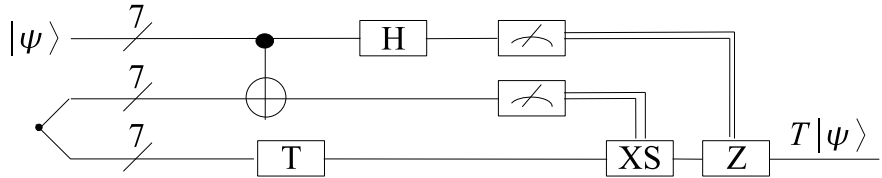}}
\end{center}
\caption{\label{Fig9} A quantum circuit for making a fault tolerant $T$ gate. See the text for the connection between $T$ gate and Toffoli gate. The first qubits are each encoded to seven qubits. The quantum operations CNOT and Hadamard are in the Clifford group and then are fault tolerant. The applied $T$ gate on the third encoded qubit is the logical $T$ gate, not fault tolerant, but has been checked to be almost fine with some reasonable error. The non-unitary parts of the circuit, namely measurements, are inevitably required. The circuit is so close to the quantum circuit for teleportation, but one of the two conditional $Z$ operation has been changed by $X$ and $S$ operations, both being in the Clifford group and then being fault tolerant. The outcome of the circuit is the fault tolerant $T$ gate. The important point, which we would like to emphasize here is that the initial second and third qubits are prepared as to be entangled states and clearly the success of the process is based on the involved entangled states.} 
\end{figure}

 The important fact, which has been shown by Shor, is that $T$ gate can be implemented using Clifford gates, measurement in the standard computational basis, $Z$, and prepared entanglement. Then, {\it entanglement makes the fault tolerant computation possible}. The quantum circuit for making a fault tolerant $T$ gate is shown in Figure \ref{Fig9}. It is clear from the Figure that, the process to make a fault tolerant $T$ gate is so close to teleportation, but slightly different. It is {\it computation by teleportation}. In this sense, we see very nice unification of the concepts which have been explained up to here, and the unification point is {\it entanglement}.
 
The other point, which has been mentioned to make a system fault tolerant is that to avoid the error propagation. The problem is serious as to how to avoid propagation of errors all during the computation and more seriously in the operational measurement or syndrome measurement. This is possible to solve the problem by making the bit-wise gates or introducing some ancilla in cat state. However, the more extended explanations may be required which would not be presented here but can be found in related papers.

 \section{Conclusion}
 We, on purpose, made this chapter in two parts. First part includes classical complexity versus quantum complexity. Some quantum algorithms are revisited. Quantum properties such as superposition and entanglement turn out to be prerequisite. The attempts for introducing quantum algorithms have been focused on attaining an exponential enhancement over classical scheme. In other words, We want a quantum computer to solve polynomially those problems which do not have any up to date known polynomial solution with classical computers. However, polynomial complexity of classical problems still might be more complicated. In this regards, measurement based model has been explained followed by parallelisibility. Exactly speaking the scheme is much closely related to entanglement. Therefore, entanglement turns out to be important in this regard, too.
 
 In the second part, we studied fault tolerant computation. Quantum system is much fragile and if it is going to be used for any computation, stability of the system should be considered  with high priority. This is a very nice result that quantum fault tolerant computation is possible even with a very strict threshold. The important point, as to our interest, is that in order to introduce a fault tolerant quantum computation, entanglement should be inevitably involved. Therefore, we conclude from this chapter, as a tour on quantum computing, that entanglement is required for making a reliable quantum computer, which can implement real quantum operations in order to truly give the desired advantages, far beyond any classical processing.

 \vspace{5cm}
\chapter{Quantum Entanglement; Characterization and Measurement}
From the time around $1900$, when quantum physics has been introduced, superposition of states has been known as the main quantum property. Namely, for a definite physical property, there are probabilities for the state to be at any possible physical superposition of the states and the total state for the property is defined by a superposition of all the possible states. This fundamental postulate of quantum physics is originally out of the classical concepts, in a sense that in classical physics the state of a physical property is assumed to be well defined even prior to the measurement. However the interaction or the measurement is the process which makes a quantum system to collapse to one of the state even though before the measurement, the state is ought to be in a superposition of all the states each might be given with a particular probability during the measurement.

This is then a kind of paradox for quantum physics. How would it be possible to prove the concept of superposition, if we accept that a quantum state just collapses to one of its eigenstates with some particular probability and we are not able to extract all the possibilities through measurement of the quantum state. One may claim that a quantum state originally had been in the state which is detected through the measurement and there would not be any practical proof. Then, there would not be possible to verify the concept of superposition.

In a sense, entanglement is a keynote in this regard. The paradoxical behaviour of some particular quantum states, namely entangled state, has been primarily pointed out by Einstein, Podolskey and Rosen \cite{D1}. Later, Bell \cite{D2} intelligently gave an approach to detect how entanglement can be verified through measuring a quantum system with a physical apparatus. Then, entanglement actually could make it possible to check the quantum fundamental concept, superposition.

We try to get close to the concept of entanglement as it is required for quantum information processing and quantum computation. There are some very important questions in this regard. Namely, what is entanglement? What is correlation? Are there some check points to detect or even to quantify entanglement?

Entanglement is known as a resource. This is important to know the reason on why we give such an outstanding role for entanglement. This can be explained briefly by giving an example. One very important topic in quantum information processing is quantum teleportation \cite{Y19}. There are so many things that one can do with teleportation. Apart from its application for sending quantum information, teleportation has been known to be generalized to some other concepts. Namely, quantum cryptography \cite{D3}; sending information in a safe manner, quantum computation and quantum error correction, which is discussed in chapter $1$. These facts show that with having only teleportation we can do lots of things and this is in a sense appreciating the concept of entanglement. Since, teleportation is a pure quantum processing and up to now there is no classical counterpart for it. {\it Teleportation is based on entanglement}. Then, having entanglement, one can make teleportation with which she can do lots of things and in this regard entanglement is a kind of physical resource.

Hence, entanglement is introduced as an important property among quantum states. In fact, we need to go further now and to find it out a way to work with this quantum property. How to detect and how to measure it quantitatively, are important topics, which are open questions still for general cases. However, for special classes of states, there are some very excellent papers and remarkable results, which we review here before introducing our new measure of entanglement. We will show a new entanglement measure, that is a proper case specially for the physical system that has been considered for our experiments on realizing quantum entanglement. In other words, we keep reviewing the known measures of entanglement and make them step toward introducing the new measure. As, we claim on the importance of entanglement, our attempts would be on realization of a true entanglement. Then, the realized entanglement in the laboratory would be measured with our homemade entanglement measure. The reason is that it is introduced in a way that it is specialized for our experiment then it is considerably easier to be measured.

We first start with general explanations in order to get more familiar with entanglement. The starting point would be entanglement in bipartite system. Later we will make the system more complicated but we will see that there would not be any complete or perfect explanation for a general case.

\section{Entanglement in Bipartite System}
Suppose that we have two identifiable systems, A and B, with finite dimension Hilbert spaces, $\mathcal{H}_A=\phi^d$ and $\mathcal{H}_B=\phi^d$. In particular, $\mathcal{H}_A=\phi^2$ and $\mathcal{H}_B=\phi^2$, for bipartite of two qubits.

Here, we would like to study the entanglement for a general bipartite system. Firstly, we consider the problem of characterization or detection of entanglement. Later, we will go around the problem on how to introduce a valid measure of entanglement and will give some examples. 
\subsection{Characterization of Entanglement; Pure State}
The given bipartite state may be a pure or a mixed state. For each case, separately we study the concept of entanglement. 

For a bipartite pure state, by definition, a state is separable if and only if it is a product state. A pure state, $|\psi\rangle$, is a product state if it can be written as 
\begin{equation}
\label{product}
|\psi \rangle =|\psi\rangle _A \otimes |\psi \rangle _B,
\end{equation}
where $|\psi\rangle_A \in \mathcal{H}_A$ and $|\psi\rangle_B \in \mathcal{H}_B$. If the state is not a product state, meaning that if it is not a separable state, then the state is an entangled state. For example
\begin{equation}
\label{phi+}
|\phi^+\rangle = \frac{1}{\sqrt 2}(|00\rangle+|11\rangle),
\end{equation}
is an entangled state. For this particular example, it is trivial to check the impossibility of writing the state $|\phi^+\rangle$ as a product of the general states $(\alpha |0\rangle +\beta |1\rangle)$ and $(\alpha^\prime |0\rangle +\beta^\prime |1\rangle)$, because it requires the following equations to be satisfied, simultaneously,
\begin{eqnarray}
\frac{1}{\sqrt 2}(|00\rangle+|11\rangle)=(a|0\rangle_A+\beta|1\rangle_A)\otimes(\alpha^\prime|0\rangle_B+\beta^\prime|1\rangle_B) \\ \nonumber
\alpha \alpha^\prime =\frac{1}{\sqrt 2}\hspace{2cm} \alpha \beta^\prime =0\\ \nonumber
\beta \alpha^\prime=0\hspace{1cm} \beta \beta^\prime =\frac{1}{\sqrt 2}.
\end{eqnarray}
Then, the state $|\phi^+\rangle$ is called an entangled state.

However, in general case, this is not as easy as the above example. In practice, we need to find the right basis vectors to be able to write the state as a product state, and only if this is shown to be impossible to find such a product elements for a given state then the state would be proved to be entangled. Schmidt decomposition is useful in this regard. In other words, the number of terms in a Schmidt decomposition of a particular state, called Schmidt rank or Schmidt number, would be useful.

\subsubsection{Schmidt decomposition}
As far as one is dealing with pure states, Schmidt decomposition is a powerful way to get the information on the entanglement and also to calculate the amount of entanglement. Schmidt decomposition is a decomposition into the biorthogonal basis, and gives the smallest possible number of terms for a product basis. Generally, any vector $|\psi \rangle$ in $\mathcal{H}_A \otimes \mathcal{H}_B$ can be expressed as follows
\begin{eqnarray}
\label{general}
|\psi \rangle&=&\sum_{i,j}C_{i,j}|a_i,b_j \rangle \\ \nonumber
             &=&\sum_k \lambda_k |a^{\prime}_k, b^{\prime}_k \rangle,
\end{eqnarray}
which is the Schmidt decomposition. The coefficients $C_{i,j}$ is the elements in a $d_A\times d_B$ matrix $C$. If $C$ is a Hermitian matrix, then always it is possible, by using the eigenvectors and eigenvalues, to find the diagonal matrix $D$ such that
\begin{equation}
C=UDU^{\dagger},
\end{equation}
and to make the Schmidt decomposition. On the other hand, if $C$ are not a Hermitian matrix still it is possible to calculate the Schmidt decomposition by the singular value method. Generally, it is possible to write 
\begin{equation}
C=UDV,
\end{equation}
$U$ is a $d_A \times d_A$ unitary matrix and $V$ is a $d_B\times d_B$ unitary matrix. Therefore, the state $|\psi\rangle$ can be written as
\begin{eqnarray}
|\psi\rangle &=& \sum_{i=1}^{d_A}\sum_{j=1}^{d_B}\sum_{k=1}^{{\rm min}(d_A, d_B)}u_{ik}d_{kk}v_{kj}|a_i\rangle\otimes|b_j\rangle\\ \nonumber
             &=& \sum_{k=1}^{{\rm min}(d_A, d_B)}d_{kk}\left( \sum_{i=1}^{d_A}u_{ik}|a_i\rangle\right)\otimes\left(  \sum_{j=1}^{d_B}v_{kj}|b_j\rangle\right)\\ \nonumber
             &=& \sum_{k=1}^{{\rm min}(d_A, d_B)} \lambda_k |a_k^{\prime}\rangle\otimes|b_k^{\prime}\rangle,
\end{eqnarray}
where $\lambda_k=d_{kk}$ and we have used the new basis vectors
\begin{equation}
|a_k^{\prime}\rangle\equiv \left(\sum_{i=0}^{d_A}u_{ik}|a_i\rangle\right),\hspace{1cm}|b_k^{\prime}\rangle\equiv \left(\sum_{j=0}^{d_B}v_{kj}|b_j\rangle\right).
\end{equation}

The coefficients $\lambda_k$ are called the Schmidt coefficients.

The Schmidt number is at most the minimum of $d_A$ and $d_B$, where $d_A$ and $d_B$ are the dimensions of subsystems $\mathcal{A}$ and $\mathcal{B}$. But, when some of the Schmidt coefficients are zero, then the Schmidt number would be less than min$(d_A, d_B)$.

In a physical point of view, the Schmidt number gives the number of degrees of freedom. For a product state, there is only a single degree of freedom and then the Schmidt number is $1$. Therefore, for a bipartite pure state, the necessary and sufficient condition for the separability of the state is that the Schmidt number being equal to $1$.

The important point is that, after writing the state based on the schmidt decomposition, then it is so easy to detect the entanglement and measure it. If the Schmidt decomposition is given then just by looking at the Schmidt decomposition it is possible to get the information on the entanglement of the states. If the Schmidt decomposition contains only one term then clearly the state is a product or separable state and then not entangled. On the other hand, if a state can not be written in a Schmidt decomposition of only one term, then the state is entangled.

\subsection{Characterization of Entanglement; Mixed State}

In a similar way, suppose there are two systems, say ${\mathcal A}$ and ${\mathcal B}$. There is an interaction between the two systems, which may raise entanglement. The state of the system is described by the density matrix $\rho$. The task here for us is to detect the existence of entanglement.

For a bipartite mixed state, by definition, the state is called separable if it can be written as a product state. The product state is given as follows
 \begin{equation}
 \sum_ip_i\rho_i^{(A)}\otimes \rho_i^{(B)},
 \end{equation}
 with $p_i\geq 0$ and $\sum_i p_i =1$. For the case of the pure state we could use this definition extensively as there is a well established approach to decompose the state, namely Schmidt decomposition. In the case of mixed state, however, there is not such a generally applicable approach. 
 
 For bipartite mixed states, there is a necessary condition for separability of states. {\it Peres-Horodecki criterion}, or PPT \cite{D4}, \cite{D5}, which stands for positive partial transpose. PPT says, if the state is separable, then the partial transpose of the density operator with respect to one subsystem is positive. PPT is a necessary condition for separability.
  
 For a chosen orthonormal product basis $\{|v_iv_j\rangle\}\equiv\{|v_i\rangle\otimes|v_j\rangle\}$ for the state $\rho$, the partial transpose $\rho^{T_B}$ is defined by its matrix elements 
 \begin{equation}
 \rho_{m\mu,n\nu}^{T_B}=\langle v_mv_\mu|\rho^{T_B}|v_nv_\nu\rangle=\rho_{m\nu,n\mu}.
 \end{equation}
 
For small dimensions, $2\times2$ and $2\times3$, the PPT-criterion is also sufficient condition for the state to be separable. However, we should say, that for a general bipartite mixed state, the situation is not still clear. There exist entangled PPT states, \cite{D6}.

The concept of entanglement for mixed state is richer compared to the previously studied pure state. There are some mixed state entanglement, which cannot be interchanged to the Bell states. The states are called Bound entanglement.
\subsection{Measurement of Entanglement; General Axioms} 
There are some axioms intuitively that are required for entanglement measures to be met. During the process to make any entanglement  measure, usually it happens that some of the conditions are discarded. Therefore, for each one of the axioms there are several modified versions. We will see that while the modified versions are mostly satisfied for the entanglement measures on mixed states, for the case of the pure states the essential properties in their strong forms are mostly adequate. The conditions on pure state entanglement measures are those specialized from the ones on the mixed states.
 
One of the most important of these conditions is that, entanglement, apart from the method for its quantification, should not increase on average by any LOCC, local operation and classical communication. This is called the criterion of the monotonicity under LOCC. Monotonicity is an essential requirement as it is even argued that in case of satisfying this condition, the other properties such as convexity and so on would also be satisfied. 
  
  The axioms for entanglement measures are summarized as follows\cite{D7}, \cite{D8}
 \begin{description}
\item[(E0a)] $E(\rho)=0$, if and only if $\rho$ is separable. This is useful, though it is very strong condition. For example, Bound entangled states have zero distillable entanglement, but they are not separable. Therefore, slightly modified version is as follows.
\item[(E0b)] $E(\rho)=0$, if $\rho$ is separable.

\vspace{2cm}

\item[(E1a)] Normalization. For a maximally entangled state in $d\times d$ dimensions, e.g. $|\phi_d^+\rangle\langle\phi_d^+|$, the amount of entanglement is quantified to be $E(|\phi_d^+\rangle\langle\phi_d^+|)=\log d$. Here,
\begin{equation}
|\phi_d^+\rangle=\sum_{i=1}^d \frac{1}{\sqrt d}(|i\rangle_A\otimes|i\rangle_B),
\end{equation}
with the two sets $|i\rangle_A$ and $|i\rangle_B$ are orthogonal bases.
\item[(E1b)] For a Bell state, $|\phi^+\rangle$, the entanglement is $E(|\phi^+\rangle\langle\phi^+|)=1$. 

\vspace{2cm}

\item[(E2a)] Monotonicity. For any LOCC operation $\Lambda$, $E(\Lambda(\rho))\leq E(\rho)$. 
\item[(E2b)] The equality in the monotonicity condition, $\Lambda$, $E(\Lambda(\rho))= E(\rho)$  is for the case that $\Lambda$ is a {\it strictly local operation}. This means that $\Lambda$ is a unitary operation or it only adds some extra dimensions.
\item[(E2c)] When $\Lambda$ is {\it strictly local unitary operation}, then $E(\Lambda(\rho))=E(\rho)$.

\vspace{2cm}

\item[(E3a)] (Continuity) Suppose two sequences of bipartite states, $\{\rho_n\}$ and $\{\sigma_n\}$, living on a sequence of Hilbert spaces, $\{\mathcal{H}_n\}$. For all these sequences which satisfy the condition $\|\rho_n-\sigma_n\|_1\to 0$, then
\begin{equation}
\frac{E(\rho_n)-E(\sigma_n)}{1+\log(\dim \mathcal{H}_n)}\to 0,
\end{equation}
where, by definition $\|\,...\,\|_1$ stands for the trace norm defined as follows
\begin{equation}
\|A\|_1=tr(\sqrt{A^\dagger A}).
\end{equation}

\item[(E3b)] Continuity condition as explained above can be weakened if the state, e.g. $\rho_n$ is a representation of pure states.

\vspace{2cm}

\item[(E4a)] (Weak additivity) For all states $\rho$ and any number $n\geq 1$, then
\begin{equation}
\frac{E(\rho^{\otimes n})}{n}=E(\rho).
\end{equation}

\item[(E4b)] (Asymptotic weak additivity). The condition of weak additivity is just modified to get the asymptotic version. Given $\epsilon>0$ and a state $\rho$, there exists an integer $N>0$, such that for all integers $n\leq N$ we have
\begin{equation}
\frac{E(\rho^{\otimes n})}{n}-\epsilon\leq E(\rho)\leq\frac{E(\rho^{\otimes n})}{n}+\epsilon.
\end{equation}

\vspace{2cm}

\item[(E5a)] (Subadditivity) For all states $\rho$ and $\sigma$, 
\begin{equation}
E(\rho\otimes\sigma)\leq E(\rho)+E(\sigma).
\end{equation}

\item[(E5b)] For all states $\rho$ and $m, n \geq 1$, then
\begin{equation}
E(\rho^{\otimes(m+n)})\leq E(\rho^{\otimes m})+ E(\rho^{\otimes n}).
\end{equation}

\item[(E5c)] (Existence of a regularisation) Finally, the weakest form for the additivity conditions, which is satisfied usually by any measure of entanglement is the one explained here. It states that for all bipartite state $\rho$, there exists a limit as follows
\begin{equation}
E^{\infty} (\rho)\equiv\lim_{n\to\infty}\frac{E(\rho^{\otimes n})}{n},
\end{equation}
for which $E^{\infty}$ is called the regularisation of $E$.

\vspace{2cm}

\item[(E6a)] (Convexity) Mixing states does not increase entanglement. Namely, for all bipartite states $\rho$ and $\sigma$
\begin{equation}
E(\lambda \rho+(1-\lambda)\sigma)\leq\lambda E(\rho)+(1-\lambda)E(\sigma),
\end{equation}
for all $0\le \lambda \le 1$. 
\item[(E6b)] For any bipartite state and any pure state realization $\rho=\sum_ip_i|\psi_i\rangle\langle\psi_i|$, $p_i\geq 0$ and $\sum_i p_i =1$, then
\begin{equation}
E(\rho)\leq \sum_ip_iE(|\psi_i\rangle\langle\psi_i|).
\end{equation}
\end{description}

\subsection{Generalized Measurement}
In general, a measurement is represented by a set of linear operators $\{M_i\}$, called measurement operators satisfying the completeness relation as follows
\begin{equation}
\sum_mM_m^\dagger M_m=1.
\end{equation}

The probability for the outcome to be $m$, $p(m)$, is 
\begin{equation}
p(m)=\langle \psi|M_m ^{\dagger} M_m |\psi\rangle.
\end{equation}

After the measurement, with result $m$, the state of the system changes into
\begin{equation}
|\psi^{\prime}\rangle \equiv \frac{M_m|\psi\rangle}{\sqrt{\langle\psi|M_m^{\dagger}M_m|\psi\rangle}}.
\end{equation}

Generalization to the mixed state is also possible. If the density matrix representing a particular mixed state, corresponds to a pure state $|\psi\rangle$, meaning that $\rho=|\psi\rangle\langle\psi|$, then generalization is as follows
\begin{equation}
p(m)=tr(M_m\rho M_m^\dagger),
\end{equation}
$p(m)$ is the probability for the outcome $m$. Then the state changes into the following state after the measurement with the outcome $m$
\begin{equation}
\rho^\prime=\frac{M_m\rho M_m^\dagger}{tr(M_m\rho M_m^\dagger}).
\end{equation}

When the density matrix is a representation for a mixed state, instead of a pure state, the generalization is still valid if we recall that
\begin{equation}
\rho=\sum_ip_i|\psi_i\rangle\langle\psi_i|=\sum_ip_i\rho_i.
\end{equation}

Therefore, the probability to get the outcome $m$ from the measurement is calculated as follows
\begin{eqnarray}
p(m)&=&\sum_ip_itr(M_m\rho_iM_m^\dagger)\\ \nonumber
    &=&tr\left(M_m\left[\sum_ip_i\rho_i\right]M_m^\dagger\right)\\ \nonumber
    &=&tr(M_m\rho M_m^\dagger).
    \end{eqnarray}
The new state after the measurement with outcome $m$ would be
\begin{eqnarray}
\rho^\prime&=&\sum_ip_iM_m\rho_iM_m^\dagger\\ \nonumber
           &=&M_m\left[\sum_ip_i\rho_i\right]M_m^\dagger\\ \nonumber
           &=&M_m\rho M_m^\dagger.
\end{eqnarray}
The projective measurement is a special class of the generalized measurement, where the observable, $V$, is a Hermitian operator and can be decomposed according to the spectral form as follows
\begin{equation}
V=\sum_mv_m|v_m\rangle\langle v_m|,
\end{equation}
 where $v_m$ and $|v_m\rangle$ are the eigenvalues and eigenvectors for $V$, respectively. The probability for an outcome $m$ is $p(m)=|\langle v_m|\psi\rangle|^2$ and the state after the measurement is $e^{i\theta} |v_m\rangle$, with an arbitrary phase $\theta$.
 
\subsection{Measures of Entanglement; Pure State}

In principle, it is possible to construct any measurement, which converts an initial state to a desired state, with a certain probability. For simplicity, suppose that two parties Alice and Bob are given a Bell state $|\phi^+\rangle$, generalization to a higher dimension would be also possible. They want to transform it into the state $\cos \theta |00\rangle +\sin \theta |11\rangle$. Note, that this is almost general example as the final state would be transformed to a general state as $\cos \theta |u_0,v_0\rangle+\sin\theta|u_1,v_1\rangle$, by local operations which do not change the amount of entanglement. 
 
 To do the task, Alice performs a measurement described by the measurement operators
 \begin{equation}
 \left(\begin{array}{@{\,}cc@{\,}}\cos \theta&0\\0&\sin\theta\\ \end{array}\right) \hspace{1cm} {\rm and} \hspace{1cm}\left(\begin{array}{@{\,}cc@{\,}}\sin\theta&0\\0&\cos\theta\\ \end{array}\right).
 \end{equation}
 
 The measurement operators clearly satisfy the completeness relation. After the measurement, we may have two states, depending on the outcome, as follows
 \begin{equation}
 |\psi_1\rangle=\cos \theta |00\rangle+\sin\theta|11\rangle, \hspace{1cm}|\psi_2\rangle=\sin\theta|00\rangle+\cos\theta|11\rangle.
 \end{equation}
 
 Therefore, by this measurement, the maximum entangled Bell state has been changed into a less or equally entangled state. If $\theta=0$ or $\pi/2$, then the output state is a product state, which is not entangled. However, if $\theta=\pi/4$ or $3\pi/4$ then the output state is a Bell state.
 
 The above example, shows generally how a maximum entangled state can be transformed to a less or equivalent desired state by means of measurement. Hence, suppose the dual problem. Meaning that the two parties, Alice and Bob, want to change the state $|\psi_1\rangle=\cos \theta |00\rangle+\sin\theta|11\rangle$ into the Bell state $|\phi^+\rangle$, by measurement. Then we need to define the measurement operators. The operator $M_1$, defined as follows will give the desired state.
 \begin{equation}
 M_1={\rm C} (\frac{1}{\cos\theta}|0\rangle\langle0|+\frac{1}{\sin\theta}|1\rangle\langle1|),
 \end{equation}
 where C is a constant. The measurement operators should satisfy the completeness relation. The new state, after measurement with operator $M_1$, is 
 \begin{equation}
 M_1|\psi_1\rangle={\rm C}(|00\rangle+|11\rangle),
 \end{equation}
 with a probability of
 \begin{equation}
 p(1)=\langle \psi_1|M_1^\dagger M_1|\psi_1\rangle={\rm C}^2(\langle00|00\rangle+\langle11|11\rangle)=2{\rm C}^2.
 \end{equation}
 
 The desired situation is maximizing this probability. However, because of the completeness relation, $M_2^\dagger M_2=1-M_1^\dagger M_1$, the maximum value should be restricted. In the valid range for $\theta$, then we conclude that the constant value would be $\sin \theta$. Then, the measurement operators are achieved as follows
 \begin{eqnarray}
 M_1&=&\tan\theta|0\rangle\langle0|+|1\rangle\langle 1|,\\ \nonumber
 M_2&=&\sqrt{1-\tan^2\theta}|0\rangle\langle0|.
 \end{eqnarray}
 The probabilities for the outcomes are
 \begin{eqnarray}
 p(1)&=&\langle\psi_1|M_1^\dagger M_1|\psi_1\rangle=2\sin^2\theta ,\\ \nonumber
 p(2)&=&\langle\psi_1|M_2^\dagger M_2|\psi_1\rangle=1-2\sin^2\theta,
 \end{eqnarray}
 and the normalized new states are
 \begin{eqnarray}
 |\psi^\prime_1\rangle&=&\frac{1}{\sqrt2}(|00\rangle+|11\rangle)=|\phi^+\rangle ,\\ \nonumber
 |\psi^\prime_2\rangle&=&|00\rangle.
 \end{eqnarray}
 Therefore, there is a certain probability to change the partially entangled state into a maximally entangled state. The probability depends on how strong is the entanglement of the initial state. As much as the initial state is entangled, $\sin\theta=\frac{1}{\sqrt 2}$ for entangled state, it is a larger probability, $p(1)=1$ for entangled state, to change the state into a maximally entangled state.
 
To conclude, we may use the above perspective in order to give the measures of entanglement. The {\it distillable entanglement} $E_d$, sometimes called {\it entanglement of distillation}, is defined as the maximum yield of Bell states that can be obtained, optimized over all possible LOCC. On the other hand, the {\it entanglement cost}, $E_c$, is defined as the minimum number of Bell states needed to create a given state by means of LOCC. It should be clear, that two separated parties cannot prepare an entangled state if they can only communicate classically. But if they have some entangled state, they can convert them to the desired entangled states. These definitions should be now clear after the above general examples on how to make the both side transformations, from Bell states to partially entangled states and vice versa.

The above definitions are given for finite regimes. However, calculations of $E_d$ and $E_c$ are hard for finite case. Only, intuitively, it is known that 
\begin{equation}
E_d\leq E_c.
\end{equation}

The reason is clear. If it would not be correct the above inequality, then it would be possible to create entanglement by means of LOCC, simply by converting Bell states to a state not satisfying the above inequality and then converting them back to the Bell state. In this process entanglement would be increased(!). 

In the asymptotic limit, the corresponding definitions are
\begin{eqnarray}
E_D(|\psi\rangle)\equiv E_d^{\infty}(|\psi\rangle)\equiv\lim_{n\to\infty}\frac{E_d(|\psi\rangle^{\otimes n})}{n},\\ \nonumber
E_C(|\psi\rangle)\equiv E_c^{\infty}(|\psi\rangle)\equiv\lim_{n\to\infty}\frac{E_c(|\psi\rangle^{\otimes n})}{n}.
\end{eqnarray}
It is shown that for the asymptotic version, both the distillable entanglement and entanglement cost are equal to the entropy of entanglement, $E_D(|\psi\rangle)=E_C(|\psi\rangle)=E_E(|\psi\rangle)$, defined as follows.

Recall the example of the Bell basis, $|\phi^+\rangle$, on a composed system of two subsystems $A$ and $B$. The information on the total state does not verify the information on each individual spaces $A$ and/or $B$. This is clear since $\rho_{A}=\rho_{B}=I$.

The fact, as mentioned intuitively with the above example, would be useful to get some idea about the measure of entanglement. Entanglement is very much related to the mixedness of the reduced density operator. Since, we know how to measure mixedness; typically by entropy; entanglement also would be measured accordingly.

Therefore, entropy seems to be a proper way for quantifying the entanglement. Entanglement of $|\psi\rangle$ is defined
\begin{eqnarray}
\label{entropy}
E(\psi) &:=&S(\rho_A)\\ \nonumber
        &=&-tr [\rho_A \log \rho_A]\\ \nonumber
        &=&-\sum_k \rho_k \log \rho_k,
\end{eqnarray}
where $S(\rho_A)$ is the von Neumann entropy of the state $\rho_A$. Von Neumann entropy for the state $\rho$ is defined as follows
\begin{equation}
S(\rho)\equiv-tr(\rho\log\rho),
\end{equation}
we have also used
\begin{equation}
\rho_A=\sum\rho_k|k\rangle\langle k|,
\end{equation}
entropy is written based on the eigenvalues.

 For example, consider the state $|\psi\rangle =|ab\rangle$. Then $\rho_A=|a\rangle \langle a|$, which gives $E(\psi )=-1\log 1=0$. On the other hand, for $|{\phi}^{+} \rangle$, we can simply calculate $E({\phi}^+)=-1/2\log 1/2-1/2\log1/2=1$. This value is actually the much that we can have for an entangled state.
 
  Therefore, after finding out the Schmidt coefficients entanglement is also quantified as 
\begin{equation}
E(\psi)=-\sum\lambda_k^2\log\lambda_k^2,
\end{equation}

  Finally, we conclude that if the state is a bipartite pure state, there is a well established way to detect the entanglement and also to quantify the amount of the entanglement.
  
\subsection{Measures of Entanglement; Mixed State}
 In contrast to pure state, for mixed states, there is no unique way to quantify entanglement, even for asymptotic cases. For pure state, Schmidt decomposition turned out to be a very powerful way for detection and measurement of entanglement. Schmidt decomposition posses an excellent property that it gives the minimum number required for decomposition of states. However for mixed state this is not always true.
 
 For example suppose an equal mixture of Bell states as follows
 \begin{equation}
 \rho=\frac{1}{2}(|\phi^+\rangle\langle\phi^+|+|\phi^-\rangle\langle\phi^-|).
 \end{equation}
 
 Each the component represented as the Bell state is entangled, however the total state would be also equivalently written as 
 \begin{equation}
 \rho=\frac{1}{2}(|00\rangle\langle00|+|11\rangle\langle11|).
 \end{equation}
 
 In this form, it is clear that the state is not entangled. therefore, even though there is a possibility to write the state based on entangled state, but this possibility does not posses the entanglement of the state. Actually, as far as there is some inexpensive, separable, way to make a state, the total state would be generally separable state valued. Even though one may try expensivly to construct the same state, but his try would be no more appreciated. 
 
 The above description can be emphasized in mathematical view point, too. The state $\rho$ is separable, not entangled, if $\exists p_k\geq0, \sum_k \lambda_k=1, |a_k\rangle\in\mathcal{H}_A, |b_k\rangle\in\mathcal{H}_B$, such that $\rho=\sum \lambda_k|a_kb_k\rangle\langle a_kb_k|$. If there is not any $p_k$ satisfying the previous statement then the state is entangled. This means that all the possible constructions to make the state should be checked and this task in general is not possible. Therefore, intuitively, we see the reason on why there is not a general and easy approach for measuring the entanglement of a mixed state.
 
\subsubsection{Entanglement Cost and Distillable Entanglement}

 The entanglement cost $E_c$ is the minimum number of Bell states required to produce the state by means of LOCC. The distillable entanglement $E_d$ is the maximum number of Bell states that can be distilled by an optimal LOCC distillation protocol. Because that a perfect conversion between mixed states in the finite regime is generally not possible, therefore for mixed state the asymptotic versions, $E_D$ and $E_C$ are considered. It is shown that for any measure $E$ which satisfies certain natural conditions for asymptotic measures, the entanglement cost and distillable entanglement provide upper and lower bounds
\begin{equation}
E_D\leq E\leq E_C.
\end{equation}

For Bound entangled states, the situation is clear as $E_D=0$ and $E_C$ is finite. Generally speaking, based on the axioms that each entanglement measure should satisfy, the upper and lower bound is defined. 
\begin{description}
\item[i)]
 If the entanglement measure satisfies $(E1a)$, $(E2a)$, $(E3a)$ and $(E4a)$, for all the states $\rho$, then
\begin{equation}
E_D(\rho)\leq E(\rho)\leq E_C(\rho).
\end{equation}

These conditions are very strong. Up to now, only the ^^ ^^ squashed entanglement" \cite{D9} has been known to satisfy them all.
\item[ii)]
 For an entanglement measure satisfying $(E1a)$, $(E2a)$, $(E3a)$ and $(E5c)$, then for all states $\rho$, the regularized version $E^{\infty}$ is bounded
\begin{equation}
E_D(\rho)\leq E^{\infty}(\rho)\leq E_C (\rho),
\end{equation}
these conditions are easier to be satisfied and they are satisfied by ^^ ^^ entanglement of formation" and ^^ ^^ relative entropy of entanglement".
\end{description}

\subsubsection{Distance Based Measures}

The class is introduced with properties that satisfy the conditions $(E0a)$, $(E2a)$ and $(E2c)$, namely the measures that are included in this class would be zero if and only if the state is separable, the expectation value does not increase under LOCC and they are left unchanged under local unitary operations.

Suppose that $D(\rho,\sigma)$ is a distance function representing the distance between two states $\rho$ and $\sigma$. Here, $\sigma \in \mathcal{S}(\mathcal{H})$, with $\mathcal{S}(\mathcal{H})$ to be the set of separable states. Then the entanglement of the state $\rho$ is defined to be as follows
\begin{equation}
E(\rho)\equiv \underset{\sigma\in\mathcal{S}(\mathcal{H})}{\rm inf}D(\rho, \sigma).
\end{equation}

{\it Relative Entropy of Entanglement:} The quantum relative entropy, $S(\rho\|\sigma)$, can be used as the distance function $D(\rho, \sigma)$. 
 \begin{equation}
 S(\rho\|\sigma)\equiv tr[\rho(\log \rho-\log\sigma)].
 \end{equation}
 
 The required conditions to be satisfied by $D(\rho, \sigma)$, now should be satisfied by $S(\rho\|\sigma)$. Relative entropy is not a metric and it is not even symmetric. But still it has some good properties. It is nonnegative, and zero for identical density operators. The same unitary operations on both states leave it unchanged. Then it can be used as a distinguishablity between the states. 
 
 Suppose that we are given a large but finite number of $n$ quantum states, all the same, being either $\rho$ or $\sigma$. Our task is to find the state by measurement. The probability for inferring from optimal measurements on the composite system, that the given state is $\rho$, while it is indeed $\sigma$ is \cite{D10}
 \begin{equation}
 P_n(\sigma\to\rho)=2^{-nS(\rho\|\sigma)}.
 \end{equation}
 
 It is clear that this expression is not symmetric. Relative entropy can be calculated from the eigenvalues and eigenvectors of density operators. Suppose that $r_i$ and $|r_i\rangle$ are the eigenvalues and eigenvectors of the state $\rho$ and $s_i$ and $|s_i\rangle$ are the eigenvalues and eigenvectors of the state $\sigma$. Then generally we have
 \begin{equation}
 S(\rho\|\sigma)=\sum_i\{r_i\log r_i-\langle s_i|\rho|s_i\rangle\log s_i\}.
 \end{equation}
 
 If the eigenvectors for the two density matrices are the same, i.e. when $\rho$ and $\sigma$ commute, then rearrangement of the eigenvectors gives result as follows
 \begin{equation}
 S(\rho\|\sigma)=\sum_i r_i(\log r_i -\log s_i),
 \end{equation}
 that is much easier for calculation point of view.
 
 Then relative entropy of entanglement is defined to be as follows \cite{D11}
 \begin{equation}
 E_r(\rho)\equiv \underset{\sigma\in\mathcal{S}(\mathcal{H})}{\rm inf}S(\rho\|\sigma).
 \end{equation}
 
 Relative entropy of entanglement satisfies the condition $(E0a)$, because that $S(\rho\|\sigma)=0$ only when $\rho=\sigma$. It also satisfies the normalization conditions $(E1a)$ and continuity on pure states $(E3b)$. It is nonincreasing under LOCC, satisfying the condition $(E2s)$ \cite{D11}, \cite{D12}, \cite{D13}, \cite{D14}, \cite{D15}.
 
 For the regularization of the relative entropy of entanglement we have
 \begin{equation}
 E_R(\rho)\equiv E_r^{\infty}=\underset{n\to\infty}{\rm lim}\frac{E_r(\rho^{\otimes n})}{n}.
 \end{equation}
 
 The above measure also satisfies automatically $(E0a), (E1a), (E2a), (E5a)$ and $(E6a)$, as these conditions extend to any regularization of a measure that satisfies them \cite{D15}. 
 
 In order to generate other distance based measures, the quantum relative entropy can be replaced by other forms of distances. In the following part, we have listed some of these measures and the properties as well.
 
 {\it Bures metric based measure:} The metric is defined as \cite{D16}, \cite{D17}, \cite{D18}, \cite{D11}
 \begin{equation}
 D_B(\rho\|\sigma)\equiv 2-2\sqrt{F(\rho,\sigma)},
 \end{equation}
 where $F(\rho, \sigma)$ is called Uhlmann's transition probability, defined as follows
 \begin{equation}
 F(\rho,\sigma)\equiv[tr\{(\sqrt\sigma\rho\sqrt\sigma)^{1/2}\}]^2.
 \end{equation}
 
 The corresponding entanglement measure satisfies $(E0a), (E1b)$ and $(E2a)$.
 
 {\it Hilbert-Schmidt distance based measure:} The distance is defined as \cite{D11}, \cite{D19}
 \begin{equation}
 D_{HS}(\rho,\sigma)\equiv\|\rho-\sigma\|^2_{HS}=tr[(\rho-\sigma)^2].
 \end{equation}
 
 however, it has been shown by Ozawa \cite{D20} that the above distance does not satisfy the sufficient conditions for a nonincreasing function under LOCC. But the distance defined as follows is shown to have the required properties \cite{D21}
 \begin{equation}
 D_T(\rho,\sigma)\equiv \|\rho-\sigma\|={\rm tr}[\sqrt{(\rho-\sigma)^2}].
 \end{equation}
 
 \subsubsection{Entanglement of Formation}
 
 Recall that in the asymptotic limit, the entanglement cost for pure state is given by the entropy of entanglement. In a sense, entanglement of formation is extension of this effect to the mixed states \cite{D22}, \cite{D23}. Then, for a given ensemble of pure states, $\varepsilon=\{p_i,|\psi_i\rangle\}$, the entanglement of formation is defined as
 \begin{equation}
 E_f(\varepsilon)\equiv\sum_ip_iE_E(|\psi_i\rangle),
 \end{equation}
 which is actually an average of the entropy of entanglement for all the states in the ensemble. 
 
 A mixed state can be realized by a multitude of pure state ensembles. The entanglement of formation for different ensembles might be different to each other. But, by definition, we consider the most economic ensemble, meaning that the one which is minimized among all the possible cases. Then the definition of the entanglement of formation for mixed state is as follows
 \begin{equation}
 E_f(\rho)\equiv \underset{\varepsilon}{\rm inf}\sum_ip_iE_E(|\psi_i\rangle),
 \end{equation}
 where the infinimum is taken over all ensembles $\varepsilon=\{p_i,|\psi_i\rangle\}$, that realizes the state $\rho$.
 
 Speaking about the conditions which can be satisfied by entanglement of formation, the definition of the entanglement of formation can be useful to show the satisfaction of several of the conditions. Namely, $(E0a)$, $(E1a)$, $(E2a)$, $(E3a)$, $(E5a)$ and $(E6a)$, all are shown to be satisfied. Additivity has been shown for some special cases to be satisfied, however for a general case still it is not proved \cite{D24}, \cite{D25}, \cite{D26}, \cite{D22}, \cite{D27}. 
 
 In the definition of the entanglement of formation, we see the inifinimum which makes calculation of the amount of the entanglement of formation hard. However, for special case of two qubits it is possible to calculate it by using the concept of {\it concurrence} \cite{D28}, \cite{D29}.
 
 By definition, spin flip for a pure state $|\psi\rangle$, is defined as
 \begin{equation}
 |\tilde{\psi}\rangle=Y|\psi^{\star}\rangle,
 \end{equation}
 where $Y$ is the Pauli operator, $|\psi^{\star}\rangle$ is the complex conjugate of $|\psi\rangle$, when it is expressed in a fixed basis such as $\{|00\rangle, |01\rangle, |10\rangle, |11\rangle\}$. For a general $\rho$ spin flip is 
 \begin{equation}
 \tilde{\rho}=(Y\otimes Y)\rho^*(Y\otimes Y).
 \end{equation}
 
 The Hermitian matrix $R$ is
 \begin{equation}
 R\equiv\sqrt{\sqrt{\rho}\tilde{\rho}\sqrt{\rho}},
 \end{equation}
 with eigenvalues, $\lambda_i$, in descending order, concurrence $C$ is defined as
 \begin{equation}
 C(\rho)\equiv {\rm max}\{0,\lambda_1-\lambda_2-\lambda_3-\lambda_4\}.
 \end{equation}
 
  Then the entanglement of formation is calculated as 
 \begin{equation}
 E_f(\rho)=\varepsilon(C(\rho)).
 \end{equation}
 
 Here, the function $\varepsilon$ is defined as
 \begin{equation}
 \varepsilon(C)\equiv h\left(\frac{1+\sqrt{1-C^2}}{2}\right),
 \end{equation}
 with $h$ standing for the binary entropy function as follows
 \begin{equation}
 h(x)\equiv -x\log x-(1-x)\log (1-x).
 \end{equation}
 
 Then, entanglement of formation, for this special class of states can be calculated. There have been also attempts on generalizing the scheme, though the complete case is still an open problem.
 
 \subsubsection{Negativity}
 
 This entanglement measure is introduced with the aim of introducing a computable measure \cite{D30}. All the entanglement measures including distillation of entanglement and entanglement cost, have been introduced up to now, include in their definition some form of optimization and therefore are difficult to be measured. However, negativity is based on the trace norm, which has been introduced before and can be calculated using standard linear algebra packages. There are two quantities. The first quantity is {\it negativity} defined as \cite{D31}, \cite{D30}
 \begin{equation}
 \mathcal{N}(\rho)\equiv\frac{\|\rho^{T_B}\|-1}{2},
 \end{equation}
 which is equal to the absolute value of the sum of negative eigenvalues of $\rho^{T_B}$. The other quantity is {\it logarithmic negativity}, defined as follows
 \begin{equation}
 E_{\mathcal{N}}(\rho)\equiv\log \|\rho^{T_B}\|_1.
 \end{equation}
 
 However, this is not strictly monotone under LOCC, only it is for a subclass of LOCC. 
 
 Some of the required conditions are satisfied by $\mathcal{N}$. It is zero for all the separable states, but it is also zero for all the PPT entangled states. Therefore, it satisfies $(E0b)$ but does not satisfy $(E0a)$. It does not satisfy the normalization criterion $(E1a)$, as it is not one but $1/2$ for the Bell states. It is monotone under LOCC but the conditions, of additivity $(E4a)$ and subadditivity $(E5a)$ are not satisfied, since the superadditivity is satisfied as follows
 \begin{equation}
 \mathcal{N}(\rho\otimes\sigma)=\mathcal{N}(\rho)+\mathcal{N}(\sigma)+2\mathcal{N}(\rho)\mathcal{N}(\sigma).
 \end{equation}
 
 On the other hand, logarithmic negativity $E_{\mathcal{N}}$ satisfies some criteria, too. Similarly, it satisfies $(E0b)$. It does not satisfy, monotonicity $(E2a)$, nor convexity $(E6a)$. But it satisfies strong additivity as follows
 \begin{equation}
 E_{\mathcal{N}}(\rho\otimes \sigma)=E_{\mathcal{N}}(\rho)+E_{\mathcal{N}}(\sigma).
 \end{equation}
 
 Then weak additivity $(E4a)$ and subadditivity $(E5a)$ are satisfied.
 
\subsubsection{Squashed Entanglement}

 As we explained in the above sections, among the entanglement measures that have been introduced up to now, there is not a perfect one, in a sense that it satisfies all the required conditions. Squashed entanglement \cite{D9} is the one which satisfies all the conditions. This measure is defined as follows
 \begin{equation}
 E_{sq}(\rho)\equiv{\rm inf}\{\frac{1}{2}I(A;B|E)|\rho^{ABE} {\rm extension\,of}\,\rho\,{\rm to}\,{\mathcal{H}}_E\},
 \end{equation}
 where $\rho\equiv\rho^{AB}={\rm tr}_E(\rho^{ABE})$, as the infinimum is taking over all extensions to a third subsystem $E$. $I(A;B|E)$ is the quantum mutual conditional information defined as
 \begin{equation}
 I(A;B|E)\equiv S(\rho^{AE})+S(\rho^{BE})-S(\rho_{ABE})-S(\rho^E).
 \end{equation}
 
 Except for $(E0a)$, which is unknown yet, and instead it satisfies $(E0b)$. Squashed entanglement satisfies the other  conditions in their strictest manner, namely $(E1a)$, $(E2a)$, $(E3a)$, $(E4a)$, $(E5a)$ and $(E6a)$. This measure of entanglement is new and still is being developed. 
 
\subsubsection{Witness Entropy of Entanglement}

 This entanglement measure is based on the concept of entanglement witness \cite{D32}, \cite{D33}. Entanglement witness is a Hermitian operator, which has positive expectation values for all separable states $\rho$, tr $(W\rho)\geq 0$, and a negative value for at least one entangled state $\sigma$, $ tr \, (W\sigma)< 0$. Therefore, entanglement is detected if a negative value is detected for the observable. Conventionally, the entanglement witness is normalized to have $tr \, (W)=1$. The entanglement witness is called to be optimal, $W_{op}$ for a state $\sigma$ if
 \begin{equation}
 {\rm tr}(W_{op}\sigma)\leq {\rm tr} (W\sigma),
 \end{equation}
 for all entanglement witnesses $W$. 
 
 The witness entropy of entanglement is defined to be as follows
 \begin{equation}
 E_\omega(\rho)\equiv\log(D/d){\rm max}[0, -{\rm tr}(W\rho)],
 \end{equation}
 where $D$ and $d$ are dimension of the total Hilbert space and the smallest of the dimensions of the Hilbert spaces of the subsystems.
 
 Except the additivity, this measure satisfies other conditions, namely $(E0a)$, $(E1a)$, $(E2a)$, $(E3a)$, $(E5a)$ and  $(E6a)$.
 
 \section{Entanglement in Multipartite System}
 Entanglement also can exist among more than two parties. Then it is called multipartite entanglement. The situation is much more complicated compared to the bipartite case and there are so many open problems which are still under progress. Nevertheless, some of the essential concepts and entanglement measures can be generalized to multipartite system from bipartite one. 

 Generally speaking multipartite entanglement complication, apart from its several reasons, is also because that multipartite system is composed of subsystems, called here partitions. Entanglement on the whole system does not pertain the entanglement on the individual subsystems. Though in some cases, it is pertained on each subsystem too. Then, the problem itself has an intrinsic complexity. 
 
 For example consider the Greenberger, Horne and Zeilinger, GHZ, state as follows
 \begin{equation}
 |{\rm GHZ}\rangle\equiv\frac{1}{\sqrt 2}(|000\rangle+|111\rangle).
 \end{equation}
 
 This state is the maximum entanglement between three subsystems. However, if any of the subsystems is traced out then the density matrix sharing by the two remaining parties is represented as follows
 \begin{equation}
 \rho=\frac{1}{2}(|00\rangle\langle00|+|11\rangle\langle 11|),
 \end{equation}
 which is separable. Thus, the maximal entanglement on three parties as represented above does not include any bipartite entanglement. 
 
 Some of the notions from bipartite system can be extended to multipartite system. Generalization of separability condition gives a corresponding condition for the separability of multipartite state. Namely, a state $\rho$ is separable if it can be written as a convex combination of product states
 \begin{equation}
 \rho=\sum_ip_i\rho_i^{(A)}\otimes \rho_i^{(B)}\otimes ... \otimes \rho_i^{(N)},
 \end{equation}
 with $p_i\geq 0$ and $\sum_ip_i=1$.
 
 A separable state may be arisen from ignorance of a subsystem of the total multipartite entangled state. Because, as it has been shown by a simple example above, an entanglement multipartite state may or may not be an entangled state if one of the subsystems is traced out. By definition, a multipartite state that loses its entanglement and becomes a separable state by tracing out a substate is called, {\it multiseparable}, in accordance to the separable state with the above definition, \cite{D34}.
 
 On the other hand, the PPT criterion for separability simply extends to the multipartite entanglement. Because, if a multipartite state is a separable, then transposing any of the substates should satisfy the positivity.
 
 However, for measurement of the entanglement, it is not always possible to get the proper measure by extension of the bipartite measures of entanglement, in general. Entanglement of distillation and entanglement cost have been introduced as two important measures for the case of bipartite state. However, the definition of these two measures are based on the Bell state, and Bell state is a bipartite state. It is not so clear how to extend the Bell state to the higher level of states. Therefore, these two entanglement measures are not easily applicable for multipartite state. Hence, there have been some attempts on this regards. For almost the same reason, entanglement of formation also can not be adequate as it is based on the entanglement cost.
 
 In order to define a measure of entanglement, in a similar manner to the bipartite case, we need to have some conditions which should be satisfied with the entanglement measures. It can be easily accepted that all the conditions have been introduced previously for the case of bipartite entanglement can be applied for multipartite case too, except that the normalization condition which imply that measure should be normalized for Bell state is not applicable for a clear reason.
 
 Distance based measures of entanglement do not have any problem in their definitions to be extended to the multipartite case. But even in this form, the intrinsic meaning of the entanglement measure is vague compared to the bipartite case. The above example on GHZ state shows that bipartite entanglement is zero for the maximally three-partite entangled state GHZ. On the other hand, there are multipartite not maximal but entangled states with non-zero bipartite entanglement. Therefore, measurement on the entanglement of the multipartite state does not have a clearly known constitute on its subsystems.
 \section{Continuous Variable}
 Almost recently, entanglement has been considered in continuous variable regime too. Our study up to now has been on finite dimensional Hilbert spaces, though. Generalization of the previous discussion to the continuous regime has not been yet completed. But it is easy to accept that for an infinite dimensional state, there are states with infinite amount of entanglement. Recall that the entropy of entanglement for a maximally entangled state is $\log d$, then for the case where the state is infinite, it gives infinite entanglement \cite{D35}, \cite{D36}.
 
 The definition of separability in a continuous variable bipartite system is as follows
 \begin{equation}
 \sum_ip_i\rho_i^{(A)}\otimes \rho_i^{(B)},
 \end{equation}
 where $p_i$ is a probability distribution. But in an infinite dimension, states only can be approximately described with a form as the above one. However, PPT criterion is still valid for infinite dimension. For some special class such as Gaussian state, it is also a sufficient condition in addition to be a necessary condition \cite{D37}, \cite{D36}, \cite{D38}, \cite{D39}, \cite{D40}, \cite{D41}.
 
 The subject of the multipartite entanglement is still an active field of research and more elaborated results may be represented in this field.
 \section{Operational Entanglement Measures}
 Barbara Terhal in an excellent review paper \cite{D42}, addresses very important note as follows, ^^ ^^ Understandably, when considering more complex physical systems, the dividing line between what is entangled and what is not entangled, may become somewhat fuzzy. The guideline in deciding these matters, I believe, should be the question: Do we have an operational form of quantum entanglement? What resource does the particular state constitute in quantum communication and computation?"
 
 The concept of operational entanglement measure is also very important. Here, it might be better to explain it in the part for Bipartite entanglement measures. But, as far as operational entanglement measure turns out to be much useful for our case as to introduce a proper entanglement measure, we would like to explain it separately.
 
 Operational entanglement measure can be explained as follows: firstly, take your favorite application of entanglement. For example, superdense coding \cite{D43}. Then, study how well or bad it performs for different sets of states, for example states involved in nuclear magnetic resonance implementation of the particular example here, superdense coding. Then, define an entanglement measure from this and check that it satisfies at least some of the conditions for entanglement, above all LOCC monotonicity \cite{D8}.
 
 We have used almost the same prescription in introducing a new entanglement measure. Superdense coding has been selected and the states for NMR experiment have been studied. However, we have modified the scheme in a sense that we demonstrate our entangled measure as based on the concept of entanglement witness. We will explain it more in the next chapter after explaining the NMR system.
 
 One slightly similar entanglement measure in this regard has been constructed based on the capacity of superdense coding \cite{D43}, by Hiroshima \cite{D44}, \cite{D45}. An asymptotic entanglement measure for any bipartite state has been introduced as based on superdense coding capacity, that has been optimized with respect to local quantum operations and classical communication, LOCC. General properties for an entanglement measure also have been proved for this particular case. 
\vspace{5cm}
\chapter{NMR Quantum Information Processing and Quantum Computation}
Nuclear magnetic resonance, NMR, is a well established field in physics, originally, and in chemistry, widely.  NMR technology goes back to 1940's and constantly has been used largely for variety of purposes. In chemistry, it is a major tool to study the structure and properties of solids, liquids and gases, \cite{S1}, \cite{S2}, \cite{S3}, \cite{S4}.

There are several good reasons on the wide attentions to NMR as a means to realize a quantum computer, \cite{S5}, \cite{S6}, \cite{S7}, \cite{S8}. In this chapter, we try to give some points on NMR as a fascinating physical system for realizing a quantum computer. We first start from the criteria which are required for any valid candidate for realization of a quantum computer. These criteria are known as DiVincenzo criteria, \cite{S9}. Then we will see NMR as a physical system, which naturally satisfies most of the required criteria. However, there are some critical bad behaviour for NMR, and we will explain them. Finally, we will prove that some NMR realization of quantum information processing, namely superdense coding, is also accessible with a classical computer. Therefore, it is vague to assume the corresponding experiment as a true quantum processing. Then we will introduce an appropriate way for detecting the entanglement. the scheme will be shown to have advantages over the conventional ways for detecting entanglement as it uses easy to access operators for NMR and determines the status of entanglement in the system, with only a single run experiment.

It turns out that repairing NMR from this disadvantages might be possible, however may not be so simple, \cite{S10}, \cite{S11}. Therefore, in the next chapter we will try another physical system, electron nuclear double resonance, ENDOR, \cite{S12}. ENDOR pertains most of the advantages of NMR system, while the existence of an electron spin also makes the required experimental conditions for constructing high spin polarized states much easier .
\section{DiVincenzo Criteria}
The minimal requirements for a quantum computer have been originally introduced by DiVincenzo. Even though after it, it has been some additional criteria added by others. Here we list the original criteria, \cite{S13}.
\subsection{A Scalable Physical System with Well Characterized Qubits}
A set of physical systems is required, in order to give a respresentation of quantum bits. The physical system should have two distinct levels as to be a representation of a two-level quantum bit. The famous examples are spin-$1/2$ particles and polarized photons. Scalability also is a very important property, which should be satisfied by the physical systems. While a quantum computer with a number of qubits up to $50$ explores $2^{50}$ computational paths in parallel, well out of reach for any classical computers, for e.g. factorization problem of a $400$ digit number, a few thousands logical qubit are required. Hence, adding up the required number of qubits for error correction, the number is almost $100$ times more. Then, the problem of scalability is always very challenging for all the candidates of a quantum computer.
\subsection{A Universal Set of Quantum Gates}
Operations in a quantum computer are defined in terms of quantum gates. Therefore, it is very important for any realization of a quantum computer to be able to give a valid realization for quantum gates. At least, a universal set of quantum gates should be possible to be performed by the corresponding physical system.

The time evolution of a quantum system is determined by its Hamiltonian. Therefore, in order to realize quantum logic gates, it must be some ability to control the Hamiltonian over time. In other words, the resulting time evolutions correspond to the computational steps of an algorithm.

Theoretically attempts have been on simplifying the required steps and also to rearrange the requirements and to make them based on the existing terms in the Hamiltonian of the system. Also, experimentally, engineering of the Hamiltonian is of a very top priority field of study.
\subsection{The Ability to Initialize the State of Qubits to a Simple Fiducial State}
This is a fact that if the input of a computation is random, the output is of little use. Then it is highly required to have the ability to prepare reliably a pure input state. For a classical computer it is so easy to make all the bits initialized, however for a quantum computer somehow it can be one of the most challenging requirements as we will see that this issue is so problematic for NMR quantum computation.
\subsection{A Qubit-specific Measurement Capability}
A computation is useful only if we can read out the result of the processing. Because of the postulates of quantum physics, it is impossible to get all the information about the state of a qubit by a measurement. However, quantum algorithms make it possible to extract the {\it required} information through the projective measurements in the basis. The basis, can be specially the computational basis as $|0\rangle$ and $|1\rangle$ or any other set of basis, because that we can interchange the basis by applying unitary operations.

This criterion sometimes seems to be challenging for NMR, because in NMR there is an ensemble state and all the measurements give an average on the state. However, it has been shown that almost all known quantum algorithms can be adapted to be implemented on (NMR) ensemble computer.
\subsection{Long Relevant Decoherence Times}
Long decoherence time, much longer than the gate operation time is required. Decoherence time of a physical system representing a qubit should be much longer than the gate operation time in order to have possibility to perform several required operations in a computational process. This requirement is important and in so many quantum physical candidates for realization of a quantum computer, it is actually very critical subject. However, in the case of NMR quantum computing and quantum information processing, we need not to be worried about decoherence time. Since nuclear spins pertain very long decoherence time usually seconds or order of seconds while the order of computational time is microsecond.
\section{Liquid State NMR Quantum Computing}
In this section, we study, very briefly, whether liquid state NMR satisfies the required criteria, as summarized in the above section. We will not go to the details and will emphasize those requirements which are important mostly for this contribution. Our interest is mainly initialization. This criterion causes some disadvantages for NMR. We will see that as far as spin polarization is very low and the initialization to a particular pure state is not possible for NMR,  the credibility of the particular quantum processing, by means of NMR is not well approved.
\subsection{Scalable and Well Characterized Physical System of Qubits}
The basic system used in NMR quantum computing is an $N$-spin system. The system is a molecule which involves $N$ magnetically distinct qubits. Usually the spins are half spins, though the higher spins are also workable.

In a liquid NMR experiment, the sample including the specific molecules are dissolved in a liquid solvent. Sample is in a special tube, called test-tube, which is specially assigned for NMR. The tube is placed in a strong static magnetic field, $B_0$. We usually call the direction of the strong magnetic field the $\hat z$ direction.

Because of the strong static magnetic field, and the spin behaviour of the particles, each spin precesses along the magnetic field. The Hamiltonian for the system of $N=2$ spins can be written as follows
\begin{equation}
{\mathcal H}=\hbar \omega_AI_{zA}+\hbar \omega_BI_{zB}+\hbar\omega_{AB}I_{zA}I_{zB}.
\end{equation}

 In the above equation, $I_z$ is the spin angular momentum in the $\hat z$ direction. $\omega_\xi=-\gamma_\xi B_0$, $\xi=A,B$, where $\gamma_\xi$ is the gyromagnetic ratio for spin $i$. The last term gives to first order, the spin-spin coupling, because that spins are also coupled to each other via a through-bond electric interaction.

In order to have some intuition about the values in the above Hamiltonian, let us give some numerical values. Suppose a molecule includes two carbon-$13$, spin half, called spins $A$ and $B$. NMR strong magnetic field is typically $9.3$ Tesla. Then, the resonance frequencies would be $\frac{\omega_A}{2\pi}\sim\frac{\omega_B}{2\pi}\sim100$ MHZ. Suppose the magnetic field to be $11.7$ Tesla. Then a proton in this field would precess with a resonance frequency of $\frac{\omega}{2\pi}\sim500$ MHz. Typically, carbon-carbon coupling is $\frac{\omega_{AB}}{2\pi}\sim100$ Hz. The proton-proton coupling is several Hertzs. 

Then in NMR, nuclei represent the two-level physical systems required for quantum computation. The characterization of the spins would be possible if we consider the resonance frequencies for each different spin. However, the weak point for NMR, in this part, is its scalability. The number of nuclei in each molecule is of course restricted and cannot be increased to infinity. There have been some attempts to resolve this disadvantage of NMR. The {\it cellular automata} has been proposed by Lloyd, \cite{S14}, which might be useful in this sense. However, still the problem of the scalability, in its real meaning as the number of qubits should be possible to be increased extendedly while the computation should be a fault tolerant, for liquid state NMR still is an open problem. Here, the main attention might not be focused on this part as we will see that even for small number of qubits still there are some points which should be considered more seriously.
\subsection{Universal Set of Quantum Gates}
In a theoretical point of view, gates or operations on qubits are fixed and qubits are introduced to the gates. However, in NMR quantum computation, qubits are being represented by spins that are fixed in molecular structure and quantum gates are defined by pulses that are selectively applied on each spin.

Generally speaking, spins are manipulated by applying much smaller radio-frequency field, {\rm rf}, ${\rm B}_1$ in the $\hat x-\hat y$ plane to excite the spins on their resonant frequencies $\omega_i$. More explicitly speaking, the applied pulse is an electromagnetic field with the strength ${\rm B}_1$, which rotates in the transverse plane at $\omega_{{\rm rf}}$, at or near the spin precession frequency $\omega_0$. The Hamiltonian of {\rm rf} is, \cite{S15}, \cite{S16}, \cite{S17}
\begin{equation}
{\mathcal H}_{{\rm rf}}=-\hbar \omega_1[\cos (\omega_{{\rm rf}}t+\phi){\rm I}_x+\sin (\omega_{{\rm rf}}t+\phi){\rm I}_y],
 \end{equation}
 where $\omega_1=\gamma {\rm B}_1$ and $\phi$ is the phase of {\rm rf} field.
 
 The motion of a nuclear spin that is subjected to both the static magnetic field ${\rm B}_0$ and a rotating magnetic field ${\rm B}_1$ is rather complex, if it is described in the usual laboratory coordinate system. However, comparably easier, we find that in the rotating frame and at $\omega_{{\rm rf}}=\omega_0$, the spin evolves under an effective field
 \begin{equation}
 \overrightarrow {\rm B}={\rm B}_1\cos (\phi)\hat x+{\rm B}_1\sin (\phi)\hat y.
 \end{equation}
 
By varying phase, $\phi$, and the magnitude of ${\rm B}_1$, the rotation angle and axis can be controlled.
 \begin{equation}
 \theta =\gamma {\rm B}_1p\omega,
 \end{equation}
 where $p\omega$ is called {\it pulse width} or {\it pulse length}. Then the state of the system is calculated to be as follows
 \begin{equation}
 |\xi(t)\rangle=e^{i\delta t \sigma_x/2}|\xi(0)\rangle.
 \end{equation}
 
  Therefore, the effect of $e^{i\delta t \sigma_x/2}$ is just a rotation of the state about the $\hat x$ axis by the angle $\delta t$, where $\delta t$ represents the integrated power of the applied resonant {\rm rf} field. As far as , only the relative phase between pulses applied to the same spin is a matter of importance then by changing the phases of the {\rm rf} field by $90^\circ$, we can similarly implement Y pulses. Typically, for the system introduced above proton pulse length is $5$ microseconds for $\theta=90^\circ$.
 
 Rotation about the $\hat z$ axis is also possible after implementing any arbitrary rotation about $\hat x$ and  $\hat y$ axes, since
 \begin{equation}
 R_z(\theta)=R_x(-\pi/2)R_y(\theta)R_x(\pi/2).
 \end{equation}
 
 However, there are more elaborated approaches to implement an arbitrary Z rotation.
 
 Operations can be applied particularly on the selected spin, by applying {\rm rf} field on resonance frequency with the particular spin. Other techniques such as {\it pulse shaping} also would be used for this purpose. 
 
 Two-qubit gates are implemented by using the pairwise interactions between spins in the same molecule. There are two dominant systems of interactions; {\it direct dipolar coupling}, and {\it indirect through-bond electronic interactions}. Dipolar coupling is described by an interaction Hamiltonian of the form
 \begin{equation}
 {\mathcal H}_{A, B}^D=\frac{\gamma_A\gamma_B\hbar}{r^3}[{\rm I}_A. {\rm I}_B-3({\rm I}_A. {\bf n})({\rm I}_B. {\bf n})],
 \end{equation}
 where {\bf n} is the unit vector in the direction joining the two nuclei, and I is the magnetic moment vector. However, dipolar interactions rapidly average away in a low viscosity liquid.
 
 Through-bond interaction, {\it J-coupling}, is an indirect interaction. The magnetic field seen by one nucleus is perturbed by the state of electronic clouds, which interacts with another nucleus through a Fermi contact interaction. The Hamiltonian is
 \begin{equation}
 {\mathcal H}_{A, B}^J=hJ{\rm I}_A{\rm I}_B= h J {\rm I}_z{\rm I}_z+\frac{hJ}{2}[{\rm I}_+{\rm I}_-+{\rm I}_-{\rm I}_+],
 \end{equation}
 where $J$ is the scalar constant ; but generally a tensor. The flip-flop term is negligible for weak coupling or heteronuclear species. Therefore, up to a very good approximation it would be as follows
 \begin{equation}
 {\mathcal H}_{AB}^J=hJ{\rm I}_{zA}{\rm I}_{zB},
 \end{equation}
 and for a duration of $t$, the evolution would be
  \begin{equation}
 V_J(t)=\exp [-i2\pi J{\rm I}_z^1{\rm I}_z^2t].
\end{equation}

Arbitrary single bit operations in addition to the CNOT gate make a universal set of gates. The ability to construct the arbitrary single bit operation has been already discussed. CNOT gate also is implemented in several ways, e.g. as shown in the followings
\begin{equation}
{\rm CNOT}_{AB}=R_{yA}(-90)R_{zB}(-90)R_{zA}(270=-90)R_{zAB}(180)R_{yA}(90).
\end{equation}

This is shown as follows, \cite{Y23},
\begin{eqnarray}
\label{CNOTNMR}
C_{AB}&=&\frac{1}{2^{5/2}}\left(\begin{array}{@{\,}cccc@{\,}}1&-1&0&0\\1&1&0&0\\0&0&1&-1\\0&0&1&1\\ \end{array}\right)\left(\begin{array}{@{\,}cccc@{\,}}1-i&0&0&0\\0&1-i&0&0\\0&0&1+i&0\\0&0&0&1+i\\ \end{array}\right)\\ \nonumber
      & &\times\left(\begin{array}{@{\,}cccc@{\,}}1-i&0&0&0\\0&1+i&0&0\\0&0&1-i&0\\0&0&0&1+i\\ \end{array}\right)\left(\begin{array}{@{\,}cccc@{\,}}1+i&0&0&0\\0&1-i&0&0\\0&0&1-i&0\\0&0&0&1+i\\ \end{array}\right)\\ \nonumber
      & &\times \left(\begin{array}{@{\,}cccc@{\,}}1&1&0&0\\-1&1&0&0\\0&0&1&1\\0&0&-1&1\\ \end{array}\right)=\sqrt 2 \left(\begin{array}{@{\,}cccc@{\,}}1&0&0&0\\0&1&0&0\\0&0&0&1\\0&0&1&0\\ \end{array}\right),
\end{eqnarray}
that is the CNOT operation up to an irrelevant overall phase.

Therefore, any arbitrary operation can be performed on the selected spin as far as we have known how to make a universal set of gates. However, the trivial operation of ^^ ^^ identity" seems to be non-trivial in NMR, since evolutions of spins are always on. This can be done using a common NMR tool known as {\it refocucing}. 

The point for NMR quantum computing and quantum information processing is that NMR is an already rich field of research going back to $50$ years ago. Then occasionally, the advanced techniques of NMR have been proved to be much useful for the quantum computational purposes specially constructing quantum operations.
\subsection{Initialization of the State}
The issue of initialization is much complicated in NMR quantum computation and quantum information processing. As much as we discuss the technology of NMR, we will see that inherently NMR suffers from the lack of a true initial state even though it has been some proposals for artificial realization of an initial state. The fact which we will discuss more extensively is that the problem of initialization for NMR makes it possible to prove the classicality of the realization of some particular NMR experiment, which has been claimed to be a quantum based processing. We discuss the issue of initialization for NMR by recalling some fundamental aspects.

Liquid state NMR quantum computation and quantum information processing have been done conventionally and mostly at a thermal equilibrium condition. The thermal equilibrium state of the NMR system is described by density matrix $\rho$ as follows
\begin{equation}
\rho=\frac{e^{-\beta {\mathcal H}}}{{\rm Tr}(e^{-\beta {\mathcal H}})},
\end{equation}
where $\beta=\frac{1}{k_BT}$.

At room temperature, $k_BT$ is much larger than the energy differences of the levels, typically $10^5$ larger. Then, the state is approximated, with a very good accuracy for approximation, as follows
\begin{equation}
\rho\sim1-\frac{\beta{\mathcal H}}{2^N},
\end{equation}
where $N$ is the number of spins. For e.g. $N=2$ and suppose for a moment that $\omega_A\sim4\omega_B$, then, \cite{Y23},
\begin{equation}
\rho\sim1-\frac{\hbar\omega_B}{4k_BT}\left(\begin{array}{@{\,}cccc@{\,}}5&0&0&0\\0&3&0&0\\0&0&-3&0\\0&0&0&-5\\ \end{array}\right).
\end{equation}

This is a mixture of all possible pure states in the computational basis.

Recall the fact that the NMR observables are traceless. Then the identity term in the above equation is not observable in NMR. The thing which is observable is only the difference in the populations of different states. 

Microscopic state for NMR might be in a large variety of states, however as far as only the ensemble averages are measured, there is no way to distinguish between them.

Notably, and we will discuss it more in details later, entanglement is a microscopic property for NMR system. Then it seems appropriate to discuss the entanglement of NMR only in case of the direct access to individual states but this is not possible in NMR. This may look as a serious paradox then how to establish the entanglement of the NMR states. One way to this purpose, which we have used for our proposal on detection of entanglement in NMR, is based on the operational entanglement measures. Any quantum processing, which is based on entanglement would not be reasonably performed successfully with a system of separable states. This fact will be shown to be much useful for NMR.
\subsection{How to Solve the Problem?}
Pseudo-pure state generally has been introduced by different people in several ways. Preparation of any pseudo-pure state inherently requires some non-unitary operations, such as summation of different experiments. 

Pseudo-pure state in any form of its realization has not been validated perfectly for our purposes for several reasons. Firstly, the number of resources, molecules or experiments, required for making a pseudo-pure state scales exponentially with the number of qubits. Therefore, an exponential number of e.g. molecules are required to make signals readable. In the current NMR experiment, with an Avogadro number of molecules and spin polarization $\sim 10^{-5}$, up to $20$-qubit is possible to be worked with. By increasing the number of qubits in a same experimental condition, would give a very weak signal, smaller than the standard deviation in reading the results. Therefore, signals cannot be detected.

Pseudo-pure state involves generally non-uniform operations for its realization hence its evolution is not clearly approved. Finally and more importantly to sake our interest, it is naive to think that pseudo-pure state gives entanglement even with low polarized spins. Pseudo-pure state might seemingly look working well, but the large amount of mixed states, any way, still remains in the NMR system. Recall, the discussion on the measure of entanglement for mixed states. There might be some way to construct a separable mixed state from entangled states. However, as far as the final mixed state is separable, the state is a separable. 

It would be all perfect if it could have been possible to make pure states directly from NMR mixed states. Direct purification of the NMR mixed states at the ground state requires very strict experimental conditions of the temperature down to milli Kelvin. Apart from the complexity of how to reach easily such a low temperature, the sample would be no more a liquid sample but would change into a solid one. Solid state sample, then needs more elaborated operations on it. Dipolar couplings would be important in the Hamiltonian and need to have special care on manipulations of spins due to the more complex Hamiltonian, whereas the strict experimental conditions also would not be stable easily. 

Therefore, instead of getting involved on preparation of the experimental condition for making a pure state, people have considered indirect ways. There are several approaches to this purpose, among them are {\it optical pumping}, {\it para-hydrogen} and {\it dynamic nuclear polarization} or DNP, \cite{S13}, \cite{S17}, \cite{S18}. 

DNP is the transfer of polarization from the electron spin in a free radical to the nuclear spins. DNP has resulted to a very high spin polarization up to $0.7$. Though the molecule was not useful directly for an NMR quantum computation. The fact used for DNP is that the magnetic moment of electron spin is $\sim 10^3$ larger than nuclear spin, then nuclear spin polarization can be enhanced by electron spins. There has not been reported any successfully demonstration of the scheme for NMR quantum computation and quantum information processing.

Finally, we summarize that the initialization is a significant problem for quantum computation and quantum information processing by means of NMR. Pseudo-pure state has intrinsically problems, specially for the problem of entanglement, as will be discussed more in details in following sections. Enhancement of polarization is possible whether directly or indirectly. While the direct approach needs very much elaborated experimental conditions, none of the indirect several techniques has been perfectly successfully done for special purpose of quantum computation, yet.
\subsection{Readout}
Measurement in NMR is done with an {\rm rf} coil placed close to the sample. In order to readout a spin, a readout pulse, $R_x(90)$ is applied on the thermal equilibrium state. As a result the nuclei are tiped from the $\hat z$ axis into the $\hat x-\hat y$ plane. Then, nuclei precess along the $\hat z$ axes and generate a magnetic induction signal known as the free induction decay, FID. FID is captured by the pickup coil. Then FID is Fourier transformed to obtain the spectrum. In the obtained spectrum, different spins, qubits, in a molecule are spectrally distinguishable via their Larmor frequencies, $\omega_i$, and the amplitude and phase of the different spectral lines give information about the corresponding spin states. The extracted information depends on the convention, typically a positive absorptive line is due to a $|0\rangle$ and negative absorptive line is due to the spin in $|1\rangle$.

Suppose the state of the system at the start of the measurement is described by $\rho$. Then, mathematically the experimentally detected signal can be calculated from the transverse magnetization for a single spin \cite{S17},
\begin{equation}
M=d\omega \hbar {\rm Tr}[\rho(i\sigma_x+\sigma_y)],
\end{equation}
here $d$ is the volume density of the detected spin. Then, for a $K$-turn solenoid coil with quality factor $Q$ and area $A$, and magnetic flux $\phi$, we have 
\begin{eqnarray}
V(t)&=&QK\frac{d\phi}{dt}\\ \nonumber
    &=&QK\frac{d}{dt}\mu_0MA\\
    &\sim& QK\gamma B\mu_0e^{-t/T_2}\{d\gamma \hbar{\rm Tr}[\rho (i\sigma_x+\sigma_y)]\}A.
\end{eqnarray}
The corresponding voltage is complex. Both the magnitude and phase are measured. Then the envelope of the FID decays as $e^{-t/T_2}$, and yields to a Lorentzian line after Fourier transform as follows
\begin{equation}
\sim \frac{1}{1+(\omega-\omega_0)^2}-\frac{i\omega}{1+(\omega-\omega_0)^2},
\end{equation}
and has a linewidth at half height of the following form
\begin{equation}
\Delta f=\frac{\Delta \omega}{2\pi}=\frac{1}{2\pi T}.
\end{equation}

The magnetic signal of a single nuclear spin is too weak to be directly detected. Therefore, NMR experiment should be done using a large ensemble of identical molecules, typically on the order of $10^{18}$, disolved in a liquid solvent. The system is an ensemble system, rather than a single $N$-spin molecule.

This raises to some difficulty. Suppose an NMR implementation of a particular quantum algorithm. Then, the information should be given by the distribution of the random numbers detected through the measurement. However, signal coming from the NMR ensemble state is an averaged over the ensemble state. Average of random numbers clearly is useless. This problem has been solved by slightly modifying quantum algorithms in a way that the averaged result also gives the required outcome.

The other problem is that a density matrix representing the state of an ensemble system is completely described if all the elements are detected and known. However, NMR measurement of the operator $-i{\rm I}_x^i-{\rm I}_y^i$ selects specific entries in the density matrix, called {\it single quantum coherence} (SQC) elements. The SQC elements connect basis states, which differ by only one quantum of energy. Therefore, in order to get the total density matrix elements, several $90^\circ$ pulses are systematically applied in order to rotate each spin around the $\hat x$ and $\hat y$ axes to make observable all of the terms in the deviation of the density matrix from the identity. The process to get the total density matrix through the measurement is called {\it state tomography}. In a general scheme, without any prior knowledge on the state, the state tomography involves many experiments, then it is impractical for experiments involving large number of qubits.

In the following sections, we will show that a proper scheme to measure the entanglement of states is the one which gives the required information on the status of entanglement with a less number of experiments as compared to the state tomography. Even though, it is always possible, for small degrees of freedom, to measure the status of entanglement after state tomography, but the scheme which employs less number of experiments is of course more appreciated. 
\subsection{Decoherence}


The decoherence process of uncoupled nuclear spins is well described by combination of two phenomena: longitudinal and transverse relaxations \cite{S17}, \cite{S2}, \cite{S4}. These two processes are closely related to generalized amplitude damping and phase damping, respectively.

Relaxation of nuclear spins is caused by fluctuations in the magnetic field experienced by the spins. Whether the magnetic field fluctuations contribute to energy exchange with the bath or phase randomization depends on the time scale of the fluctuations. Roughly speaking, two processes are happened as follows. Fluctuations at resonance frequency of the nuclear spins lead to efficient energy exchange with the spins. However, slow fluctuations or fluctuations at zero frequency, give rise to phase randomization.

The issue of decoherence generally is very critical for any realization of a quantum computer. However, NMR is very much appreciated on this matter. Therefore, we are not going to get involved in this topic. The extended discussion might be found in literature, \cite{S17}, \cite{S2}, \cite{S4}. We only recall as stated above that two main relaxation processes have been explained as $T_1$, spin-lattice relaxation time and $T_2$, spin-spin relaxation time. They are measurable quantities e.g. by standard inversion recovery and Carr-Purcell-Meiboom-Gill pulse sequences, \cite{S15}.

Generally speaking $T_2\leq T_1$. For liquids with low viscosity $T_2\sim T_1$ and for viscous liquids $T_2\ll T_1$, due to long range of spin-spin interactions. In common organic liquids $T_2\sim T_1\sim 1$, while in noble gases $T_1$ is several hours. In a liquid NMR experiment, e.g. with carbon-13 labeled chloroform, proton and carbon represent $T_2< T_1$, of several seconds.

\section{What is the Problem With NMR?}
Through the above sections, we understand the reason that why NMR has been introduced as one of the most viable methods for realization of a quantum computing. liquid state NMR has been used extensively for implementing even relatively complicated quantum algorithms, \cite{S19}, \cite{S20}, \cite{S21}, \cite{S22}, \cite{S23}.

However, the status of NMR has not been completely approved compared to the other candidates, such as photons and ion traps, in that it treats an ensemble system composed of a large number of molecules. Under the current experimental conditions, the states are mixed rather than quantum pure states, \cite{S24}, \cite{S25}, \cite{S26}, \cite{S27}.

 Accordingly, nearly all previous attempts to demonstrate NMR quantum information processing (QIP) and quantum computations (QC) relied on pseudo-initialization of highly mixed states. However, with or without employing the pseudo-pure states, mathematical arguments refute the credibility of NMR QIP and QC except at very low spin temperatures by verifying separability of the states.

 The issue of entanglement of the states, has been the main focus of so many researches. Nevertheless, still the exact and unified solution has not been accepted widely. The separability of NMR mixed states, density matrix, is mathematically proved. But, there are also some discussions from the other side. For example, it has been claimed that a certain mixed state would be possible to be explained by several density matrices. More particularly, suppose a system of mixed states. Conventionally, we know that all the density matrix representations for this system are equivalent. Therefore, if a density matrix among them all has been proved to be separable, then the system of mixed state also is separable. This is because that the different representations cannot be distinguished from each other by any physical means. However, there has been some discussions also to show that this can not be always true. The different density matrices can be distinguished by showing the fluctuation, through measuring some particular observables. Then separability of a particular density matrix representation for a certain mixed state does not verify completely the separability of the mixed state, \cite{LZJ}.
 
 One more reliable way to solve the discrepencies, we believe that would be to show that any particualar implementation of a quantum algorithm by means of NMR is also accessible for a classical system. Therefore, in this manner, there would be no quantum beneficial to use the NMR system because the corresponding processing would be classical. The exact proof for all the cases is not still done. However, particularly, we can show that an implementation of a quantum algorithm, namely superdense coding, is possible with classical system.
 
 We show that with NMR mixed states, even if the states are definitely separable, still there is a considerably high probability to detect signals with appropriate signs which apparently imply transfer of two-bit message even though we encode only one qubit.  The reason for detection of the signals is proved to be because of the large number of molecules being involved in the NMR ensemble state. Therefore, a system composed of a huge number of molecules should be necessarily prepared in order to realize the transfer of two bits of information. Then, taking account the required number of resources, molecules, it is totally non-sense to claim on the quantum advantage over the classical counterpart. However, for an exact demonstration of SDC with NMR, in the sense that it can realize enhancement over the classical communication, nuclear spin polarization should be increased above a certain threshold, and this threshold coincides with the mathematical criterion for non-separability of the density matrix.
 \subsection{Ideal SDC with Nuclear Spins}
\begin{figure}
\begin{center}
\scalebox{0.6}
{\includegraphics[0cm,21cm][20cm,27cm]{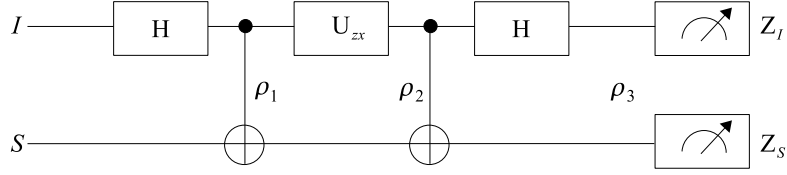}}
\end{center}
\caption{\label{Figure. 1} A quantum circuit for superdense coding. Two nuclear spins $I$ and $S$ are involved. After the first Hadamard and controlled-NOT, the spin $I$ is given to Bob and $S$ to Alice. Bob encodes the spin $I$ by applying the unitary transformation ${\rm U}_{zx}$ and then sends it off to Alice. Alice has now the two nuclear spins in her disposal. She applies the second controlled-NOT and Hadamard gates and measures the spin magnetizations. As a result, she obtains the encoded message, meaning that two-bit message is transferred while only a single spin has been encoded.}
\end{figure}

Let us consider a pair of nuclear spins $I=1/2$ and $S=1/2$ placed in a static magnetic field $B_0$, and suppose for a moment that the system is initially in a pure state $|\psi_0\rangle = | 00 \rangle$. Then, the procedure of SDC, whose quantum circuit is described in Figure \ref{Figure. 1}, is as follows. 
Firstly, the entangling operation is applied on the state of the nuclear spins. The entangling operation is composed of a Hadamard gate (H) on the $I$ spin and then a controlled-NOT gate (${\rm U_{cn}}$) whose control and target qubits are the $I$ and $S$ spins, respectively. The quantum state $ |\psi_1 \rangle $ after the entangling operation is represented as
\begin{equation}
\label{1}
  |\psi_1\rangle = {\rm U}_{\rm ent} |\psi_0\rangle =\frac{1}{\sqrt{2}} (|00\rangle + |11\rangle)= |\beta_{00}\rangle,
\end{equation}
where we have defined ${\rm U}_{\rm ent}={\rm U_{cn}} ({\rm H}_I\otimes {\rm I}_S)$. Here, $ | \beta_{00}\rangle$ is known as one of the four Bell states
\begin{equation}
\label{2}
|\beta _{zx}\rangle\equiv\frac{|0,x\rangle+(-1)^z|1,\bar{x}\rangle}{\sqrt 2},
\end{equation}
where $z, x=0, 1$ and $\bar{x} = 1-x$. Now, suppose that the nuclear spin $I$ is given to Bob and the other nuclear spin to Alice. Bob encodes a two-bit classical message $zx$ by applying the unitary operation ${\rm U}_{zx}={\rm Z}^z {\rm X}^x$ on the $I$ spin and then sends off the encoded qubit to Alice. The effect of the unitary transformation by Bob is to toggle $|\psi_1\rangle = |\beta_{00}\rangle$ into the other Bell state. That is,
\begin{eqnarray}
\label{3}
  |\psi_2\rangle = {\rm U}_{zx} |\beta_{00}\rangle
                 = |\beta_{zx}\rangle.
\end{eqnarray}
Then, Alice applies the disentangling operation, which is composed of a ${\rm U_{cn}}$ followed by ${\rm H}$. The state $|\psi_3\rangle$ after the disentangling operation is
\begin{equation}
\label{4}
  |\psi_3\rangle = {\rm U}_{\rm Bell}|\psi_2\rangle=|zx\rangle,
\end{equation}
where we defined ${\rm U}_{\rm Bell}=({\rm H}_I\otimes {\rm I}_S) {\rm U_{cn}}$. Finally, she performs measurement of the resultant magnetizations, ${\rm Z}_I$ and ${\rm Z}_S$ and extracts the results as $(-1)^z$ and $(-1)^x$.

If it was possible to execute all the above process in an NMR apparatus, then the signal intensities detected through the measurement in this ideal case of the fully polarized initial state would have been maximum and information on the choice of $ {\rm U}_{zx} $ could be extracted with a success probability of unity.   In this case,  the maximum entanglement is shared between Alice and Bob, which causes transfer of two classical bits of information from Bob to Alice via the $I$ spin alone. Communication by SDC is efficient by a factor of two as compared to the classical communication.

\subsection{Transfer of Two-Bit Message by Encoding a Single Spin in NMR}
NMR is a spectroscopy, manipulating the nuclear spins placed in a strong, typically 10 Tesla, static magnetic field $B_0$. The corresponding total spin Hamiltonian in an isotropic liquid state is composed of several terms, but the most important parts are considered as the Zeeman Hamiltonian and spin-spin interaction parts. So that we take
\begin{equation}
H=- \sum_{i=0}^{m-1} \frac{\hbar \omega_{i}}{2} {\rm Z}_i +\frac{1}{4} \sum_{i=0}^{m-2} \sum_{j>i} 2 \pi \hbar J_{ij} {\rm Z}_i {\rm Z}_j,
\end{equation}
where $\omega_{i}$ is the Larmor frequency and ${\rm Z}_i$ is the Pauli operator for the ${\rm z}$ component of the $i$th spin and $J_{ij}$ is the coupling constant between $i$th and $j$th spins in a molecule. In the high field approximation, which is valid for most practical cases, in the thermal equilibrium at temperature $T$, the polarization $\epsilon_i$ is given by
\begin{equation}
\label{epsilon}
\epsilon_i={\rm tanh} ( \frac{-\gamma_i\hbar B_0}{2k_{\rm B}T} ),
\end{equation}
where $\gamma_i$ is gyromagnetic ratio. Accordingly, the probabilities $p_i$ and $q_i$ of finding the $i$th spin in the states $|0\rangle$ and $|1\rangle$, respectively, are given by
\begin{equation}
\label{probability}
p_i=\frac{1+\epsilon_i}{2},\, \hspace{1cm} q_i=\frac{1-\epsilon_i}{2}.
\end{equation}

Then, the density matrix in thermal equilibrium is given by
\begin{equation}
\label{ron}
\rho_{\rm m-qubit}= \rho^1 \otimes ... \otimes \rho^m=\otimes_{i=1}^m\rho^{i},
\end{equation}
where $\rho^{i}$ is the density matrix for the $i$th nuclear spin in a molecule and is written as
\begin{equation}
\label{roi}
\rho^{i} =  p_i | 0 \rangle \langle 0 | + q_i | 1 \rangle \langle 1 |.
\end{equation}

At room temperature, the energy splitting between the ground state and the excited state is much smaller than the thermal energy and as a result, large number of spins (qubits) are in statistical mixture with low spin polarization. Accordingly, the information processing should be evaluated statistically.

For nuclear spins $I=1/2$ and $S=1/2$ the initial state is a mixed state described by partially polarized density matrix with the form of (\ref{ron}) and (\ref{roi}) with ${\rm m=2}$ as
\begin{eqnarray}
\label{roin}
\rho_{0}&=&(p_I|0\rangle\langle 0|+q_I|1\rangle\langle 1|)\otimes (p_S|0\rangle\langle 0|+q_S|1\rangle\langle 1|)\\ \nonumber
        &=&p_Ip_S|00\rangle\langle00|+p_Iq_S|01\rangle\langle01|+q_Ip_S|10\rangle\langle10|+q_Iq_S|11\rangle\langle11|.
\end{eqnarray}
After performing the entangling operation the state of the two qubits (\ref{roin}) changes to a general Bell diagonal state of the form
\begin{eqnarray}
\label{roshared}
\rho_{1}&=& {\rm U}_{\rm ent} \rho_0 {\rm U}^\dagger_{\rm ent}\\ \nonumber
        &=& p_Ip_S |\beta _{00}\rangle \langle \beta_{00}|+p_I q_S|\beta _{01}\rangle \langle \beta_{01}|\\ \nonumber 
        &+& q_Ip_S|\beta_{10}\rangle \langle \beta_{10}|+q_Iq_S|\beta _{11}\rangle \langle \beta_{11}|.
\end{eqnarray}
This is the shared state between Alice and Bob. Suppose that Bob owns the state of the nuclear spin $I$. Then, Bob applies the unitary operation ${\rm U}_{zx}$. The state $\rho_2$, after the encoding operation is still a general Bell diagonal state as follows
\begin{eqnarray}
\label{encoded}
\rho_{2}&=& {\rm U}_{zx} \rho_1 {\rm U}^\dagger_{zx}\\ \nonumber
        &=& p_Ip_S |\beta _{z, x}\rangle \langle \beta_{z, x}|+p_Iq_S|\beta _{z, {\bar x}}\rangle \langle \beta_{z, {\bar x}}|\\ \nonumber
        &+& q_Ip_S|\beta _{{\bar z}, x}\rangle \langle \beta_{{\bar z}, x}|+q_Iq_S|\beta _{{\bar z}, {\bar x}}\rangle \langle \beta_{{\bar z}, {\bar x}}|.
\end{eqnarray}
He hands over the nuclear spin to Alice, who applies the disentangling operation.  Then, the state $\rho_2$ changes into
\begin{eqnarray}
\rho_{3}&=& {\rm U}_{\rm Bell} \rho_2 {\rm U}^\dagger_{\rm Bell}\\ \nonumber
        &=& p_Ip_S |z, x\rangle \langle z, x|+p_Iq_S|z, {\bar x}\rangle \langle z, {\bar x}|\\ \nonumber
        &+& q_Ip_S|{\bar z}, x\rangle \langle {\bar z}, x|+q_Iq_S|{\bar z}, {\bar x}\rangle \langle {\bar z}, {\bar x}|\\ \nonumber
        &=& (p_I|z\rangle \langle z|+q_I|{\bar z}\rangle \langle {\bar z}|)\otimes (p_S|x\rangle \langle x|+q_S|{\bar x}\rangle \langle {\bar x}|).
\end{eqnarray}
Measurement of the spin magnetizations $ {\rm Z}_I $ and $ {\rm Z}_S $ is done on the total ensemble state composed of $n$ molecules. Then, measurement on the product state $ \otimes_{i=0}^n \rho^{(i)}_3$ is composed of separate but simultaneous measurements on the spin magnetizations and is represented as $\sum_{i=1}^n {\rm Z}_I^{(i)}\otimes {\rm Z}_S^{(i)}$. Note, that $\rho^{(i)}_3$ stands for the density matrix $\rho_3$ of the $i$th molecule. Measurement of the spin magnetizations gives results as binomial probability distributions over $(-n, -n+1, ..., -1, 0, 1, ..., n-1, n)$, with the mean values to be as follows
\begin{eqnarray}
\label{mean}
\mu_I&=&(-1)^z n p_I+(-1)^{\bar z} n q_I=(-1)^z n\epsilon_I,\\ \nonumber
\mu_S&=&(-1)^x n p_S+(-1)^{\bar x} n q_S=(-1)^x n\epsilon_S.
\end{eqnarray}
Let us make an assumption that $z, x=0$, which does not influence the generality of the discussion. The corresponding variances are characterized by
\begin{eqnarray}
\label{var}
\sigma_I^2&=&4np_Iq_I=n(1-\epsilon_I^2),\\ \nonumber
\sigma_S^2&=&4np_Sq_S=n(1-\epsilon_S^2).
\end{eqnarray}
Recall that $\sigma_I$ and $\sigma_S$ are essentially the widths of the range over which the outcomes are distributed around the mean values (\ref{mean}). The relative widths of distributions are characterized by
\begin{eqnarray}
\frac{\sigma_I}{\mu_I}=\frac{\sqrt{n(1-\epsilon_I^2)}}{n\epsilon_I}\approx \frac{1}{\epsilon_I\sqrt n},\\ \nonumber
\frac{\sigma_S}{\mu_S}=\frac{\sqrt{n(1-\epsilon_S^2)}}{n\epsilon_S}\approx \frac{1}{\epsilon_S\sqrt n}.
\end{eqnarray}
Then it is clear that the relative widths decrease as $\sqrt n$ with increasing  the number of molecules $n$. Thus, the greater the number of molecules, the more likely it is that an observation gives a result which is relatively close to the mean values (\ref{mean}).

Now, it is also required to calculate the corresponding error probability of detection of a negative value for $z=0$. For simplicity, let us consider only ${\rm Z}_I^{(i)}$, which is the measurement on the spin magnetization of the nuclear spin $I$ in the $i$th molecule. It should be clear at the moment that results are applicable to the signals detected through measurement on the other spin magnetization in the molecule, ${\rm Z}_S^{(i)}$. The error probability is defined as
\begin{equation}
P_e=P(\sum_{i=1}^n{\rm Z}^{(i)}_I<0|z=0).
\end{equation}

We would like to calculate the error probability to show that for a range of $n$ of the current NMR experiments, this quantity is negligible. From the DeMoivre and Laplace theorem, \cite{S28}, if $n$ is large enough then generally we have
\begin{equation}
P\left\{\alpha<\frac{\sum_i^n{\rm Z}^{(i)}_I-{\mu_I}}{{\sigma_I}}<\beta\right\}\approx\frac{1}{\sqrt{2\pi}}\int_{\alpha}^{\beta}e^{-\frac{x^2}{2}}dx.
\end{equation}

For the error probability $P_e$, therefore
\begin{eqnarray}
P_e&=&P\left\{-\infty<\frac{\sum_i^n{\rm Z}^{(i)}_I-{\mu}_I}{\sigma_I}<-\frac{\mu_I}{\sigma_I}\right\}\\ \nonumber
   &\approx&\frac{1}{\sqrt{2\pi}}\int_{-\infty}^{-\mu_I/\sigma_I}e^{-\frac{x^2}{2}}dx\\ \nonumber
   &\approx&\frac{1}{\sqrt{2\pi}}e^{-\frac{{\mu_I}^{2}}{2{\sigma_I}^2}}\frac{{\sigma_I}}{\mu_I}\\ \nonumber
   &\approx&\frac{1}{\sqrt{2\pi}}e^{-\frac{({n\epsilon_I^2})}{2}}\frac{1}{({\sqrt n \epsilon_I})}.
\end{eqnarray}
If the number of molecules are large enough, the result of the measurement on the spin magnetizations $ {\rm Z}_I $ and $ {\rm Z}_S $ on the ensemble state gives results very close to the mean values (\ref{mean}) with negligible error probabilities. In a conventional NMR experiment, the number of molecules are $n\sim 10^{18}$, with low spin polarization $\epsilon\sim 10^{-5}$. Then we calculate $P_e\ll 10^{-100}$, to be negligible. Therefore, in an NMR apparatus, if signals can be detected despite of the low spin polarization the choice $zx$ can be evaluated with probability of almost one through the signs of the signals (\ref{mean}). 

Detection of signals depends on the NMR apparatus. The NMR signal intensity is defined by amplitude
\begin{equation}
\label{S}
V_{\rm S}=\frac{1}{4}\sqrt{(Q/V)\mu_0R\omega_I}\hbar \gamma_In\epsilon_I,
\end{equation}
and the noise is determined through Nyquist formula, \cite{S29}
\begin{equation}
\label{N}
V_{\rm N}=\sqrt{4k_{\rm B}TR\Delta\nu},
\end{equation}
where $Q$ is the quality factor of the resonance coil, $V$ is the volume of the coil, $R$ is the resistance, $\omega_I$ is the Larmor frequency of the nuclear spin $I$, and $\Delta\nu$ is the amplifier bandwidth. The noise which is defined by (\ref{N}) is entirely classical noise generated by the equilibrium fluctuations of the electric current inside an electrical conductor. Nevertheless, for an NMR experiment with two qubit liquid ensemble at room temperature with $V=1{\rm cm}^3$ and $Q=10^3$, the number of required molecules is bounded by $N>10^{16}$, \cite{S30}.

Therefore, in the case which we are facing, $n$ is large to get strong enough signal intensities to be detectable with an NMR apparatus. Deviations are very small (\ref{var}) and the signals are very close to the mean values (\ref{mean}) with the error probabilities to be negligible. Thus the signs of the detected signals give information on the choice of $zx$, and hence the encoded two-bit message, regardless of the separability due to the very low polarized initial states. Then we emphasize the results of the measurement of the spin magnetizations ${\rm Z}_I$ and ${\rm Z}_S$ to be as follows
\begin{equation}
\label{result}
\langle {\rm Z}_I \rangle=(-1)^{z}\epsilon_I,\hspace{1.5cm} \langle{\rm Z}_S \rangle=(-1)^{x}\epsilon_S.
\end{equation}

We note that this does not necessarily mean that the experiment described here is a demonstration of SDC with NMR mixed states, even though the signs of the detected signals appear to give the two-bit message while the $I$ spin alone has been applied by encoding pulses. SDC relies on entanglement of states in order to realize transfer of two-bit information by only encoding a single qubit. In the case of the NMR experiment, as explained here, the large number of molecules are inevitably required for detection of encoded two-bit message. Then taking account the required number of molecules it is in principle erroneous to call the corresponding NMR experiment a demonstration of SDC.
 \section{Appropriate Entanglement Witness for NMR}
Given the density matrix (\ref{encoded}), we obtain a probability for a successful SDC to be $p_Ip_S$, with which a two-bit message can be transferred inside any individual single molecule. Therefore, for this experiment to represent the quantum information advantage and to outperform the classical one-bit communication through one-qubit channel, the success probability, $p_Ip_S$, must exceed $50\%$. This is because two-bit information can be obtained by one-bit classical communication and a one-bit of random guess, which comes true with a probability of $1/2$. Then SDC is beyond the classical achievements only if the inequality
\begin{equation}
\label{16}
p_Ip_S>1/2
\end{equation}
is satisfied. It is worth noting that this condition exactly coincides with the condition for the non-separability of $\rho_2$ derived from e.g. the negativity criterion.

Therefore, only when $F$, to be defined as follows, imposes a negative value, the NMR SDC is successful and there exists entanglement.
\begin{equation}
\label{17}
F \equiv 1/2 - p_I p_S.
\end{equation}

Detection of entanglement through finding a negative value for an observable is reminiscent of entanglement witness. Entanglement witness is a Hermitian operator $W=W^{\dagger}$ which has positive mean values for all separable states $\rho$, $\mathrm{Tr}(W\rho)>0$, but a negative mean value for at least one entangled state $\sigma_{\rm ent}$, $\mathrm{Tr}(W\sigma_{\rm ent})<0$. In other words, entanglement is detected if a negative mean value is obtained through the measurement of the entanglement witness.

Using the spin polarizations, we rewrite $F$ as
\begin{eqnarray}
\label{20}
F&=&\frac{1}{2} - \frac{1}{4} (1+\epsilon_I)(1+\epsilon_S)\\ \nonumber
 &=&\frac{1}{2} - \frac{1}{4} (1+|\langle {\rm Z}_I \rangle |)(1+|\langle {\rm Z}_S \rangle | ),
\end{eqnarray}
where we used (\ref{result}). The absolute values are required for the evaluation of the function $F$ for different choices of ${zx}$.

Measurement on the state $\rho_3$ with the observables ${\rm Z}_I$ and ${\rm Z}_S$ is equivalent to measurement on the state $\rho_2$ (or $\rho_1$ in the special case of $x, z=0$) in the Bell basis, because
\begin{eqnarray}
\label{23}
\langle {\rm Z}_I \rangle&=&\mathrm{Tr} \rho_3 ( {\rm Z}_I \otimes {\rm I}_S)\\ \nonumber
                         &=&\mathrm{Tr} \rho_2 ( {\rm X}_I \otimes {\rm X}_S)=\langle W_1 \rangle 
\end{eqnarray}
\begin{eqnarray}
\label{24}
\langle {\rm Z}_S \rangle&=&\mathrm{Tr} \rho_3 ( {\rm I}_I \otimes {\rm Z}_S)\\ \nonumber
                         &=&\mathrm{Tr} \rho_2 ( {\rm Z}_I \otimes {\rm Z}_S)=\langle W_2 \rangle ,
\end{eqnarray}
where the two observables $W_1$ and $W_2$ are defined as follows
\begin{equation}
\label{25}
W_1={\rm U}_{\rm Bell}^{\dagger} ( {\rm Z}_I \otimes {\rm I}_S) {\rm U}_{\rm Bell}={\rm X}_I \otimes {\rm X}_S, 
\end{equation}
\begin{equation}
\label{26}
W_2={\rm U}_{\rm Bell}^{\dagger} ( {\rm I}_I \otimes {\rm Z}_S) {\rm U}_{\rm Bell}={\rm Z}_I \otimes {\rm Z}_S.
\end{equation}

From (\ref{20}), (\ref{23}) and (\ref{24}), then $F$ is further rewritten as
\begin{equation}
\label{27}
F \equiv f(\langle W_1 \rangle,\langle W_2 \rangle)=\frac{1}{2} - \frac{1}{4} (1+|\langle W_1 \rangle |)(1+|\langle W_2 \rangle | ).
\end{equation}

Measurement of the two observables $W_1$ and $W_2$ on the ensemble system described by $\rho_2$ is related to the measurement on the state $\rho_3$ of the spin magnetization, which affords information on the spin polarizations. Therefore, separate and simultaneous measurement of the observables $W_1$ and $W_2$ is possible in a single experiment, and tells the existence of entanglement. That is, if $\langle W_1 \rangle$ and $\langle W_2 \rangle$ satisfy
\begin{equation}
\label{28}
F \equiv f(\langle W_1 \rangle,\langle W_2 \rangle)<0,
\end{equation}
then the state is entangled. In this sense, $W_1$ and $W_2$ are regarded as a new class of entanglement witnesses. These observables are easily measurable in a single run NMR experiment and give information on the status of entanglement through quantitative evaluation of the function $F$ (\ref{27}).

We also note, that entanglement can be detected in principle through measuring the conventional entanglement witness.  For $\rho_2$, entanglement witness is derived by the conventional approach as
\begin{equation}
\label{32}
W=\frac{1}{4} ({\rm I}_I \otimes {\rm I}_S+(-1)^{\bar z}{\rm X}_I \otimes {\rm X}_S +(-1)^{\bar z}(-1)^{\bar x}{\rm Y}_I \otimes {\rm Y}_S +(-1)^{\bar x}{\rm Z}_I \otimes {\rm Z}_S).
\end{equation}

If we assume the ability of implementation of any form of the unitary transformations with NMR pulse sequences, then the conventional entanglement witness as defined by (\ref{32}) also can be measured in a single run and by measuring the spin magnetizations (see appendix A). We emphasize here, that the scheme which is introduced in this contribution for detection of entanglement, still has advantage in the sense that for different choices of $z$ and $x$ it is applicable by just evaluation the function with the absolute values. In other words, there is no need to change the experimental operations. However, the conventional entanglement witness even though is proved here to be measurable in a single run NMR experiment by only measuring the spin magnetizations, still requires different pre-applied unitary transformations, which depend on the choices of $z$ and $x$.
\subsection{Single Run Detection of the Conventional Entanglement Witness}
\paragraph{Theorem:}
Suppose that we have the observable 
$\tilde W=U^\dagger \tilde W_o U=U^\dagger(aZ_I\otimes I_S+bI_I\otimes Z_S+cI\otimes I)U$ with arbitrary coefficients $a,b,c \in \mathbf{R}$ and a unitary transformation $U\in \mathrm{U}(4)$, and have the conventional entanglement witness
$W=\frac{1}{4}(I_I\otimes I_S + (-1)^{\bar z}X_I\otimes X_S + (-1)^{\bar z}(-1)^{\bar x}Y_I\otimes Y_S+(-1)^{\bar x}Z_I\otimes Z_S)$. Then there exist sets of $a,b,c$, and $U$ such that $\mathrm{Tr} \rho_2(p_I,p_S) \tilde W = \mathrm{Tr} \rho_2(p_I,p_S) W$ for all possible $p_I, p_S\ge 0.5$ when we specify $(z,x)$ to one of $(0,0),(0,1),(1,0),(1,1)$.

\paragraph{Proof:}
Let us write $U=\sum_{k,l = 0,0}^{3,3}u_{kl}|k\rangle\langle l|$.
In this proof, we make use of the following equations: for
$x=z=0$, we have
\begin{equation}
 W=\frac{1}{2}
 \begin{pmatrix}
0&0&0&-1\\0&1&0&0\\0&0&1&0\\-1&0&0&0
\end{pmatrix}.
\end{equation}
$\tilde W_o$ is written as
\begin{equation}
 \tilde W_o=\begin{pmatrix}
a+b+c&0&0&0\\0&a-b+c&0&0\\0&0&-a+b+c&0\\0&0&0&-a-b+c
\end{pmatrix}.
\end{equation}

In the following sentences, we restrict the problem to the case of $x=z=0$.

The eigenvalues of $W$ are different to the ones for $\tilde W$. Then the two observables are essentially different to each other. However they can still give the same value after calculating the traces.

The problem to be solved is as follows:
\paragraph{Problem for $x=z=0$:}
Suppose that we have the observable 
$\tilde W=U^\dagger \tilde W_o U=U^\dagger(a W_1 + b W_2 + c)U=
U^\dagger(aZ_I\otimes I_S+bI_I\otimes Z_S+cI\otimes I)U$
with coefficients $a,b,c \in \mathbf{R}$
and a unitary transformation $U\in \mathrm{U}(4)$, and have the 
conventional entanglement witness
$W=\frac{1}{4}(I_I\otimes I_S - X_I\otimes X_S + 
Y_I\otimes Y_S - Z_I\otimes Z_S)$.
Then find a set of $a,b,c$, and $U$ such that
$\mathrm{Tr} \rho_2(p_I,p_S) \tilde W =
\mathrm{Tr} \rho_2(p_I,p_S) W$ for all possible $p_I, p_S\ge 0.5$.
\paragraph{Solution:}
Firstly, consider a maximally mixed state 
$\rho_2(\frac{1}{2},\frac{1}{2})=I\otimes I/4$ for the input. We have
\begin{equation}
 \mathrm{Tr}\rho_2(\frac{1}{2},\frac{1}{2})W=\frac{1}{4},
\end{equation}
and 
\begin{equation}
 \mathrm{Tr}\rho_2(\frac{1}{2},\frac{1}{2})\tilde W
=\mathrm{Tr}U\rho_2(\frac{1}{2},\frac{1}{2})U^\dagger\tilde W_o
=c.
\end{equation}

Therefore $c=1/4$ holds. We have
\begin{equation}
 \tilde W_o=\begin{pmatrix}
\alpha&0&0&0\\0&\beta&0&0\\0&0&-\beta&0\\0&0&0&-\alpha
\end{pmatrix}+\frac{1}{4}I\otimes I,
\end{equation}
where $\alpha=a+b$ and $\beta=a-b$.

Secondly, we change the problem to an equivalent problem of finding eigenvalues. Consider the unitary transformation
\begin{equation}
 U_{HCN}=\frac{1}{\sqrt{2}}\begin{pmatrix}
1&0&0&0\\0&1&0&0\\0&0&0&1\\0&0&1&0
\end{pmatrix}\begin{pmatrix}
1&0&1&0\\0&1&0&1\\1&0&-1&0\\0&1&0&-1
\end{pmatrix}=\frac{1}{\sqrt{2}}\begin{pmatrix}
1&0&1&0\\0&1&0&1\\0&1&0&-1\\1&0&-1&0
\end{pmatrix}.
\end{equation}

With this transformation, we change the basis:
\begin{equation}
 \mathrm{Tr}\rho_1(p_I,p_S)W'=\mathrm{Tr}\rho_2(p_I,p_S)W=\frac{1}{2}-p_Ip_S
\end{equation}
with
\begin{equation}
 \rho_1(p_I,p_S)=U_{HCN}^\dagger\rho_2(p_I,p_S)U_{HCN}
=\begin{pmatrix}p_Ip_S&0&0&0\\0&p_Iq_S&0&0\\
0&0&q_Ip_S&0\\0&0&0&q_Iq_S\end{pmatrix},
\end{equation}
and
\begin{equation}
 W'=U_{HCN}^\dagger W U_{HCN}=\begin{pmatrix}
-\frac{1}{2}&0&0&0\\0&\frac{1}{2}&0&0\\0&0&\frac{1}{2}&0\\
0&0&0&\frac{1}{2}
\end{pmatrix}.
\end{equation}

We also have
\begin{equation}
 \mathrm{Tr}\rho_1(p_I,p_S)V^\dagger\tilde W_oV=
\mathrm{Tr}\rho_2(p_I,p_S)\tilde W
\end{equation}
with $V=UU_{HCN}$. We write that $V=\sum_{kl}v_{kl}|k\rangle\langle l|$.

Now the problem is to find a set of $\alpha,\beta$ and $V$ that satisfy
\begin{equation}
 \mathrm{Tr}\rho_1(p_I,p_S)(W'-V^\dagger \tilde W_oV)=0.
\end{equation}

Because $\rho_1(p_I,p_S)$ is a diagonal matrix, the above equality hold 
for $\forall p_I,p_S$ if and only if the diagonal elements of
$(W'-V^\dagger \tilde W_oV)$ are $0$. This condition is expressed in 
the next equation:
\begin{equation}
 W'-V^\dagger \tilde W_oV = A,
\end{equation}
where 
\begin{equation}
 A=\begin{pmatrix}
0&a_{01}&a_{02}&a_{03}\\
a_{01}^*&0&a_{12}&a_{13}\\
a_{02}^*&a_{12}^*&0&a_{23}\\
a_{03}^*&a_{13}^*&a_{23}^*&0
\end{pmatrix}.
\end{equation}

This leads to that
\begin{equation}\begin{split}
 \begin{pmatrix}-\frac{3}{4}&0&0&0\\0&\frac{1}{4}&0&0\\
0&0&\frac{1}{4}&0\\0&0&0&\frac{1}{4}
\end{pmatrix}
-V^\dagger
\begin{pmatrix}
\alpha&0&0&0\\0&\beta&0&0\\0&0&-\beta&0\\0&0&0&-\alpha
\end{pmatrix}
V=A,\\
\begin{pmatrix}
-\frac{3}{4}&-a_{01}&-a_{02}&-a_{03}\\
-a_{01}^*&\frac{1}{4}&-a_{12}&-a_{13}\\
-a_{02}^*&-a_{12}^*&\frac{1}{4}&-a_{23}\\
-a_{03}^*&-a_{13}^*&-a_{23}^*&\frac{1}{4}
\end{pmatrix}
=V^\dagger
\begin{pmatrix}
\alpha&0&0&0\\0&\beta&0&0\\
0&0&-\beta&0\\0&0&0&-\alpha
\end{pmatrix}V.
\end{split}\end{equation}
Therefore the problem has been led to an equivalent problem:\\
\paragraph{Equivalent problem for $x=z=0$:} Find an Hermitian matrix
$B=\begin{pmatrix}
-3&b_{01}&b_{02}&b_{03}\\
b_{01}^*&1&b_{12}&b_{13}\\
b_{02}^*&b_{12}^*&1&b_{23}\\
b_{03}^*&b_{13}^*&b_{23}^*&1
\end{pmatrix}$
whose eigenvalues are written as
$\pm 4\alpha,\pm 4\beta$ where
$\alpha,\beta\in\mathbf{R}$.

It is obvious that this problem has solutions because
$\mathrm{Tr}B=\mathrm{Tr\,diag}(4\alpha,4\beta,-4\beta,-4\alpha)=0$
and any unitary transformation in $\mathrm{U(4)}$ preserves the trace.\\

To find an explicit values of $a$, $b$, and the elements of $U$,
we solve the equation:
\begin{equation}\begin{split}
 B&=V^\dagger \mathrm{diag}(4\alpha,4\beta,-4\beta,-4\alpha) V\\
&=\sum_kA_k\begin{pmatrix}
|v_{k0}|^2&v_{k1}v_{k0}^*&v_{k2}v_{k0}^*&v_{k3}v_{k0}^*\\
v_{k0}v_{k1}^*&|v_{k1}|^2&v_{k2}v_{k1}^*&v_{k3}v_{k1}^*\\
v_{k0}v_{k2}^*&v_{k1}v_{k2}^*&|v_{k2}|^2&v_{k3}v_{k2}^*\\
v_{k0}v_{k3}^*&v_{k1}v_{k3}^*&v_{k2}v_{k3}^*&|v_{k3}|^2
\end{pmatrix},
\end{split}\end{equation}
where $A_0=4\alpha,A_1=4\beta,A_2=-4\beta,A_3=-4\alpha$.
This leads to the equations:
\begin{align}
&4\alpha|v_{00}|^2+4\beta|v_{10}|^2-4\beta|v_{20}|^2-4\alpha|v_{30}|^2=-3,\\
&4\alpha|v_{01}|^2+4\beta|v_{11}|^2-4\beta|v_{21}|^2-4\alpha|v_{31}|^2=1,\\
&4\alpha|v_{02}|^2+4\beta|v_{12}|^2-4\beta|v_{22}|^2-4\alpha|v_{32}|^2=1,\\
&4\alpha|v_{03}|^2+4\beta|v_{13}|^2-4\beta|v_{23}|^2-4\alpha|v_{33}|^2=1.
\end{align}
Suppose that $\alpha\ge\frac{3}{4}$ and $\beta=0$. Then we have the system of
algebraic equations:
\begin{align}
&4\alpha|v_{00}|^2-4\alpha|v_{30}|^2=-3,\\
&4\alpha|v_{01}|^2-4\alpha|v_{31}|^2=1,\\
&4\alpha|v_{02}|^2-4\alpha|v_{32}|^2=1,\\
&4\alpha|v_{03}|^2-4\alpha|v_{33}|^2=1.
\end{align}
As far as we only want to show the possibility of the measurement of the conventional entanglement witness by only measuring the spin polarizations in a single run NMR experiment, it is enough to give an example and to avoid the unnecessary complexities due to a general proof.

An example $V_{ex}(\alpha)$ of $V$, as a function of $\alpha$, is written as
\begin{equation}
 V_{ex}(\alpha)=\begin{pmatrix}
0&\frac{1}{\sqrt{3}}&
\frac{1}{\sqrt{3}}e^{i\pi2/3}&\frac{1}{\sqrt{3}}e^{-i\pi2/3}\\

0&\frac{1}{\sqrt{3}}&
\frac{1}{\sqrt{3}}e^{-i\pi2/3}&\frac{1}{\sqrt{3}}e^{i\pi2/3}\\

-\frac{\sqrt{4\alpha - 3}}{2\sqrt{\alpha}}&\frac{1}{2\sqrt{\alpha}}&
\frac{1}{2\sqrt{\alpha}}&\frac{1}{2\sqrt{\alpha}}\\

\frac{\sqrt{3}}{2\sqrt{\alpha}}&\frac{\sqrt{4\alpha - 3}}{2\sqrt{3\alpha}}&
\frac{\sqrt{4\alpha - 3}}{2\sqrt{3\alpha}}&\frac{\sqrt{4\alpha - 3}}{2\sqrt{3\alpha}}
\end{pmatrix}.
\end{equation}

Therefore an example of the set of $a$, $b$, and $U$ as functions of
$\alpha$ is that
\begin{equation}
 a_{ex}(\alpha)=b_{ex}(\alpha)=\frac{\alpha}{2}, U_{ex}(\alpha)=V_{ex}(\alpha)U_{HCN}^\dagger.
\end{equation}

Suppose that the state to be measured is a maximally entangled state. Measurement with the conventional witness observable $W$ gives $\langle W \rangle =-1/2$. We impose the observable $\tilde W$ to give the same value. We observe that for the maximally entangled state, ${\rm Z}_I$ and ${\rm Z}_S$ should take $\pm 1$. Recall that generally $c=1/4$. Then, as an example, we can choose $a_{\rm ex}=b_{\rm ex}=3/8$ and the corresponding unitary transformation is obtained as follows
\begin{equation}
 V_{ex}(\frac{3}{4})=\begin{pmatrix}
0&\frac{1}{\sqrt{3}}&
\frac{1}{\sqrt{3}}e^{i\pi2/3}&\frac{1}{\sqrt{3}}e^{-i\pi2/3}\\

0&\frac{1}{\sqrt{3}}&
\frac{1}{\sqrt{3}}e^{-i\pi2/3}&\frac{1}{\sqrt{3}}e^{i\pi2/3}\\

0&\frac{1}{\sqrt{3}}&
\frac{1}{\sqrt{3}}&\frac{1}{\sqrt{3}}\\

1&0&0&0
\end{pmatrix}.
\end{equation}

For different choices of ${zx}$, the elements of the matrix $A$ should be changed in a way that the unsimilar diagonal element $-3/4$ stands for the matrix component $A_{x+2z, x+2z}$. Then similar calculation shows that the components of ${\rm U}_{\rm ex}$ for different choices of ${zx}$ interchange according to each case.

Therefore, we showed that it is also possible to decompose the conventional entanglement witness to the separate but simultaneous measurement on the spin magnetizations. However, we emphasize that still the entanglement witness, which we introduced in this work is more handy as it covers all the cases for different $z$ and $x$, by just evaluation of the function $F$ with the absolute values without additional requirements on the different experimental operations. Whereas, as far as the conventional entanglement witness is concerned, different sets of $a$, $b$ and more over different unitary transformations are required for different choices of $z$ and $x$. In other words, by only changing the signs of the absolute values $a$ and $b$ but with a fixed unitary transformation, generally it is impossible to get the equality for the expectation values of $W$ and $\tilde W$. This can be proved by contradiction.

\paragraph{Proposition:}
We already have the density matrix
$\rho_1=\mathrm{diag}(p_Ip_S,p_Iq_S,q_Ip_S,q_Iq_S)$
and the conventional entanglement witness family 
${W_1}'=\mathrm{diag}(-\frac{1}{2},\frac{1}{2},\frac{1}{2},\frac{1}{2})$,
${W_2}'=\mathrm{diag}(\frac{1}{2},-\frac{1}{2},\frac{1}{2},\frac{1}{2})$,
${W_3}'=\mathrm{diag}(\frac{1}{2},\frac{1}{2},-\frac{1}{2},\frac{1}{2})$,
${W_4}'=\mathrm{diag}(\frac{1}{2},\frac{1}{2},\frac{1}{2},-\frac{1}{2})$
for $\rho_1$. Suppose that there is a specified-and-fixed unitary 
transformation $U$ and coefficients $a,b\in\mathrm{R}$ that satisfy
\begin{equation}\label{eqw1}
 \mathrm{Tr}\rho_1{W_1}'=\mathrm{Tr}\rho_1(aU^\dagger Z\otimes IU
+bU^\dagger I\otimes ZU+\frac{1}{4}I\otimes I).
\end{equation}

Then, it is impossible to satisfy the equality
\begin{equation}\label{eqw2}
 \mathrm{Tr}\rho_1{W_k}'=\mathrm{Tr}\rho_1(a'U^\dagger Z\otimes IU
+b'U^\dagger I\otimes ZU+\frac{1}{4}I\otimes I)
\end{equation}
for $\forall k\in\{2,3,4\}$ by using the coefficients 
$a'=\pm a$ and $b'=\pm b$.

\paragraph{Proof:} We prove the proposition by using a proof by
contradiction.

Assume that both of Eqs. (\ref{eqw1}) and (\ref{eqw2})
hold. Let us denote the diagonal elements of $U^\dagger Z\otimes IU$
by $x_{00},x_{11},x_{22},x_{33}$ and the diagonal elements of
$U^\dagger I\otimes ZU$ by $y_{00},y_{11},y_{22},y_{33}$.
We have the system of equations led from Eq. (\ref{eqw1}):
\begin{align}
&\frac{1}{4}+ax_{00}+by_{00}=-\frac{1}{2},\\
&\frac{1}{4}+ax_{11}+by_{11}=\frac{1}{2},\\
&\frac{1}{4}+ax_{22}+by_{22}=\frac{1}{2},\\
&\frac{1}{4}+ax_{33}+by_{33}=\frac{1}{2}.
\end{align}
Setting $k=4$ in Eq. (\ref{eqw2}) leads to the system of equations:
\begin{align}
&\frac{1}{4}+a'x_{00}+b'y_{00}=\frac{1}{2},\\
&\frac{1}{4}+a'x_{11}+b'y_{11}=\frac{1}{2},\\
&\frac{1}{4}+a'x_{22}+b'y_{22}=\frac{1}{2},\\
&\frac{1}{4}+a'x_{33}+b'y_{33}=-\frac{1}{2}.
\end{align}
These two systems of equations lead to that
\begin{align}
&(a+a')x_{00}+(b+b')y_{00}=-\frac{1}{2},\label{eqw3}\\
&(a+a')x_{11}+(b+b')y_{11}=\frac{1}{2},\\
&(a+a')x_{22}+(b+b')y_{22}=\frac{1}{2},\\
&(a+a')x_{33}+(b+b')y_{33}=-\frac{1}{2}.
\end{align}
The candidates of $(a',b')$ for $k\in\{2,3,4\}$ are 
$(a,-b)$, $(-a,b)$, and $(-a,-b)$. Hence each of the candidates must
satisfy the above system of equations if the assumption is correct.
Nevertheless, when $(a',b')=(-a,-b)$, Eq. (\ref{eqw3}) does not hold.
The assumption has led to a contradiction.

We conclude as follows. The conventional entanglement witness is measurable in a single NMR experiment and by only measuring the spin magnetizations, provided that any required unitary transformation can be applied prior to the measurement. However, the unitary transformation has to be changed in accordance to the choices of $z$ and $x$ and in this sense the new entanglement witness, which we introduced in this contribution is proved to have its favorite advantage.
\section{Conclusion}
Liquid state NMR with very low nuclear spin polarizations prohibits the existence of entanglement. Then, with this physical system, it is absolutely impossible to demonstrate a faithful non-local quantum information processing, which requires entangled states. Although two-bit information is correctly detected in NMR SDC experiment, it is not based on the existence of entanglement but relys on the large number of molecules.

For a completely reliable demonstration of NMR SDC within a single molecule, spin polarization should inevitably be enhanced over a certain threshold and this threshold coincides with the condition for non-separability of the states. According to the results, we introduced a new class of entanglement witnesses, with several advantages particularly for NMR. The introduced entanglement witness is measurable in a single run experiment and generally is applicable for all the states without any requirement on any extra experimental operation. Detection of entanglement through the conventional entanglement witness is also proved to be possible in a single NMR experiment, however for different states under investigation, it requires different pre-application of somehow complicated unitary transformations.

\vspace{5cm}
\chapter{ENDOR-Based Quantum Computing}
\section{Introduction}
In this thesis, the main purpose for our study has been mainly understanding  the concept of entanglement. As a theoretical point of view entanglement has been studied in chapters 1 and 2. Furthermore, we focused on a true realization of entanglement. Current NMR, as it is usually used for QC is studied in chapter 3 and has been shown to be far beyond the proper apparatus that can give, at least easily, quantum entanglement. Enhancement of the spin polarizations in NMR seems to be in reach considering the amount of researches in this field, for example see \cite{Anwar} and references therein. Though, we are looking for a rather bit different approach for realizing quantum entanglement. In order to get entanglement, spin polarizations should be enhanced. Electron spin, because of larger gyromagnetic ratio, three orders of magnitude larger than nuclear spin polarization in equilibrium, can be very useful in this regard. At the same time we would like to keep nuclear spins because of their inevitable advantages  for QC that have been proved in NMR QC. The spin technology which satisfies our interests and involves electron and nuclear spins is called Electron Nuclear DOuble Resonance, ENDOR.

Recall that NMR is a spectroscopy with high resolution, but rather low sensitivity. On the other hand, electron paramagnetic resonance, EPR is a highly sensitive spectroscopy but it does not represent much high resolution. ENDOR is a double resonance technique that combines the high resolution and nuclear selectivity of an NMR experiment with the high sensitivity of an EPR experiment.

In ENDOR, first developed by Feher in 1956, \cite{S12}, a change in the intensity of an EPR signal is observed as an RF frequency on nuclear resonance applied to the system. In order to get ENDOR, we first perform EPR. Then, the magnetic field is set and fixed at the resonance position of the EPR absorption signal. For the cw mode of the experiment, high power microwave is applied, giving rise to partial saturation of the EPR signal. For low power experiments, usually the EPR signal grows with the square root of the microwave power. But, if the power of microwave increases, the signal diminishes. This is called saturation and is important to be fulfilled to get ENDOR. Actually, ENDOR is desaturation of EPR by applying RF irradiation. As a result of desaturation, EPR signal intensity is recovered. The corresponding changes in the observed EPR signal is called ENDOR, \cite{H1}, \cite{H2}, \cite{H3}, \cite{H4}. 

Indeed, the double resonance experiment, such as ENDOR, requires more efforts than simpler EPR or NMR experiments, but ENDOR comes with excellent reward on your further efforts that is giving you more insight into spin dynamics of electron and nuclear systems in the sample. ENDOR in terms of spin technology is based on a ^^ ^^ quantum transformation of detection" from nuclear precession to electronic one. Particularly, pulse-based electron-nuclear multiple resonance is a rapidly developing field in terms of spin technology.

We study ENDOR aiming on realizing an entangled state with ENDOR systems. To some extent, we will examine the implementation of quantum operations with ENDOR. In other words, full quantum computing with ENDOR has been  aiming to be established. We get results of selective manipulations of spins, implementations of quantum operations, and measurements. However, quantum entanglement is just to be realized as the required experimental conditions are fulfilled and there is no fundamental restriction for it, as will be explained in the following section.

In the following sections, some of the experiments have been done for the sake of {\it sample measurements} for quantum computing and quantum information processing, {\it decoherence time} measurements and so on. The experimental evidences on quantum information processing and quantum computation with ENDOR will be given. We are mainly interested on pulsed ENDOR for QC. However, continuous wave (cw) ENDOR and pulsed EPR will also be used for sample measurements and decoherence time measurements. For sample preparation and measurements in order to get the required information of the particular sample for quantum computation, we first start with cw ENDOR. Strictly speaking, in terms of magnetic resonance spectroscopy, pulse-based ENDOR has never been almighty, but cw and pulsed ENDOR technologies are complementary.

For any physical system as a candidate for the realization of a quantum computer, there are some fundamental criteria, known as DiVincenzo criteria that should be met. Molecule-based ENDOR systems are also expected to meet these criteria in order to be a realistic physical system for QIP and QC; see Table \ref{T0} for a list of DiVincenzo criteria and properties of the ENDOR system, in this regard.

In the molecule-based ENDOR QIP and QC, molecular electron spins in addition to nuclear spins have been introduced as quantum bits (qubits). In a thermal equilibrium, the relative populations in the ground states with molecular electron spins are more than $10^3$ times larger than the corresponding excited states in the presence of a static magnetic field or the ones with zero-field splitting if system involves more than two electrons. Therefore, with ENDOR systems, achievement of the required experimental conditions for preparing the initial state for QIP and QC, compared with NMR systems, seems to be substantially easier with the current technology.

Any physical system for QIP and QC should be chemically stable during computational processing. We have prepared robust organic open-shell entities against long and high-power irradiations of radio frequency and microwave even at ambient temperature, as some among those exemplified in this contribution. In addition, the corresponding decoherence time of the qubits is expected to be long enough compared with the computational or operational time. As a result of the existence of the nuclear spins in molecular open-shell entities, there is a wide possibility to work with samples with long decoherence times. Proper samples with long decoherence time and their synthetic procedure should be considered, in advance. In this context, realistic molecule-based multiple-qubit systems require component synthetic chemistry combined with knowledge of molecular magnetism. Long enough decoherence times for several samples have been measured during the course of this work.

One important point which we will discuss in a separate section in the following is that, for a matter of initial state we still use a pseudo-pure state. For spin with low spin polarization, the pseudo-pure state has been pointed out to have several problems. It does not solve the problem of the separability of the states. Generally, it is not scalable because of the exponential increase in the required number of operational steps for realizing a pure state. However, if polarization of the spins increases to some extent, then the states would be entangled. Entanglement can exist for high polarized states although the state might be prepared as pseudo-pure states. The scalability problem for making pseudo-pure states is a different issue than entanglement and we might consider solving the scalability problem some other time specially when dealing with experiments with a rather larger number of qubits. For a large number of qubits, several proposals would be examined e.g. algorithmic cooling, which very recently has been shown to be useful for NMR QC, \cite{Baug}, \cite{Fernandez}.

ENDOR systems are essentially useful as we will show in some following section. We will introduce the required steps for ENDOR QC as sample preparation and measurement of magnetic properties, decoherence time controlling and measurements, readout. We will also give experimental results supporting the discussion for each section, as well as finally some quantum computation experiments with ENDOR will be given.

 \newpage
 \begin{table}
\begin{center} 
  \caption{\label{T0} DiVincenzo criteria for ENDOR Quantum Computing.}
\small
\begin{tabularx}{\linewidth}{|X|X|X|}
\hline
... & DiVincenzo Criteria  & ENDOR\\\hline
 Qubit & Identifiable, well characterized qubits are required. & Molecule-based electron and nuclear spins in molecular open-shell entities based on chemical syntheses and identifications. \\ \hline
 Initialization & Possibility to be initialized to a simple and fiduciary state. & System of electron and nuclear spins would be cooled down in a very strong magnetic field, that the conditions for entanglement would be achieved. For preliminary experiments, pseudo-pure states will be used. For large number of qubits, the other approaches, e.g. algorithmic cooling, will be used. \\ \hline
 Decoherence time & Long relevant decoherence times, much longer than the gate operation time is necessary. & Long decoherence times of the nuclear spins and electron spin have been available for demonstration of quantum operations between the two spins. Proper molecular entities with long decoherence times for multi-qubit operations are also possible, for which stable isotope-labeled open-shell molecules have been designed and synthesized. \\ \hline
 Quantum operation & Universal set of quantum gates is required. & Quantum gates between a single electron and a single nuclear spin have been demonstrated. Electron spin bus molecule for ENDOR QC is prepared for which more general quantum operation can be applicable. \\ \hline
 Measurement & The ability of measurements on qubits to obtain the result of the computation is required. & The current measurement scheme is an ensemble one, for which modification of quantum algorithms are required. This is not a problem as it has been already solved out for NMR QC. However, we may use also STM-based electron magnetic resonance detection.\\ \hline
\end{tabularx}
\end{center}
\end{table}

\begin{figure}
\begin{center}
\scalebox{0.79}
{\includegraphics[0cm,0cm][9cm,6cm]{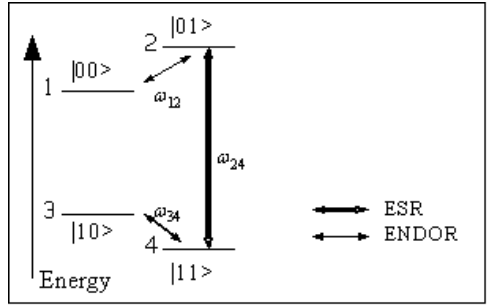}}
\end{center}
\caption{\label{ENDORlevel} Energy levels and corresponding EPR/ESR and ENDOR resonance transitions in the presence of a static magnetic field. The definition for the level is as follows; $|00\rangle$ and $|10\rangle$ denote $|M_s =+1/2, M_I =+1/2\rangle$ and $|M_s = -1/2, M_I =+1/2\rangle$, respectively. Similarly, $|01\rangle$ and $|11\rangle$ denote $|M_s = +1/2, M_I = -1/2\rangle$ and $|M_s = -1/2, M_I = -1/2\rangle$, respectively. Also, see the text for the use of the notation of the sublevels.}
\end{figure}

In QC-ENDOR, manipulation and processing on the qubits as well as the readout processing can be realized by introducing radio frequency pulses on the nuclear spins or by the microwave frequency pulses on the electron spins (see Figure \ref{ENDORlevel}).

 It is known that realization of a particular class of quantum gates known as universal gates can be enough for implementation of any other quantum gates. One-qubit gates in addition to a non-trivial two-qubit gate, e.g. Controlled-NOT gate (CNOT), give a universal set of quantum gates. For implementations of quantum gates, it is possible to perform this task by introducing the relevant pulses. From the experimental side, for the QC-ENDOR system particular quantum gates have been demonstrated by means of the two qubits, as it is described in the following section. Multi-qubit gates for molecular systems involving a larger number of qubits should be considered for the relevant physical system in terms of the particular form of the spin Hamiltonian for the corresponding sample. There are several typical types of quantum operations based on multi-qubit-gates, depending upon spin Hamiltonians for realistic QC-ENDOR experiments.
 
 For instance, we have implemented quantum gates between a single electron and a single nuclear spin. For rather more general quantum operations, proper sample, electron spin bus molecule, has been selected. For this form of samples, even if qubits are not directly coupled to each other, for instance nuclei, still it is possible to perform universal computation. Electron spin is a degree of freedom for this class of samples, for which it can be coupled to any one nuclear spin, in a controlled manner. Therefore, two-qubit gates can be defined and easily generalized to other pairs of qubits. The original idea of coupling via a bus is generally applicable foe any other degree of freedom in a sample for quantum computing. This can be typically a photon that is coupled to several quantum dots embedded in a semiconductor structure. We will use this idea for rather more upcoming advanced experiment for a complete quantum computing with ENDOr. However, for preliminary experiments, also for a matter of implementation of quantum entanglement, quantum gates can be easily demonstrated between electron and nuclear spin and essentially the situation is very simple as the number of qubits are not very large.
 
In the ENDOR based QC, the readout or the measurement part has been done through electron spin echo detection. For detection of entanglement, as far as the global information on the state of electron and nuclear spins is required, modified detection scheme, Time Proportional Phase Incrementation, TPPI has been used. In the following sections these points will be explained in more details. In virtue of TPPI, we will be dealing with spin manipulation of electron-nuclear systems.

It has been proved that for the case of pure states, entanglement is the necessary requirement for quantum exponential speed-up over the classical counterpart. For mixed states the statement has not been proved yet, but highly believed to be correct. Therefore, one very important issue which should be examined for any physical system for QIP and QC is the entanglement status.

Realization of the entanglement between an electron spin and a nuclear spin has been reported in an ENDOR experiment by using pseudo-pure states, \cite{n4}. However, it should be stated that the entanglement they achieved is ^^ ^^  pseudo-entanglement". It means that the states actually are not entangled but the quantum operations for making entanglement have been applied and if the required experimental conditions i.e. low enough temperature below critical temperature and high magnetic field, are satisfied, then quantum entanglement would be established.

In order to determine the required experimental conditions for establishment of entanglement, we should consider the Hamiltonian of the system and determine density matrix which explains the system of interest. However, the full Hamiltonian of the system might be very complicated. Simplified Hamiltonian would not give all the correspondence  between entanglement and experimental conditions, but still might be interesting to some extent. We calculate the conditions of temperature and magnetic field in order to establish entanglement. The measure that we use for calculation of entanglement is the one which we introduced previously, in chapter 3. Recall that through our discussion on SDC, states for two spins $I$ and $S$ are entangled iff $p_Ip_S>1/2$. Here $p_I$ and $p_S$ are the probabilities for nuclear and electron spins, respectively, to be at the zero state and are related to the spin polarizations as follows
\begin{equation}
p_I=\frac{1+\epsilon_I}{2}, \hspace{2cm} p_S=\frac{1+\epsilon_S}{2}.
\end{equation}

Now, substituting the corresponding spin polarizations, we get the required conditions for entanglement as follows
\begin{equation}
\label{ENTtemp}
{\rm Ent.}\equiv \epsilon_I\epsilon_S+\epsilon_I+\epsilon_S>1.
\end{equation}

Here $\epsilon_I$ and $\epsilon_S$ are the spin polarizations for nuclear and electron spins. Recall that the above inequality for entanglement is achieved for the case that a Hadamard gate and a CNOT gate are applied to the state (not pseudo-pure state) of the system. However, as far as a pseudo-pure state e.g. effective pure state can be written as a convex combination transformed thermal states, and any measure of entanglement is a convex function therefore entanglement is reduced under convex combination, then an individual transformed thermal state may possess more entanglement than the combination, \cite{YU}. For any other particular class of quantum operation, entanglement might be calculated e.g. scheme that has been explained in chapter 2.

\begin{figure}
\begin{center}
\scalebox{0.69}
{\includegraphics[0cm,7.5cm][21cm,20.5cm]{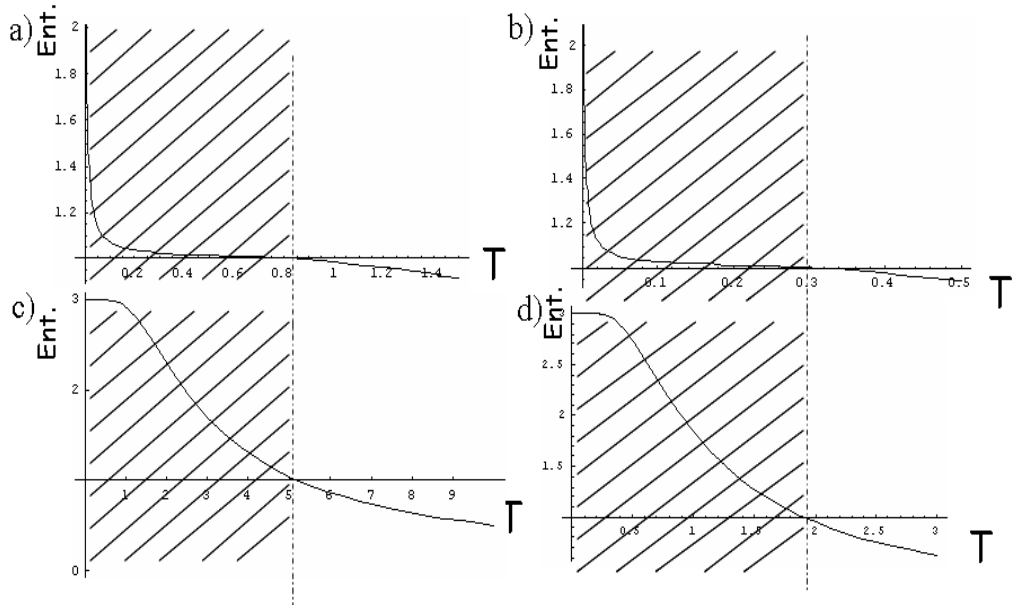}}
\end{center}
\caption{\label{threshold} Entanglement diagram as a function of temperature. Vertical axis is the amount of entanglement defined by \ref{ENTtemp}. The area in which the states are entangled is in shadow. a) W-band b) Q-band. c)W-band under the condition of high polarization for nuclear spins d) Q-band under the condition of high polarization for nuclear spins.}
\end{figure}

Referring to the above measure of entanglement, the experimental condition for existence of entanglement for a magnetic field of 95GHz, W-band experiment, is a temperature of 0.83 K. However, if a transfer of spin polarization from electron spin to nuclear spin is performed the required temperature is 5.17 K, which is well in reach with current technology with a W-band ENDOR spectrometer operating at liquid Helium temperature. Figure \ref{threshold} shows the required temperature for existence of entanglement, given by inequality \ref{ENTtemp} after substituting temperature in spin polarizations, for magnetic fields of 95 GHz, W-band, and 35 GHz, Q-band experiments, for two cases of the experiments without and with transfer of spin polarization. But, how to transfer electron spin polarization to nuclear spin. This is a matter that is discussed in the following section. It will be clear that it is an advantage of pulsed ENDOR spectroscopy that the scheme for transfer of electron spin polarization to nuclear spin is performed comparably easier and by applying relevant pulses.

\section{Transfer of Spin Polarization from Electron to Nuclear Spin}
\begin{figure}
\begin{center}
\scalebox{0.69}
{\includegraphics[0cm,0cm][21cm,23cm]{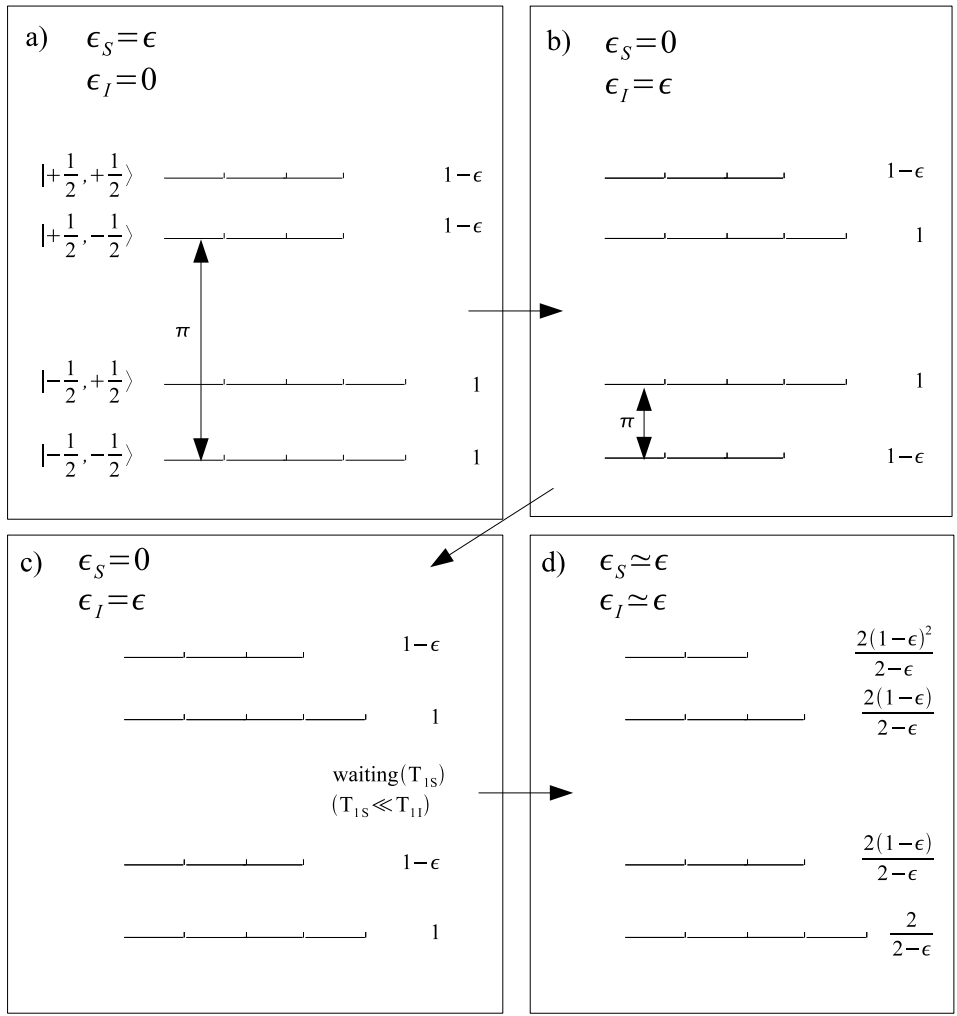}}
\end{center}
\caption{\label{transfer} A scheme on how to make nearly equal spin polarizations on nuclear spin and electron spin, equal to the initial electron spin polarization.}
\end{figure}

In the case of working with {\it pulsed ENDOR}, transfer of spin polarization is easily done by applying appropriate pulses. Following, we explain a very simple approach with which by applying some particular pulses the final nuclear spin polarization becomes almost as large as electron spin polarization. The approach that is explained here express the possibility of achieving almost equal spin polarization on electron and nuclear spins. The approach is shown in Figure \ref{transfer} and works as follows.

Suppose an isolated system composed of an electron spin $S=\frac{1}{2}$ and a nuclear spin $I=\frac{1}{2}$. The corresponding four energy levels are shown in Figure \ref{transfer}(a). In order to make a nearly equal spin polarization on two electron and nuclear spins, we propose the following pulse sequence, \cite{n5}.

In the Figure \ref{transfer}, the horizontal length of each energy level refers to the corresponding population of that level. Nuclear spin polarization is neglected in comparison with the electron spin polarization.
\begin{eqnarray}
\frac{N_4}{N_1}=\exp (\frac{-g\beta H}{k_B T})&\equiv& 1-\frac{g\beta H}{k_B T} \\ \nonumber
                                              &=& 1-\epsilon.
\end{eqnarray}
Figure \ref{transfer} shows the polarization of states using the above naming for $\epsilon=\frac{g\beta H}{k_B T}$. Initially, electron spin polarization, $\epsilon_S$, is equal to $\epsilon$, while nuclear spin polarization is zero, $\epsilon_I=0$. In  this Figure $\epsilon$ is exaggerated for the sake of easy understanding. Firstly, apply an on-resonance microwave $(\omega_{13})$ $\pi$ pulse. The state after the pulse would be as shown in Figure \ref{transfer}(b). Now, the polarizations are interchanged as follows
\begin{eqnarray}
\epsilon_S=0\\ \nonumber
\epsilon_I=\epsilon.
\end{eqnarray}
Applying an rf $\pi$ pulse $(\omega_{12})$ would interchange the population between two energy levels, one and two, but may not affect spin polarizations. Now, we make an assumption, that can be satisfied by carefully choosing proper samples. Suppose that spin-lattice relaxation time of electron, $T_{1e}$, is less than the relaxation time for nuclear spin, $T_{1n}$. Therefore, after a waiting time of $T_{1e}$ the system may go back to equilibrium on electron level and may yield the energy levels as shown in Figure \ref{transfer}(d). For instance, finally we have electron and nuclear spin polarizations equal to $\frac{2\epsilon}{2-\epsilon}$ and $\frac{2\epsilon}{2-\epsilon}(1-\epsilon)$. Hence, we have
\begin{eqnarray}
\epsilon_S \cong \epsilon \\ \nonumber
\epsilon_I \cong \epsilon.
\end{eqnarray}
The above simplified approximation for calculating the population is based on a rate-equation method. The above spin polarizations are the values that are required in much higher energy levels as shown in Figure \ref{transfer}. There might be some more elaborated pulse sequences to get high spin polarizations on both electron and nuclear spins, without a serious assumption on the sample properties. The above explained approach, however is an experimentally easy accessible approach for the current technology of ENDOR spectroscopy, and as far as the pulse sequence for QC with ENDOR is not complicated would work properly.

\section{Providing Some Theoretical Concepts}
 We have been mainly engaged in two experimental tasks for the realization of a quantum computer by molecule-based ENDOR. One has been attempts for the preparation of the experimental requirements for demonstrating quantum entanglement between a molecular electron spin and a nuclear spin by the use of a simple organic radical in the single crystal or the electron delocalized systems. High spin polarizations on both the electron and nuclear spins are essentially required to achieve entanglement between the two spins in the molecular frame.

While the preparation of all the experimental requirements is in progress, the efforts still can be maintained on some other aspects of the research. The other aspects are materials challenge to design and synthesize stable open-shell molecular entities suitable for QIP/QC ENDOR experiments. Sample selection and measurements are very heavy experimental tasks. Novel molecular open-shell systems with stable isotope labels suitable for our purposes have also been designed and synthesized. The critical temperature can be tuned by invoking stable high-spin molecular entities in near the future. 

 ENDOR consists of EPR, electron paramagnetic resonance, followed by NMR. EPR is a spectroscopic technique, which detects species that have unpaired electrons. Materials, which have unpaired electrons are various, including free radicals, many transition metal ions, and defects in materials.
 
Here we would like to first explain theoretical fundamentals of EPR/ENDOR, \cite{n6}, \cite{n7}. It should be noted that almost the same explanation goes for NMR. However, differences in the magnitudes and signs of the magnetic interaction involved for NMR and EPR/ENDOR, naturally lead to divergences in the experimental techniques that have been employed. I would like to organize a comprehensive discussion, including NMR too, in this section, but because of the space and the main interest topic for this work, I would better keep going with the most important points from EPR/ENDOR, that are required in order to understand the following experiments on ENDOR QC including material search and so on.

Electron with $S=1/2$ has a magnetic moment ${\bf \mu}_e$, which is proportional to the magnitude of the spin, that is 
\begin{equation}
{\bf \mu}_e = -g \beta  {\bf S}.
\end{equation}

Here, $g$ is a dimensionless constant, generally a tensor, called the electron $g$ factor. $\beta$ is the electronic Bohr magneton defined as $\beta=\frac{e \hbar}{2mc}$, with $-e$ and $m$ representing, respectively, the charge and mass of the electron. $\hbar${\bf S} is the spin angular momentum vector.

If we apply a steady magnetic field ${\bf H}$, there is an interaction between the field and the magnetic moment ${\bf \mu}_e$ as follows
\begin{equation}
\mathcal{H}= -{\bf \mu}_e \cdot {\bf H}.
\end{equation}

If the applied static magnetic field is in the z direction then 
\begin{equation}
\mathcal{H}=g\beta H S_{\rm z}.
\end{equation}

   \begin{figure}
\begin{center}
\scalebox{0.99}
{\includegraphics[0cm,0cm][10cm,4cm]{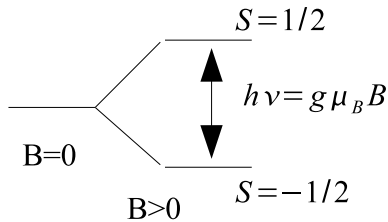}}
\end{center}
\caption{\label{EPRenlev} EPR energy levels.}
\end{figure}

As far as $S=1/2$, there are two allowed orientations of spin, parallel or antiparallel to {\bf H}, and are represented by $|\beta \rangle$ corresponding to $m_s=-1/2$ and $|\alpha \rangle$ corresponding to $m_s=1/2$, Figure \ref{EPRenlev}.

In order to introduce transitions between the two nuclear spin levels, an oscillating electromagnetic field is applied to the system. Absorption of energy occurs providing the magnetic vector of the oscillating field is perpendicular to the steady magnetic field {\bf H} and also provided the frequency $\nu$ of the oscillating field satisfies the resonance conditions
\begin{equation}
h\nu=g\beta H.
\end{equation}

The resonance frequencies for different magnetic fields are listed in Table \ref{T1}.
  \begin{table}
 \begin{center}
  \caption{\label{T1} Different EPR spectrometers with corresponding frequencies.}
\small
\begin{tabularx}{70mm}{|X|X|X|}
\hline
   & Frequency/ GHz & Magnetic Field/ G(g=2) \\ \hline
 L                 &1-2                   &500                \\ 
 S                 &2-4                   &1000               \\ 
 X                 &9-10                  &3400               \\ 
 K                 &24                    &8500               \\ 
 Q                 &34                    &12100              \\ 
 W                 &94                    &33500              \\ \hline   
 \end{tabularx}
 \end{center}
 \end{table}

It should be stated that the oscillating field is equally likely to produce transitions up or down. However, due to the fact that the population of $\beta$ electron is slightly larger than $\alpha$ in a thermal equilibrium, then we can only detect absorption of energy from the radiation field.

There are two possibilities. Conventionally, we could make the magnetic field, $B_0$ constant and vary the frequency, microwave until the signal is detected on the condition of resonance. However, because of the restriction on the Microwave technology, this option is not practical. Instead, we apply a constant microwave frequency and vary the magnetic field close to the resonance condition.
 
 Suppose the magnetic energy level around $3000 G$. Then, in a thermal equilibrium, from the Boltzmann distribution, we have
 \begin{equation}
 \frac{N_{up}}{N_{down}}=\exp (\frac{-g\mu_BH}{kT}).
 \end{equation}
 
 There would be a net alignment of the spins, however very tiny. For example, in a magnetic field of $9.8$ GHz, X-band, and at room temperature, with a sample which contains $10,000$ spins, on average $5,004$ spins would be parallel and $4996$ spins would be antiparallel. The resulting difference population would be $8$, giving the spin polarization almost $10^{-3}$. Then, levels are almost but not exactly, equally populated. If a microwave magnetic field is applied perpendicular to the static magnetic field, some transitions are acquired between the magnetic levels. Microwave stimulates transitions in both the directions. Since the population in the lower level is slightly larger than the upper level then there will be a net observed EPR signal. Under the steady condition the process would be soon stopped as the population would be equalized. The system is subject to interactions between spins of the surrounding described by characteristic $T_1$ and $T_2$, spin lattice relaxation process and spin-spin relaxation process, respectively. These make the longitudinal and transverse transitions.
 
After saturation of EPR, ENDOR is performed by monitoring the EPR signal intensity, while sweeping an RF signal to drive the NMR transitions in the cw conventional mode of experiments. Shortly speaking, there are some main steps to acquire an ENDOR spectrum.
 \begin{description}
 \item[1.] Acquire an spectrum and determine the magnetic filed values at which the EPR signals occur.
 \item[2.] Set the magnetic field to the resonance position of the absorption EPR signals.
 \item[3.] Increase the microwave power and partially saturate the EPR signal.
 \item[4.] Monitor change in the EPR signal, while sweeping the RF. For instance, it should be considered as the signals are too weak and need RF amplifier to make them detectable.
 \end{description}
 
 \subsection{The Magnetic Hamiltonian, EPR/ENDOR Spectra}
In this part, we consider a simple system of one proton and one electron spin. This can be illustrated by Hydrogen atom. We first derive expressions for the energy levels of this atom in a magnetic field. Then, possible spin transitions and EPR/ENDOR spectra will be calculated.

Let us give an overview of what will come in following. First of all an appropriate wave function should be defined that can describe the four energy levels from one electron and one proton. These energy levels are degenerate in the case that there is no magnetic field. Then applying magnetic field ${\bf H}$ makes the system to involve the interaction between ${\bf H}$ and ${\bf S}$ in addition to ${\bf H}$ and ${\bf I}$. We would also study coupling between spins, hyperfine coupling. This particular simple example possesses most of the complications that we would  require to understand and make it clear how to read EPR/ENDOR spectrum. We should be careful however that the simplification due to this particular example might be dangerous. Specially because that the hydrogen atom is spherical, then the spin Hamiltonian for the hydrogen is not anisotropic, though we need to study the anisotropical behaviour of {\bf g} and {\bf A} tensors. Therefore, later we will extend this discussion in a  way that can include the anisotropic terms, too, \cite{H5}, \cite{H6}, \cite{n6}, \cite{n7}.

\subsubsection{The Zeeman Energy}
The nucleus and the electron both interact with the steady magnetic field and yield the following energy
\begin{equation}
\mathcal{H}_0=g \beta H S_z-g_N\beta_N H I_z
\end{equation}
$g$ and $g_N$ are the $g$ factors for the electron and the nucleus. The second term, the nuclear energy is much smaller than the electronic, first term, part.

\subsubsection{The Isotropic Hyperfine Coupling}
The magnetic moments of electron and nuclei are coupled via the so-called contact interaction, which represents the energy of nuclear moment in the magnetic field produced at the nucleus by electric currents that are associated with the spinning electron. it has the form
\begin{equation}
\mathcal{H}_1=a{\bf I} \cdot {\bf S} =a(I_xS_x+I_yS_y+I_zS_z).
\end{equation}

There is a magnetic coupling between the magnetic moments of the electron and nucleus which is entirely analogous to the classical dipolar coupling between two bar magnets. But the dipolar Hamiltonian averaged out to zero whenever the electron cloud is spherical.

Totally the complete Hamiltonian is 
\begin{equation}
\mathcal{H}=g\beta HS_z-g_N\beta_NHI_z+a{\bf S} \cdot {\bf I}.
\end{equation}

It is the task of the experimentalists to find the proper Hamiltonian with parameters.To make it more general we should consider that Hamiltonian includes terms as ${\bf H} \cdot {\bf S}$, ${\bf H} \cdot {\bf I}$ and ${\bf S} \cdot {\bf I}$. The first two come in the $\mathcal{H}_0$ and the last term in $\mathcal{H}_1$. In order to solve the corresponding hamiltonian, we would require perturbation theory. The eigenfunctions of $\mathcal{H}_0$, $\phi_1, ..., \phi_4$, correspond to the ^^ ^^ zero-order" unperturbed energy values, $\epsilon_1,...,\epsilon_4$. Here $\mathcal{H}_1$ is the perturbation term that yield
\begin{eqnarray}
\psi_n&=&\phi_n-\sum_{m\neq n} \frac{\langle m | \mathcal{H}_1 | n \rangle}{\epsilon_m-\epsilon_n} \phi_m\\ \nonumber
E_n&=&\epsilon_n +\langle n|\mathcal{H}_1|n\rangle - \sum_{m\neq n}\frac{\langle m|\mathcal{H}_1|n\rangle\langle H_1|m\rangle}{\epsilon_m-\epsilon_n}
\end{eqnarray}
the last equation for energy includes the first-order correction in its second term and the second-order correction in its last term. it is clear that in the first order of perturbation the energy of $\phi_n$ is changed without changing the form of $\phi_n$, however in the second-order of perturbation the changes of energy also mixes $\phi_n$ with other function $\phi_n$. This will be clear in the experiment with DPNO that we will show in the following sections.

\subsubsection{Zero-Order Energy}
Recall that we denoted the two possible spin function of an electron $S=1/2$, with $|\alpha_e\rangle$ and $|\beta_e\rangle$ corresponding to $m_s=+1/2$ and $m_s=-1/2$, respectively. These come from the eigenequations as follows
\begin{eqnarray}
S_z|\alpha_e\rangle&=&+1/2|\alpha_e\rangle,\\ \nonumber
S_z|\beta_e\rangle&=&-1/2|\beta_e\rangle.
\end{eqnarray}
Similar fact is going on for nuclear spin $I=1/2$, as follows
\begin{eqnarray}
I_z|\alpha_N\rangle=+1/2|\alpha_N\rangle, \\ \nonumber
I_z|\beta_N\rangle=-1/2|\beta_N\rangle.
\end{eqnarray}
Therefore, at this level, the appropriate basis functions are 
\begin{eqnarray}
\phi_1&=&|\alpha_e \alpha_N \rangle, \hspace{2cm}  \phi_2=|\alpha_e \beta_N \rangle, \\ \nonumber
\phi_3&=&|\beta_e \alpha_N \rangle, \hspace{2cm} \phi_4=|\beta_e \beta_N \rangle.
\end{eqnarray}
Initially the four states are degenerate, but after applying the steady magnetic field represented by Hamiltonian $\mathcal{H}_0$ then the states will separate as follows, e.g.
\begin{eqnarray}
\mathcal{H}_0|\alpha_e \beta_N\rangle &=& g \beta H S_z|\alpha_e\rangle |\beta_N\rangle-g_N\beta_NH|\alpha_e\rangle I_z|\beta_N\rangle \\ \nonumber
&=& \frac{1}{2} g \beta H |\alpha_e\rangle |\beta_N\rangle +\frac{1}{2} g_N\beta_N H |\alpha_e \rangle|\beta_N\rangle \\ \nonumber
&=& (\frac{1}{2} g \beta H +\frac{1}{2} g_N H) |\alpha_e \beta_N\rangle .
\end{eqnarray}
Then the zero-order energy values are
\begin{eqnarray}
\epsilon_1&=&\frac{1}{2} g \beta H - \frac{1}{2} g_N \beta_N H \hspace{3cm} |\alpha_e \alpha_N\rangle, \\ \nonumber
\epsilon_2&=&\frac{1}{2} g \beta H + \frac{1}{2} g_N \beta_N H \hspace{3cm} |\alpha_e \beta_N\rangle, \\ \nonumber
\epsilon_3&=&-\frac{1}{2} g \beta H - \frac{1}{2} g_N \beta_N H \hspace{2.4cm} |\beta_e \alpha_N\rangle, \\ \nonumber
\epsilon_4&=&-\frac{1}{2} g \beta H + \frac{1}{2} g_N \beta_N H \hspace{2.4cm} |\beta_e \beta_N\rangle.
\end{eqnarray}

\subsubsection{The First-Order Energies}
At this level $\mathcal{H}_1$ works as a perturbation on the zero-order and it gives results as follows
\begin{equation}
\mathcal{H}_1= a {\bf S} \cdot {\bf I} =a(S_zI_z+S_xI_x+S_yI_y),
\end{equation}
for the first-order, the first term is only involved and the latter terms are for the second-order and will be discussed later. hence, we have
\begin{eqnarray}
\langle \alpha_e \alpha_N |aS_zI_z|\alpha_e\alpha_N\rangle&=&\frac{1}{4}a, \\ \nonumber
\langle \beta_e \alpha_N |aS_zI_z|\alpha_e\beta_N\rangle&=&-\frac{1}{4}a, \\ \nonumber
\langle \beta_e \alpha_N |aS_zI_z|\beta_e\alpha_N\rangle&=&-\frac{1}{4}a, \\ \nonumber
\langle \beta_e \beta_N |aS_zI_z|\beta_e\beta_N\rangle&=&\frac{1}{4}a.
\end{eqnarray}

\subsubsection{The Second-Order Energies}
This is due to the term $a(S_xI_x+S_yI_y)$. We recall that shift operators are defined as follows
\begin{eqnarray}
S^+&=&S_x+iS_y, \\ \nonumber
S^-&=&S_x-iS_y.
\end{eqnarray}
Therefore, considering also the relation for shift operators, we have 
\begin{equation}
S_xI_x+S_yI_y=\frac{1}{2}(S^+I^-+S^-I^+).
\end{equation}

The nonvanishing matrix elements are
\begin{eqnarray}
\langle\alpha_e\beta_N|S^+I^-|\beta_e\alpha_N\rangle&=&1, \\ \nonumber
\langle\beta_e\alpha_N|S^-I^+|\alpha_e\beta_N\rangle&=&1.
\end{eqnarray}
The complete matrix of the operator $\mathcal{H}_1$ is 
\begin{equation}
 a{\bf I}\cdot{\bf S} = \frac{1}{4} a \left(\begin{array}{@{\,}cccc@{\,}}1&0&0&0\\0&-1&2&0\\0&2&-1&0\\0&0&0&1\\ \end{array}\right).
 \end{equation}
 
 Then perturbation theory changes the states into new states as follows
 \begin{eqnarray}
\label{perturb}
 \psi_2&=&|\alpha_e \beta_N\rangle+\frac{a}{2(g\beta H+g_N\beta_NH)}|\beta_e\alpha_N\rangle, \\ \nonumber
 \psi_3&=&|\beta_e \alpha_N\rangle-\frac{a}{2(g\beta H+g_N\beta_NH)}|\alpha_e\beta_N\rangle.
\end{eqnarray}
Corresponding to the energies
\begin{eqnarray}
E_2&=&(\frac{1}{2} g\beta H+\frac{1}{2}g_N\beta_NH)-\frac{1}{4}+\frac{a^2}{4(g\beta H+g_N\beta_NH)},\\ \nonumber
E_3&=&-(\frac{1}{2} g\beta H+\frac{1}{2}g_N\beta_NH)-\frac{1}{4}-\frac{a^2}{4(g\beta H+g_N\beta_NH)}.
\end{eqnarray}
We just remind that for the simple case as the hydrogen atom, we would solve the problem even without using perturbation theory.

\subsubsection{The First-Order Spectrum}
Suppose that an oscillating magnetic field of strength $2H_1\cos \omega t$ acts on the atom. For transitions applying a proper e.g. microwave frequency may cause allowed transitions like $\alpha_e\alpha_N \to \beta_e \alpha_N$, or forbidden transition, $\alpha_e \beta_N \to \beta_e \alpha_N$, with very small probability. On the other hand, RF pulse causes allowed transition $\alpha_e\alpha_N\to \alpha_e\beta_N$.

If the oscillating magnetic field is applied along the z direction, it will not cause any energy absorption, because it just modulate the energy levels of the spin system. On the other hand, if it is applied across the steady field $H$ in the x direction we have 
\begin{eqnarray}
V(t)&=& 2(g\beta H_1S_x-g_N\beta_NH_1I_x)\cos \omega t \\ \nonumber
    &=& 2V\cos \omega t.
\end{eqnarray}
The transition probability from state $n$ to the state $m$ is equal to 
\begin{equation}
P_{nm}=\frac{2\pi}{\hbar^2}|\langle n|V|m\rangle|^2\delta(\omega_{mn}-\omega).
\end{equation}

If $H_1$ corresponds to the electron spin then resonance transitions of electron spins are aquired as follows
\begin{equation}
P_{nm}=\frac{2\pi}{\hbar^2}g^2\beta^2H_1^2|\langle n|S_x|m\rangle|^2\delta(\omega_{mn}-\omega).
\end{equation}

If we use the shift operators, then we have 
\begin{equation}
\langle \alpha_e \alpha_N |S_x|\beta_e\alpha_N \rangle=\frac{1}{2}\langle\alpha_e|S^+|\beta_e\rangle=\frac{1}{2},
\end{equation}
with the transition probability 
\begin{equation}
P=\frac{\pi}{2\hbar^2}g^2\beta^2H_1^2\delta(\omega_{mn}-\omega).
\end{equation}

We replace $\gamma=g\beta$. Also, to be closer to the real situation, we write the above equation as based on the general line shape function $g(\omega)$.
\begin{equation}
P=\frac{1}{2} \pi \gamma^2 H_1^2 g(\omega).
\end{equation}

And the frequency for the transition would be
\begin{eqnarray}
h\nu_a&=&(\frac{1}{2}g\beta H-\frac{1}{2}g_N\beta_NH+\frac{1}{4}a)-(-\frac{1}{2}g\beta H-\frac{1}{2}g_N\beta_NH-\frac{1}{4}a)\\ \nonumber
      &=&g\beta H +\frac{1}{2}a.
\end{eqnarray}
\begin{eqnarray}
h\nu_b&=&(\frac{1}{2}g\beta H+\frac{1}{2}g_N\beta_NH+\frac{1}{4}a)-(-\frac{1}{2}g\beta H+\frac{1}{2}g_N\beta_NH+\frac{1}{4}a)\\ \nonumber
      &=&g\beta H -\frac{1}{2}a.
\end{eqnarray}
Then the first order EPR spectrum consists of two lines separated by a, the hyperfine constant. Now, the magnetic field is set to the top for the EPR. Then microwave power is increased to saturate the EPR signal. ENDOR comes by monitoring EPR signals, while sweeping the RF irradiation. For instance it should be considered as the signals are too weak and need RF amplifier to make them detectable. Suppose that the microwave is fixed on the lower state with energy $-1/2g\beta H$ and the RF is applied. The resonance frequency would be given in a same manner as above but this time for ENDOR as follows
\begin{eqnarray}
h\mu_a&=&(-\frac{1}{2}g\beta H +\frac{1}{2}g_N\beta_NH+\frac{1}{4}a)-(-\frac{1}{2}g\beta H-\frac{1}{2}g_N\beta_NH-\frac{1}{4}a)\\ \nonumber
      &=& g_N\beta_NH+\frac{1}{2}
\end{eqnarray}
\begin{eqnarray}
h\mu_b&=&(-\frac{1}{2}g\beta H +\frac{1}{2}g_N\beta_NH-\frac{1}{4}a)-(-\frac{1}{2}g\beta H-\frac{1}{2}g_N\beta_NH+\frac{1}{4}a)\\ \nonumber
      &=& g_N\beta_NH-\frac{1}{2}
\end{eqnarray}

\subsubsection{The Second-Order Forbidden Transitions}
The transition in which both electron and nuclear spin states are changed is strictly forbidden in first-order, but in the second-order, \ref{perturb}, it is weakly allowed, provided the oscillating magnetic field is polarized parallel to the static field, rather than perpendicular. In the following calculation we used $\lambda=\frac{a}{2(g\beta H+g_N\beta_NH)}$. We used the states which we already calculated for the first order perturbation, then we have
\begin{eqnarray}
\langle \psi_2|S_z|\psi_3\rangle&=&\langle \alpha_e\beta_N+\lambda \beta_e\alpha_N|S_z|\beta_e\alpha_N-\lambda\alpha_e\beta_N\rangle\\ \nonumber
                                &=&\langle\alpha_e\beta_N|S_z|\beta_e\alpha_N\rangle+\lambda\langle\beta_e\alpha_N|S_z|\beta_e\alpha_N\rangle\\ \nonumber
                                &-&\lambda\langle\alpha_e\beta_N|S_z|\alpha_e\beta_N\rangle-\lambda^2\langle^2\langle\beta_e\alpha_N|S_z|\alpha_e\beta_N\rangle\\ \nonumber
                                &=&0-\frac{1}{2}\lambda-\frac{1}{2}\lambda-0\\ \nonumber
                                &=&-\lambda.
\end{eqnarray}
$\lambda$ is small, then $\lambda^2$ is neglected.

The corresponding transition probability is
\begin{equation}
P=2\lambda^2\pi\gamma^2H_1^2g(\omega).
\end{equation}

We would like to close this part here and move to more precise explanation particularly for liquid and solid state sample as will be directly used for sample study for ENDOR QC. But first we would like to note that there are much more corrections which we neglected here. Also, generally $g$ and $a$ are anisotropic and represented by tensors. This will be clear by studying the orientational speciment (single crystal). the principal values of the $g$ tensor and/or $a$ tensors can only be determined from solid state study.
\subsection{Liquid State Sample Study}
Here, we concentrate our attention on liquid state sample study. We shall still use the general discussion that is given above and only add some more important points. We will see that incase the number of nuclei is large, ENDOR gives much more clear spectra as compared to EPR.

Electron nuclear hyperfine interaction, in general, is represented as follows ${\bf S}\cdot{\bf A}\cdot.{\bf I}$, where ${\bf A}$ is the hyperfine tensor. This tensor is decomposed into the sum of two terms $a{\bf S}\cdot{\bf I}+{\bf S}\cdot{\bf A}\cdot{\bf I}$, where $a$ is the isotropic constant part, and ${\bf A}$ is the magnetic dipolar tensor. In solid state it gives the anisotropic term that shall be discussed in the following section. However, the average value of ${\bf A}$ is zero and {\it the observed hyperfine splitting gives the isotropic coupling constant directly.}

 The spin Hamiltonian for a system composed of an electron spin and one nuclear spin can be written as follows, in a very simple form. We have assumed that $g$ and $A$ are isotropic scalars as would be the case for a liquid phase sample, \cite{H1}, \cite{H8}. Also, we have the approximation of the first order perturbation theory or high field limit. In other words, the strong field is taken in the $z$ direction and $m_S$ and $m_I$ are valid quantum numbers. This approximation ignores second order hyperfine terms, $h^2A^2/g\beta H$. 
 \begin{equation}
 \mathcal{H}= g\beta HS_z + SAI - g_N\beta_NHI_z.
 \end{equation}
 
 The first term is electron Zeeman effect and actually it is the most dominant term in the Hamiltonin. This interaction determines the magnetic field at which the EPR signal is centered. The information on the electronic structure of the radical can be extracted through this term. However, some important information such as the molecular structure, spin distribution, or dynamics of the sample would not be possible to be measured by this term. Nuclear Zeemann interaction is represented with the last term. It gives information on the structure, or identity, of the nuclear spin. This term is comparably small. Usually, it is not possible to be measured in a typical EPR experiment because of the selection rule and more adequate ENDOR experiment would make it detectable.
 
 The second term in the above Hamiltonian is due to the interaction between electron and nuclear spins, hyperfine interaction. Very essential information on the structure, identity and number of atoms involved in the radical, their distances from the unpaired electron and electron spin densities are available through the hyperfine couplings if some quantum chemical considerations are made. The other term, not appearing in the above equation,  is the electric quadropolar interaction for the nuclear spin. P is the quadropolar matrix, generally a tensor and vanishing in solution. The nucleus with a spin larger than $1/2$ gives a non-zero quadropolar term. 
 
\begin{figure}
\begin{center}
\scalebox{0.69}
{\includegraphics[0cm,11cm][18cm,26.5cm]{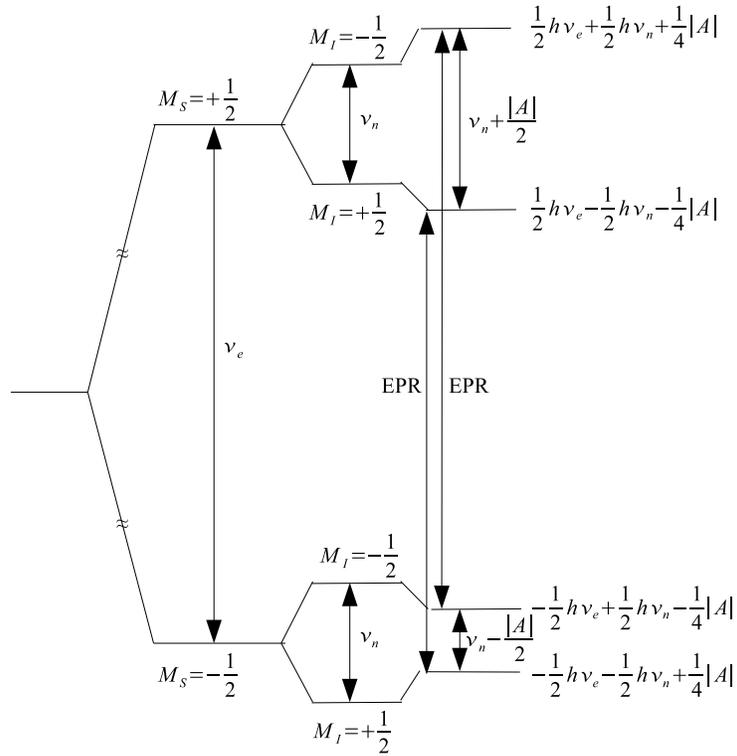}}
\end{center}
\caption{\label{lev1} Energy levels for a system composed of one electron spin $m_S=1/2$ and a nuclear spin $m_I=1/2$. Here, $\nu_n>|A|/2$, and $A<0$ and $g_n>0$.}
\end{figure}

\begin{figure}
\begin{center}
\scalebox{0.69}
{\includegraphics[0cm,11cm][18cm,26.5cm]{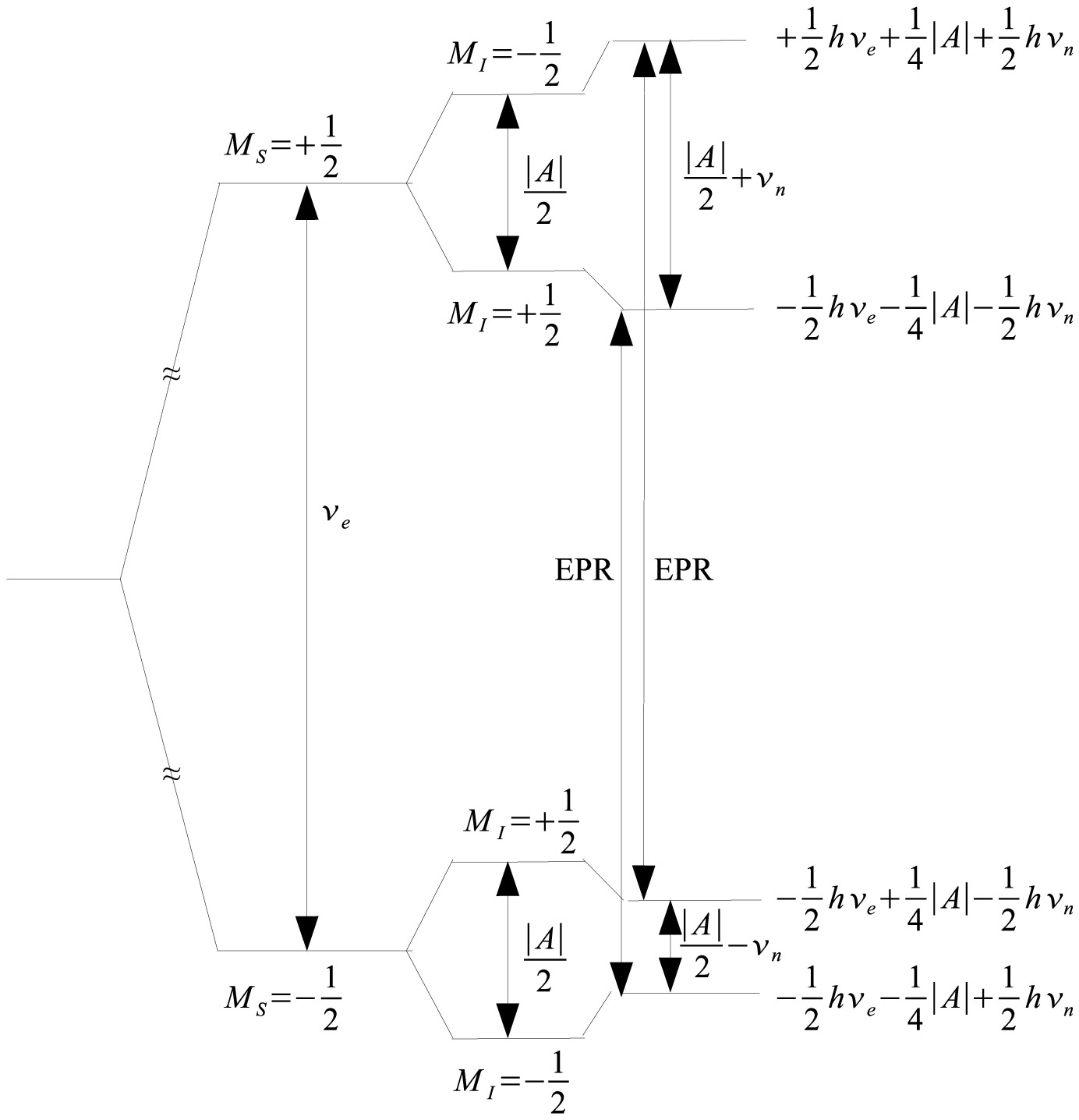}}
\end{center}
\caption{\label{lev2} Energy levels for a system composed of one electron spin $m_S=1/2$ and a nuclear spin $m_I=1/2$. Here, $|A|>2\nu_n$, and $A<0$ and $g_n>0$.}
\end{figure}

 We define the electron and nuclear spin resonances as follows
 \begin{eqnarray}
 \nu_e&=&g\beta H/h\\ \nonumber
 \nu_n&=&g_N\beta_N H/h
 \end{eqnarray}
 $\nu_e$ and $\nu_n$ are the electron spin resonance and nuclear spin resonance, respectively. Then, energy eigenstates are calculated as follows, see Figs.\ref{lev1} and \ref{lev2}
 \begin{equation}
 E_{m_S,m_I}/h=\nu_em_S-\nu_nm_I+am_Sm_I.
 \end{equation}
 
 Selection rules give the allowed transitions for EPR and ENDOR as follows
 \begin{equation}
 \Delta m_S=\pm 1, \Delta m_I=0,
 \end{equation}
 for EPR, and
  \begin{equation}
 \Delta m_S=0, \Delta m_I=\pm 1,
 \end{equation}
 for NMR.

\begin{figure}
\begin{center}
\scalebox{0.69}
{\includegraphics[0cm,8cm][21cm,22cm]{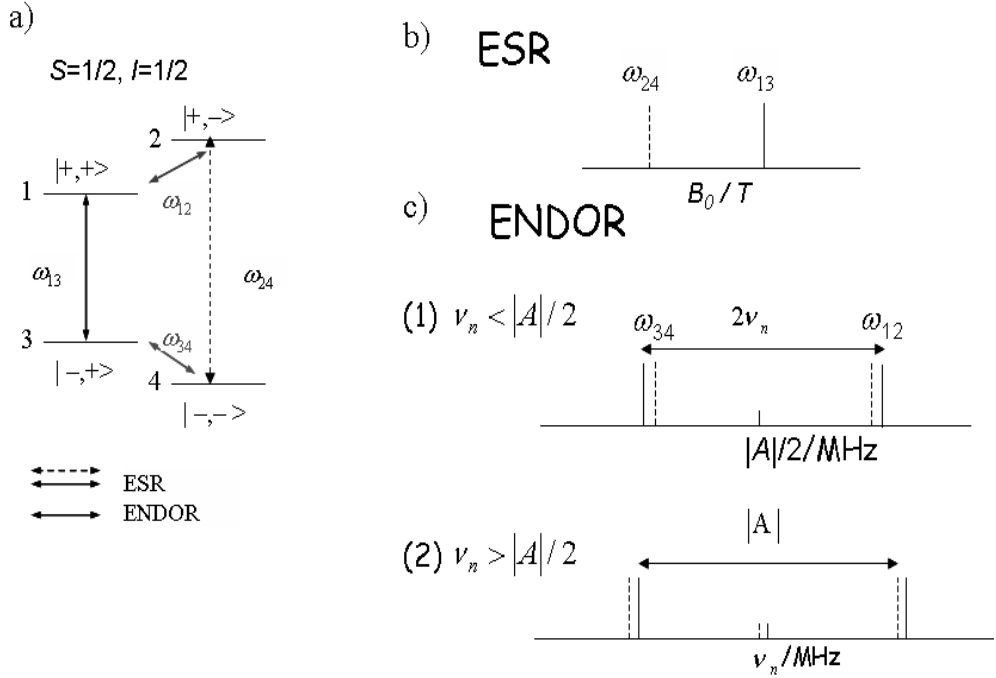}}
\end{center}
\caption{\label{spectra} A simple illustration of ENDOR Spectra.}
\end{figure}

 The typical EPR spectrum for this system, composed of an electron and a nuclear spin $1/2$ is two lines being centered at $\nu_e$ and separated by $|A|$, see Figure \ref{spectra}. For NMR, there are two possibilities. If $\nu_n>|A|/2$ then NMR signals are two picks, centered at $\nu_n$ and separated by $|A|$. If $\nu_n<|A|/2$ then the two lines are centered at $|A|/2$ and separated by $2\nu_n$. This is a direct explanation of energy eigenstates, as given by the above equation. This is shown in Figure \ref{spectra}
and the energy levels are shown in Figs. \ref{lev1}-\ref{lev2}. Through the calculation, parameters for any molecule being as the sample can be measured. We show in the following section a particular experiment, which is done in liquid phase.

In order to show the difference between EPR and ENDOR for the case that the number of nuclei is large let us assume, for simplicity, the situation in which a free radical involves two nuclei. The Hamiltonian of the system could be
\begin{equation}
\mathcal{H}=g\beta HS_z-g_N\beta_NH(I_{1z}+I_{2z})+a_1{\bf S}.{\bf I}_1+a_2{\bf S}.{\bf I}_2.
\end{equation}

For EPR the terms describing the nuclear Zeeman interactions can be omited because they do not affect EPR. However, they do affect ENDOR. Up to the first order perturbation, for EPR, we have
\begin{equation}
\mathcal{H}=g\beta HS_z+S_z(a_1I_{1z}+a_2I_{2z}).
\end{equation}

The energies are proportional to
\begin{equation}
E=m_s[g\beta H+a_1m_1+a_2m_2].
\end{equation}

Usually, the frequency is fixed and field is swiped in a small range on either side of the field that corresponds to the condition on resonance for electron $H^*$. We may convert the coupling constants $a_1$ and $a_2$ into Gauss. (1 Gauss$=2.80$ MHz, for $g$ of free electron.)

Then, we have
\begin{eqnarray}
\mathcal{H}&=&\mathcal{H}^*-a_1m_1-a_2m_2\\ \nonumber
           &=&\mathcal{H}^*-\sum_ia_im_i.
\end{eqnarray}
In EPR spectrum, for two nuclear spin sample, there are four lines. However, it often happens that several nuclei are equivalent to each other, usually occupying symmetrically equivalent positions in the molecule. Then, particularly in the above example we have only three lines, and then relative intensities are $1:2:1$, different to the previously equivalent intensity. Generally speaking, for one electron and $n$ equivalent e.g. protons, the EPR would show $n+1$ lines whose relative intensities are proportional to the coefficients of binomial expansion of $(1+x)^n$. For more than one electron also there would be some corresponding overlaping. For instance, naphtalene negative ion contains two sets of four identical protons. One set would give EPR with five lines with relative intensities as $1:4:6:4:1$. The other set would separate lines. Hence, complete spectrum contains $25$ lines and its analysis yields the two hyperfine couplings. By the intensity pattern the number of equivalent nuclei would be known.

However, ENDOR, as a double resonance technique, offers superior resolution to the hyperfine coupling constants compared to EPR. The nuclear $g$ values, $g_N$ is also measured in ENDOR, fascillating the identification of nucleous. The only information lost because of the fewer number of signals is the number of equivalent nuclei. But this can be achieved in advance through EPR.

The dramatic resolution enhancement achieved by ENDOR results to a large extent from the fact that two resonance conditions have to be fulfilled simultaneously, as to be clear from the Hamiltonian of the system. Then, it can highly simplify the spectrum.

\subsubsection{The Unpaired Spin Density}
The unpaired electron density or spin density $\rho({\bf r}_N)$ at the nucleus or at some other point in space is a probability density, measured in electrons/(angestroms)$^3$. In a real situation, a molecule contains many electrons, all of whose spins are couples together, and it is not in general possible to say there is just one ^^ ^^ unpaired electron" in a certain orbital, while the other electrons are perfectly paired to each other.

The correct contact Hamiltonian for molecule is represented by the operator as follows
\begin{equation}
H_c=\frac{8\pi}{3}g\beta g_N\beta_N\sum_k\delta({\bf r}_k-{\bf r}_N){\bf S}_k . {\bf I},
\end{equation}
where the sum runs over all the electrons, and the value of the splitting constant is found by averaging over the full many-electron wave function. The result can be written as follows
\begin{equation}
a=\frac{4\pi}{3}g\beta g_N\beta_N\rho({\bf r}_N),
\end{equation}
where $\rho({\bf r}_N)$, the unpaired electron density at the nucleus is defined as
\begin{equation}
\rho({\bf r}_N)=\int\phi^*\sum_k2S_{zk}\delta({\bf r}_k-{\bf r}_N)\psi d\tau,
\end{equation}
and the spin component of the state $\psi$ is taken to be $S_z=1/2$. $\rho({\bf r}_N)$ is simply the difference between the average number of electrons at the nucleus which have spin $\alpha$ and the number with spin $\beta$; the operator $2S_{zk}$ gives a factor $\pm 1$ in the above equation depending on the spin, while the $\delta$ function ensured that the electron is at the nucleus. If the nucleus is in a position where an excess of electrons have $\beta$ spin the value of $\rho({\bf r}_N)$ will be negative, and one may speak of a negative spin density.

On the other hand, the unpaired electron, in a molecular radical is often distributed over many atoms and gives proportionally smaller splitting. Thus, the probability that the electron is on the particular nuclear spin is quite small. Then, the spin density in an orbital, $\rho_H$, is a number, representing the functional population of unpaired electron, on an atom. There is a direct proportionality between the spin density and the hyperfine splitting. For instance, McConnel's relation is a good, not perfect, relation to provide a means of interpreting the spectra of virtually all aromatic free radicals. We will see some relations on this regard in the section for experiment with liquid state sample.

\subsection{Solid State Sample Study}
Many unstable radicals can be trapped in solids, and give highly interesting results. The radicals may be formed by ultraviolet irradiation, electron bombardment, or exposure to nuclear radiations. In this work we consider single crystals. The radicals are almost always regularly oriented relative to the crystal axes, and this provides the opportunity to rotate the crystal in the magnetic field and measure the anisotropic magnetic interactions. Although the presence of anisotropic tensors makes the analysis and interpretation of spectra more difficult than in solution, the results give valuable information about the structure and orientation of the unstable molecules.

Here we would like to have an overview on how to read spectra in a single crystal and to find out the anisotropic g and hyperfine tensors. We have used \cite{n6} and \cite{n7}, however other textbooks might be useful too.

In a particular crystal sample study, generally there would be no prior information on the principle axes of hyperfine and {\bf g} tensors and these would be determined through the experiment. The elements of {\bf g} and {\bf A} might be evaluated relative to an orthogonal system that can be identified by obvious structural features of the crystal. Then, through diagonalization of the {\bf g} and {\bf A} tensors, principle axes would be determined. Therefore, some considerations on the crystal systems and corresponding angles would be helpful, and might be found in some related text books.

\subsubsection{Evaluation of the {\bf g} Tensor}
In an experiment as you may see in the following section, {\bf g} tensor evaluation for interested sample, DPNO, has been done through EPR measurement. To the first order, one can obtain g-only experimental curves by taking measurements at the center of the symmetrical hyperfine multiplets. 

Second order effects can, however, skew the center of the hyperfine patterns, when the hyperfine splitting is large. In case the first order is just contributed the analyses would be as follows. Measurement of the sample would be done in different planes. For instance, suppose measurements are taken in the xy plane where  {\bf H} forms an angle $\theta$ with the $x$ axis, the effective $g_{\theta}^2$ for various $\theta$ values is given as follows
\begin{eqnarray}
g_{\theta}^2&=&(g_{xx}\cos\theta+g_{yx}\sin\theta)^2+(g_{xy}\cos\theta+g_{yy}\sin\theta)^2 \\ \nonumber
            &+& (g_{xz}\cos\theta+g_{yz}\sin\theta)^2 \\ \nonumber
            &=& g_x^2\cos^2\theta+g_y^2\sin^2\theta+2(g_{xx}g_{yx}+g_{xy}g_{yy}+g_{xy}g_{yz})\sin\theta\cos\theta,
\end{eqnarray}
where 
\begin{equation}
g_x^2=g_{xx}^2+g_{xy}^2+g_{xz}^2,
\end{equation}
\begin{equation}
g_y^2=g_{yy}^2+g_{yx}^2+g_{yz}^2,
\end{equation}
and where $g_x$ and $g_y$ are the effective $g$ values along the $x$ and $y$ axes. The above equation can be rewritten as the following form which is more convenient for plotting.
\begin{equation}
\label{gtheta}
g_{\theta}^2=K_1(x,y)+K_2(x,y)\cos2\theta+K_3(x,y)\sin2\theta,
\end{equation}
where
\begin{equation}
K_1(x,y)=\frac{1}{2}(g_x^2+g_y^2),
\end{equation}
\begin{equation}
K_2(x,y)=\frac{1}{2}(g_x^2-g_y^2),
\end{equation}
\begin{equation}
K_3(x,y)=g_{xx}g_{yx}+g_{xy}g_{yy}+g_{xz}g_{yz}.
\end{equation}

Expressions for the other coordinate planes can be obtained by obvious exchange of the coordinates.

Measurements along the three axes give the three effective values $g_x$, $g_y$, and $g_z$, and measurements in the three planes away from the axes at any $\theta$ can give only three additional constants. Because the tensor is symmetric, $g_{xy}=g_{yx}$, and so on, there are only six unknown tensor elements to be evaluated, and there are six undependent equations linking these unknowns to the measurable constants. Thus all elements of the {\bf g} tensor are determinable.

The effective $g_{\theta}$ values to be used in the \ref{gtheta} may be obtained from the relation
\begin{equation}
g_\theta=\frac{\nu_0}{\beta H_{\theta}},
\end{equation}
where $H_\theta$ are the resonance field values at the fixed frequency $\nu_0$ for the different orientations $\theta$.

In case that the matrix elements of ${\bf g}$ expressed in the arbitrary system $x$, $y$, $z$, as described in the preceding paragraphs, the resulting tensor can be diagonalized by a rotational transformation of the tensor to the principal axes. Let $l_{ix}$, $l_{iy}$, and $l_{iz}$ be the direction consines between the principal axis $i$ and the refrence axes $x$, $y$, and $z$, respectively, and let $g_i$ be the corresponding principal value. Since the off-diagonal elements vanish in the principal system, the transformation may be expressed as
\begin{equation}
\left(
\begin{array}{@{\,}ccc@{\,}}g_{xx}&g_{xy}&g_{xz}\\g_{yx}&g_{yy}&g_{yz}\\g_{zx}&g_{zy}&g_{zz}\\ \end{array}
\right)
\left(
\begin{array}{@{\,}c@{\,}}l_{ix}\\l_{iy}\\l_{iz} \\ \end{array}
\right)
= g_i
\left(
\begin{array}{@{\,}c@{\,}}l_{ix}\\l_{iy}\\l_{iz} \\ \end{array}
\right).
\end{equation}

This is the problem of finding the eigenvalues and eigenvectors. Solution for this problem gives three values of $g_i$, which are the principal values of {\bf g}. We use the designations $g_u$, $g_v$, and $g_w$, where $g_u$ is the smallest, $g_v$ is the intermediate and $g_w$ is the largest principal element of {\bf g} and where $u$, $v$, and $w$ signify the principal axes. Substitution of each of the principal values in turn for $g_i$ yields a set of three equations involving the three direction cosines for the particular principal axis. These equations with the auxiliary equation $l_{ix}^2+l_{iy}^2+l_{iz}^2=1$ may be solved for the direction cosines for the particular principal value $i=u, v, w$ that is substituted.

As a method of procedure, it is usually simpler to obtain the elements $(g^2)_{ij}$ of the squared $g$ matrix from \ref{gtheta} than those of the unsquared $g$. One can then diagonalize the squared matrix and take the square root of the diagonal elements to obtain the principal $g$ values. Since the $g$ values are always positive, this procedure causes no ambiguity in sign. The relationship of the squared to the unsquared elements is easily found from the matrix product rule:
\begin{equation}
(g^2)_{ij}=\langle i|g^2|j \rangle=\sum_{k=x,y,z}\langle i|g|k \rangle \langle k|g|j \rangle =\sum_k g_{ik}g_{ki}.
\end{equation}

The resulting expressions are simplified because the $g$ matrix is symmetric, that is, $g_{ij}=g_{ji}$. For example,
\begin{equation}
(g^2)_{xx}=g_{xx}g_{xx}+g_{xy}g_{yx}+g_{xz}g_{zx}=g_{xx}^2+g_{xy}^2+g_{xz}^2=g_x^2.
\end{equation}

Similarly, $(g^2)_{yy}=g^2_y$, $(g^2)_{zz}=g_z^2$, and
\begin{equation}
(g^2)_{xy}=g_{xx}g_{xy}+g_{xy}g_{yy}+g_{xz}g_{zy}=(g^2)_{yx}=K_3(x,y)=K_3(y,x).
\end{equation}

With the corresponding relationship for the other coordinates, $g$ square is expressed in terms of the observed constants by
\begin{equation}
\left|
\begin{array}{@{\,}ccc@{\,}}(g^2)_{xx}&(g^2)_{xy}&(g^2)_{xz}\\(g^2)_{yx}&(g^2)_{yy}&(g^2)_{yz}\\(g^2)_{zx}&(g^2)_{zy}&(g^2)_{zz} \\ \end{array}
\right| =
\left|
\begin{array}{@{\,}ccc@{\,}}g^2_x&K_3(x,y)&K_3(x,z)\\K_3(y,x)&g_y^2&K_3(y,z)\\K_3(z,x)&K_3(z,y)&g_z^2\\ \end{array}
\right|,
\end{equation}
where the $g_x^2$, $g_y^2$, and $g_z^2$ values along the respective axes are related to the $K$ constants by
\begin{equation}
g_x^2=K_1(x,y)+K_2(x,y)=K_1(z,x)-K_2(z,x),
\end{equation}
\begin{equation}
g_y^2=K_1(y,x)+K_2(y,x)=K_1(x,y)-K_2(x,y),
\end{equation}
\begin{equation}
g_z^2=K_1(z,y)+K_2(z,y)=K_1(y,z)-K_2(y,z).
\end{equation}

Then solving the eigenvalue problem gives the three values of $g_i^2$ that are the squared principal elements, $g_u^2$, $g_v^2$ and $g_w^2$, of the $g$ tensor.

The direction cosines, or eigenvectors, for the principal elements are, of course, the same as those for the squared principal elements and hence may be obtained from solution of the set of equations
\begin{equation}
[K_1(x,y)+K_2(x,y)-g_i^2]l_{ix}+K_3(x,y)l_{iy}+K_3(x,z)l_{iz}=0,
\end{equation}
\begin{equation}
K_3(y,x)l_{ix}+[K_1(y,x)+K_2(y,x)-g_i^2]l_{iy}+K_3(y,z)l_{iz}=0,
\end{equation}
\begin{equation}
K_3(z,x)l_{ix}+K_3(z,y)l_{iy}+[K_1(z,x)+K_2(z,y)-g_i^2]l_{iz}=0,
\end{equation}
where $i=u,v,w$ and where the values $g_u^2$, $g_v^2$, and $g_w^2$ obtained from eigenvalue problem as above, are in turn substituted for $g_i^2$. The solution of these equations yields only the ratios of the direction cosines; the addition $l_{ix}^2+l_{iy}^2+l_{iz}^2=1$ is required to give their specific value. For instance, the analysis of the DPNO, we used the both ways as described above.
\subsubsection{Evaluation of Hyperfine Coupling Tensor}
In a similar manner to that explained for the evaluation of the $g$ tensor as explained above, here also we would scape the theoretical background that might be found in text books and will show how to evaluate hyperfine tensor from the detected spectrum. We neglect the effect of $g$ tensor anisotropy. This assumption is valid for most free radicals. For these cases, $g$ tensor anisotropy is sufficiently small that its effect on the nuclear coupling tensor is negligible. Then, almost the same process as for $g$ tensor would be performed for $A$.

We start calculation for the effective coupling in the $xy$ reference plane. 
\begin{equation}
A_\theta^2=A_x^2\cos^2\theta+A_y^2\sin^2\theta+2(A_{xx}A_{yx}+A_{xy}A_{yy}+A_{xz}A_{yz})\sin\theta\cos\theta,
\end{equation}
where $A_x$ and $A_y$ are the effective coupling along the $x$ and $y$ axes, which are expressed in terms of the squared tensor elements by
\begin{equation}
A_x^2=A_{xx}^2+A_{xy}^2+A_{xz}^2,
\end{equation}
\begin{equation}
A_y^2=A_{yy}^2+A_{yx}^2+A_{yz}^2.
\end{equation}

Note that these equations have the same form as those for the $g$ tensor and hence may be expressed as
\begin{equation}
A_\theta^2(x,y)=C_1(x,y)+C_2(x,y)\cos 2\theta+C_3(x,y)\sin 2\theta,
\end{equation}
where $\theta$ is the angle between the $x$ axis and the applied field, and where
\begin{equation}
C_1(x,y)=\frac{1}{2}(A_x^2+A_y^2),
\end{equation}
\begin{equation}
C_2(x,y)=\frac{1}{2}(A_x^2-A_y^2),
\end{equation}
\begin{equation}
C_3(x,y)=A_{xx}A_{yx}+A_{xy}A_{yy}+A_{xz}A_{yz}.
\end{equation}

Similar expressions for the other coordinate planes may be obtained by obvious changes of the coordinates.

By measuring the hyperfine splittings in the three coordinate planes and by squaring the resulting values, one can evaluate the six independent $C$ constants. The elements of the squared hyperfine coupling may be obtained from these measurable constants by use of the relations
\begin{equation}
(A^2)_{xx}=A_x^2=C_1(x,y)+C_2(x,y)=C_1(z,x)-C_2(z,x),
\end{equation}
\begin{equation}
(A^2)_{yy}=A_y^2=C_1(y,x)+C_2(y,x)=C_1(x,y)-C_2(x,y),
\end{equation}
\begin{equation}
(A^2)_{zz}=A_z^2=C_1(z,y)+C_2(z,y)=C_1(y,z)-C_2(y,z),
\end{equation}
\begin{equation}
(A^2)_{xy}=(A^2)_{yx}=C_3(x,y)=C_3(y,x),
\end{equation}
\begin{equation}
(A^2)_{xz}=(A^2)_{zx}=C_3(x,z)=C_3(z,x),
\end{equation}
\begin{equation}
(A^2)_{yz}=(A^2)_{zy}=C_3(y,z)=C_3(z,y),
\end{equation}
which are similar to those for squared $g$ values. One substitutes the numerical values of squared hyperfine coupling and solves for the three eigenvalues $A_i^2=A_u^2, A_v^2, A_w^2$, which are the squared principal elements. The direction cosines for each of the principal axes $i=u,v,w$ would be the eigenvectors. The direction cosines of the principal axes of the nuclear tensor are often the same as those of the $g$ tensor, but not generally.

 \section{Practice of EPR/ENDOR}
 
 It might be easier to understand EPR/ENDOR, if we have a look at the components for an EPR/ENDOR spectrometer. The heart of the EPR/ENDOR is the resonant cavity, which contains the sample and directs the microwave to and from the sample. The absorption of microwave by the sample is monitored, amplified and recorded by the detection with modulation systems. To provide a resonant condition requires a stable, linearly variable, and homogeneous magnetic field of arbitrary magnitude.

\begin{figure}
\begin{center}
\scalebox{0.8}
{\includegraphics[0cm,13cm][21cm,27cm]{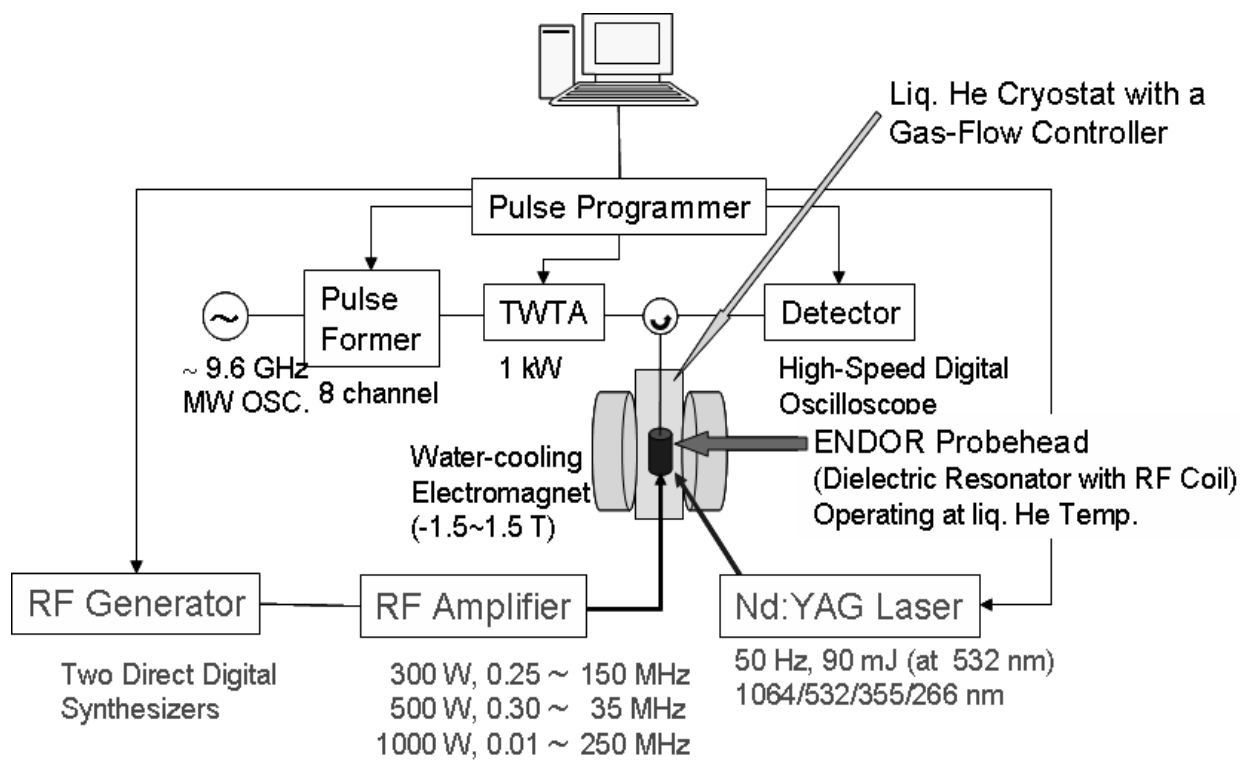}}
\end{center}
\caption{\label{aparatus} Block diagram of (pulsed QC-)ENDOR setup, X-band (10 GHz).}
\end{figure}

The main components of an ENDOR spectrometer is shown in Figure \ref{aparatus}.
 The first significantly visible part is a magnet, which turns the electric energy levels. Then, the important part is a microwave resonant cavity, an ENDOR RF coil, which are metal box in order to amplify weak signals from the sample. Sample is placed in the cavity. In our cw experiment, a TM mode cavity has been used. The important part is bridge that includes the electromagnetic radiation source, microwave, and the detector. In addition, similar to other spectroscopies, console contains signal processing and control electronics and a computer, \cite{H1}, \cite{H6}, \cite{H7}.
 
 Microwave bridge contains the microwave source and the detector. The microwave radiation is originated by a vacuum tube called a Klystron. In our experiment a GUN oscillator has been used. The reason why we fix the frequency of microwave and change the magnetic field is that microwave resources can be tuned only over a very small frequency range comparing with magnitudes of interactions appearing in the spin Hamiltonian.
 
 The microwave is applied to the sample inside the cavity through a small hole in the cavity called an iris, \cite{H1}, \cite{H6}. Then iris should be properly adjusted in a way that the microwave power changes at the detector only because of the absorption of microwave energy by the sample. In order to adjust the level of microwave power, an attenuator is placed in between the output of the microwave source and the sample. Attenuator controls the microwave power by blocking the flow of microwave radiation. There is also a circulator inside the bridge, which determines the trajectories for which each input and output microwave should follow. In order to detect the signal, there is a Schottkey barrier diode, which converts the microwave power to an electric current. For an accurate qualitative signal intensity measurements as well as optimal sensitivity, the diode should be fixed to the linear working range, meaning the range where the current is proportional to the square root of the microwave power. This happens at higher levels, greater than 1 milliWatt.
 
 A microwave resonant cavity is made specially with dimensions, that is on resonance with the fundamental frequency of the microwave. For example, for a cavity dimensions of $1.7\times 0.9\times 0.45$ in$^3$ , the fundamental frequency occurs at a microwave frequency of $9.5$ GHz. Out of resonance, no microwave would be derived. Hence, any absorption of the fundamental resonance of the cavity by sample will disturb the resonance frequency, which can in turn be detected by the proper circuitry.
 
 Sample size also may turn out eventually to be of particular importance. EPR/ENDOR signal intensity depends on the total number of spins in the active part of the cavity. Then, it will be better to use the largest sample tube in order to get higher signal intensity.
 
 For particular practice of ENDOR, only there are some other experimental equipments that are added to EPR. DICE ENDOR is for generating RF field. Then, a proper RF power amplifier is also required. In Figure \ref{aparatus}, the laser part is for our future work of high spin polarization and selective addressing.
 
 The work that is explained here is summarized as following steps.
 \begin{description}
 \item[1.] We measure cw EPR and ENDOR of liquid and solid state samples to see whether there could be any signal from those samples, also to get sample information that is required for QC. The information should be composed of elaborate analyses of molecular magnetic properties of the open-shell entity under study.
 
 \item[2.] Then throughout the pulsed EPR experiment, we measure $T_1$ and $T_2$ to see whether the sample can be a proper choice for quantum computing and quantum information processing for a matter of decoherence time. Recall that one of the conditions for a proper sample is the long decoherence time.
 \item[3.] Finally, the experiments on implementations of quantum operations, making pseudo-pure state and measurement issue by means of pulsed ENDOR will be explained and more importantly, the experiment on entanglement preparation between spins involved in ENDOR experiment will be discussed.
 \end{description}
 
 We first start with experiments on samples in liquid phase. Samples in liquid phase, which enjoy rich nuclear qubits are for future usage for solution-ENDOR QC experiments. More elaborated examinations of the magnetic property such as linewidth of ENDOR signals as a function of temperature, etc. are required.Then we will give some experimental results on solid state samples, particularly appropriate for low temperature ENDOR QC, for experiment with quantum entanglement.
 

 \subsection{Experiment on 2-(N-aminoxyl-N-t-Bu)-benzimidazole, BABI:}
\begin{figure}
\begin{center}
\scalebox{0.99}
{\includegraphics[6cm,20cm][15cm,24cm]{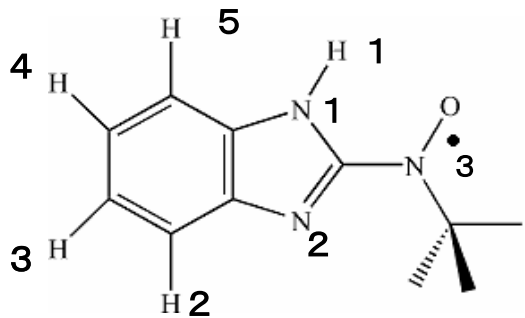}}
\end{center}
\caption{\label{BABI} Molecular structure of 2-(N-aminoxyl-N-t-Bu)-benzimidazole, BABI.}
\end{figure}

\begin{figure}
\begin{center}
\scalebox{0.79}
{\includegraphics[0cm,0cm][13cm,8.5cm]{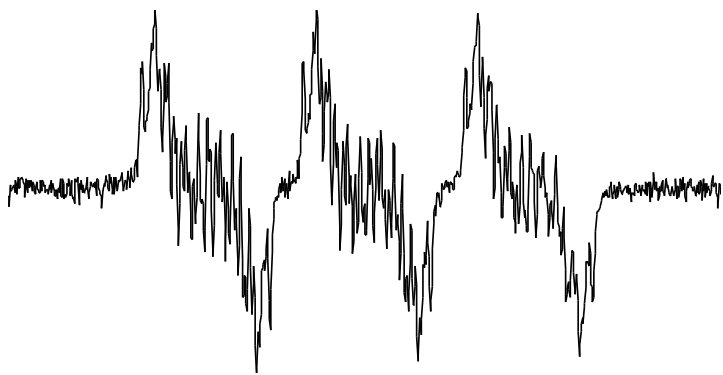}}
\end{center}
\caption{\label{BABIepr} EPR spectra of 2-(N-aminoxyl-N-t-Bu)-benzimidazole, BABI.}
\end{figure}

\begin{figure}
\begin{center}
\scalebox{0.39}
{\includegraphics[0cm,5.5cm][21cm,21cm]{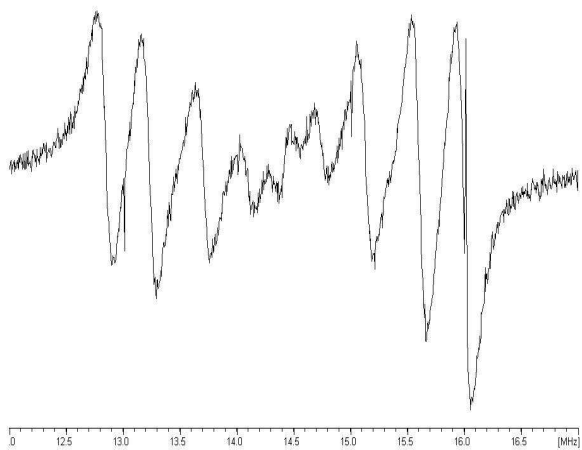}}
\end{center}
\caption{\label{BABIendor} ENDOR spectra of 2-(N-aminoxyl-N-t-Bu)-benzimidazole, BABI.}
\end{figure}

 The particular sample, which we have studied is 2-(N-aminoxyl-N-t-Bu)-benzimidazole, BABI, in liquid phase. The molecular structure is shown in Figure \ref{BABI}.
 This molecule is an open shell molecule with hydrogen bonding in its crystalline state. From solution EPR and $^1$H-ENDOR spectra observed in mineral oil at ambient temperature, all the $^{14}$N- and $^1$H-hyperfine coupling parameters of their compounds were determined. An EPR spectrum is shown in Figure \ref{BABIepr}, and the ENDOR spectra is shown in Figure \ref{BABIendor}.

In order to discuss electronic and molecular structures of their compounds, theoretical calculations based on the density functional theory were carried out. The observed hyperfine coupling constants have been compared with those by the DFT, density functional theory, calculations, Table\ref{T2}. 

 \begin{table}
 \begin{center}
  \caption{\label{T2} Experimentally determined hyperfine coupling constants compared with those by DFT calculations, $a/$MHz. Previous work refers to [].}
\small
\begin{tabularx}{\linewidth}{|X|XX|XX|}
\hline
 Site  & This Work  &                & Previous Study &               \\  \hline
       & Experiment & UB3LYP/ cc-pVDZ & Experiment     & UB3LYP/ cc-pVDZ \\ \hline
 N(1)  &1.164       & 1.134          & 2.162          & 0.881          \\ 
 N(2)  &6.040       & 7.865          & 5.980          & 6.387          \\
 N(3)  &28.752      & 25.990         & 28.525         & 19.791         \\ 
 H(1)  &-1.524      & -1.735         & 2.471          & -1.775         \\ 
 H(2)  &-2.446      & -2.192         & 1.432          & -1.977         \\ 
 H(3)  &0.231       & 0.537          & 1.039          & -0.174         \\ 
 H(4)  &-3.170      & -2.843         & 0.590          & -2.774         \\  
 H(5)  &0.676       & 0.701          &0.309           & 0.076         \\\hline   
 \end{tabularx}
 \end{center}
 \end{table}

\begin{figure}
\begin{center}
\scalebox{0.99}
{\includegraphics[0cm,0.5cm][10cm,6cm]{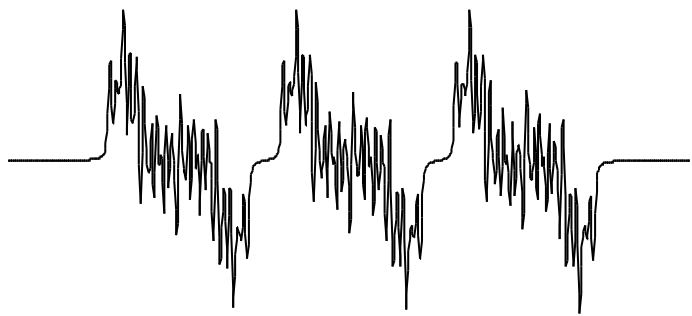}}
\end{center}
\caption{\label{BABIsim} Simulation of EPR spectra of 2-(N-aminoxyl-N-t-Bu)-benzimidazole, BABI.}
\end{figure}

The simulated EPR spectrum which is based on the calculated parameters is shown in Figure \ref{BABIsim}. Very well coincidence exists between the two EPR spectra, one from experiment and the other from simulation.

\begin{figure}
\begin{center}
\scalebox{0.69}
{\includegraphics[0cm,0cm][10cm,6cm]{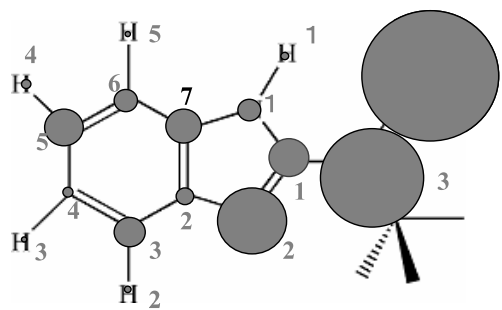}}
\end{center}
\caption{\label{BABIspin} Spin densities on each site.}
\end{figure}

 Spin density can be calculated through McConnell's equation as follows
 \begin{eqnarray}
 &a&({\rm H})=1426.2\times\rho({\rm H})\\ \nonumber
 &a&({\rm N})=84.2\times\rho({\rm N})\\ \nonumber
 &a&(\pi{\rm C-H})=-61.8\times\rho(\pi{\rm C-H})
 \end{eqnarray}
 The spin densities calculated for each site is shown in Table\ref{T3}. 
  \begin{table}
 \begin{center}
  \caption{\label{T3} Spin densities from experiment and calculation.}
\small
\begin{tabular}{|c|c|c|}
\hline
 Site  & Experiment  &  UBLYP/cc-pVDZ    \\ \hline
 N(1)  &0.0138       & 0.0240          \\ 
 N(2)  &0.0717       & 0.1293          \\
 N(3)  &0.3414       & 0.3494          \\ 
 C(1)  &             & -0.0698         \\ 
 C(2)  &             & -0.0248         \\ 
 C(3)  &0.0396       & 0.0368          \\ 
 C(4)  &-0.0037      & -0.0120         \\  
 C(5)  &0.0514       & 0.0476          \\
 C(6)  &-0.0110      & -0.0176         \\  
 C(7)  &             & 0.0358          \\
 H(1)  &-0.0011      & -0.0020         \\ 
 H(2)  &-0.0017      & -0.0014         \\ 
 H(3)  &0.0002       & 0.0005          \\ 
 H(4)  &-0.0022      & -0.0018         \\  
 H(5)  &0.0005       & 0.0004          \\
 O     &             & 0.5026          \\ \hline   
 \end{tabular}
 \end{center}
 \end{table}
 The spin density on each site is shown in Figure \ref{BABIspin} by circles with areas representing the magnitude.

 \subsection{Solid State Sample Study for ENDOR QC; Malonyl Radical}
  
\begin{figure}
\begin{center}
\scalebox{0.59}
{\includegraphics[0cm,19cm][21cm,24cm]{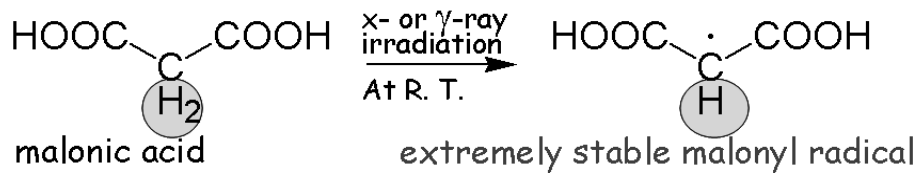}}
\end{center}
\caption{\label{malonyl} Generation of malonyl radical by irradiating the sample.}
\end{figure}

Malonyl radicals incorporated in the single crystal of malonic acid were generated by X-ray irradiation at ambient temperature as shown in Figure \ref{malonyl}. Spin Hamiltonian parameters of malonyl radical under study, as summarized in Table \ref{T4}, have been reported by McConnell and coworkers.

 \begin{table}
 \begin{center}
  \caption{\label{T4} Magnetic tensors parameters of malonyl radical.}
\small
\begin{tabular}{|c|ccc|ccc|}
\hline
    S & {\bf G}$_{xx}$ &{\bf G}$_{yy}$ &{\bf G}$_{zz}$& $^a$A$_{xx}$/MHz & $^a$A$_{yy}$/MHz & $^a$A$_{zz}$/MHz\\ \hline
 1/2  &2.0026        & 2.0035        &  2.00331     &  -29          &  -61          &  -91   \\ \hline  
 \end{tabular}
 \end{center}
 \end{table}

  \begin{figure}
\begin{center}
\scalebox{0.79}
{\includegraphics[0cm,0cm][7cm,4cm]{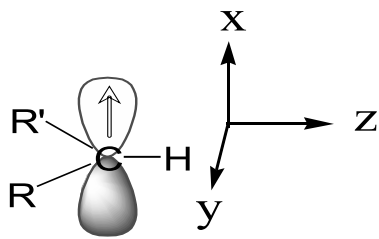}}
\end{center}
\caption{\label{Fig11} Principal axes of malonyl acid.}
\end{figure}

  \begin{figure}
\begin{center}
\scalebox{0.79}
{\includegraphics[0cm,0cm][11cm,6.5cm]{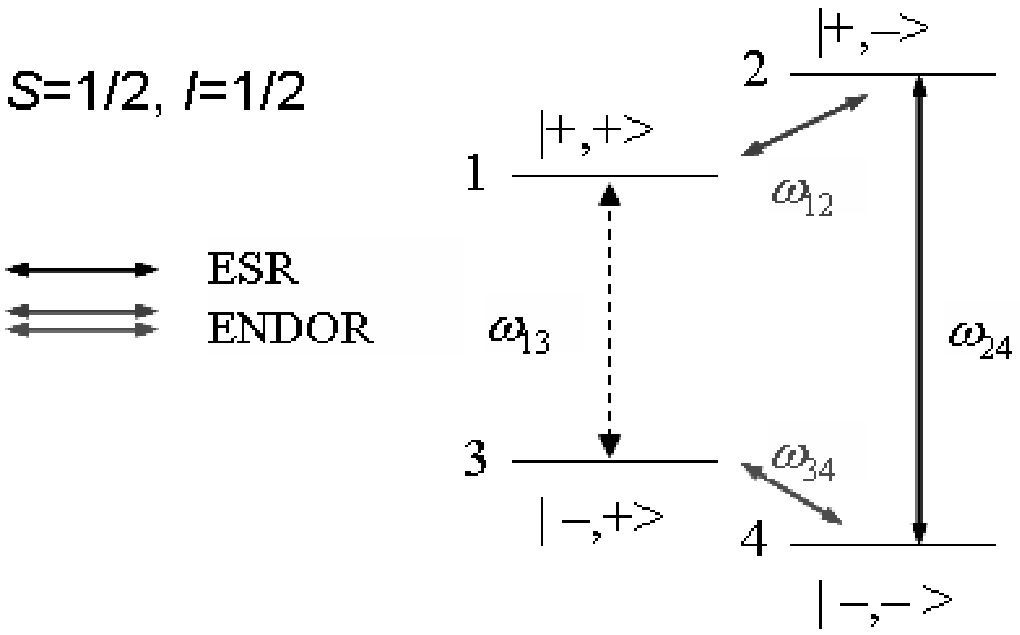}}
\end{center}
\caption{\label{MalonylX} Energy levels for malonyl, X-band.}
\end{figure}

  \begin{figure}
\begin{center}
\scalebox{0.79}
{\includegraphics[0cm,0cm][11cm,6.5cm]{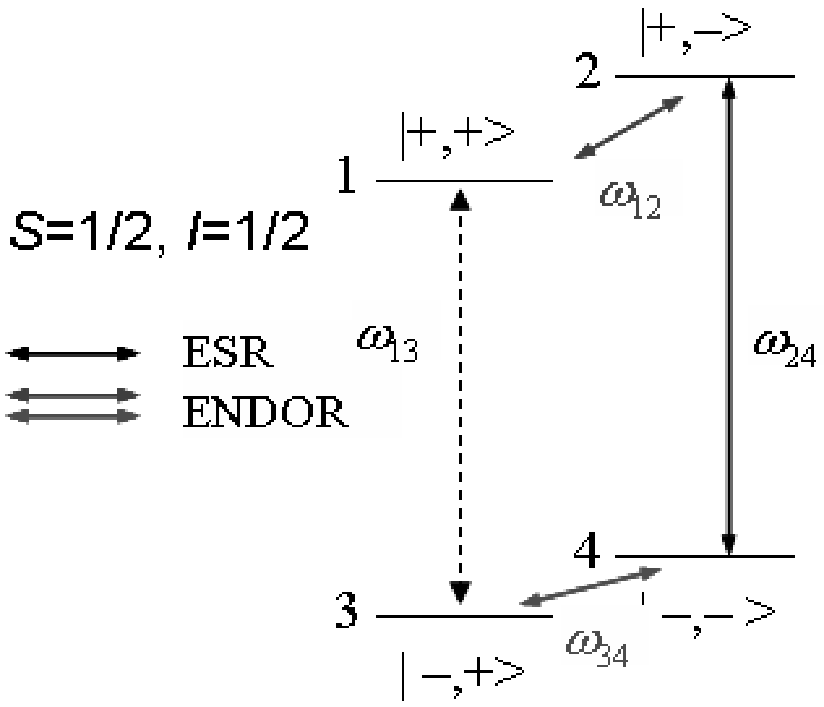}}
\end{center}
\caption{\label{MalonylQ} Energy levels for malonyl, Q-band.}
\end{figure}

  \begin{figure}
\begin{center}
\scalebox{0.79}
{\includegraphics[0cm,0cm][12cm,7cm]{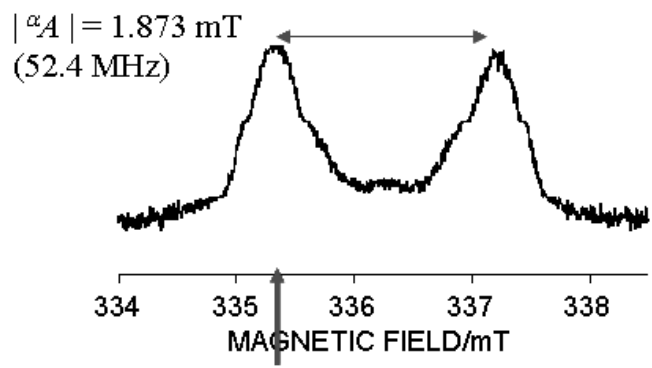}}
\end{center}
\caption{\label{MalonylXepr} Pulsed EPR spectrum of malonyl radical in the single crystal, X-band. The arrow indicates the static magnetic field for the ENDOR measurements.}
\end{figure}

  \begin{figure}
\begin{center}
\scalebox{0.79}
{\includegraphics[2cm,18.5cm][19cm,25cm]{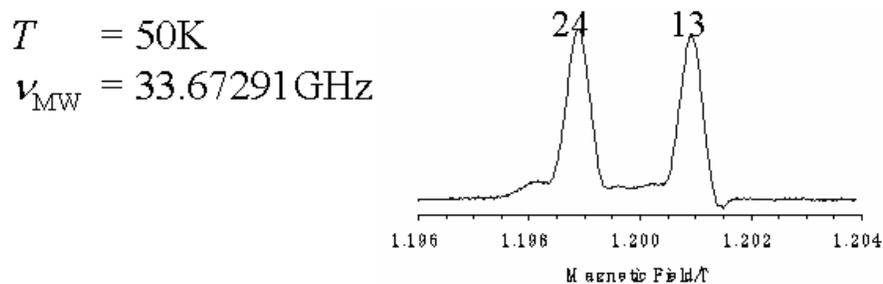}}
\end{center}
\caption{\label{MalonylQepr} Pulsed EPR spectrum of malonyl radical in the single crystal, Q-band. }
\end{figure}

  \begin{figure}
\begin{center}
\scalebox{0.69}
{\includegraphics[0cm,0cm][20cm,6cm]{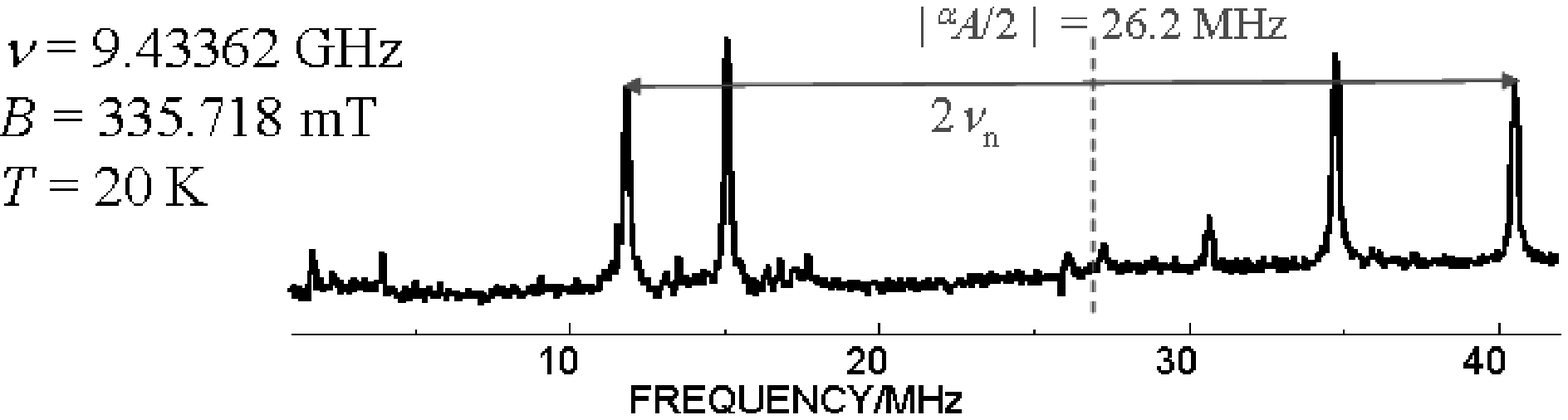}}
\end{center}
\caption{\label{MalonylXendor} Pulsed ENDOR spectrum from malonyl radical with the magnetic field set as in Figure \ref{MalonylXepr}, X-band}
\end{figure}

  \begin{figure}
\begin{center}
\scalebox{0.69}
{\includegraphics[0cm,0cm][21cm,9cm]{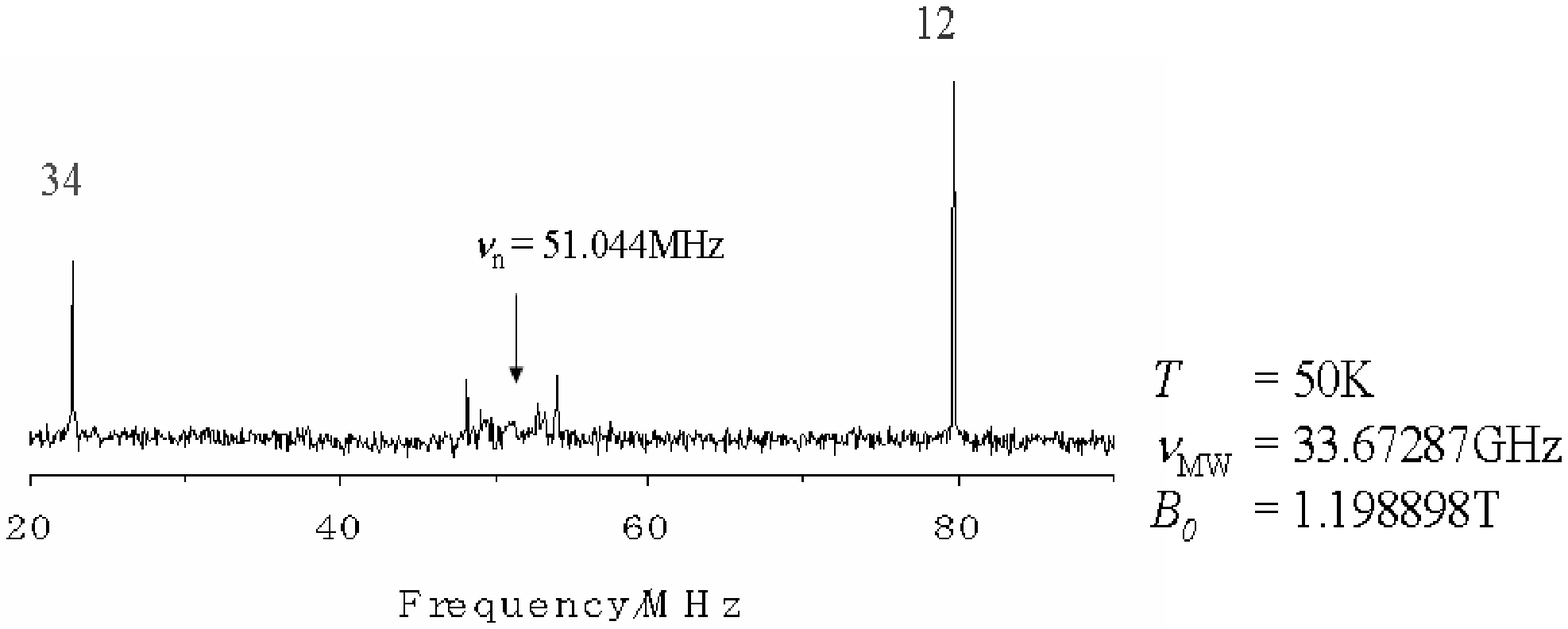}}
\end{center}
\caption{\label{MalonylQendor} Pulsed ENDOR spectrum from malonyl radical with the magnetic field set as in Figure \ref{MalonylQepr}, Q-band.}
\end{figure}
 Figs. \ref{MalonylX}-\ref{MalonylXendor} show the experimental results on malonyl radical at X-band and Q-band. The corresponding energy levels and pulsed EPR spectra and Pulsed ENDOR spectra are shown to be different particularly in case of pulsed ENDOR for X-band and Q-band. The magnetic field for Q-band is more than three times larger than X-band. Then the nuclear magnetic field becomes three times lager and its correspondance relation with the hyperfine coupling changes and yields to a different form of ENDOR spectra.

 \subsection{Solid State Sample Study for ENDOR QC; Diphenylnitroxide}
 Open-shell molecular entities as molecular devices such as quantum computing and quantum information processing have emerged in interdiciplinary areas, giving us new aspects of molecular spinics or molecular spin science. During the coarse of implementation of pulsed ENDOR based quantum computing and quantum information processing, we have studied extremely stable diphenylnitroxide, DPNO, profiles which are necessary for molecular designing for the above molecular devices.
 
 One nice feature of DPNO, which makes it potentially a proper sample for quantum information processing and quantum computation is the decoherence time. Through the measurement of decoherence time, it turns out that DPNO is extremely stable sample and even more importantly it is possible rather simply to control the decoherence time by controlling the dilution of the sample. Therefore we have been trying to get full analysis of DPNO.
 
 DPNO has been studied previously by several groups. To our knowledge, the first pioneering work has been Deguchi, 1961. This work is an ESR work both on pure DPNO single crystals and on DPNO diluted in benzophenone single crystals, even before the x-ray structural analysis appeared in 1968. The exchange line broadening was analyzed in this work. Then, later high resolution hyperfine EPR spectra in solution were analyzed by Yamauchi, 1967. The {\bf g} tensor for DPNO diluted in benzophenone lattices and the hyperfine tensor of the nitrogen nucleus of the nitroxide site were determined by conventional EPR spectroscopy as early as 1970's by Tien-S. Lin. The principle values for {\bf g} tensor of DPNO and hyperfine tensor {\bf A} for nitrogen have been reported to be, $g_{xx}=2.0092,\,g_{yy}=2.0056,\,g_{zz}=2.0022,\,A_{xx}=1.9\,{\rm G},\,A_{yy}=3.6\,{\rm G},\, A_{zz}=23.8\,{\rm G}$. Brustolon's group revisited DPNO in the benzophenone lattice by cw ENDOR spectroscopy in 1988. According to their paper, the {\bf g} tensor has the principle values of $g_{xx}=2.0079,\,g_{yy}=2.0040,\,g_{zz}=2.0014$. Only preliminary results of the nitrogen hyperfine coupling and quadropole tensors in the abc system were given, though the tensor analysis was not given. Also, $^1$H ENDOR/general-TRIPLE  spectroscopy was carried out in isopropanol at 210 K and radical pairs were detected from the concentrated mixed crystals. In between the two latter introduced works, Yamauchi et al. did proton ENDOR in ethylbenzene at 203 K, 1987. The substituted ortho-methyl effect was influential.  Nevertheless, complete analyses of the magnetic tensors of DPNO from the experimental side have never been documented yet because of inhomogeneously broadened EPR lines in the solid state due to the existence of protons. 

 There are several reasons on our interest on studying DPNO. We have selected Diphenylnitroxide as an {\it electron-spin bus QC/QIP molecule} as a central hardware for a molecule-based quantum spin computer. This is due to possibility of the sample for controlling the decoherence time. DPNO is selected also because that it is a stable molecular-spin physical system with {\it well characterized magnetic tensors} under strong microwave and radio frequency irradiations. The irradiation in this regard is sometimes highly required for quantum information processing and quantum computation. DPNO is very easy-to-dilute molecular spin in diamagnetic molecular lattices which makes it a quite perfect sample to "control" decoherence time. We have prepared mixed single crystals of a desired host-guest with variety of concentration ratios. DPNO is also selected as it is feasible to have measurements in a {\it wide range of temperature}. This point specially turns out to be important if we consider the critical experimental conditions for realization of a quantum entanglement as it is required to go down to a few Kelvin. Furthermore, it is also possible to extend the ENDOR based quantum computation with this sample to {\it larger number of qubits} as this sample is a proper sample for extension to multi-nuclear spin-bus systems by stable isotopic labeling and $A$ tensor engineering. All the points indicated here, if collected to each other make it a valuable task to study and extract full analysis of DPNO.  
 
\begin{figure}
 \begin{center}
\scalebox{0.69}
{\includegraphics[0cm,11.5cm][21cm,17cm]{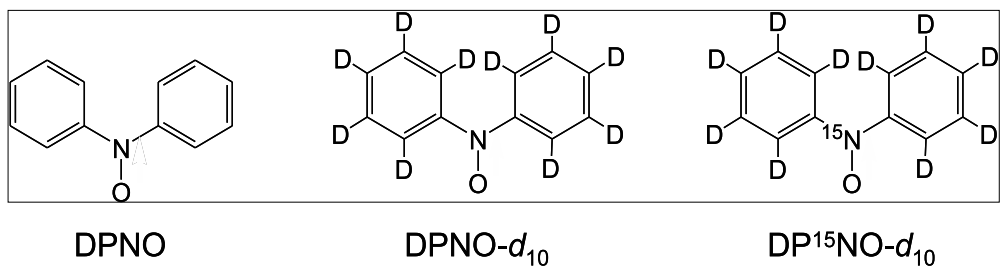}}
\end{center}
\caption{\label{DPNO2} Different samples that have been used for acquiring data on magnetic tensors for DPNO.}
\end{figure}

 The accurate magnetic tensors of DPNO in the benzophenone lattice have been acquired, for the first time, in this work, by single crystal cw ENDOR spectroscopy, X-band, at ambient temperature. The analyses include an accurate determination of the quadropolar tensor of the nitrogen nucleus.  In order to come over the broadening problem due to existence of protons, we have synthesized DPNO, DPNO-$d_{10}$ and stable isotope labeled ones, including DP$^{15}$NO-d$_{10}$, according to the method modifying the procedure described in the literature,  Figure \ref{DPNO2}.

\begin{figure}
 \begin{center}
\scalebox{0.69}
{\includegraphics[0cm,0cm][9cm,10cm]{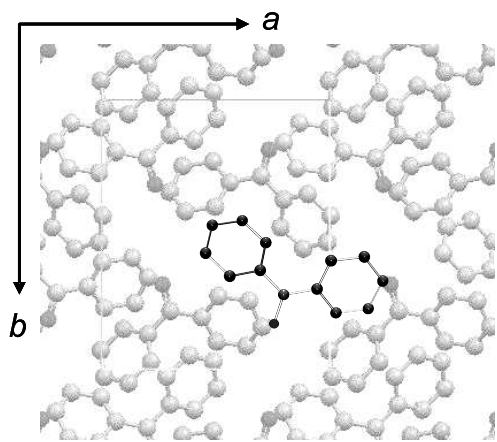}}
\end{center}
\caption{\label{DPNO3} Benzophenone molecule is isostructurally substituted by DPNO .}
\end{figure}

\begin{figure}
 \begin{center}
\scalebox{0.69}
{\includegraphics[3cm,10cm][18.5cm,18cm]{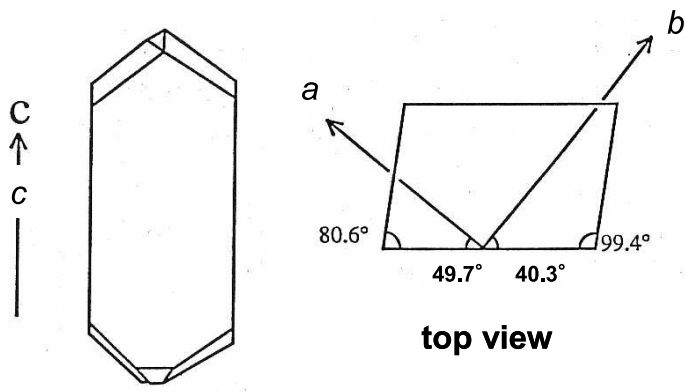}}
\end{center}
\caption{\label{DPNO4} External form of benzophenone single crystal.}
\end{figure}

 For sample preparation, DPNO has been magnetically diluted in benzophenone single crystals. benzophenone molecule is isostructurally substituted by DPNO at a desired concentration, see Figure \ref{DPNO3}. 
 External form of benzophenone single crystals, as a host molecule, is shown in Figure \ref{DPNO4}. 
The crystal symmetry of benzophenone is orthorhombic with $a = 10.30\,\overset{\circ}{\rm A}$, $b = 12.15\,\overset{\circ}{\rm A}$ and $c =   8.00\,\overset{\circ}{\rm A}$, with four molecules in each unit cell.

\begin{figure}
 \begin{center}
\scalebox{0.69}
{\includegraphics[0cm,1cm][21cm,11cm]{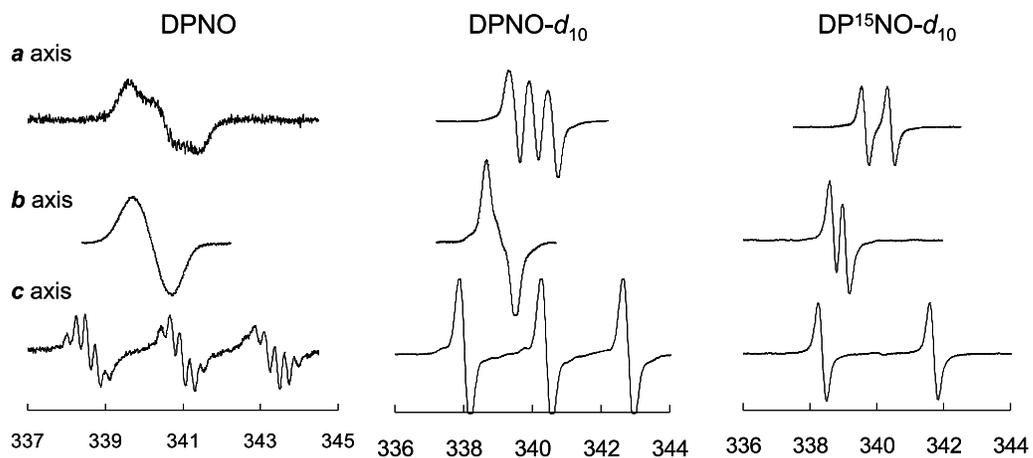}}
\end{center}
\caption{\label{DPNO5} Single-crystal EPR spectra of DPNO, DPNO-$d_{10}$, and DP$^{15}$15NO-$d_{10}$ in a benzophenone single crystal. The horizontal axes are magnetic field in milli Tesla}
\end{figure}

All the angular dependencies of EPR and ENDOR spectra were carried out in the crystallographic abc coordinate system at ambient temperature with a Bruker ESP350 spectrometer (a TM mode cavity and helical coil for RF irradiation). In each crystalographic plane, two of the molecules are equivalent. Therefore, two sets of molecules exist while for measurements on axis, all the four molecules of a unit cell are equivalent and then there is not any site splitting due to different sites, see Figure \ref{DPNO5}. 
We have used this property in order to define the axis during our experiment. For DPNO, as will be shown later site splitting is not so clear because of the proton broadening of the spectra, while this fact for each plane is more clear in case of DP$^{15}$NO-$d_{10}$. Note that eventhough we have made measurements at ambient temperature, the ENDOR spectra can be observed at other desired temperature in the range of 350 K to liquid helium temperature. 
 
\begin{figure}
 \begin{center}
\scalebox{0.69}
{\includegraphics[0cm,11cm][21cm,25cm]{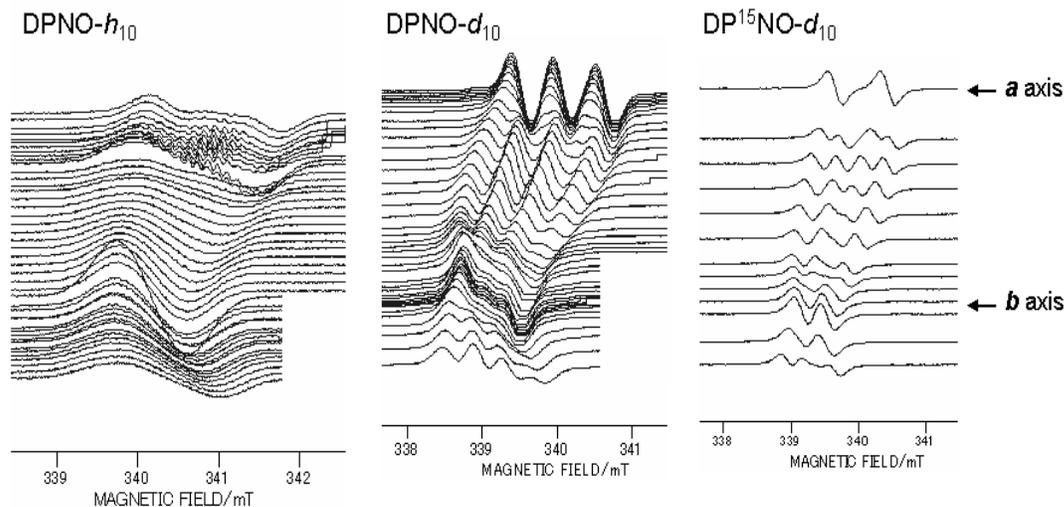}}
\end{center}
\caption{\label{DPNO6} Angular dependence of single-crystal EPR spectra of DPNO, DPNO-$d_{10}$, and DP$^{15}$NO-$d_{10}$ in the crystallographic {\it ab} plane.}
\end{figure}

\begin{figure}
 \begin{center}
\scalebox{0.69}
{\includegraphics[0cm,13cm][21cm,22cm]{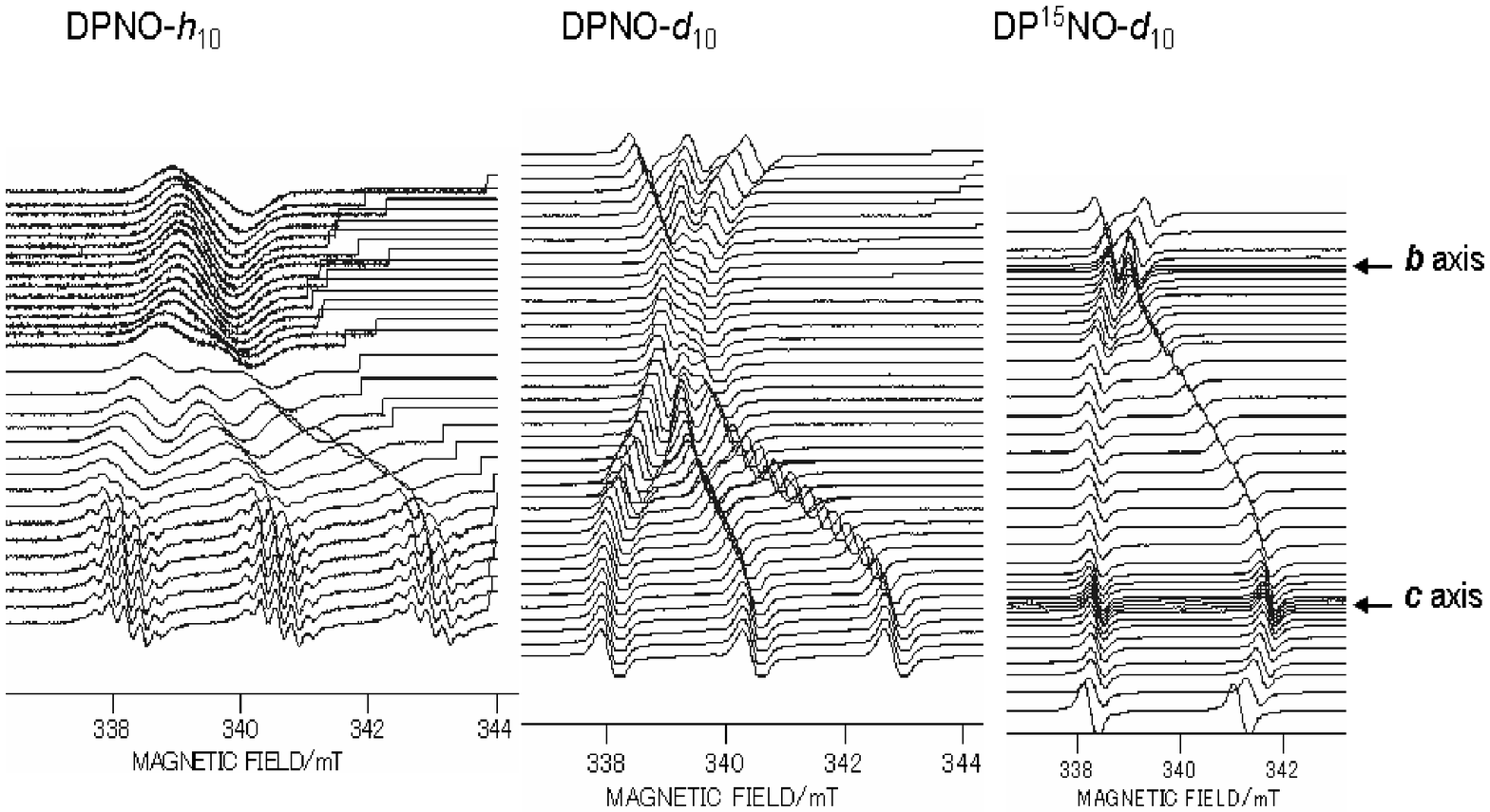}}
\end{center}
\caption{\label{DPNO7} Angular dependence of single-crystal EPR spectra of DPNO, DPNO-$d_{10}$, and DP$^{15}$NO-$d_{10}$ in the crystallographic bc plane.}
\end{figure}

\begin{figure}
 \begin{center}
\scalebox{0.69}
{\includegraphics[0cm,0cm][16cm,10cm]{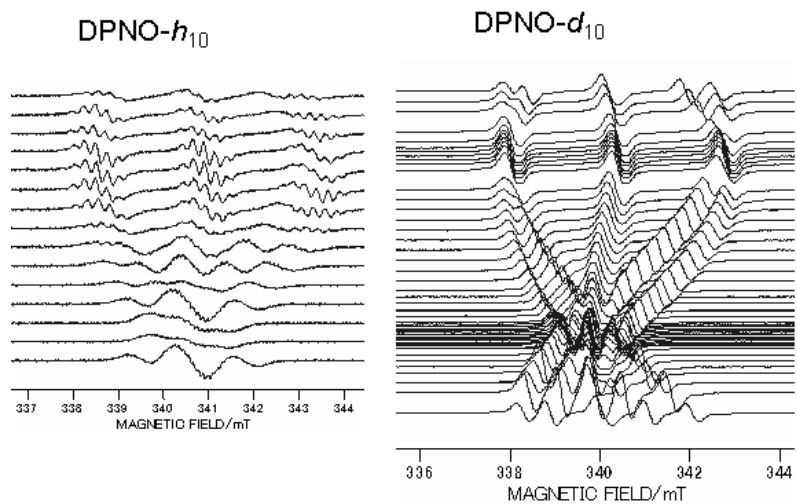}}
\end{center}
\caption{\label{DPNO8} Angular dependence of single-crystal EPR spectra of DPNO and DPNO-$d_{10}$ in the crystallographic ca plane.}
\end{figure}

For a complete analyses for finding the {\bf g} tensor, each transition has been assigned based on the general solution for the resonance fields and ENDOR frequencies as given by the second-order perturbation theory. The {\bf g} tensor was acquired from DP$^{15}$NO-$d_{10}$. Figure \ref{DPNO6}-Figure \ref{DPNO8} show the angular dependencies of single-crystal EPR spectra of DPNO, DPNO-$d_{10}$, and DP$^{15}$NO-$d_{10}$ in the crystallographic ab, bc and ca planes. 

\begin{figure}
 \begin{center}
\scalebox{0.79}
{\includegraphics[0cm,4cm][21cm,17cm]{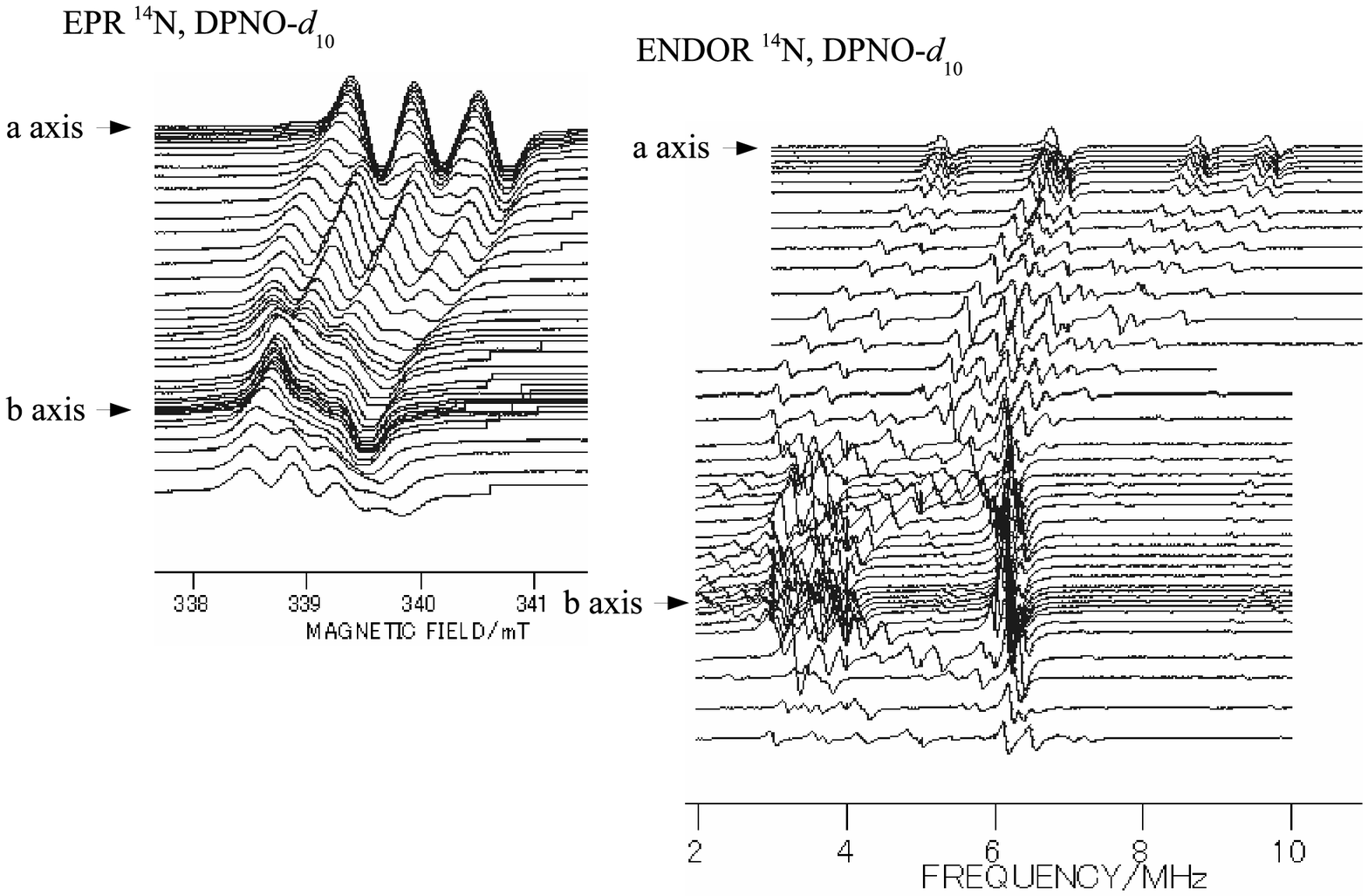}}
\end{center}
\caption{\label{DPNO9} EPR spectrum of DPNO-$d_{10}$ and ENDOR of nitrogen$^{14}$ of the same molecule in {\it ab} plane. Around $b$ axis, the perturbation treatment is broken.}
\end{figure}

\begin{figure}
 \begin{center}
\scalebox{0.69}
{\includegraphics[0cm,9cm][21cm,20cm]{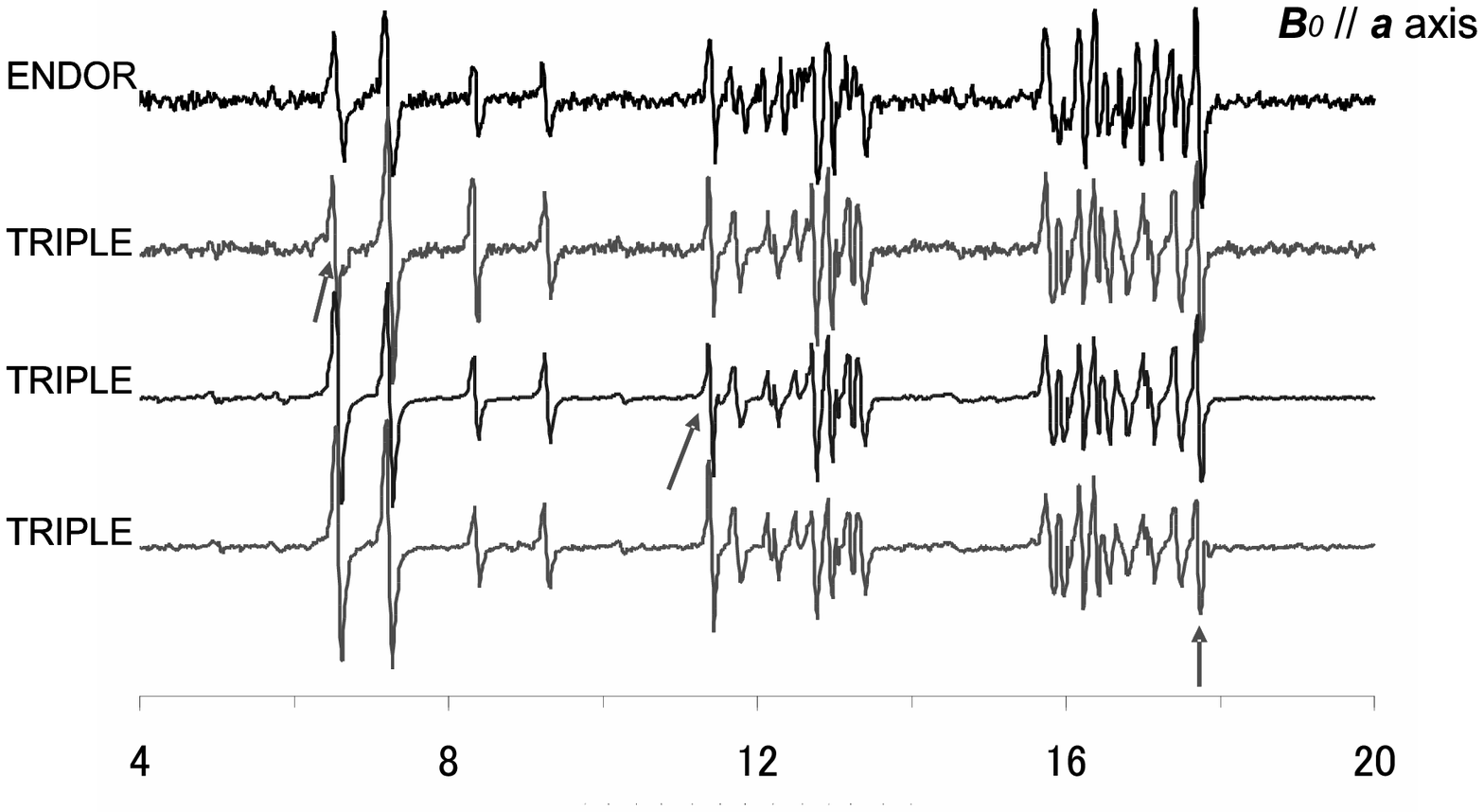}}
\end{center}
\caption{\label{DPNO10} Typical $^1$H-ENDOR and TRIPLE spectra of DPNO-$h_{\rm 10}$ in a benzophenone-$h_{\rm 10}$ single crystal. Arrow indicates the second frequency for pumping.}
\end{figure}

\begin{figure}
 \begin{center}
\scalebox{0.69}
{\includegraphics[0cm,9cm][21cm,20cm]{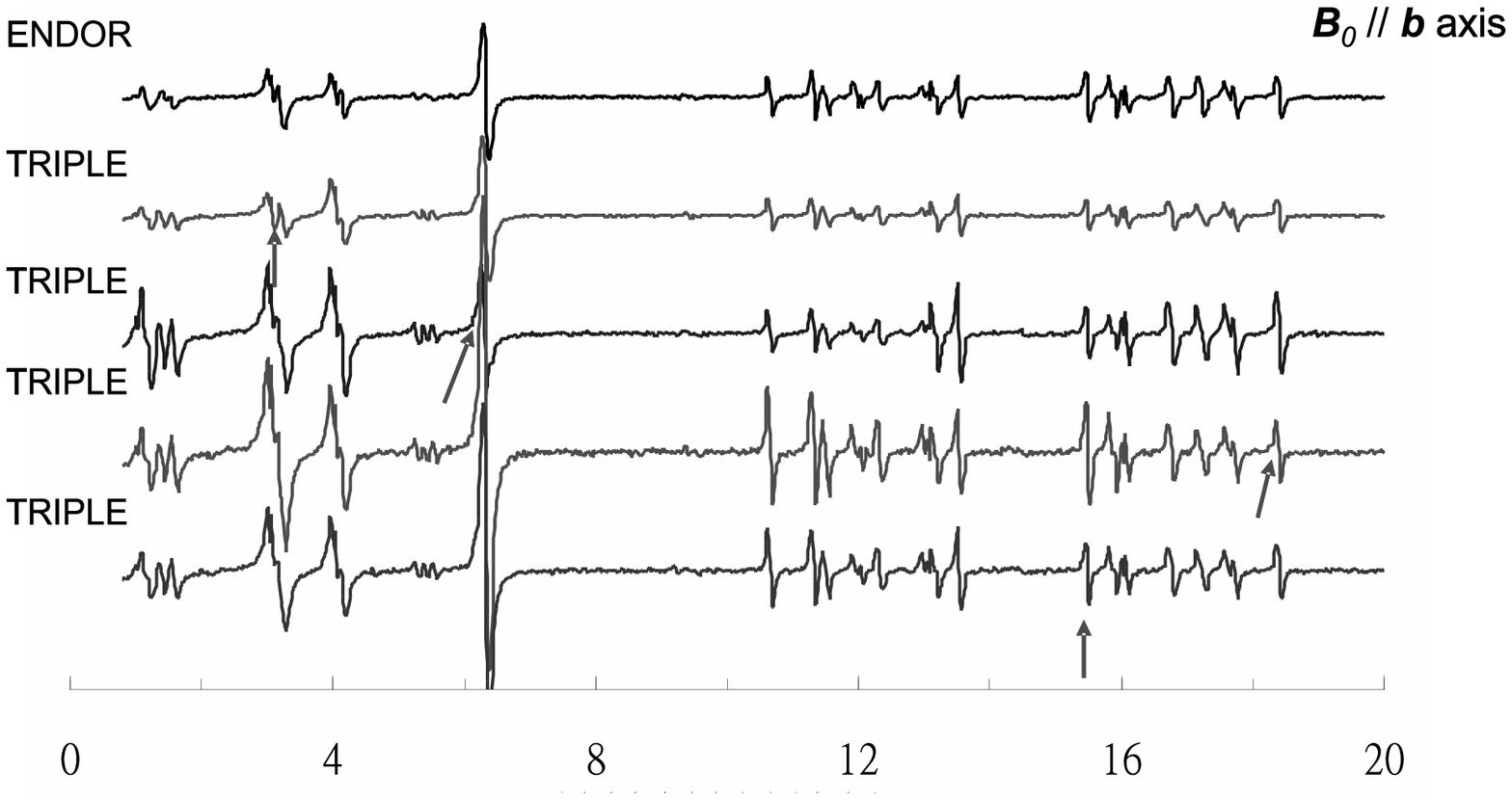}}
\end{center}
\caption{\label{DPNO11} Typical $^1$H-ENDOR and TRIPLE spectra of DPNO-$h_{\rm 10}$ in a benzophenone-$h_{\rm 10}$ single crystal. Arrow indicates the second frequency for pumping.}
\end{figure}

Complete ENDOR measurement in each crystallographic plane, for all the samples has been carried out. Figure \ref{DPNO9} shows the EPR and ENDOR angular dependency measurements on DPNO-$d_{10}$.
 Furthermore, $^1$H ENDOR/general-TRIPLE  spectroscopy has been carried out, see Figure \ref{DPNO10} and Figure \ref{DPNO11}.

After assigning all the transitions for the resonance fields and ENDOR frequencies of the observed spectra in the abc crystallographic systems for the {\bf g} tensor and nitrogen {\bf A} and {\bf Q} tensors, then the least square fitting has been done. Finally, SIMPLEX optimization of the observed ENDOR frequencies has been performed in the abc crystallographic systems for the {\bf g}, nitrogen {\bf A} and {\bf Q} tensors, comparing with the theoretical ones obtained by the numerical diagonalization based on the hybrid eigenfield approach.

Attempts to calculate the isotropic {\bf g}-value of DPNO by quantum chemical computations of high level have been appearing, recently. We have employed the scheme with different basis sets in order to calculate the parameters of this particular sample. Table \ref{T6} shows the experimental and quantum chemical calculational results for {\bf g} tensor values and Table \ref{T6} shows the nitrogen $A$ and $Q$ tensors in comparison to the calculated ones.
  \begin{table}
 \begin{center}
  \caption{\label{T6} g tensor obtained from the experiment as compared with the calculated one. $*$:ADF2005.01, basis set: QZ4P, LDA: VWN and XC: Becke Perdew.}
\begin{tabular}{|c|c|c|c|c|}
\hline
                   & g$_{iso}$ &  g$_{xx}$ &  g$_{yy}$  &  g$_{zz}$  \\ \hline\hline 
 Experiment        &2.0069     & 2.0110   &  2.0065   &  2.0033   \\ \hline
 Calculation       &2.0062     & 2.0109   &  2.0057   &  2.0021   \\\hline   
 \end{tabular}
 \end{center}
 \end{table} 
 
\begin{figure}
 \begin{center}
\scalebox{0.69}
{\includegraphics[0cm,0cm][6cm,9cm]{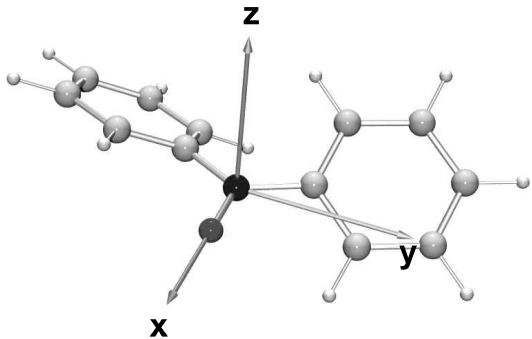}}
\end{center}
\caption{\label{DPNO12} Molecular structure for diphenylnitroxide in Benzephenone.}
\end{figure}

 For this results, the molecular structure in accordance to the crystallographic axis is shown in Figure \ref{DPNO12}.

   \begin{table}
\begin{center}
\caption{\label{Table 5}  A and Q tensors Obtained from the experiment as compared with the calculated ones for different basis set. Also the spin densities on nitrogen and oxygen have been shown.}
\small
\begin{tabularx}{\linewidth}{|X|X|X|X|X|X|X|X|X|X|X|}
\hline
Basis         &     & 6-31g(d)& 6-311g(d,p)& EPR-II  &  EPR-III  &  DZVP  &{\bf DZVP2}  &  TZVP  &  Exp.   \\ 
Set           &     &         &            &         &           &        &             &        &         \\ \hline
A/            &  xx &56.195   &51.386      & 55.084  &  58.360   & 66.592 &{\bf 68.643} & 57.490 & 68.16   \\
MHz           &  yy &3.744    &-5.107      & -1.665  &  -2.037   & 6.488  &{\bf 7.018}  & -2.366 &  8.55   \\
              &  zz &2.830    &-6.097      & -2.430  &  -2.731   & 5.796  &{\bf 6.362}  & -3.121 &  6.56   \\ \hline
A$_{iso}$/MH  &     &20.923   &13.394      & 16.996  &  17.864   & 26.292 &{\bf 27.341} & 17.335 &  27.13  \\
A$_{aniso}$/  &  xx &35.272   &37.992      & 38.088  &  40.496   & 40.300 &{\bf 41.302} & 40.155 &  40.87  \\
MHz           &  yy &-17.179  &-18.501     & -18.661 &  -19.901  & -19.804&{\bf -20.32} & -19.700& -19.33  \\
              &  zz &-18.093  &-19.491     & -19.426 &  -20.595  & -20.496&{\bf -20.97} & -20.455& -21.53  \\ \hline
Q/            &  xx &-2.481   &-2.744      & -2.758  &  -2.610   & -3.110 &{\bf -3.344} & -2.820 &  -2.48  \\
MHz           &  yy &1.834    &2.054       & 2.043   &  2.014    & 2.245  &{\bf 2.434}  & 2.160  &  1.89   \\
              &  zz &0.647    &0.690       & 0.715   &  0.596    & 0.866  &{\bf 0.910}  & 0.660  &  0.59   \\ \hline
$\rho$        &  N  &0.331    &0.317       & 0.333   &  0.295    & 0.327  &{\bf 0.322}  & 0.328  &         \\
$\rho$        &  O  &0.495    &0.497       & 0.478   &  0.497    & 0.486  &{\bf 0.484}  & 0.482  &         \\  \hline 
 \end{tabularx}
 \end{center}
 \end{table}
 
For theoretical calculation of {\bf g} tensor, various correction terms have been introduced to the isotropic {\bf g} value of the free electron.
\begin{equation}
{\bf g} = {\bf g}_e + \Delta{\bf g}^{Rel} + \Delta{\bf g}^{Dia} + \Delta{\bf g}^{Para},
\end{equation}
${\bf g}_e$ is the g value of the free electron, ${\bf g}_e = 2.0023$.  $\Delta{\bf g}^{Rel}\sim -0.0002$ is relativistic correction including $\Delta{\bf g}^{KE}$, $\Delta{\bf g}^{MV}$, etc. Relativistic correction is tiny here as we are dealing with nitrogen, proton and carbon, that none of them has a large atomic number. Anisotropic corrections, diamagnetic correction, $\Delta{\bf g}^{Dia}\sim 0.0001$ and paramagnetic correction, $\Delta{\bf g}^{Para}\sim 0.004$ also have been included. Diamagnetic term is small due to considerably small nuclear gyromagnetic ratio as compared to the paramagnetic term. The latter one is more influential and is calculated as follows for molecular orbital (MO) spins
\begin{equation}
\Delta{\bf g}^{para}=\sum_k \xi \frac{|\langle n|L|k \rangle|}{\epsilon_n-\epsilon_k},
\end{equation}
where $\xi$ is the spin-orbit coupling constant.

\begin{figure}
 \begin{center}
\scalebox{0.69}
{\includegraphics[0cm,7cm][21cm,21cm]{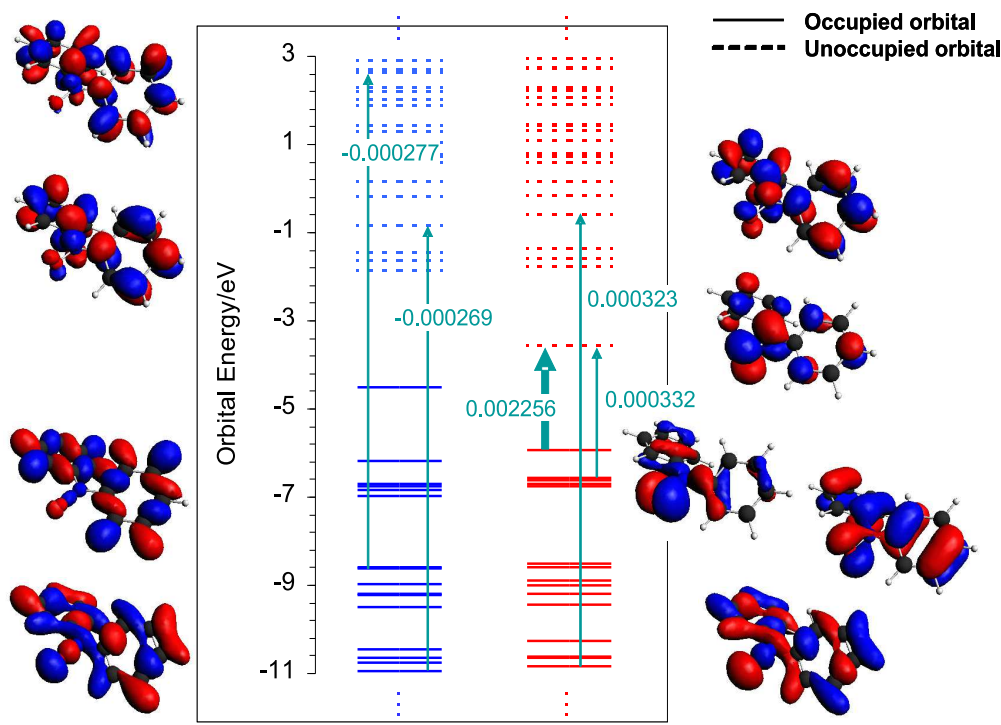}}
\end{center}
\caption{\label{DPNOeleden} Contribution of the paramagnetic terms to {\bf g} tensor of DPNO; idea for {\bf g} tensor engineering. Spin densities on each spin site is shown in colors, red for positive spins and blue for negative spins. Radiuses are for the magnitude of the densities.}
\end{figure}

Then there is a notable point here that with quantum chemistry, it is possible to control the {\bf g} tensor through location of the excited states, {\bf g} tensor engineering. For ENDOR quantum computation and even more elaborated schemes such as ELDOR, Electron-Electron DOuble Resonance quantum computation and specially the case for the larger number of qubits, it is highly required to introduce spins with different {\bf g} tensors as to be distinguishable from each other. Then the concept of {\bf g} tensor engineering turns out to be very important. For this particular molecule, the status is shown in Figure \ref{DPNOeleden}. 
It is then clear that a transition of a $\beta$ spin gives a considerable contribution to the {\bf g} tensor. This effect can be used for engineering spins with different tensorial values. 

We emphasize here that the computational chemistry is very important specially for {\bf g} tensor engineering. Although the results reported in the above table compared to the experimental values have been selected after calculations with several basis sets. The proper basis set then has been selected as the one which yields to values closer to the experimental results. But the basis as reported here as the proper basis set, can not be generalized as it is special case of the molecule and for other samples the situation might be reported to be completely different. For theoretical calculations of {\bf A} and {\bf Q} tensors, we have considered the corresponding equations. 
   
\begin{figure}
 \begin{center}
\scalebox{0.89}
{\includegraphics[0cm,0cm][10cm,10cm]{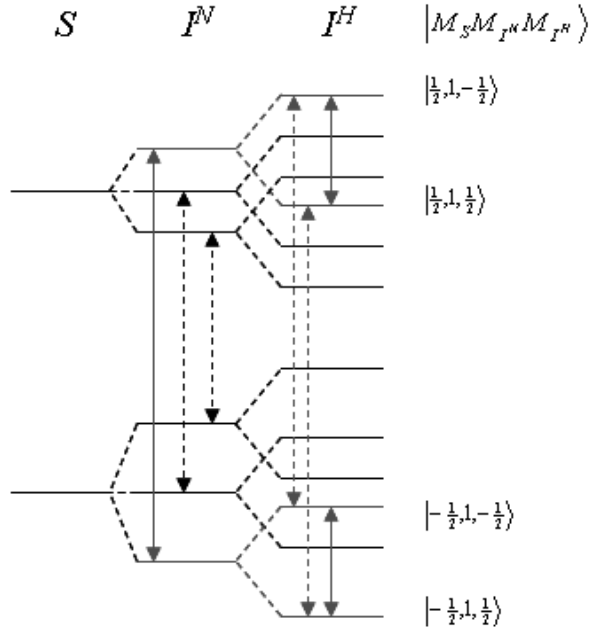}}
\end{center}
\caption{\label{DPNOQ} The energy levels for DPNO.}
\end{figure}

\begin{figure}
 \begin{center}
\scalebox{0.89}
{\includegraphics[0cm,0cm][14cm,6cm]{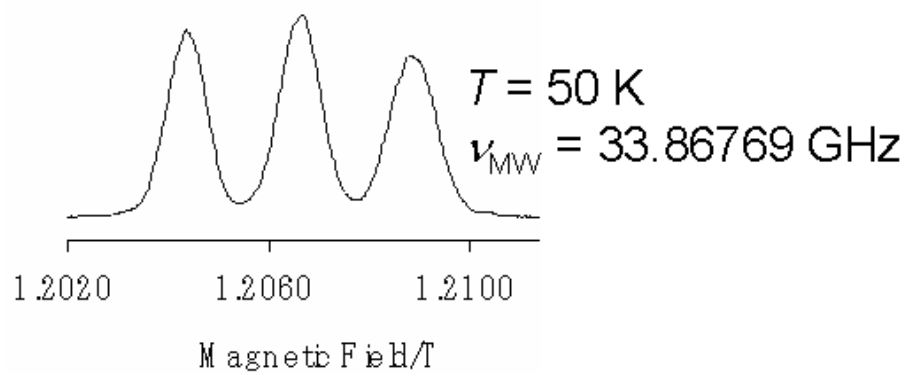}}
\end{center}
\caption{\label{DPNOepr} Pulsed EPR measurement of DPNO.}
\end{figure}

\begin{figure}
 \begin{center}
\scalebox{0.79}
{\includegraphics[0cm,0cm][15cm,8cm]{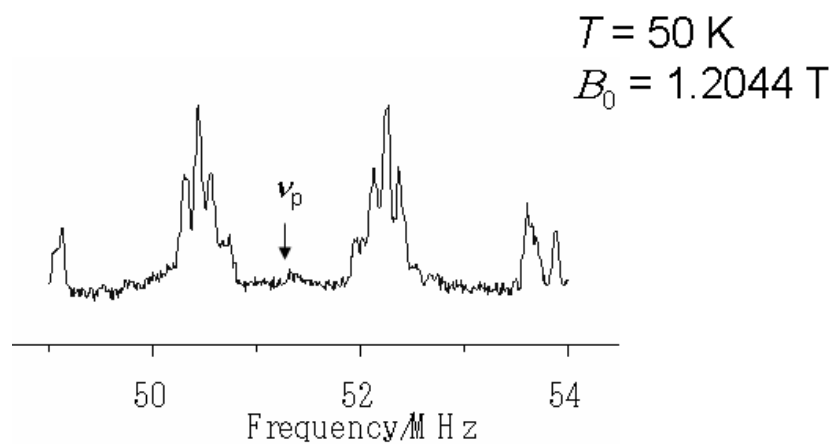}}
\end{center}
\caption{\label{DPNOendor} Pulsed ENDOR measurement of DPNO.}
\end{figure}

  We are ready for measurement of DPNO for pulsed ENDOR QC. Experiment has been done at X-band and Q-band, at different temperatures. Energy levels for DPNO at Q-band is shown in Figure \ref{DPNOQ}.  Figure \ref{DPNOepr} shows the pulsed EPR measurement of DPNO at Q-band and 50 K. Pulsed (Davis) ENDOR spectrum of DPNO is shown in Figure \ref{DPNOendor} 
  
  
\begin{figure}
 \begin{center}
\scalebox{0.89}
{\includegraphics[0cm,0cm][17cm,6.5cm]{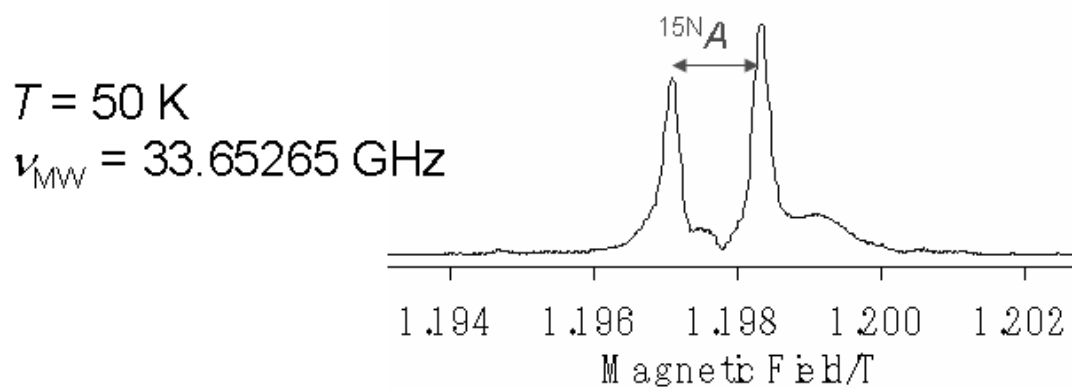}}
\end{center}
\caption{\label{DP15NOepr} Pulsed EPR measurement of DP$^{15}$NO$-{\it d}_{10}$.}
\end{figure}

  We did also experiment with DP$^{15}$NO$-{\it d}_{10}$. Pulsed EPR is shown in Figure \ref{DP15NOepr} and pulsed ENDOR (Mims) is shown in Figure \ref{DPNOendor}. 
\begin{figure}
 \begin{center}
\scalebox{0.79}
{\includegraphics[0cm,17.5cm][21cm,26cm]{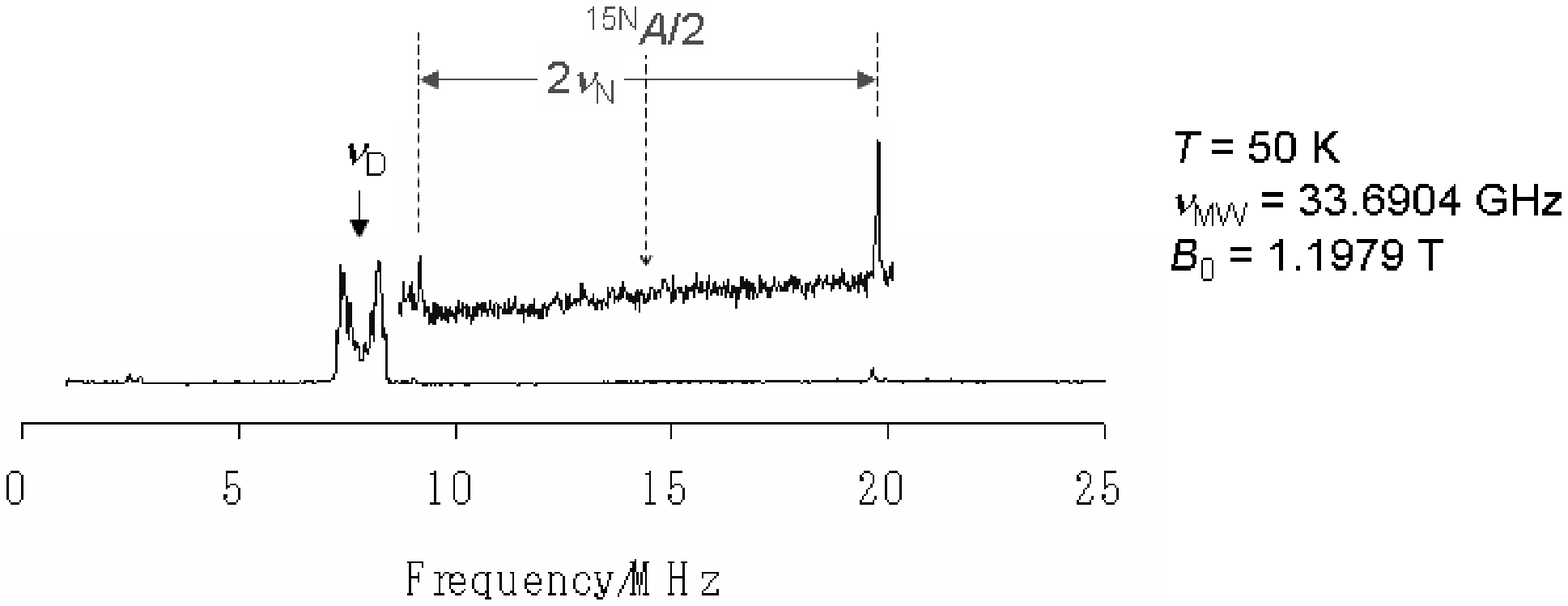}}
\end{center}
\caption{\label{DP15NOendor} Pulsed ENDOR measurement of DP$^{15}$NO$-{\it d}_{10}$.}
\end{figure}
This molecule provides three qubits and to some extent proves partly our claim that this molecule is an electron spin bus molecule and is possible to increase the number of qubits for this molecule. However, in Figure \ref{DP15NOendor} the spectrum is not so clear in comparison to the previous ones. The reason for this is also easy to understand. See the table \ref{T7} that shows nuclear frequencies for X-band and Q-band, for some spins. Nuclear spin frequencies of $^{15}N$ is much smaller than e.g. proton. Therefore, in order to manipulate $^{15}$N we need a very high power rf amplifier, more than two times of the one that we had for this experiment. Unfortunately, at the time of the experiment, the required rf amplifier has not been available. 
  \begin{table}
 \begin{center}
  \caption{\label{T7} NMR frequencies of various nuclei.}
\begin{tabular}{|c|c|c|}
\hline
   Nucleus           & $\nu_n$ at 9.5 GHz (3390 G) & $\nu_n$ at 35 GHz (12,488 G) \\ \hline
    $^1$H   & 14.4 MHz  & 53.2 MHz  \\ 
    $^2$H              & 2.22  & 8.16\\ 
    $^{14}$N    & 1.04   & 3.83\\ 
    $^{15}$N    & 1.46   & 5.39\\ 
    $^{19}$F  & 13.6   & 50.0\\   \hline
 \end{tabular}
 \end{center}
 \end{table}
     
  In order for quantum computation, decoherence time profiles for several sample have been measured by pulsed-based EPR. Spin-lattice relaxation time, $T_1$ for DPNO is 392/msec and for DP$^15$NO-$d_{10}$ is 42.5/ msec, at 10K. measurement on spin-spin relaxation time $T_2$ gives results for DPNO to be $0.777/\mu$sec and for DP$^15$NO-{\it d}$_{10}$ to be $0.489/\mu$sec. These values are quite fine for introducing quantum computation and information with this sample.
    
 \section{Decoherence Time Measurement}
 Decohrence time measurement is done with pulsed EPR. Here, we give a brief introduction on pulsed EPR as it is also preliminary step to pulsed ENDOR and will be used frequently for pulsed ENDOR based quantum computation. 
 
 Remind that NMR most generally is performed with employing a pulse FT technique. The reasons are simple as NMR is slow and mostly it contains very narrow lines spread over a wide frequency range, as compared to the linewidth. Then NMR in its pulsed performance because that it reduces the measurement time by moving to a pulse FT technique gives essentially very nice advantages. However, EPR spectra is wide. Then, it seems useless to move to pulsed EPR scheme. Although, it is still some more advantages through pulsed scheme compared to continuous wave scheme. In a pulsed performance it is possible to get more information by measuring the time domain and using the multi-dimensional techniques.
 
 With pulsed EPR e.g. it is possible to measure relaxation times more directly as compared to continuous wave EPR. With several pulsed techniques, variety of information can be accessed in a particular experiment. For example, ESSEM, Electron Spin Echo Envelope Modulation gives information on interactions of electron spin with the surrounding nuclei.
 
 Pulse EPR spectroscopy is performed by applying a short microwave pulse, less than $20$ ns, but intense, more than $300 W$. Then the microwave signals which have been generated by the sample is measured and Frourier transformed and gives the experimental results.
 
 Let us assume the laboratory frame as usually is defined with the magnetic field, $B_0$, parallel to the $z$ axis, the microwave magnetic field, $B_1$, parallel to the $x$ axis, and $y$ axis to be orthogonal to the $x$ and $z$ axes. Same as the processing which goes on in NMR, here also the physical background of the system can be explained simply by assuming a spin which is placed in a magnetic field. For EPR, a torque is exerted on the electron spin and causes the magnetic moment to precess about the magnetic field with the angular frequency named as Larmor frequency defined as follows
 \begin{equation}
 \omega_L=-\gamma B_0,
 \end{equation}
 where $\omega_L$ is the Larmor frequency. For a free electron which has a $\gamma/2\pi$ value of approximately $-2.8$ MHz/Gauss, the Larmor frequency is about $9.75$GHz for a field of $3480$ Gauss. Suppose that a large number of electron spins are placed in a strong magnetic field, $B_0$. Then the electron spins are precessing about the magnetic field, however their orientation should be completely random in $xy$ plane, since there is not any preferable direction in this plane. Now, suppose that a rotating magnetic field, $B_1$, is applied, let's say along the $+x$ direction. $B_1$ is also assumed to be much smaller than $B_0$. In order to study the effect of $B_1$ on the electron spins, we would be better to move to a frame rotating with the microwave frequency. Then, in this frame $B_1$ looks as if it is a stationary field. The assumption here is also that we are working at resonance, meaning that $\omega_L=\omega_0$, where $\omega_0$ is the microwave frequency. 
 
 Resulting effect from above is that the magnetization would be precessing about $B_1$ at a frequency as follows
 \begin{equation}
 \omega_1=-\gamma B_1,
 \end{equation}
 $\omega_1$ is called Rabi frequency. The magnetization will rotate about $B_1$, $+x$ as assumed here, as long as the microwave are applied and the rotation angle is calculated as follows
 \begin{equation}
 \alpha=\gamma|B_1|t_p.
 \end{equation}
 
 The angle $\alpha$ is called the {\it tip angle} and $t_p$ is the {\it pulse length}. It is then clear from the bove equation that the tip angle depends on both the magnitude of $B_1$ and the length of the pulse, in other words how strong pulse for how long is applied on the sample.
 
 Now let's assume the original problem. For an on resonance condition and $B_1$ along $+x$ direction, the magnetization after a $\pi/2$ pulse would have a direction along $-y$ direction. In laboratory frame, as introduced above, the magnetization then has  a rotating motion in $xy$ plane at the Larmor frequency. This rotation actually generates current and voltage in the resonator and gives a signal which is called FID, free induction decay.
 
 The condition above for resonance, which means that the Larmor frequency being equal to the microwave frequency, clearly can not be satisfied generally as there are many different frequencies in an EPR spectrum. Therefore, most the cases experience nonresonance condition. Then, the magnetization after applying the microwave pulse would have a rotation in $xy$ plane which might be added up to be canceled partly by the rotating field $B_1$. The net effect is chosen as follows
 \begin{equation}
 \Delta \omega =\omega-\omega_0.
 \end{equation}
 
 If $\Delta \omega>0$ then the magnetization is gaining and will rotate in a counter-clockwise fashion. Conversely, if $\Delta \omega<0$ then the magnetization is lagging and will rotate in a clockwise fashion. Then, different resonance frequencies in the EPR spectrum are discriminated as they have different $\Delta \omega$ and this is encoded in the FID of the sample.
  
 The above discussion, however has been realistically true and in real world it can not be fulfilled completely. Magnetization after applying any pulse does not remain in a particular direction for a long time, for reasons as the spins pertain interactions with each other and with the environment. This effect is determined with two constants, $T_1$ and $T_2$. By definition, $T_1$ is the spin lattice relaxation time and describes how quickly the magnetization returns to alignment with the $z$ axis. Also, $T_2$ is the transverse relaxation time and describes how quickly the magnetization in the $xy$ plane disappears.
   \begin{figure}
\begin{center}
\scalebox{0.69}
{\includegraphics[3cm,16.5cm][17cm,29.5cm]{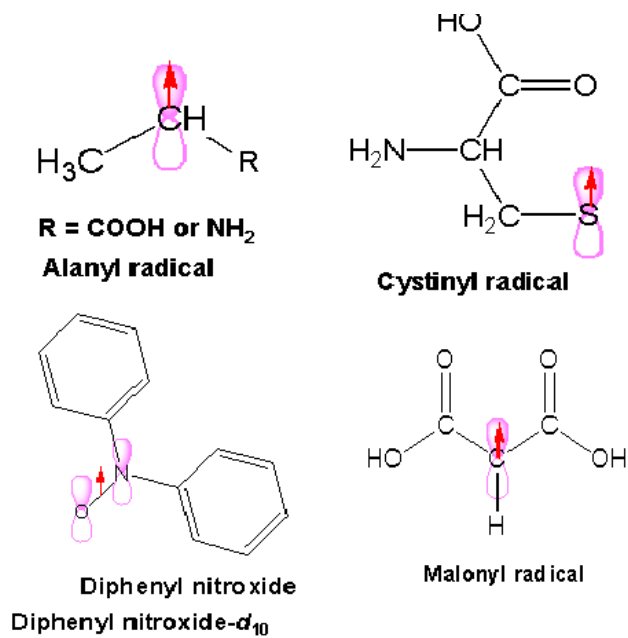}}
\end{center}
\caption{\label{T1samples} Decoherence time has been measured for different samples.}
\end{figure}

 Thermal equilibrium is for before applying any pulses. After applying a pulse, then there will be no longer the thermal equilibrium. The equation which explains the evolution is then explained as follows
 \begin{equation}
 m_z(t)=m_0\times (1-e^{-(1/T_1)}).
 \end{equation}
 
 The above equation is for the case that we apply a $\pi$ pulse and it will be slightly different for other pulses, e.g. $\pi/2$. therefore, the rate constant at which the magnetization recovers to thermal equilibrium is $T_1$ and can be calculated as above after getting the signal which has to be taken by the FID of the corresponding experiments.
 
\begin{figure}
\begin{center}
\scalebox{0.55}
{\includegraphics[2cm,12cm][17cm,16cm]{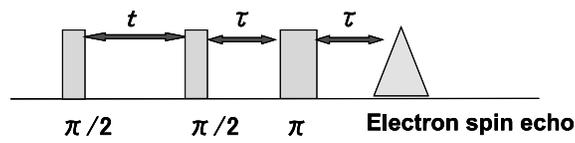}}
\end{center}
\caption{\label{T1pulses} Pulses for $T_1$ measurement with saturation recovery.}
\end{figure}

\begin{figure}
\begin{center}
\scalebox{0.55}
{\includegraphics[2cm,10cm][18cm,18cm]{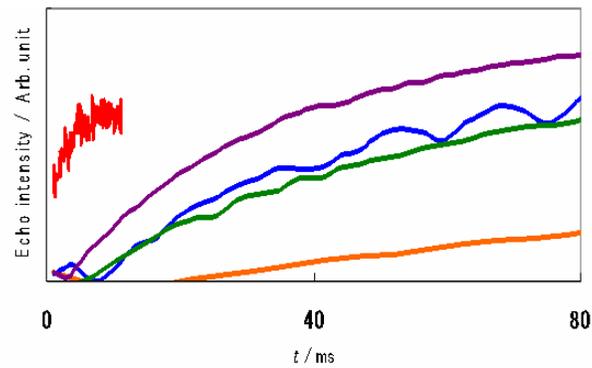}}
\end{center}
\caption{\label{T1recovery} Saturation recovery curves at 10 K. Malonyl radical: green, DPNO: yellow, DPNO-$d_{10}$: purple, Alanyl radical: blue and cystinyl radical: red.}
\end{figure}

\begin{figure}
\begin{center}
\scalebox{0.49}
{\includegraphics[3.5cm,8cm][17cm,20.5cm]{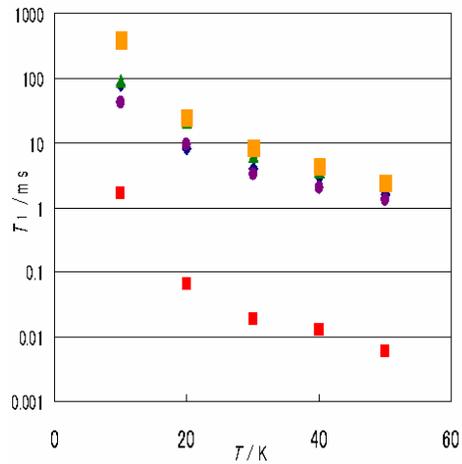}}
\end{center}
\caption{\label{T1temperature} Spin-lattice relaxation times as a function of temperature. Malonyl radical: green, DPNO: yellow, DPNO-$d_{10}$: purple, Alanyl radical: blue and cystinyl radical: red.}
\end{figure}

 We did measurement of $T_1$ for some particular samples, Figure \ref{T1samples}. The pulse sequence for measurement of $T_1$ is shown in Figure \ref{T1pulses}. The acquired spin lattice relaxation time profiles are given in Table \ref{T8}. The experimental detected saturation recovery curves are given in Figure \ref{T1recovery}. $T_1$ is temperature dependent. Figure  \ref{T1temperature} shows the dependencies for particular sample of our interest.
 
  \begin{table}
 \begin{center}
  \caption{\label{T8} Spin-lattice relaxation times at 10 K.}
\begin{tabular}{|c|c|}
\hline
    Radical           & $T_1/$mS \\ \hline
    Malonyl radical   & 91.5    \\ 
    DPNO              & 392.0   \\ 
    DPNO-$d_{10}$    & 42.5    \\ 
    Alanyl radical    & 76.4    \\ 
    Cystinyl radical  & 2.2    \\   \hline
 \end{tabular}
 \end{center}
 \end{table}
  
 Measurement of $T_2$ also can be understood in almost the same manner. The spins interact with each other, resulting in mutual and random spin flip-flop. Molecular motion can also contribute to this relaxation. These random fluctuations contribute to a faster fanning out of the magnetization. The decay of the transverse magnetization or FID from this mechanism is in general exponential with relation to the spin-spin relaxation time, and then gives a way to measure $T_2$, through the following equation.
 \begin{equation}
 m_{-y}=e^{(t/T_2)}.
 \end{equation}
 
\begin{figure}
\begin{center}
\scalebox{0.55}
{\includegraphics[0cm,0cm][16cm,3cm]{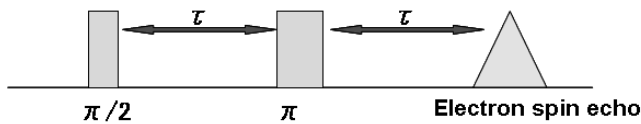}}
\end{center}
\caption{\label{T2pulses} Pulses for $T_2$ measurement with two-pulse echo decay.}
\end{figure}

\begin{figure}
\begin{center}
\scalebox{0.55}
{\includegraphics[3cm,9.5cm][18.5cm,19cm]{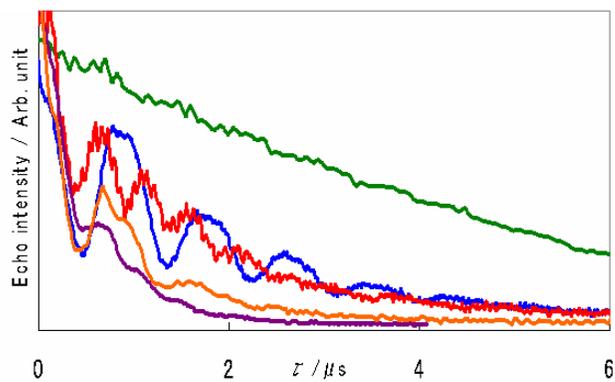}}
\end{center}
\caption{\label{T2decay} Echo decay curves at 10 K. Malonyl radical: green, DPNO: yellow, DPNO-$d_{10}$: purple, Alanyl radical: blue and cystinyl radical: red.}
\end{figure}

\begin{figure}
\begin{center}
\scalebox{0.49}
{\includegraphics[0cm,0cm][13.5cm,13.5cm]{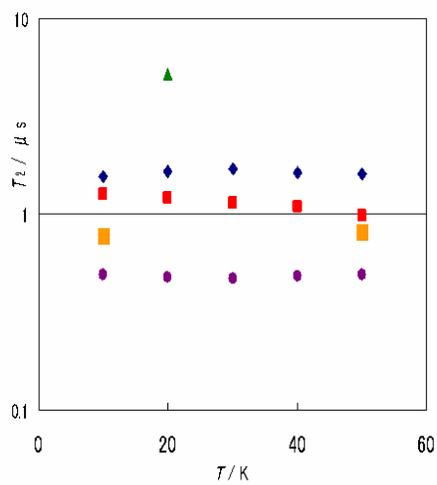}}
\end{center}
\caption{\label{T2temperature} Spin-spin relaxation times as a function of temperature. Malonyl radical: green, DPNO: yellow, DPNO-$d_{10}$: purple, Alanyl radical: blue and cystinyl radical: red.}
\end{figure}

 The pulse sequence to acquire information of $T_2$ is given in Figure \ref{T2pulses}, with which threshold for $T_2$ of samples are given in Table \ref{T9}. Detected experimental results of echo decay curves are shown in Figure \ref{T2decay} and temperature dependencies in Figure \ref{T2temperature}. $T_2$ is not temperature dependent and this fact is clear from our experimental results, too.
 
  \begin{table}
 \begin{center}
  \caption{\label{T9} Spin-spin relaxation times at 10 K. $T_2$ for malonyl is given at 20 K.}
\begin{tabular}{|c|c|}
\hline
    Radical           & $T_2/ \mu$S \\ \hline
    Malonyl radical   & 5.200    \\ 
    DPNO              & 0.777   \\ 
    DPNO-$d_{10}$    & 0.489    \\ 
    Alanyl radical    & 1.560    \\ 
    Cystinyl radical  & 1.267    \\   \hline
 \end{tabular}
 \end{center}
 \end{table}

\section{Generation of a Pseudo-pure State for ENDOR Quantum Computing}
In our experiments, pseudo-pure states are acquired by applying some particular pulse sequence and waiting time. For instance, we give an example on how to get a pseudo-pure state. Initially, the Larmor frequency of the nuclear spins has been ignored due to that it is smaller than the Larmor frequency of the electron spins. For making pseudo-pure state , \cite{n4}, the initial state is the truncated Boltzmann density matrix, $\rho^\prime=-\frac{3}{2} {\bf S}_{\bf z}$, from the original Boltzmann spin density matrix for high temperature as $\rho_{\bf B}=(1-K_{\bf B})\frac{1}{4}{\bf I}_4+K_{\bf B}\cdot \rho_{\bf B}$, with $K_{\bf B}=\mu_{\bf B}B_0/3k_BT$ for $g=2$ and $\rho_p=1/4 {\bf I}_4-3/2 {\bf S}_{\bf z}$. Pseudo-pure state is prepared by providing some particular microwave and radiofrequency  pulses in addition to a waiting time for making transverse components decay for equalizing different populations, \cite{n4}.

\begin{figure}
\begin{center}
\scalebox{0.69}
{\includegraphics[0cm,0cm][19.5cm,3.5cm]{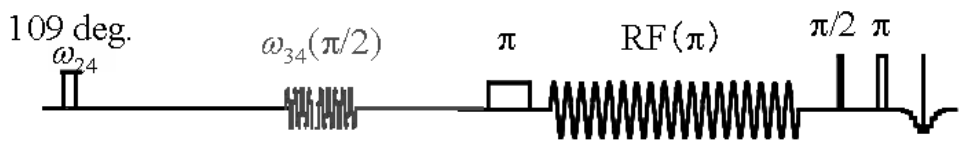}}
\end{center}
\caption{\label{pstatepulses} The pulse sequence that has been used for generation of pseudo-pure state as shown in Figure \ref{pstate}}
\end{figure}

\begin{figure}
\begin{center}
\scalebox{0.69}
{\includegraphics[0cm,0cm][18cm,10cm]{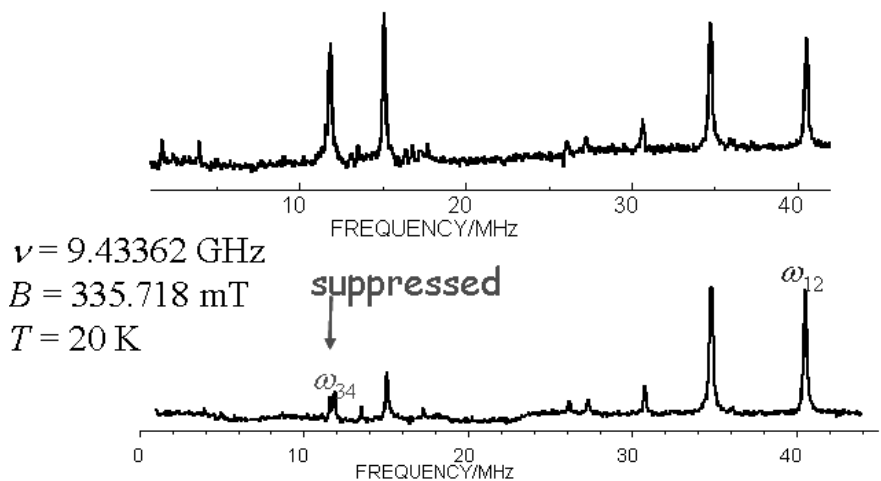}}
\end{center}
\caption{\label{pstate} ENDOR measurement before and after applying the pulse sequence for making pseudo-pure state.Pseudo-pure state is originated after applying the pulses.}
\end{figure}

In a rather similar approach as stated above, we implemented pseudo-pure state in an experiment. Figure \ref{pstate} gives the initial state in comparison to the pseudo-pure state after applying the pulse sequence Figure \ref{pstatepulses} for making pseudo-pure state. It is clear that the state corresponding to $\omega_{34}$ is suppressed giving us the pseudo-pure state on the state corresponding to $\omega_{12}$. We should note that in the pulse sequence Figure \ref{pstatepulses} the first 109 deg. microwave pulse of $\omega_{24}$ and $\pi/2$ radiofrequency pulse of $\omega_{34}$ are for pseudo-pure state generation and the other pulses are applied for (Davies) ENDOR detection. Pseudo-pure state to the other energy levels also would be acquired by applying some other form of pulse sequences.
More details on how to make different pseudo-pure states, by interchanging the rf and microwave pulses, would be clear through our experiments on ENDOR quantum computation in the following parts of this chapter.
 \section{Universal Set of Gates with ENDOR}
\begin{figure}
\begin{center}
\scalebox{0.79}
{\includegraphics[0cm,0cm][18cm,3.5cm]{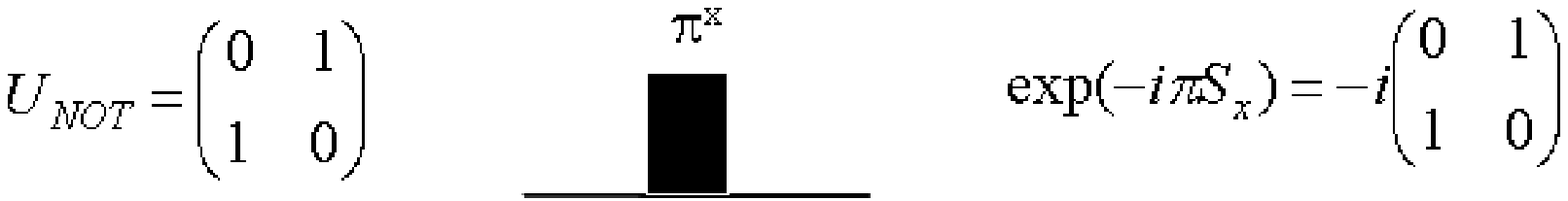}}
\end{center}
\caption{\label{xnotgate} Not-gate is performed by applying a $\pi$ pulse.}
\end{figure}

 Quantum gates are applied by Microwave pulses on electron spin and radiofrequency pulses on nuclear spins. For making a universal set of gates, an arbitrary one-qubit gate and a non-trivial two-qubit gate are required. One-qubit gate is easy to be performed in pulsed ENDOR quantum computation. For instance, NOT-gate is done with a $\pi$ pulse applied in e.g. x direction, Figure \ref{xnotgate}.
 
\begin{figure}
\begin{center}
\scalebox{0.79}
{\includegraphics[0cm,0cm][18cm,4cm]{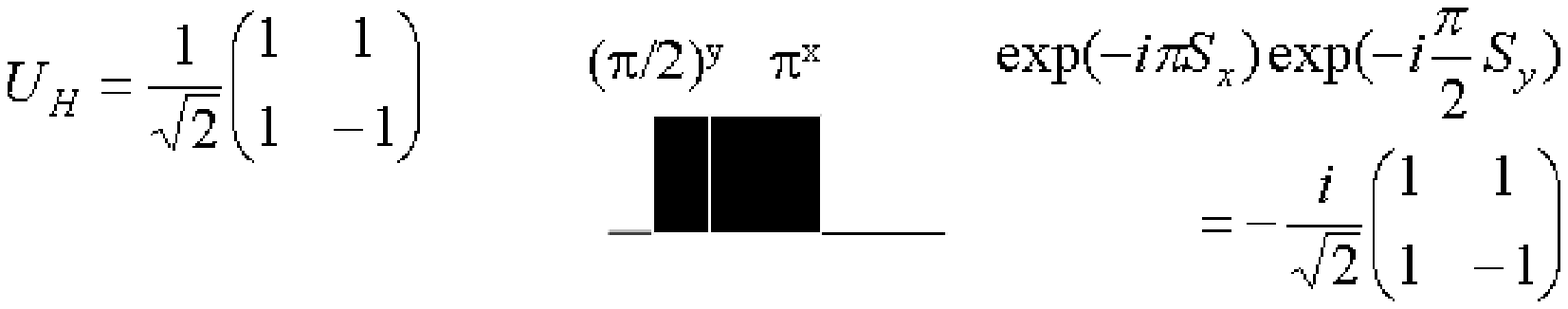}}
\end{center}
\caption{\label{hadamardgate} The pulse sequence for making a Hadamard gate, up to an overall phase.}
\end{figure}

 Hadamard gate is performed, up to an overall phase,  with a $\pi/2$ pulse in y direction followed by a $\pi$ pulse in x direction, Figure \ref{hadamardgate}.
 
\begin{figure}
\begin{center}
\scalebox{0.79}
{\includegraphics[0cm,0cm][21cm,4cm]{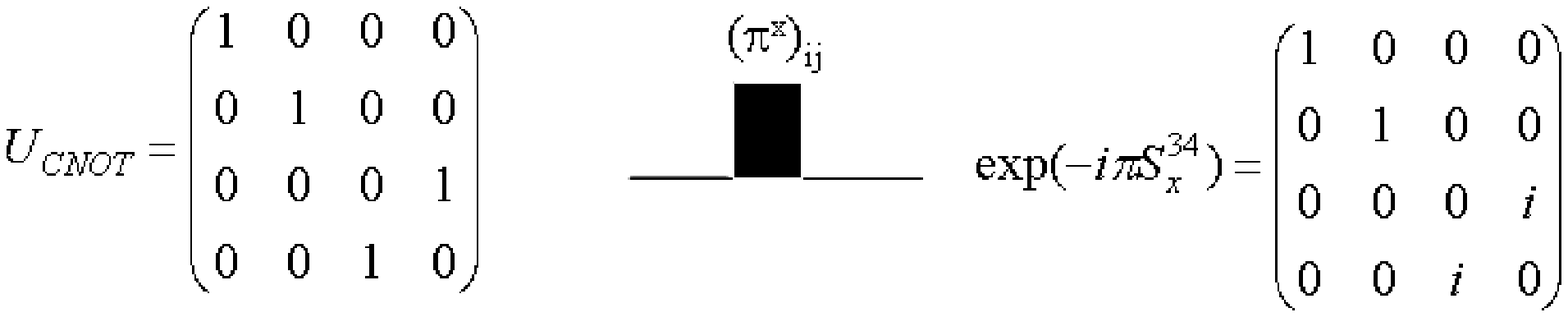}}
\end{center}
\caption{\label{CNOTgate} The pulse sequence that we have used for making CNOT in the experiment for making entanglement. For a general CNOT gate the pulse sequence is more complicated. See text.}
\end{figure}

 CNOT gate is also performed by applying a $\pi$ radiofrequency pulse in x direction, Figure \ref{CNOTgate}.
There would be a phase in this performance for making CNOT gate, however as far as we are working in a sequence of gates, for our special interest quantum circuit for making entanglement, this would be possible to eliminate the effect of this phase, as we did and results are shown in further coming experiments. Also, it is possible to apply some more complicated pulse sequences that will give the exact form of the CNOT gate. The corresponding part on NMR QC, in Chapter 3, has been useful in this regard. For example, the pulse sequence that has been introduced for making CNOT, \ref{CNOTNMR}, up to a total phase, is essentially applicable for pulsed ENDOR QC, too. 

Experimental evidences on the establishment of different quantum gates will be clear in the last part of this chapter that we present the experiments on making entanglement with ENDOR and implementation of quantum operations for superdense coding.
  \section{Readout of Quantum Computation with ENDOR}
  Readout in ENDOR experiment is done conventionally through electron spin echo. We use the conventional scheme for detection of states in addition to other scheme that will be explained in this section. First, we give a brief introduction to the electron spin echo detection in ENDOR. Then, we will explain additional detection scheme, TPPI, time proportional phase incrementation for quantum computation. 
  
 \subsection{Electron Spin Echo:}
\begin{figure}
\begin{center}
\scalebox{0.79}
{\includegraphics[0cm,20cm][14cm,24cm]{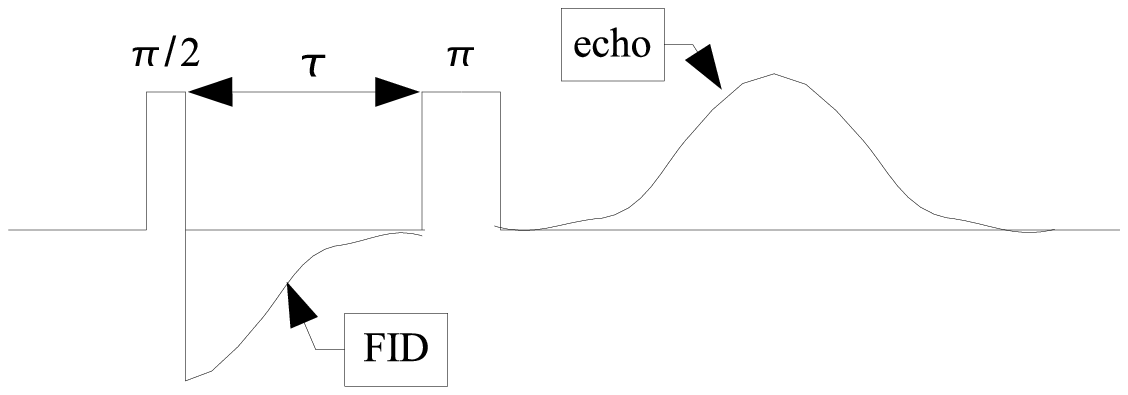}}
\end{center}
\caption{\label{ESE} A Hahn echo.}
\end{figure}

 After that we apply a microwave pulse, it produces a signal that decays away and gives the FID. However, if the spectrum is broadend enough it is possible to recover this effect by applying another microwave that produces Hahn echo. The processing for echo is simply represented in Figure \ref{ESE}. Here, the transverse relaxation leads to an exponential decay in echo height as follows
 \begin{equation}
 {\rm echo}\, {\rm height}\sim e^{\frac{-2\tau}{T_2}}.
 \end{equation}

Result of the measurement is expressed with the echo intensity that is the signal intensity detected after applying the pulse sequence that includes a $\frac{\pi}{2}$ followed by a $\pi$ pulse in $x$ direction in addition to a final echo pulse in $y$ direction. Detection through electron spin echo is the generally used measurement scheme for any pulsed ENDOR experiment. We have used almost the same approach. Information on the nuclear spins might exchange and be detected through electron spin echo. However, for some special cases, e.g. detection of entanglement that a global information on the state of the two spins is required electron spin echo may not be completely adequate and rather more elaborated scheme is required that is explained in the following section.

 \subsection{Time Proportional Phase Incrementation, TPPI:}
 TPPI is known in ENDOR for separation of multiple quantum coherences. However, we used the scheme in a way that it can give a global information on the state of electron and nuclear spins. This is for resolving the restriction of ENDOR spectroscopy that any measurement can only be done via electron spin echo. Specially for the case that information on the state of entanglement is required, TPPI is an essential part for detection of resolving the entangled state from the simple superposition states. 
 
 In TPPI, the phase shifts are implemented by {\it incrementing} the phase frequency of the individual {\it detection pulses} in {\it consecutive experiments}. What we mean by detection pulses will be clear after that we represent our experiments for entanglement with pulsed ENDOR. Meanwhile, this term might be understood if we recall the part for entanglement witness derived from NMR SDC, see chapter 3. We introduced some unitary back operations for transferring the state to the computational state. Hereafter, the corresponding unitary back operation would be called detection pulses.
 
 Firstly, we apply detection pulses with arbitrary phase frequencies, $\nu_j$ on $j$th spin. Therefore, we have
 \begin{equation}
 \Delta \omega_j=2\pi\Delta \nu_j.
 \end{equation}
 
 Then consecutive experiments are performed for a time $\Delta t$. We have
 \begin {equation}
  \phi_j=\Delta \omega_j \Delta t.
  \end{equation}
  
  The phase shifts are detected through the experiment and give information on the state. Let us give an example. Suppose that the state in the experiment, is a Bell state of the following form
  \begin{equation}
  |\psi\rangle =\frac{1}{\sqrt 2}(|01\rangle-|10\rangle).
  \end{equation}
  
  Detection pulses that we apply are a microwave $\pi$ pulse followed by a radiofrequency $\pi/2$ pulse, with phases of $\phi_1$ and $\phi_2$, respectively. Then the detection unitary operation might be written as
  \begin{equation}
  U_d=\omega_{34}(\pi/2,\phi_2)\omega_{24}(\pi, \phi_1).
  \end{equation}
  
  The first pulse is a microwave $\pi$ pulse $(\omega_{24}(\pi))$ and the second pulse is a $\pi/2$ radiofrequency pulse $(\omega_{34}(\pi/2))$.
  
  Then, measurement is performed with electron spin echo. The intensity would be as follows
  \begin{eqnarray}
  S_d^{\psi}(\phi_1,\phi_2)&=&tr\{S_z^{24}U_d|\psi\rangle\langle\psi|U^\dagger_d\}\\ \nonumber
                           &=& -\frac{1}{2}[1-\cos(\phi_1-\phi_2)].
\end{eqnarray}
Therefore, through detection by TPPI we would get the phase shift as follows
\begin{equation}
           \phi_1-\phi_2=\Delta (\omega_1-\omega_2)\Delta t.
           \end{equation}
           
Here, for the angular frequencies, $\omega_j$, we have
\begin{equation}
\Delta(\omega_1-\omega_2)=2\pi \Delta(\nu_1-\nu_2),
\end{equation}
$\nu_j$ is the phase frequency on the $j$th spin. Detection of a pulse with a phase shift of $\phi_1-\phi_2$ demonstrate that the state has been $|\psi\rangle$ and not the separate states of spins. Because, if the state under measurement is a separable state of the spins, after applying detection unitary operations with $\nu_1$ and $\nu_2$, phase frequencies on spins then TPPI measurement would give results as only $\phi_1$ and $\phi_2$ and not $\phi_1\pm \phi_2$.

\section{Pseudo-entanglement with ENDOR at Q-band and Low Temperature}
In the following part we will give report on our experiment as how the phase control has been acquired and furthermore the results will show nicely a strong evidence on realizing the (pseudo) entanglement. Higher magnetic field and lower temperature and probably other experimental properties would give quantum entanglement.

 
In our experiment, spins have been selectively addressed by their resonance frequencies. We did experiment either with pseudo-pure state or without pseudo-pure state. Detection has been performed by electron spin echo while phase manipulation has been done by TPPI.

The experiment that is explained here is performed on two different samples. The first choice has been malonyl radical. The other sample, has been DPNO, the molecule which we have extensively studied and believe that can be a proper electron spin bus molecule for quantum computation. For both the cases, we got some reasonably good results on implementation of (pseudo-)entangled state. However, we emphasize the spin manipulations by means of phase incrementation. This is quite useful point for further progress in this field.

 Pseudo-pure state has been made in a similar manner as discussed in the previous section and the final state of the system is then $|10\rangle$, after a $109.47^{\circ}$ microwave $\omega_{24}$ followed by a $\frac{\pi}{2}$ radio frequency $\omega_{12}$. Slightly different pulse sequences give the other choices for the initial state. Then, the initial state changes into a Bell state if pulses are applied as follows
 \begin{eqnarray}
 |10\rangle & & \xrightarrow{\omega_I^{(34)}(\frac{\pi}{2})} \frac{1}{\sqrt 2}(|10\rangle+|11\rangle) \\ \nonumber
            & & \xrightarrow{\omega_S^{(24)}({\pi})} \frac{1}{\sqrt 2}(|10\rangle-|01\rangle).
\end{eqnarray}
 This is equivalent to have a Hadamard gate on the nuclear spin followed by a CNOT gate on the state of the electron and nuclear spins.
 
 Then, detection is done by unitary back operations. Entanglement is a global property. Then, a single local measurement as opposed by electron spin echo would not generally give any information on the status of entanglement. Therefore, we have used the phase incrementations on both microwave and radio frequency to extract the required information. We apply phase dependent pulses, $\pi$ microwave $\omega_{24}$ with phase $\phi_1$ followed with a $\pi/2$ radio frequency $\omega_{34}$ with phase $\phi_2$, that gives the phase dependent echo intensity of the following form
 \begin{equation}
 {\bf I}= -\frac{1}{4}[1-\cos (\phi_1- \phi_2)].
 \end{equation}
 
\begin{figure}
\begin{center}
\scalebox{0.69}
{\includegraphics[0cm,0cm][13.5cm,13.5cm]{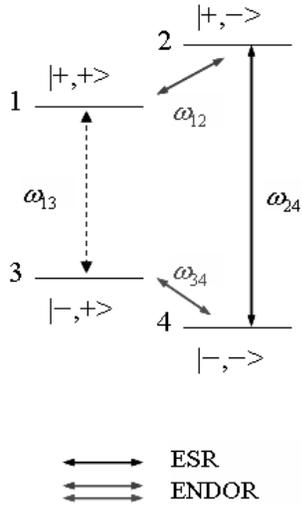}}
\end{center}
\caption{\label{entlevels} Energy levels for a system composed of two spin $1/2$. The transitions naming are in accordance to the ones indicated in the pulse sequence for making entanglement Figure \ref{entpulses}}
\end{figure}

\begin{figure}
\begin{center}
\scalebox{0.69}
{\includegraphics[0cm,0cm][19cm,4cm]{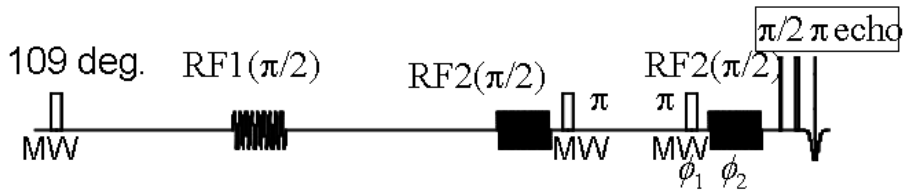}}
\end{center}
\caption{\label{entpulses} The pulse sequence that has been used for generation of entanglement state.}
\end{figure}

 To be noted here, in case that the initial state has been some other state, then the other Bell state would be generated after applying the corresponding pulses for making entangled states. After the phase dependent pulses for detection, we would get other combination of the phases. This is shown in Table \ref{T10} for several choices of microwave and radiofrequencies for pulse sequence for making entanglement, Figure \ref{entpulses}, with resonance frequencies shown in Figure \ref{entlevels}.
 
  \begin{table}
 \begin{center}
  \caption{\label{T10} Different Bell states derived by different pulse sequences.}
\small
\begin{tabular}{|c|c|c|c|c|}
\hline
 MW            & RF1          & RF2           & Bell state                               & $\nu_{TPPI}$ \\ \hline
 $\omega_{24}$ & $\omega_{12}$& $\omega_{34}$  & $\frac{1}{\sqrt 2}(|10\rangle-|01\rangle)$ & $\phi_1-\phi_2$ \\ \hline 
 $\omega_{24}$ & $\omega_{34}$& $\omega_{12}$  & $\frac{1}{\sqrt 2}(|00\rangle+|11\rangle)$ & $\phi_1+\phi_2$ \\ \hline 
 $\omega_{13}$ & $\omega_{12}$& $\omega_{34}$  & $\frac{1}{\sqrt 2}(|00\rangle-|11\rangle)$ & $\phi_1+\phi_2$ \\ \hline 
 $\omega_{13}$ & $\omega_{34}$& $\omega_{12}$  & $\frac{1}{\sqrt 2}(|10\rangle+|01\rangle)$ & $\phi_1-\phi_2$ \\ \hline 
 \end{tabular}
 \end{center}
 \end{table}
 
 Also, we would like to emphasize that the scheme which has been explained here is for detection of the entangled state from the simple superposition of the state. On the other hand, we always get one out of the two possible combinations of the phases and this is not sufficient to fully determine the particular Bell state. However, it is sufficient to determine the entanglement of the state. Then, the other matter, meaning that determination of the particular Bell state would be possible by considering the initial state and the pulses which have been applied in order to make the entangled state.

\begin{figure}
\begin{center}
\scalebox{0.89}
{\includegraphics{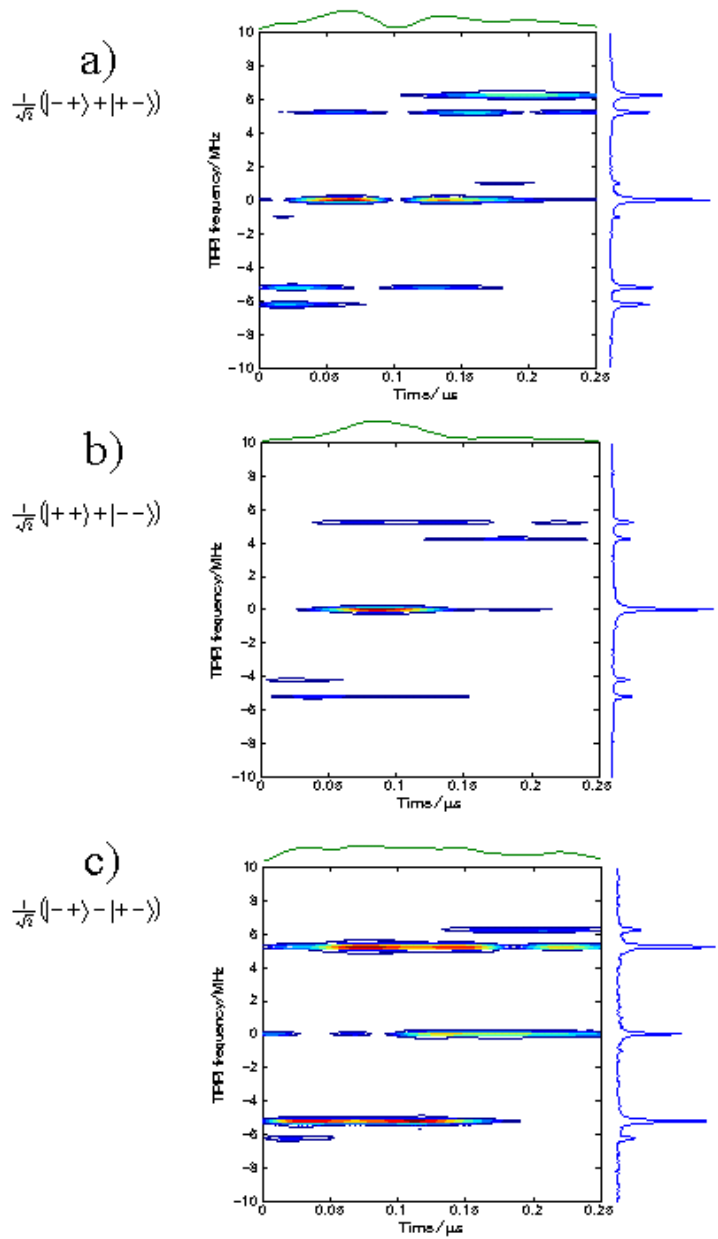}}
\end{center}
\caption{\label{13entangle} 2-dimensional spectra, TPPI frequency against time for detecting the spin echo intensity through phase incrementing, in consecutive experiments on malonyl. The interferogram of the spectra shown here is Fourier transformed to get the information on the individual phases as well as the combination of the phases, in case of existence of any entanglement. Particularly, for a) this process is explained in more details in text. Also, see Figure \ref{InterferogramFT}}
\end{figure}

 The artificial phase frequencies in our experiments are $\Delta \omega_j=2\pi\Delta\nu_j$, $\nu_1=1.0$ MHz and $\nu_2=5.2$ MHZ, as arbitrary values. The resultant spectrum is a 2D spectra, the TPPI frequency against time. See Figure \ref{13entangle} for an example on TPPI detected 2D spectra for different pulse sequences i.e. different Bell states.
  \begin{figure}
\begin{center}
\scalebox{0.69}
{\includegraphics[0cm,0cm][21cm,23cm]{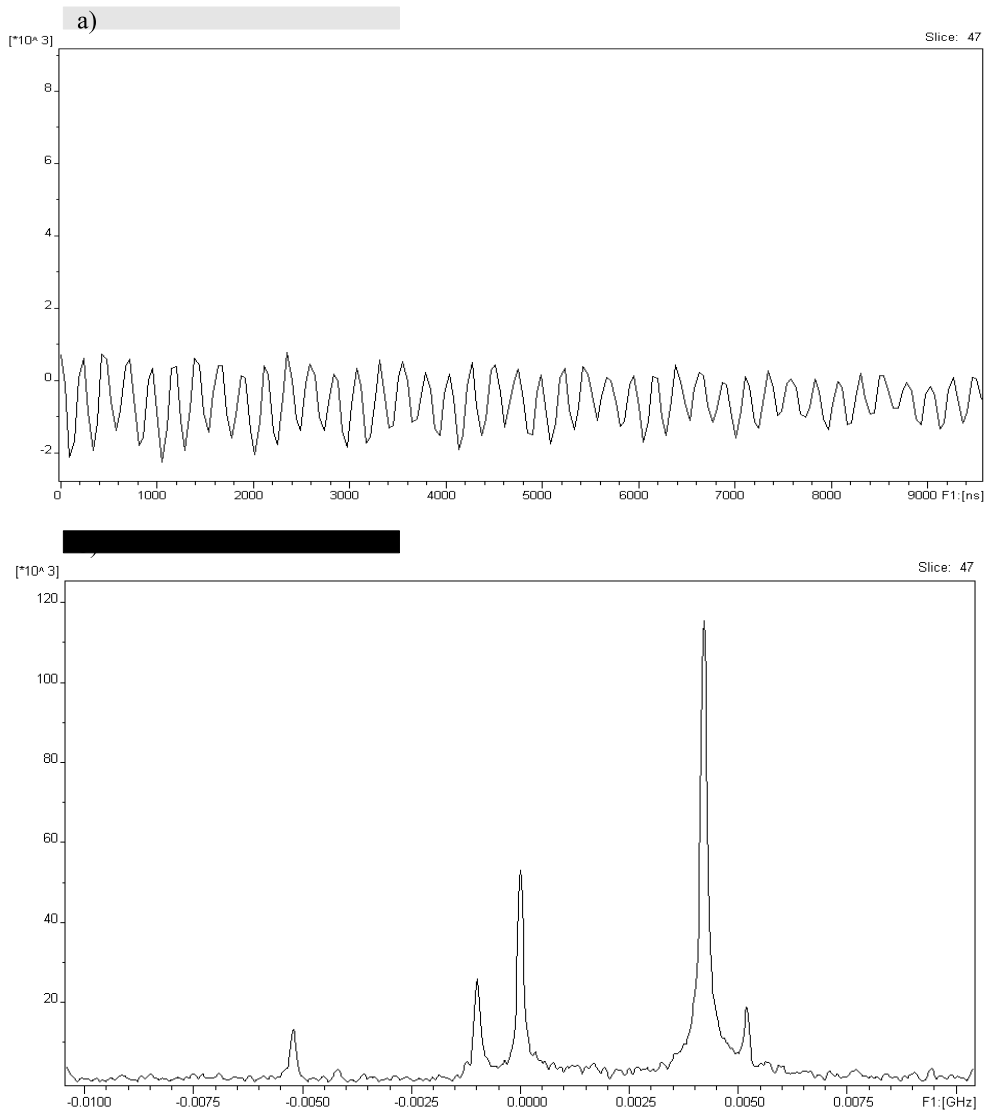}}
\end{center}
\caption{\label{InterferogramFT} This figure is to show rather more in details the scheme which has been used for detection and analyse of the phases in order to determine demonstration of (pseudo-)entanglement. a) is the interferogram (one slice) of the TPPI as shown in Figure \ref{13entangle}(a). b) is the Fourier transform of the spectrum shown in (a).}
\end{figure}
  \begin{figure}
\begin{center}
\scalebox{0.69}
{\includegraphics[0cm,0cm][18cm,11cm]{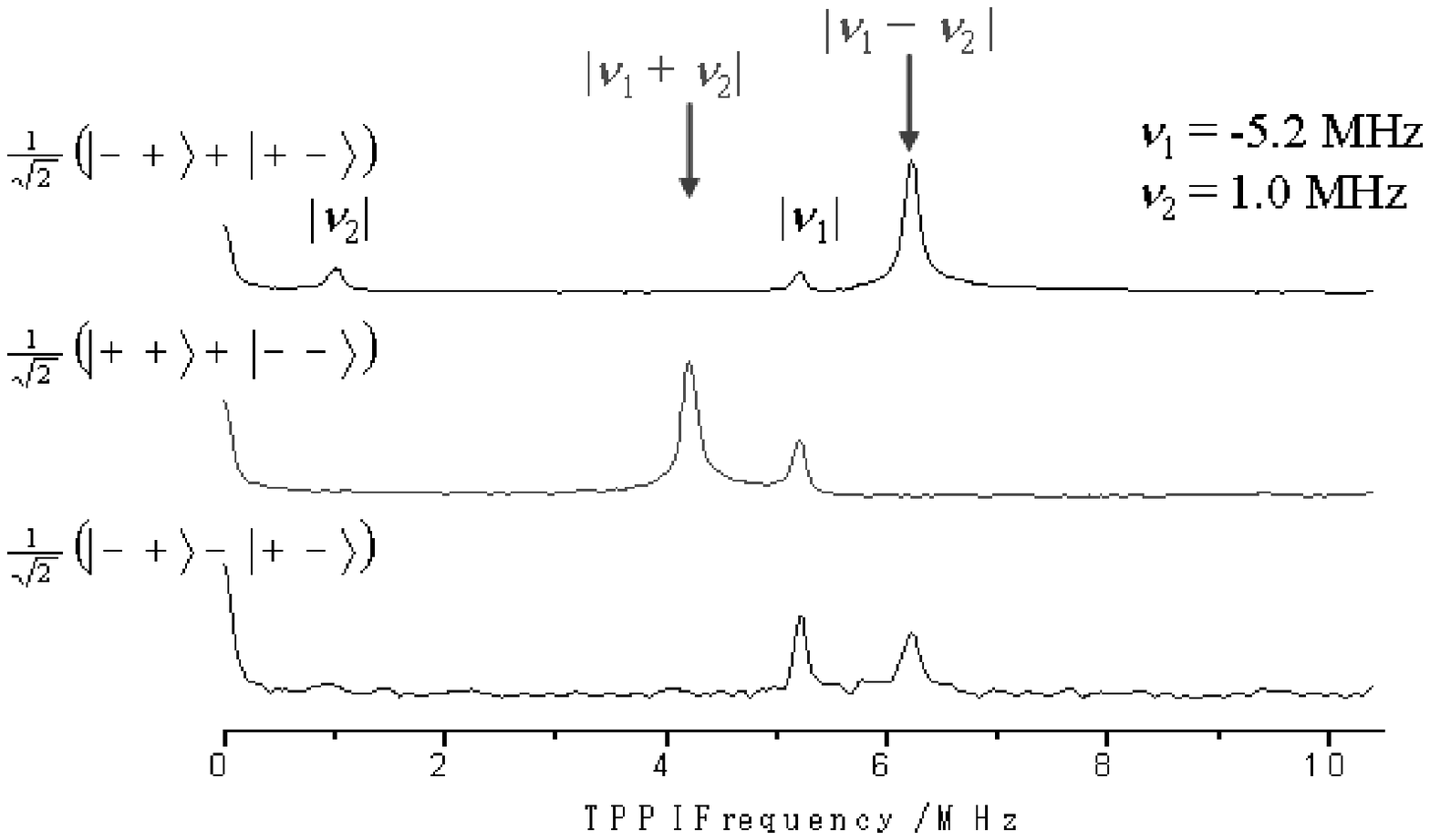}}
\end{center}
\caption{\label{entanglemaldpno} Experimental results of demonstration of (pseudo-)entanglement as the Fourier transform of the phase interfrograms. Phase frequencies are $\nu_1=1$ MHz and $\nu_2=5.2$ MHz. The bottom two correspond to the experiment on malonyl radical, while the upper one is due to the experiment on DPNO.}
\end{figure}

 Then the phase interferogram against time is Fourier transformed. See Figure \ref{InterferogramFT} for an example on an interferogram which corresponds to the TPPI detected spectra of Figure \ref{13entangle}(a). The combination of the phases, whether to be addition or subtraction, gives evidence on the phase of the Bell state and then proves the entanglement of the state. See Figure \ref{entanglemaldpno} for experimental detection of (pseudo-) entanglement for malonyl radical.
 
\begin{figure}
\begin{center}
\scalebox{0.89}
{\includegraphics{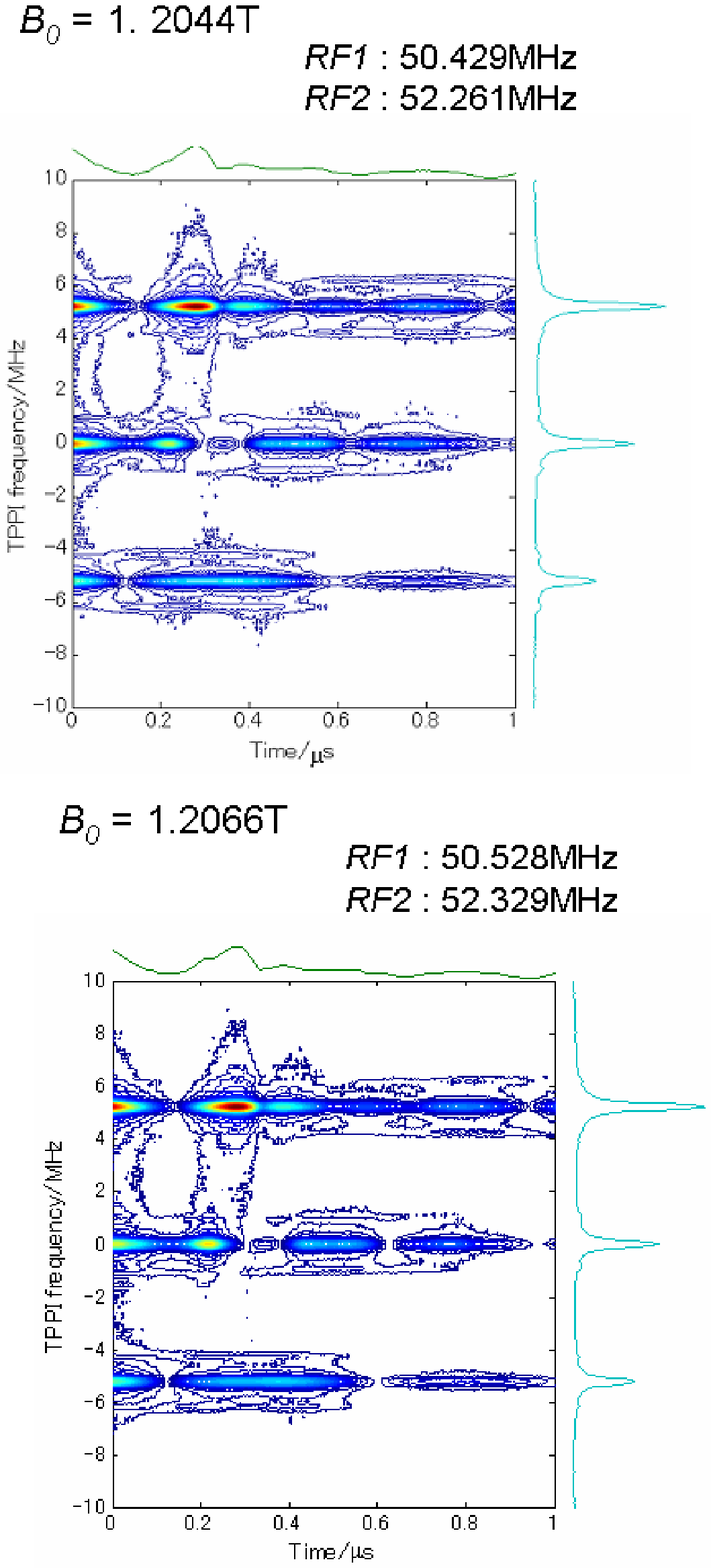}}
\end{center}
\caption{\label{DPNOTPPI} 2Dimensional spectra, TPPI frequency against Time for detecting the spin echo intensity through phase incrementing, in consecutive experiments on DPNO.}
\end{figure}

\begin{figure}
\begin{center}
\scalebox{0.69}
{\includegraphics[0cm,0cm][16cm,14cm]{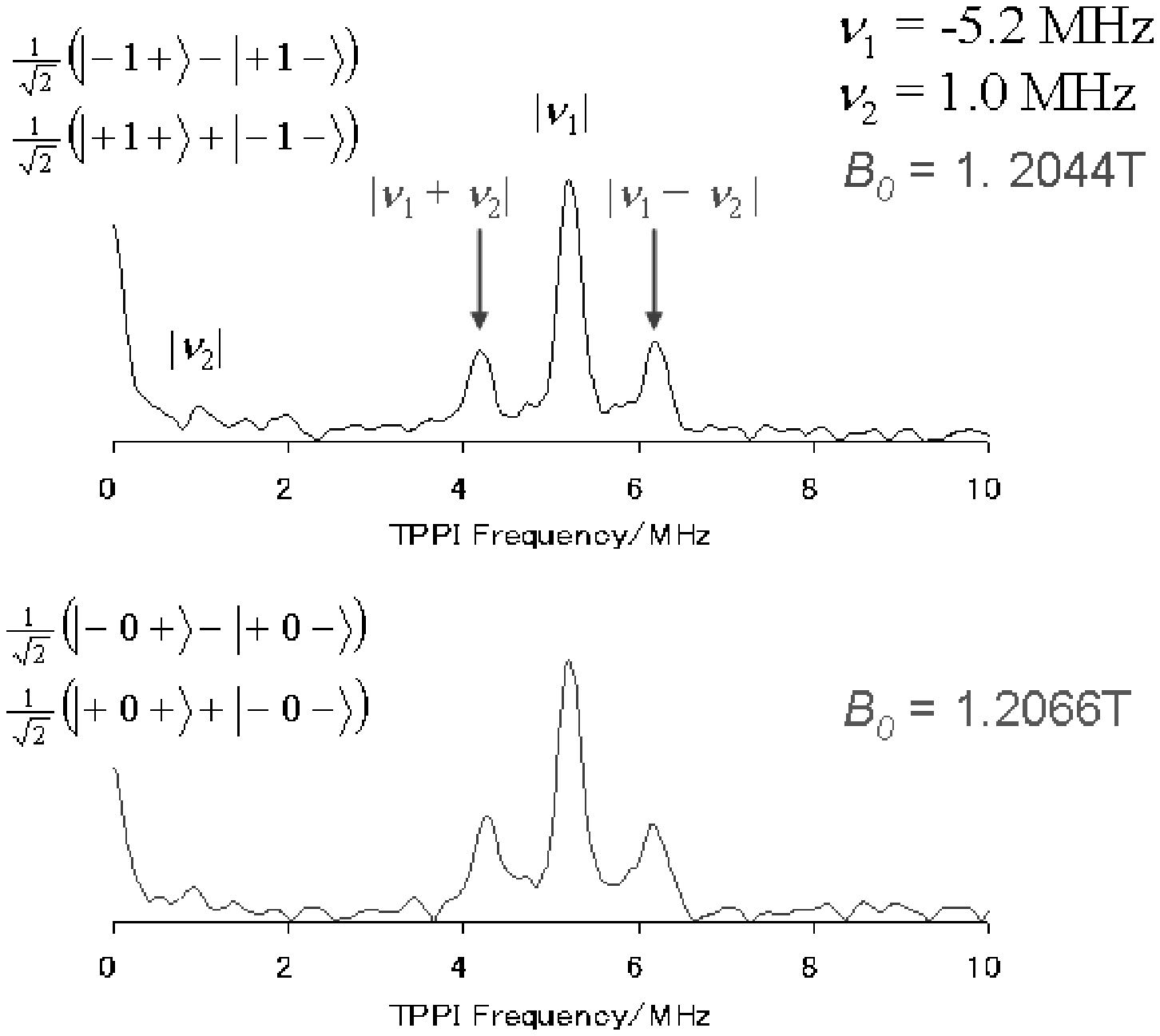}}
\end{center}
\caption{\label{DPNOFT} Experimental results of demonstration of (pseudo-)entanglement as the Fourier transform of the phase interfrograms of DPNO.}
\end{figure}

 The same experiment is done for DPNO. Two-dimensional TPPI frequency against time spectra are shown in Figure \ref{DPNOTPPI}. Furier transformed experimetal results are given in Figure \ref{DPNOFT}. The experiment with DPNO involves three qubits. However, because of the reasons that we explained before on low power radiofrequency amplifier, nitrogen could not manipulated. Therefore, the established entanglement is a bipartite entanglement as between proton and electron spin. Nitrogen is a basis for descriminating proton energy levels. However, we are planning to use this molecule for more elaborated quantum computation after providing a high power radiofrequency amplifier.

\section{Implementation of Quantum Operations for Superdense Coding; X-Band}
Superdense coding introduced by Bennett and Wiesner is a non-local quantum algorithm in which two classical bits of information have been transformed from Alice to Bob by sending only a single qubit. The scheme, explained in chapter 3, is based on the fact that the entangled initial state has been shared between the two involved parties. See chapter 3 for explanation of SDC and corresponding materials.

 \begin{figure}
\begin{center}
\scalebox{0.79}
{\includegraphics[2cm,21cm][17cm,25cm]{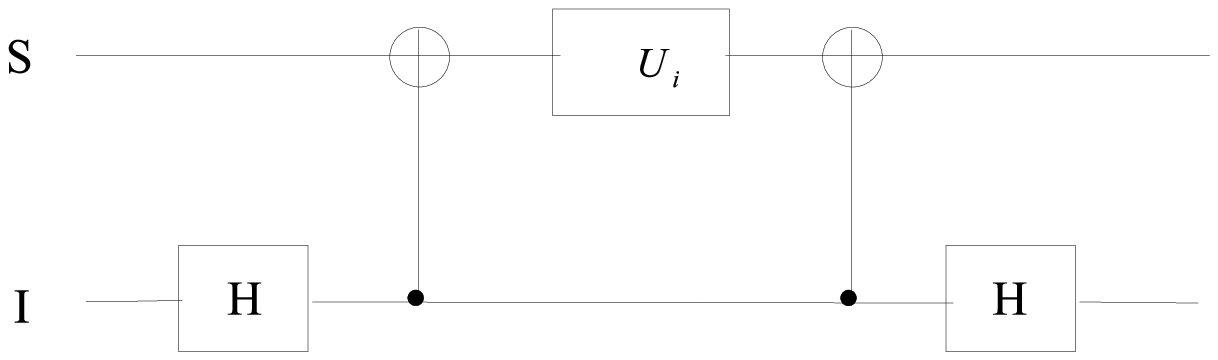}}
\end{center}
\caption{\label{endorSDCgate} Quantum circuit implementing SDC.}
\end{figure}

Concepts for ENDOR-based experimental setup for two-qubit superdense coding are depicted in Figure \ref{endorSDCgate}, in which S and I denote the electron-spin part and nuclear spin one. The quantum circuit for SDC, Figure \ref{endorSDCgate}, includes a Hadamard and controlled NOT (CNOT) gates. $U_i$ stands for one of the unitary transformation for encoding; out of the choices $\{X, Y, Z, I\}$. The first Hadamard and CNOT gates generate an entangled state between the electron and nuclear spins. Following the unitary transformation $U_i$, the CNOT and Hadamard gates (unitary back operations) the entangled state extract the encoded information. Except for a phase factor which dose not affect any signal, selective $\pi /2$ and $\pi$ pulses are available for the Hadamard gate for single qubit and the CNOT gate for two-qubit in pulsed magnetic resonance spectroscopy, respectively.  

Here we give a report on the implementation of the quantum operations for SDC by ENDOR. We do not call this experiment as an implementation of SDC because that the entanglement is actually pseudo-entanglement. If the experimental conditions, as discussed to be just in hand, are provided as it will be in a near future, then the same experiment with similar processing would be given as an implementation of SDC with ENDOR.

Here, the main idea is testing the ENDOR system for QIP giving a testing ground for QIP and QC to molecule-based ENDOR.  In our experiment, states have been prepared as pseudo-pure states in a similar approach as explained before in this chapter.

\begin{figure}
\begin{center}
\scalebox{0.79}
{\includegraphics[3cm,10cm][19cm,20cm]{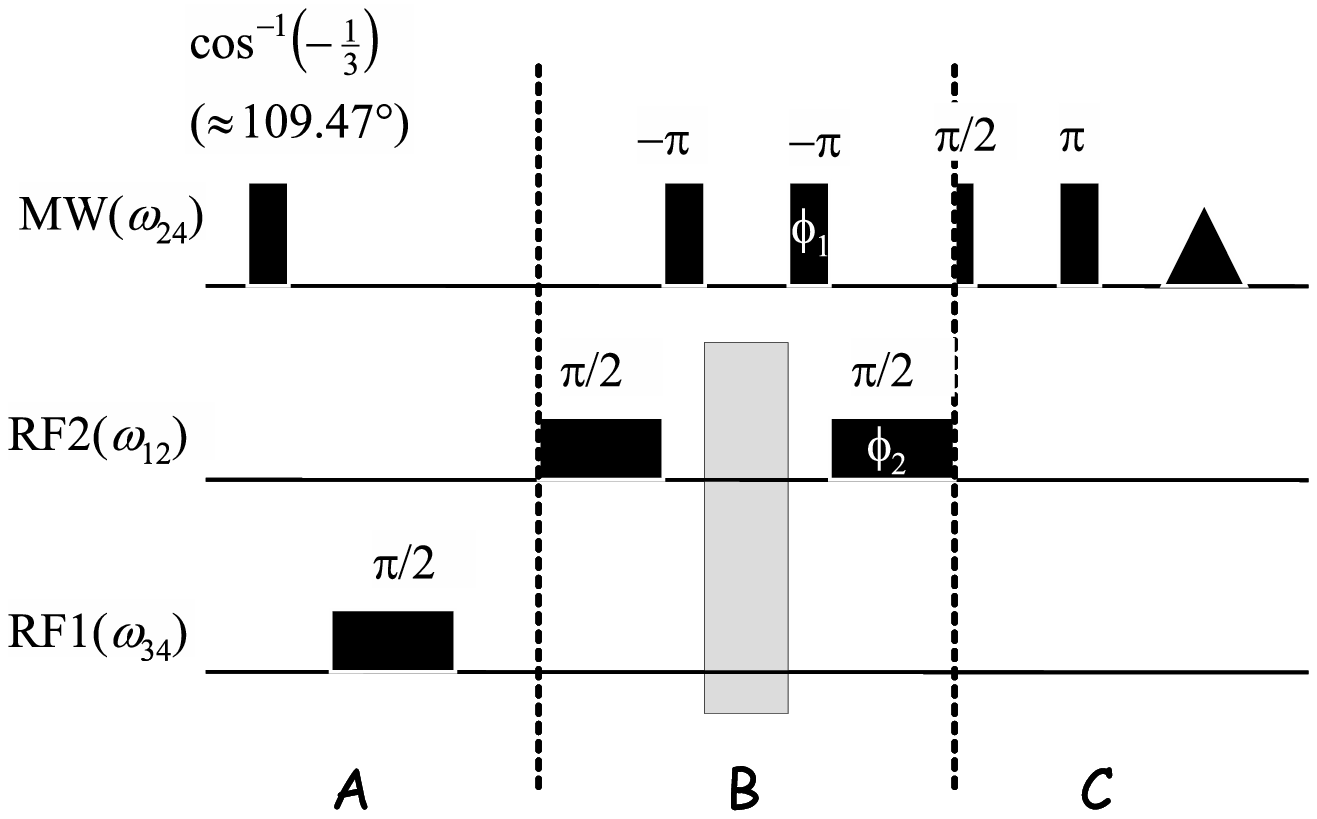}}
\end{center}
\caption{\label{endorSDCpulses} Pulse sequence for implementation of superdense coding by molecule-based ENDOR.}
\end{figure}

The pulse sequence that has been used is represented in Figure \ref{endorSDCpulses}. There are three main parts of the sequences, i.e., the preparation of the pseudo-pure states, manipulation and finally detection according to the customs in magnetic resonance spectroscopy.

The outermost left part labeled by ^^ ^^ A" is for preparation of the pseudo-pure state. Two pulses on electron and nuclear spins, with additional waiting times in order to make the off diagonal term of the density matrix vanishing, are required to acquire the pseudo-pure state. The first two pulses in the central part of the sequence are for entangling and also the last two pulses are for detection of the entanglement. Two phases of $\phi_1$ and $\phi_2$ for the pulses in the detection part are required for clearing up the entangled states from the simple superposition states explained in previous sections. In the central part between the entangling and detecting the entanglement one of the qubits, nuclear spin in our experiment, is encoded by randomly applying one of the four pulses of {I, X, Y, Z}. The necessary pulses for encoding are described in Table \ref{T11}. 

 \begin{table}
 \begin{center}
  \caption{\label{T11} The unitary operation and corresponding pulse sequences for encoding.}
\small
\begin{tabular}{|c|c|c|c|c|}
\hline
       & Initial State & Necessary Operation  & Required Pulses   &  Encoded State  \\ \hline
 U=I &$\frac{1}{\sqrt 2}(|00\rangle +|11\rangle)$ &                  &                             &$\frac{1}{\sqrt 2}(|00\rangle +|11\rangle)$ \\  
 U=X & $\frac{1}{\sqrt 2}(|00\rangle +|11\rangle)$ & $\exp(-i\pi I_x)$ & $P_x^{34}(\pi)P_x^{12}(\pi)$ & $\frac{1}{\sqrt 2}(|01\rangle +|10\rangle)$ \\  
 U=Y & $\frac{1}{\sqrt 2}(|00\rangle +|11\rangle)$ & $\exp(-i\pi I_y)$ & $P_x^{34}(2\pi)P_x^{34}(\pi)P_x^{12}(\pi)$ &$\frac{1}{\sqrt 2}(|01\rangle -|10\rangle)$ \\ 
 U=Z & $\frac{1}{\sqrt 2}(|00\rangle +|11\rangle)$ & $\exp(-i\pi I_x)$ & $P_x^{34}(2\pi)$ & $\frac{1}{\sqrt 2}(|00\rangle -|11\rangle)$ \\\hline   
 \end{tabular}
 \end{center}
 \end{table}
 
  \begin{table}
 \begin{center}
  \caption{\label{T12} Detection through the angular dependent electron spin echo intensity.}
\small
\begin{tabularx}{\linewidth}{|X|X|X|}
\hline
       & Angular dependant radio frequency pulses for encoding & Detected angular dependant echo intensity \\ \hline
 U=I &                           &$\frac{1}{4}(-1+\cos [\phi_1-\phi_2])$ \\  \hline
 U=X & $P_x^{34}(\theta)P_x^{12}(\phi)$ & $\frac{1}{16}(-3\cos \theta-\cos\phi+4\cos\frac{\theta}{2}\cos{\phi}{2}\cos [\phi_1-\phi_2])$ \\  \hline
 U=Y & $P_x^{34}(2\pi+\theta)P_x^{12}(\phi)$ & $\frac{1}{16}(-3\cos \theta-\cos\phi+4\cos\frac{\theta}{2}\cos{\phi}{2}\cos [\phi_1-\phi_2])$ \\ \hline
U=Z & $P_x^{34}(\theta)$ & $\frac{1}{16}(-1-3\cos \theta+4\cos\frac{\theta}{2}\cos [\phi_1-\phi_2])$ \\\hline   
 \end{tabularx}
 \end{center}
 \end{table}

\begin{figure}
\begin{center}
\scalebox{0.79}
{\includegraphics[0cm,12cm][21cm,28cm]{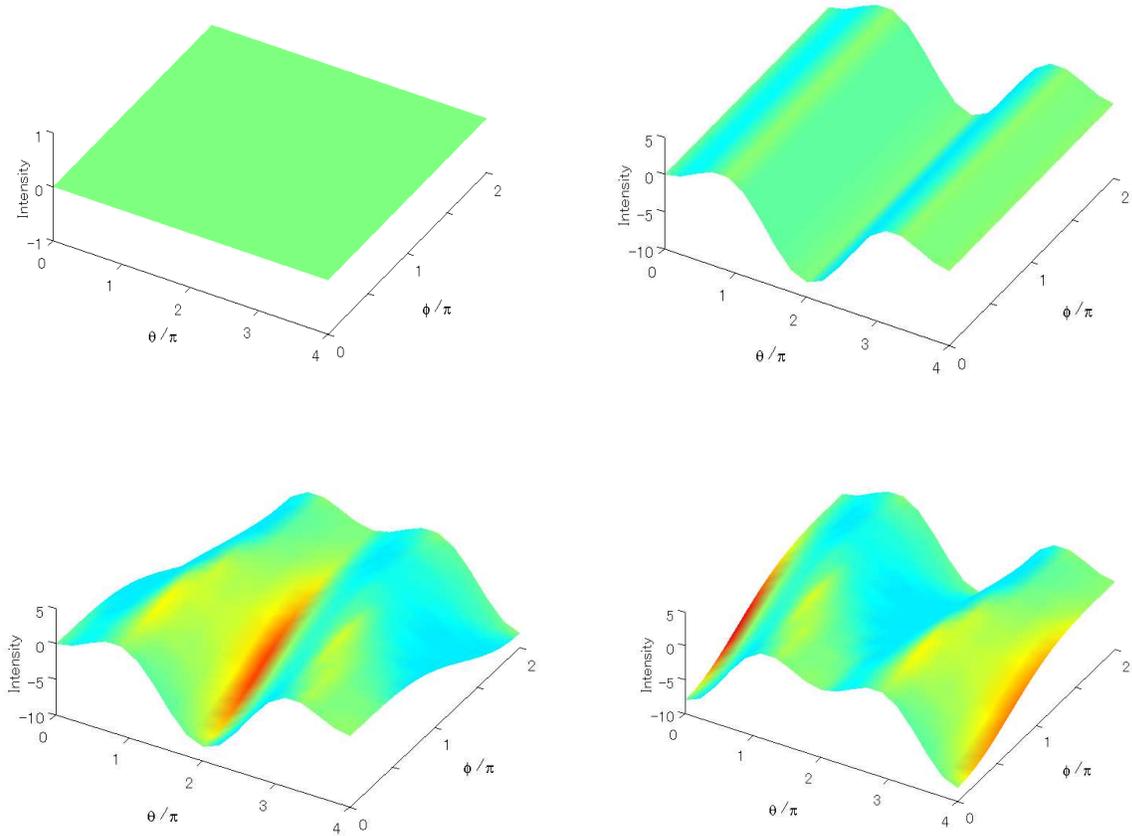}}
\end{center}
\caption{\label{endorSDCsim}Detection through the angular dependent pulses. This scheme for detection has been used in case that full analysis based on arbitrary change on microwave and radiofrequency phases have not been available. Though later the scheme has been improved as it is reported in text, following sections.}
\end{figure}

Finally, there are pulses for detection by an electron spin echo. For the measurement part, the situation has been modified for some detection considerations. This considerations have been much improved and modified for our following experiment at Q-band and low temperature. There, the phases of microwave and radio frequecies could be changed as aribitrary. Then detection has been done by TPPI scheme as will be explained later. But here we would like to represent the first proposal on detection. In this study, we have used the electron spin echo detection. The echo intensities have been detected for different angular dependencies of the pulses in the encoding part. As a result, there are four sets of angular dependencies for radio frequency pulses which have been used for encoding, (see Table \ref{T12}), see Figure \ref{endorSDCsim}.

\begin{figure}
\begin{center}
\scalebox{0.79}
{\includegraphics[0cm,0cm][11cm,9cm]{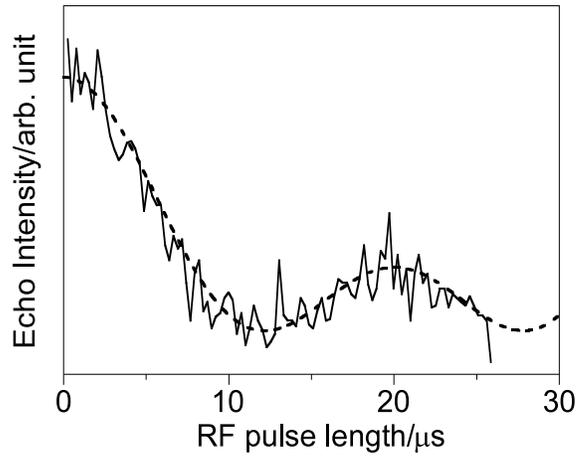}}
\end{center}
\caption{\label{4pi}4$\pi$ period dependence of the electron spin echo detection scheme. 
Microwave frequency:   = 9.38725 GHz. Static magnetic field set for ENDOR: B = 335.918 mT. Temperature: T = 20 K
.}
\end{figure}

Table \ref{T12} shows the angular dependence of the electro-spin-echo intensities in the detection part, depending also on experimental conditions for measurements. In the SDC experiments, the detected echo intensities incorporate the terms characteristic of $4\pi$ period, as shown in Table \ref{T12}. We have detected this salient behavior of the intensities, exemplifying the case for $U_i = X$ as given in Figure \ref{4pi}. In contrast to the $2\pi$ period of the population, the observed $4\pi$ period originates from the spinor property as the intrinsic nature of spins under study with the experimental condition of the selective microwave excitation.  The sign difference appearing between Figure \ref{4pi} and Table \ref{T12} is due to the difference between the experimental setup. Therefore, we have introduced pulsed ENDOR based approach to make quantum operations, pseudo-pure states and detection for SDC.
 
 \section{Conclusion}
 Our efforts have been made to develop entangling unitary operations in order to check the credibility of the molecule-based ENDOR QIP from both the theoretical and experimental sides. Angular dependent electron-spin echo intensities in the readout measurements have been derived in simple analytical expressions for SDC-ENDOR experiments. TPPI has been used for entanglement detection.
 
 Regarding the concept of entanglement, this work is much closer to the existence of quantum entanglement as compared to the previous work, \cite{n4}. In addition to the experiments at X-band and Q-band, we did also W-band at very low temperature, around $3.3$ K. Under the experimental conditions, the distance to the quantum entanglement is much less than the one that has been reported previously.
 
 The other important fact is the proper sample for quantum computing with ENDOR, that can give even quantum entanglement at low temperature and strong magnetic field. The previous work has been with malonyl redical. However, we have studied DPNO, heavily, and showed that for high magnetic field and low temperature this sample is the proper sample. Malonyl radical is not workable at this experimental conditions due to very long relaxation time. It does not give any clear ENDOR spectra. However, DPNO that has been used for very wide range of experimental conditions has been workable and gives very nice ENDOR spectra. 
 
 Also, about DPNO, it is possible to increase the number of qubits, rather simply by isotope labeling and in this regard we did several sample analytical experiments. Nitrogen and flourine labeled samples of DPNO have been synthesized, for instances. This sample also has been used for liquid state study quantum computing, due to the fact that it gives very clear ENDOR spectra for any purposes.
 
 With ENDOR, we have been throughly engaged in a complete establishment of quantum computing. This work is still going on, however we can say that the preliminary experimental results are well representing the feasibility of the ENDOR quantum information processing.

 \newpage
 
\vspace{5cm}
\chapter{Conclusion}
We have studied the concept of quantum entanglement for quantum information processing and quantum computation. Through the theoretical study, we conclude that specially for exponential enhancement, entanglement is a prerequisite quantum property. However, entanglement for ensemble quantum information processing and quantum computation is not so clear. Firstly we studied implementation of superdense coding by means of nuclear magnetic resonance, NMR. We have proved that NMR with low spin polarization, is not a quantum system that can give a quantum enhancement over classical processing. For instance, it has been proved that NMR implementation of superdense coding is erroneous, in the sense that the acquired experimental results are in reach with any classical system. 

We have also introduced a new measure of quantum entanglement, system oriented entanglement witness, that is specially applicable for NMR system.This measure of entanglement has been shown to have several advantages over any conventional entanglement witness. Firstly, it is possible to detect entanglement in a single run experiment and with a simple to implement quantum operations. This measure of entanglement is also useful for other ensemble system such as electron nuclear double resonance, ENDOR.

ENDOR has been introduced for quantum computation and it has been shown that the system is more appropriate because of the existence of an electron spin. The electron spin polarization, under the same experimental conditions, is three orders of magnitude larger than nuclear spin polarizations. This makes the required conditions for existence of entanglement much easier as compared with the case that the system only includes nuclear spins.

Quantum computation by means of ENDOR has been studied step by step. Sample study is shown to be very critical step as to detect sample with long decoherence time and well characterized magnetic tensors. As a result, molecular-spin bus systems for ENDOR QC are found. This molecule has been shown to be appropriate for elaborated experiments that may come later.

In the other step for a full quantum computation by means of ENDOR, quantum operations for ENDOR QC are demonstrated. Different magnetic field, X-band, Q-band, and W-band from the range of 9.5 GHz to 95 GHz have been used for demonstration of quantum operations at different temperatures from up to $\sim 300$K to down to $\sim 3$K. The experimental results are shown to be very much promising on reliability of ENDOR physical system for quantum computation and quantum information processing.

The other very important step, has been measurement for quantum computation. Because of some technological restriction in ENDOR spectroscopy, measurement, at a glance may seem to be problematic point for ENDOR quantum computing. However, we have modified ENDOR technology in a way that is very well organized for quantum computing. For instance, quantum state characterization by TPPI is established and has been used widely for investigation of entanglement. This technique is applicable to solution ENDOR QC experiments if suitable molecular open-shell entities are prepared. W-band ENDOR experiments with the required conditions for entanglement will give evidence of establishment of entanglement in a more straightforward manner, disclosing nature of entanglement and its new aspects possibly applicable to advanced electron magnetic resonance spectroscopy.

Finally, it should be pointed out that in the context of molecular-spin based ENDOR QC fine structure terms appearing in the spin Hamiltonian are useful for acquiring an effective temperature cooling down effect because the zero-field splitting due to the fine structure interaction enormously enhances the energy gap between two electronic sublevels involving EPR transitions.

\newpage

\end{document}